\documentclass[aps,prd,twocolumn,floatfix,superscriptaddress,
groupedaddress]{revtex4}   

\usepackage{amsmath}    
\usepackage{graphicx}   
\usepackage{longtable}
\begin{document}

\title{Search for charged massive long-lived particles at {\bf$\sqrt{s}=$} 1.96 TeV}
%
\affiliation{LAFEX, Centro Brasileiro de Pesquisas F\'{i}sicas, Rio de Janeiro, Brazil}
\affiliation{Universidade do Estado do Rio de Janeiro, Rio de Janeiro, Brazil}
\affiliation{Universidade Federal do ABC, Santo Andr\'e, Brazil}
\affiliation{University of Science and Technology of China, Hefei, People's Republic of China}
\affiliation{Universidad de los Andes, Bogot\'a, Colombia}
\affiliation{Charles University, Faculty of Mathematics and Physics, Center for Particle Physics, Prague, Czech Republic}
\affiliation{Czech Technical University in Prague, Prague, Czech Republic}
\affiliation{Center for Particle Physics, Institute of Physics, Academy of Sciences of the Czech Republic, Prague, Czech Republic}
\affiliation{Universidad San Francisco de Quito, Quito, Ecuador}
\affiliation{LPC, Universit\'e Blaise Pascal, CNRS/IN2P3, Clermont, France}
\affiliation{LPSC, Universit\'e Joseph Fourier Grenoble 1, CNRS/IN2P3, Institut National Polytechnique de Grenoble, Grenoble, France}
\affiliation{CPPM, Aix-Marseille Universit\'e, CNRS/IN2P3, Marseille, France}
\affiliation{LAL, Universit\'e Paris-Sud, CNRS/IN2P3, Orsay, France}
\affiliation{LPNHE, Universit\'es Paris VI and VII, CNRS/IN2P3, Paris, France}
\affiliation{CEA, Irfu, SPP, Saclay, France}
\affiliation{IPHC, Universit\'e de Strasbourg, CNRS/IN2P3, Strasbourg, France}
\affiliation{IPNL, Universit\'e Lyon 1, CNRS/IN2P3, Villeurbanne, France and Universit\'e de Lyon, Lyon, France}
\affiliation{III. Physikalisches Institut A, RWTH Aachen University, Aachen, Germany}
\affiliation{Physikalisches Institut, Universit\"at Freiburg, Freiburg, Germany}
\affiliation{II. Physikalisches Institut, Georg-August-Universit\"at G\"ottingen, G\"ottingen, Germany}
\affiliation{Institut f\"ur Physik, Universit\"at Mainz, Mainz, Germany}
\affiliation{Ludwig-Maximilians-Universit\"at M\"unchen, M\"unchen, Germany}
\affiliation{Fachbereich Physik, Bergische Universit\"at Wuppertal, Wuppertal, Germany}
\affiliation{Panjab University, Chandigarh, India}
\affiliation{Delhi University, Delhi, India}
\affiliation{Tata Institute of Fundamental Research, Mumbai, India}
\affiliation{University College Dublin, Dublin, Ireland}
\affiliation{Korea Detector Laboratory, Korea University, Seoul, Korea}
\affiliation{CINVESTAV, Mexico City, Mexico}
\affiliation{Nikhef, Science Park, Amsterdam, the Netherlands}
\affiliation{Radboud University Nijmegen, Nijmegen, the Netherlands}
\affiliation{Joint Institute for Nuclear Research, Dubna, Russia}
\affiliation{Institute for Theoretical and Experimental Physics, Moscow, Russia}
\affiliation{Moscow State University, Moscow, Russia}
\affiliation{Institute for High Energy Physics, Protvino, Russia}
\affiliation{Petersburg Nuclear Physics Institute, St. Petersburg, Russia}
\affiliation{Instituci\'{o} Catalana de Recerca i Estudis Avan\c{c}ats (ICREA) and Institut de F\'{i}sica d'Altes Energies (IFAE), Barcelona, Spain}
\affiliation{Uppsala University, Uppsala, Sweden}
\affiliation{Lancaster University, Lancaster LA1 4YB, United Kingdom}
\affiliation{Imperial College London, London SW7 2AZ, United Kingdom}
\affiliation{The University of Manchester, Manchester M13 9PL, United Kingdom}
\affiliation{University of Arizona, Tucson, Arizona 85721, USA}
\affiliation{University of California Riverside, Riverside, California 92521, USA}
\affiliation{Florida State University, Tallahassee, Florida 32306, USA}
\affiliation{Fermi National Accelerator Laboratory, Batavia, Illinois 60510, USA}
\affiliation{University of Illinois at Chicago, Chicago, Illinois 60607, USA}
\affiliation{Northern Illinois University, DeKalb, Illinois 60115, USA}
\affiliation{Northwestern University, Evanston, Illinois 60208, USA}
\affiliation{Indiana University, Bloomington, Indiana 47405, USA}
\affiliation{Purdue University Calumet, Hammond, Indiana 46323, USA}
\affiliation{University of Notre Dame, Notre Dame, Indiana 46556, USA}
\affiliation{Iowa State University, Ames, Iowa 50011, USA}
\affiliation{University of Kansas, Lawrence, Kansas 66045, USA}
\affiliation{Kansas State University, Manhattan, Kansas 66506, USA}
\affiliation{Louisiana Tech University, Ruston, Louisiana 71272, USA}
\affiliation{Northeastern University, Boston, Massachusetts 02115, USA}
\affiliation{University of Michigan, Ann Arbor, Michigan 48109, USA}
\affiliation{Michigan State University, East Lansing, Michigan 48824, USA}
\affiliation{University of Mississippi, University, Mississippi 38677, USA}
\affiliation{University of Nebraska, Lincoln, Nebraska 68588, USA}
\affiliation{Rutgers University, Piscataway, New Jersey 08855, USA}
\affiliation{Princeton University, Princeton, New Jersey 08544, USA}
\affiliation{State University of New York, Buffalo, New York 14260, USA}
\affiliation{University of Rochester, Rochester, New York 14627, USA}
\affiliation{State University of New York, Stony Brook, New York 11794, USA}
\affiliation{Brookhaven National Laboratory, Upton, New York 11973, USA}
\affiliation{Langston University, Langston, Oklahoma 73050, USA}
\affiliation{University of Oklahoma, Norman, Oklahoma 73019, USA}
\affiliation{Oklahoma State University, Stillwater, Oklahoma 74078, USA}
\affiliation{Brown University, Providence, Rhode Island 02912, USA}
\affiliation{University of Texas, Arlington, Texas 76019, USA}
\affiliation{Southern Methodist University, Dallas, Texas 75275, USA}
\affiliation{Rice University, Houston, Texas 77005, USA}
\affiliation{University of Virginia, Charlottesville, Virginia 22904, USA}
\affiliation{University of Washington, Seattle, Washington 98195, USA}
\author{V.M.~Abazov} \affiliation{Joint Institute for Nuclear Research, Dubna, Russia}
\author{B.~Abbott} \affiliation{University of Oklahoma, Norman, Oklahoma 73019, USA}
\author{B.S.~Acharya} \affiliation{Tata Institute of Fundamental Research, Mumbai, India}
\author{M.~Adams} \affiliation{University of Illinois at Chicago, Chicago, Illinois 60607, USA}
\author{T.~Adams} \affiliation{Florida State University, Tallahassee, Florida 32306, USA}
\author{G.D.~Alexeev} \affiliation{Joint Institute for Nuclear Research, Dubna, Russia}
\author{J.~Alimena} \affiliation{Brown University, Providence, Rhode Island 02912, USA}
\author{G.~Alkhazov} \affiliation{Petersburg Nuclear Physics Institute, St. Petersburg, Russia}
\author{A.~Alton$^{a}$} \affiliation{University of Michigan, Ann Arbor, Michigan 48109, USA}
\author{A.~Askew} \affiliation{Florida State University, Tallahassee, Florida 32306, USA}
% ...
%
%
\collaboration{The D0 Collaboration\footnote{with visitors from
$^{a}$Augustana College, Sioux Falls, SD, USA,
$^{b}$The University of Liverpool, Liverpool, UK,
$^{c}$UPIITA-IPN, Mexico City, Mexico,
$^{d}$DESY, Hamburg, Germany,
$^{e}$SLAC, Menlo Park, CA, USA,
$^{f}$University College London, London, UK,
$^{g}$Centro de Investigacion en Computacion - IPN, Mexico City, Mexico,
$^{h}$ECFM, Universidad Autonoma de Sinaloa, Culiac\'an, Mexico
and
$^{i}$Universidade Estadual Paulista, S\~ao Paulo, Brazil.
}} \noaffiliation
\vskip 0.25cm

\date{November 11, 2012}
\begin{abstract}

We present a search for charged massive long-lived particles (CMLLPs) 
that are pair produced in $p\bar{p}$ collisions at $\sqrt{s}=$ 1.96
 TeV collected by the D0 
experiment at the Fermilab Tevatron collider. Our result is a combination 
of two searches where either one or both CMLLPs are reconstructed in the detector. We select events with muon-like particles that have both speed and ionization 
energy loss (${\rm d} E/{\rm d} x$) different from muons produced in $p\bar{p}$ collisions. 
In the absence of evidence for CMLLPs corresponding to 6.3 ${\mathrm{fb}}^{-1}$ of integrated luminosity, we set limits on the CMLLP masses in several 
supersymmetric (SUSY) models, excluding masses below 278 GeV for long-lived gaugino-like charginos, and masses below 
244 GeV for long-lived higgsino-like charginos at the 95\% C.L. We also set limits on the cross section for pair production of long-lived scalar tau leptons that range from 0.04 pb to 0.008 pb for scalar tau lepton masses of 100 to 300 GeV. 

\end{abstract}

\maketitle

\section{Introduction}

Several extensions of the standard model (SM) including some SUSY models predict the existence of massive long-lived particles (MLLP)~\cite{cosmo1}. Their existence could explain the origin of  
dark matter. Primordial lithium abundance is not 
described by the current model of big bang nucleosynthesis, but it can be 
satisfactorily explained by the existence of a MLLP that decays during or 
after big bang nucleosynthesis~\cite{BBN}. MLLPs could have color or electric charge. They appear as R-hadrons (bound states 
of squarks or gluinos with SM quarks), as sleptons, or as charginos. MLLPs are relatively slow moving at the collision energy 
of $\sqrt{s} = 1.96$~TeV and for MLLP masses of 100 GeV or greater considered in this article. Charged MLLPs also have large ionization energy loss (${\rm d} E/{\rm d} x$) due to their slow speeds.
 These characteristics are different from other particles studied at high energy colliders, and thus the identification 
of such particles is simplified by the corresponding small amount of background. We therefore
 search for charged massive long-lived particles (CMLLPs) at the Tevatron.

Searches for CMLLPs were performed previously by the D0~\cite{Mike thesis, Yunhe thesis, D0 CMSP PRL Run IIa}, CDF~\cite{CDF Run1 CHAMP, CDF Champ PRL}, LEP~\cite{LEP_CMLLP}, 
CMS~\cite{CMS_CMLLP}, and ATLAS~\cite{ATLAS_CMLLP} collaborations. We present limits on masses of CMLLPs by combining data from a search 
for pair produced CMLLPs performed with 1.1 ${\mathrm{fb}}^{-1}$ integrated luminosity~\cite{D0 CMSP PRL Run IIa} 
with an analysis based on 5.2 ${\mathrm{fb}}^{-1}$ integrated luminosity~\cite{D0 CMSP PRL Run IIb}. The second analysis includes searches for either a pair of CMLLPs or a single CMLLP signature in an event. This article provides greater detail
on the analysis and results published in~\cite{D0 CMSP PRL Run IIb}. 
    
In this study ``long-lived'' refers to particles that traverse
the entire detector before decaying. Although cosmological observations place 
severe limits on stable massive particles~\cite{cosmo1, cosmo2}, these 
limits do not rule out the particles predicted by models studied here. 
We are sensitive to CMLLPs with lifetimes longer than 25 ns, with best sensitivity for lifetimes longer than 1 ${\mu}$s.

We compare the results with predictions of several SUSY models. Models with gauge-mediated SUSY-breaking (GMSB) always contain a light 
gravitino/goldstino as the lightest SUSY particle (LSP)~\cite{GMSB1, GMSB2}.
The next-to-lightest SUSY particle (NLSP) could be 
the lightest scalar tau lepton (stau) or the lightest neutralino,
depending on the model parameters~\cite{NLSP1, SUSYPrimer}.
The GMSB parameters assumed in this paper make the stau lepton 
the NLSP.
If stau lepton decays to the gravitino/goldstino are suppressed
(the effective coupling to the gravitino/goldstino is a free parameter
in the model), then the stau lepton can live long enough to escape the detector
and be a candidate CMLLP~\cite{NLSP2, NLSP3}.

Long-lived charginos can occur in models with anomaly mediated SUSY breaking and in SUSY models that do not have gaugino mass unification, 
provided the difference between the masses of the lightest chargino and the lightest neutralino 
 is less than approximately 150 MeV~\cite{AMSB1, AMSB2}. The chargino can be mostly 
higgsino or mostly gaugino. We treat these two cases separately. The analysis strategy is the same as that for the stau lepton search.

In addition, some SUSY models predict long-lived top squark NLSPs that hadronize into mesons and baryons with long enough lifetimes to be CMLLP candidates~\cite{GMSB3}. Hidden 
valley models predict scenarios where the top squark acts like the 
LSP and has a long lifetime~\cite{HiddenValley1, HiddenValley2}. In these models the top squark forms hadrons that are CMLLP candidates. Any SUSY
 scenario where the top squark is the lightest colored 
SUSY particle can have a hadron formed with a top squark that is a CMLLP. 
Colored CMLLPs will hadronize and experience charge 
exchange during nuclear interactions. This effect is taken into account in the analyses reported here.

A brief description of the D0 detector is 
given in Sec.~II, which is followed in Sec.~III by a description of the trigger and the data used. Section IV describes the theory and the 
signal generation. Section V presents the strategies and techniques 
used in these analyses. 
Section VI describes the search for pairs of CMLLPs and Sec.~VII
 the search for single CMLLPs with an integrated luminosity of 5.2 ${\mathrm{fb}}^{-1}$. ``Pair'' and ``single'' refer to the number of detected particles. In the models we consider, CMLLPs are always produced in pairs. Section VIII summarizes 
the earlier search with an integrated luminosity of 1.1 ${\mathrm{fb}}^{-1}$. The combined results are presented in Sec.~IX. Section X summarizes this study.

\section{Detector}

\begin{figure*}
\begin{center}
\scalebox{0.75}{\includegraphics{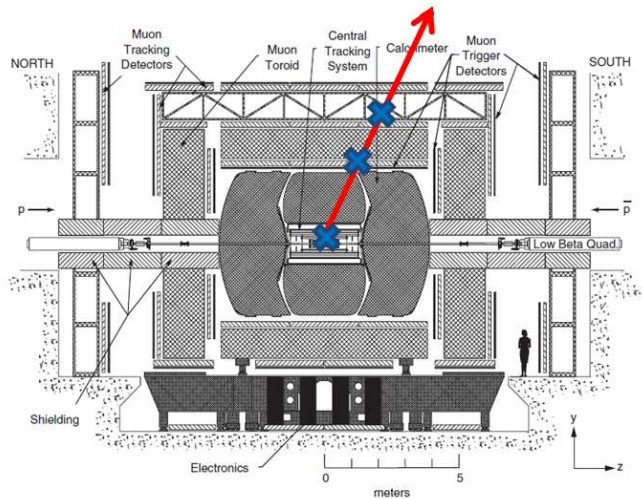}}
\caption{\label{fig:D0_detector}(color online) Diagram of the D0 detector showing the locations (blue 
crosses) where a top squark hadron must be measured as charged to be selected as a CMLLP candidate.}
\end{center}
\end{figure*}

Figure~\ref{fig:D0_detector} shows the details of the D0 detector~\cite{D0NIM} which consists of three 
primary systems: a central tracking system, calorimeters, and a muon spectrometer. 
 The polar angle $\theta$ is defined 
such that $\theta=0$ is the $+z$ direction, which is the direction of the proton beam.
 The azimuthal angle $\phi$ is defined 
such that $\phi=0$ lies along the horizontal $+x$ axis, pointing outwards from 
 the center of the Tevatron ring and $\phi= {\pi}/2$ in the $+y$ direction. The pseudorapidity of a particle is 
defined as $\eta=-\ln{[}\tan\left(\theta/2\right)${]}.

The silicon microstrip tracker (SMT) is the innermost part of the tracking 
system and has a six-barrel longitudinal structure, where each barrel 
consists of a set of four layers arranged axially around the beampipe to 
measure the $r$-$\phi$ coordinates of charged particles. A new layer of SMT sensors was installed 
near the beampipe in 2006. The data recorded before this addition are referred to as Run IIa and the subsequent data are referred 
to as Run IIb. Twelve radial disks, interspersed between the barrel 
segments, provide position measurement in the $r$-$z$ and $r$-$\phi$ 
planes. The SMT provides a spatial resolution of approximately 10~$\mu$m 
in $r$-$\phi$ and approximately 100~$\mu$m in $r$-$z$ and covers a pseudorapidity range $|\eta| <$ 3. The SMT is also used to measure ionization energy loss 
(${\rm d} E/{\rm d} x$) of tracks. The central fiber tracker (CFT) surrounds the SMT and 
consists of eight concentric carbon fiber barrels holding doublet layers of 
scintillating fibers (one axial and one small-angle stereo layer) with the 
outermost barrel covering the region $|\eta| <$ 1.7. A superconducting 
solenoidal magnet  surrounds the CFT and provides a uniform 1.9 T axial magnetic field. 

 A liquid argon/uranium calorimeter measures both electromagnetic and hadronic 
energy and is housed in three cryostats, with the central calorimeter covering 
the region $|\eta| <$ 1.1 and two end calorimeters covering the region 1.5 $<
|\eta|<$ 4.2. The calorimeter is made of pseudo projective towers consisting of an absorber 
plate and a signal board. Liquid-argon, the active material of the calorimeter, 
fills the gap. There are about 10 hadronic interaction lengths in the calorimeter at $\eta=0$.

The muon system is the outermost part of the D0 detector and covers the region 
$|\eta|<2$~\cite{d0muon}. It comprises drift tubes and scintillation counters arranged in three layers (A, B, and C). Between layers A and B, there is magnetized 
steel (6 interaction lengths at $\eta=0$) generating a 1.8 T toroidal field.
  In the central layers ($|\eta|<1$) multiwire proportional drift tubes (PDT) and in the forward layers ($1<
|\eta|<2$), mini drift tubes (MDT) are used for tracking. Scintillation counters covering the region ($|\eta|<2$) are used for triggering on muons.

The PDTs are typically 2.8 $\times$ 5.6 m$^2$, with cells that are 10 cm in diameter. 
Typical chambers are built of three or four layers of 24 cell wide planes. Each cell has an anode wire at its center. Vernier cathode pads are located on both sides of
the wires to provide information on the hit position along the wire. 
The chambers are filled with a gas mixture of 84\% argon, 8\% CF$_4$, and 8\% CH$_4$ with a drift velocity of approximately 10 cm$/{\mu}$s.

Scintillation counters are installed on the top, the sides and the bottom of the outer 
layers of the central muon PDTs. They provide a fast signal to associate a 
muon in a PDT with the appropriate bunch crossing and hence are used in 
muon triggers. They also help to discriminate against the 
cosmic ray background and to reject out-of-time particles scattered from accelerator and detector components at high $\eta$. The time resolution is approximately 2 ns for A-layer counters 
and approximately 4 ns for B, and C-layer counters. Detection efficiency is 
close to 100\% in all counters.

In the forward region, MDTs with a drift time of $\sim 90$ ns provide  
good coordinate resolution of less than 1 mm, radiation hardness, high segmentation, 
and low occupancy. Each MDT layer is divided into octants. An MDT consists 
of eight cells, each with a $9.4\times9.4\ \mathrm{mm}^2$ internal cross section and uses a 
fast gas mixture of CF$_4$/CH$_4$ (90\%:10\%). There are 4214 scintillation counters 
in the forward region, arranged in three layers (A, B, and C). The segmentation is $4.5^\circ \times 0.12\ (0.07)$ in $\phi \times \eta$ 
for the first nine inner (last three) rows of counters. The scintillation counters 
are 1.3 cm thick with various cross sections ranging from 60 $\times$ 106 cm$^2$ to 17 $\times$ 24 cm$^2$. The time resolution is approximately 2 ns and the detection 
efficiency is above 99.9\%. The CMLLPs considered in this analysis would be identified as muons in
the D0 detector as they penetrate the material of the calorimeter and the toroid and leave hits in
the muon system. An accurate measurement of the time of flight (TOF) of a charged particle reaching a scintillation counter is obtained from the position of the counter and the recorded time of the hit. Particle tracks are reconstructed in the muon system using hits from scintillation 
counters and drift tubes. A muon candidate is qualified as a good muon if it has hits in scintillation counter layers A and either B or C, and multiple drift tube hits in different detector layers. These \emph{local muons} reconstructed by the muon spectrometer are then matched to charged 
particle tracks in the central tracking system originating at the $p{\bar{p}}$ interaction vertex.  The muon candidate is 
rejected if no match is found. Otherwise, the measurement of the momentum component transverse to the beam line ($p_T$) of the muons is taken 
from the parameters of the central track. To discriminate between muons produced in hadronic decays 
(which tend to be surrounded by other charged particles and calorimeter energy deposits) and
 isolated muons, two different isolation quantities are calculated. 
\emph{Track isolation} is the sum of the $p_T$ of all other tracks in the 
central tracking system in a cone of radius $\Delta R = \sqrt{(\Delta \eta)^2 + (\Delta \phi)^2} 
< 0.5$ around the central track matched to the muon. \emph{Calorimeter isolation} is the sum of all energy deposits in the calorimeter in an annulus of 
$0.1 < \Delta R < 0.4$ around the muon trajectory. Throughout this article, isolated
muons will be those with track and calorimeter isolation less than 2.5 GeV each.

\section{Trigger}
The D0 trigger system is designed with three distinct levels with each succeeding 
level examining fewer events in greater detail so that the final trigger rate is 
low enough for the data to be recorded without causing excessive dead time. The 
first stage (L1) comprising a collection of hardware elements that selects events
based on features such as momentum, energy and particle type, provides 
an accept rate of about 2 kHz. In the second stage (L2) microprocessors associated with specific subdetectors provide information 
to a global processor to construct a trigger decision based on individual objects 
as well as object correlations. The L2 system has an accept rate of approximately 
1 kHz. Candidates accepted by L1 and L2 are sent to the third level (L3) of the trigger 
system where the data is processed by algorithms on a computing 
farm to reduce the rate to about 200 Hz. Events that pass all three trigger levels are recorded for offline reconstruction. 

As described earlier, the CMLLPs in this study have
muon-like signatures with regards to their penetration characteristics
in the detector. The $p_T$ threshold of the muon triggering the event varies from 8 GeV to 13 GeV. The earlier analysis performed with an integrated luminosity of 1.1 ${\mathrm{fb}}^{-1}$ from Run IIa searched for a pair of muons in an event using a
triggering condition that requires at least two muons to be present in the event (dimuon trigger). In the recent analysis with an integrated luminosity of 5.2 ${\mathrm{fb}}^{-1}$ of Run IIb data, the search is expanded to include events with only one 
CMLLP candidate. Hence, triggering conditions requiring the presence of at least one muon (single muon triggers) are used. 

\begin{figure*}
\begin{center}
\scalebox{0.40}{\includegraphics{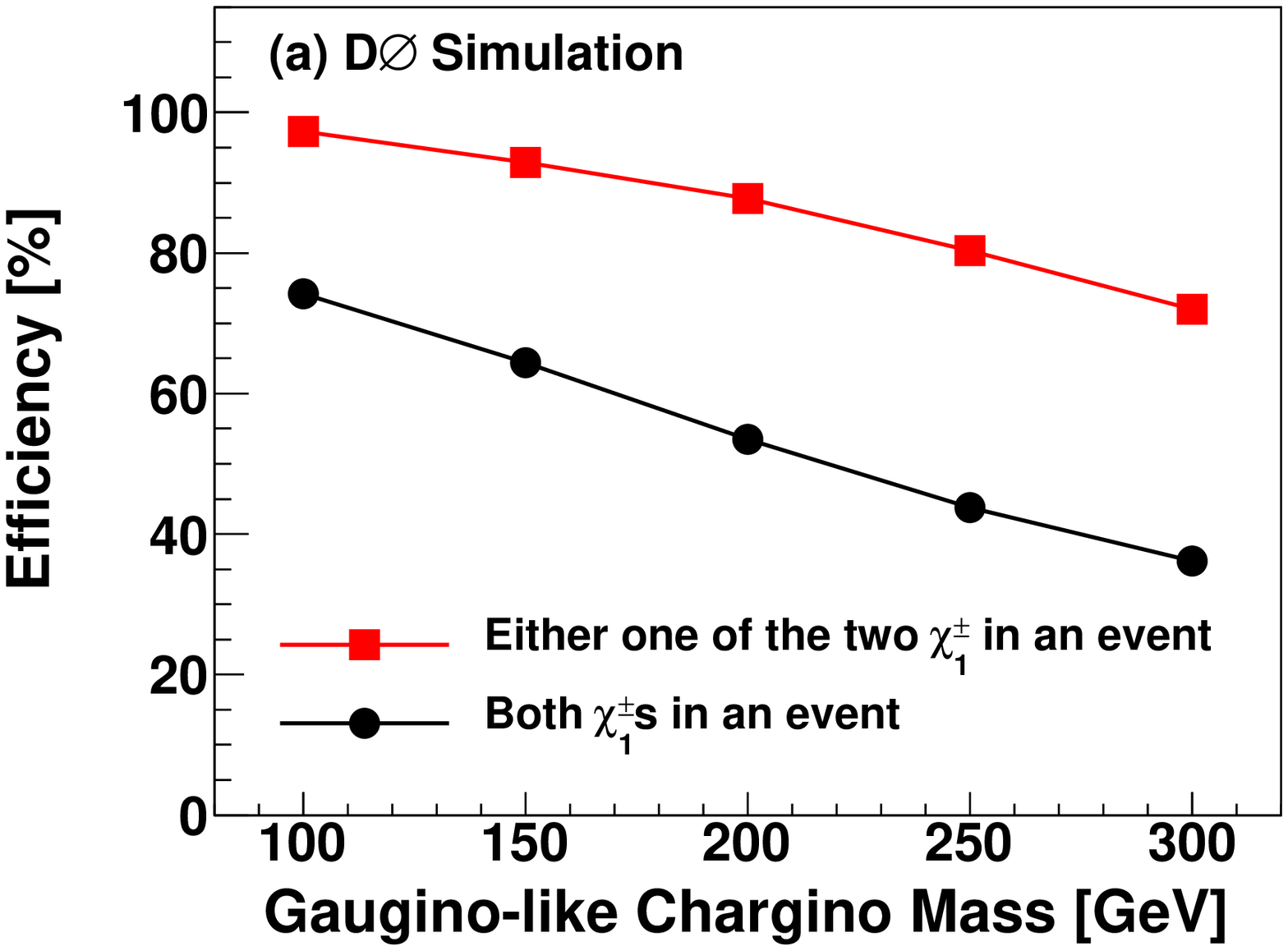}}
\scalebox{0.40}{\includegraphics{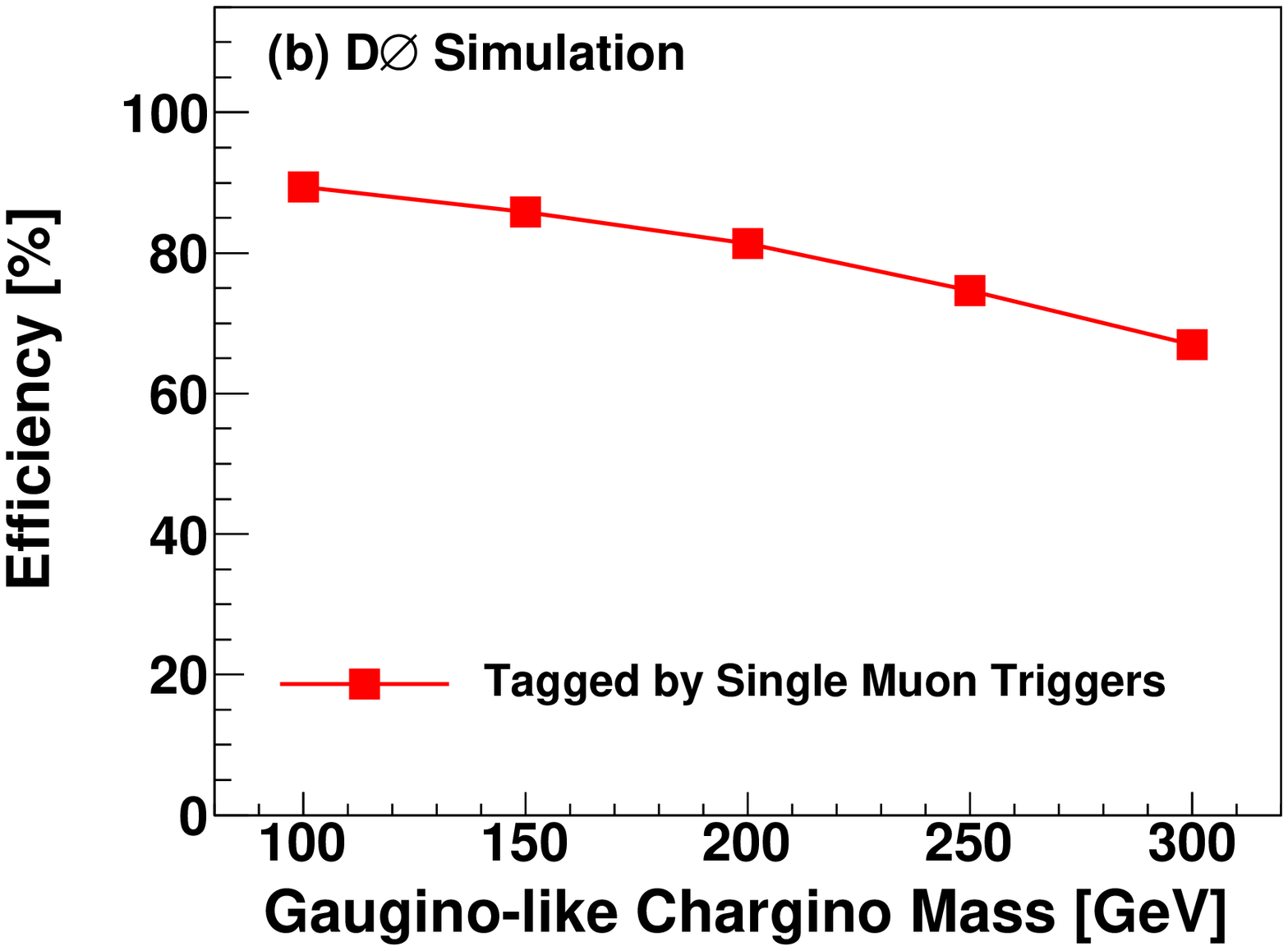}}
\caption{\label{fig:charginog_mass_trig_eff}(color online) Efficiency for slow-moving gaugino-like charginos of various masses to arrive within the L1 muon trigger gates. All events in 
this MC sample contain two gaugino-like charginos. (a) The black and red  lines show the efficiencies for a pair of charginos, or a single chargino  respectively, to be within the trigger gate.  
(b) Overall efficiency, which is a product of the trigger gate efficiency and the efficiency for single muon triggers, for the selection of single charginos.}
\end{center}
\end{figure*}

Since these triggers are designed for muons traveling close to the speed of 
light ($\beta \approx$1), there is a loss of acceptance for CMLLP candidates with $\beta<1$. The muon triggers
 impose a time window (trigger gate) on the muon scintillation counter hits. The trigger
 gate opens 15--30 ns before particles from a collision  traveling at the speed of light reach a particular muon layer and closes 15--40 ns after that time. 
%
The effect of the trigger gate on events with a pair of CMLLPs is more pronounced in di-muon triggers, where both slow moving particles must arrive within the trigger gate; see Fig.~\ref{fig:charginog_mass_trig_eff}(a). The muon triggers are simulated by applying trigger efficiencies measured in $Z\rightarrow\mu\mu$ decays in data using a tag and probe method. Trigger efficiencies for CMLLPs are calculated using Monte Carlo (MC) simulations for different CMLLP masses. 
The overall selection efficiency, which is a product of the trigger gate efficiency and the efficiency for single muon triggers, for slow moving CMLLPs is higher for single muon triggers than that for dimuon triggers. We therefore use single muon triggers in the searches for a pair of CMLLPs as well as for a single CMLLP. The overall selection efficiency, including the efficiency of the trigger gate, for a single CMLLP is shown in Fig.~\ref{fig:charginog_mass_trig_eff}(b).

A second time window (readout gate) is imposed on the muon signals during digitization. This gate opens 15--30 ns before particles from a collision  traveling at the speed of light reach a particular muon layer and closes 70--90 ns after. In the search for a pair of CMLLPs both particles must be within the
readout gate for the information to be recorded even though only one
muon needs to be within the trigger gate for the trigger to be satisfied. 

\section{Models and Signal Generation}

Signal samples with direct production of a pair of CMLLPs are simulated using {\sc pythia} 6.409~\cite{Pythia}. Data events collected from random beam crossings are overlaid on simulated events to simulate additional interactions and detector noise. Production of CMLLPs through cascade decays from heavier new particles (such as squarks) is model dependent and has not been considered here. A GMSB model with a long-lived stau lepton NLSP, model line D in Ref.~\cite{stau-NLSP}, is used to generate stau lepton pairs.  The model parameters are given in Table~\ref{tab:stau_p}.
The minimal supersymmetric standard model is used for generating long-lived 
gaugino-like charginos, higgsino-like charginos, and top squarks. The corresponding model parameters are given in Table~\ref{tab:chargino_p}, where $M_1$, $M_2$, and $M_3$ are the mass parameters for U(1), SU(2), 
and SU(3) gauginos respectively, tan$\beta$ is the ratio of the vacuum expectation 
values of the two Higgs doublets, and $\mu$ is the corresponding mass parameter. Long-lived top squarks are generated with {\sc pythia} and hadronized by linking with an algorithm external to {\sc pythia}. This algorithm~\cite{stop code} is applicable to any SUSY model that features a long-lived top squark.
\begin{table*}
\begin{center}
\begin{ruledtabular}
\begin{tabular}{c c c}
Parameter & Description & Value \\
\hline
${\Lambda}_m$ & Scale of SUSY breaking & 19 to 100 GeV \\
$M_m$ & Messenger mass scale   & 2 ${\Lambda}_m$ \\
$N_5$ & Number of messenger fields  & 3 \\
$\tan\beta$  & Ratio of Higgs field vacuum expectation values & 15 \\
sign of $\mu$ & Sign of Higgsino mass parameter & $+1$ \\
\end{tabular}
\end{ruledtabular}
\end{center}
\caption{\label{tab:stau_p} GMSB model parameters for stau lepton production.}
\end{table*}
\begin{table*}
\begin{center}
\begin{ruledtabular}
\begin{tabular}{c c c c c c c}
Model & $\mu$ (GeV)& $M_1$ (GeV)& $M_2$ (GeV)& $M_3$ (GeV)& $\tan\beta$& Squark Mass (GeV)\\
\hline
Top squark & 10,000 & 100 & 200 & 500 & 15 & 800 \\
Gaugino-like chargino& 10,000& 3$M_2$ &  100 to 300 & 500& 15 & 800 \\
Higgsino-like chargino& 100 to 300& 100,000 & 100,000 & 500 & 15 & 800  \\
\end{tabular}
\end{ruledtabular}
\end{center}
\caption{\label{tab:chargino_p} Model parameters for top squark and chargino production.}
\end{table*}
A set of 50,000 events is generated for each model and each mass point. A 
{\sc geant}-based detector simulation models the detector response for the
 MC samples~\cite{D0gstar}. We have modified {\sc geant} to treat 
our long-lived signal particles as heavy muons for purposes of tracking and 
estimating the ${\rm d} E/{\rm d} x$ of signal particles in the detector. Therefore, CMLLPs in MC samples have muon-like lifetimes. The detector 
geometry is described in detail in the simulation, which uses information on the 
position of the scintillation counters to evaluate the timing information of the 
hits in the counters. This information is used to calculate the TOF of the muons. Simulation of muon timing in the standard simulation 
software is corrected using information from data as described in 
 Sec.~\ref{tof_section}. After the simulation of the detector response, 
a simulation of the electronics and digitization is performed. The 
simulated samples are then passed through the same reconstruction software that is used to reconstruct data.

    Theoretical values of masses and couplings for different types of CMLLPs are calculated using {\sc SoftSUSY}~\cite{SoftSUSY}. This information is provided as input to {\sc PROSPINO}~\cite{Prospino} for the calculation of production cross sections and their uncertainties for different types and masses of CMLLPs. 

\section{Analysis Strategy and Techniques, and Selection Variables}

\begin{figure*}
\begin{center}
\scalebox{0.4}{\includegraphics{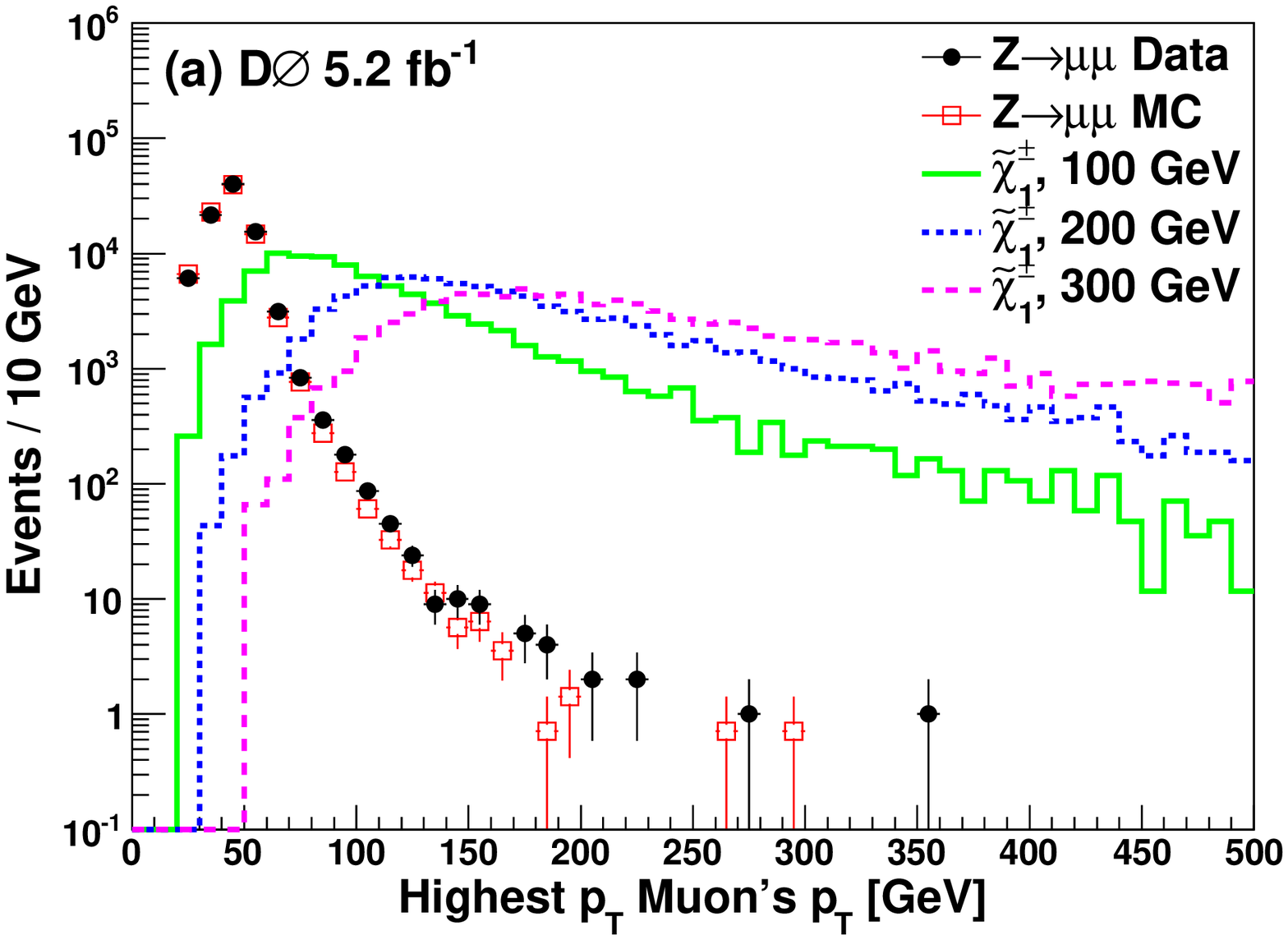}}
\scalebox{0.4}{\includegraphics{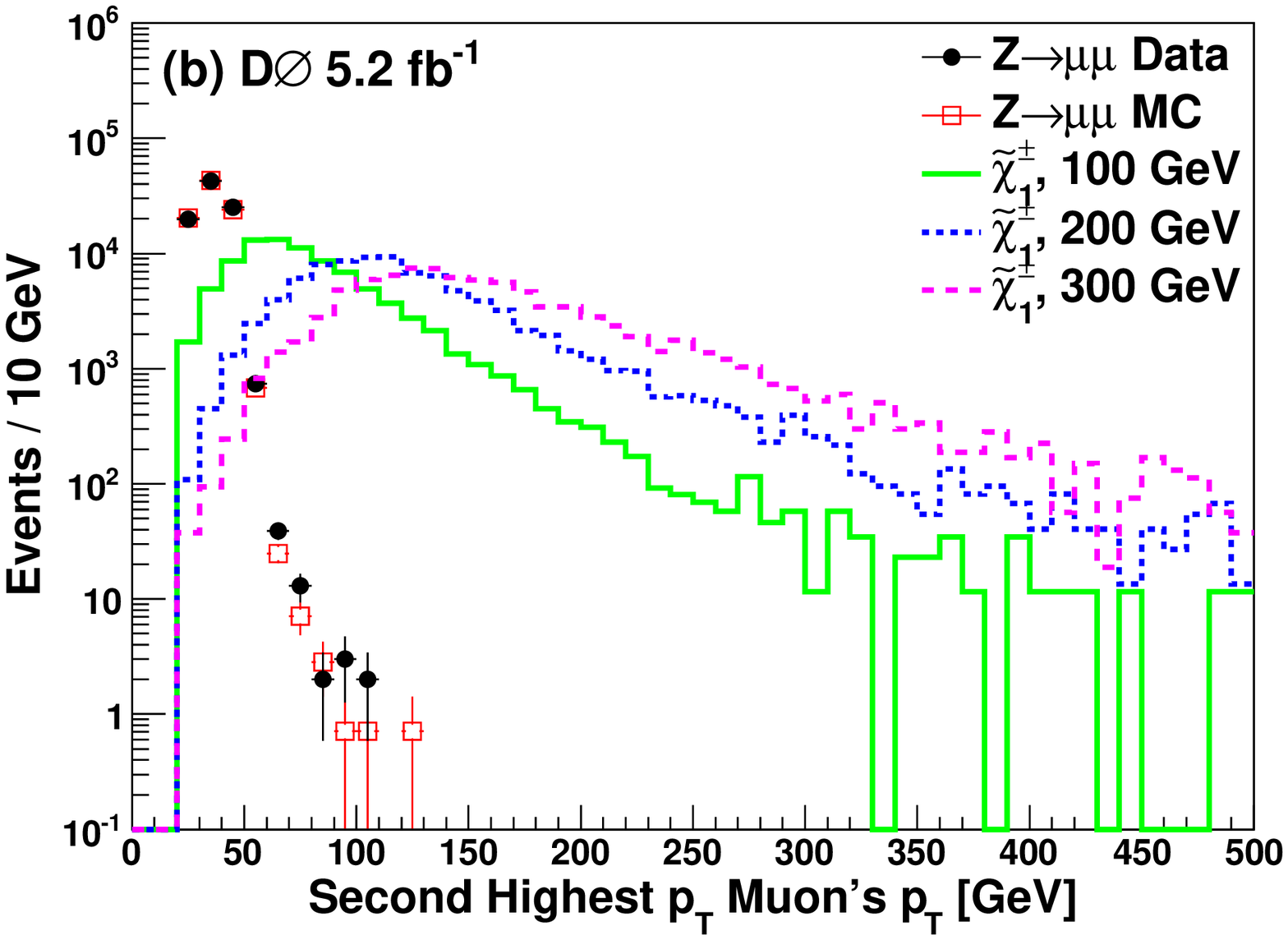}}
\caption{\label{fig:pt_muons}(color online) Distributions of $p_T$ of the  (a) highest $p_T$ and the (b) second-highest 
$p_T$ muon in an event for $Z\rightarrow\mu\mu$ data and MC, and for simulated gaugino-like charginos with masses of 100 GeV, 200 GeV, and 300 GeV.}
\end{center}
\end{figure*}

 With respect to their production at the primary vertex 
and penetration characteristics, CMLLPs are similar to prompt muons produced in ${p}\bar{p}$ collisions, 
but they travel at $\beta \approx$ 0.6 to 0.8. The momenta of  CMLLPs 
are distinctly higher than that of prompt muons despite their lower $\beta$, 
as shown in Fig.~\ref{fig:pt_muons}. In this figure we compare the highest and 
the second-highest $p_T$ CMLLPs (simulated gaugino-like charginos of masses 100 GeV, 200 GeV, and 300 GeV) in an event with the highest and the second-highest $p_T$ muons from $Z\rightarrow\mu\mu$ decays in data and MC events. Events are required to have 
two isolated muons with ${p_T}>$ 20 GeV. The muons are required to be 
inconsistent with cosmic ray muons. All distributions are normalized to the same number of events. For $Z\rightarrow\mu\mu$ events, we require 70 $<M_{\mu\mu}<$ 110 GeV. The data and MC are compared in this mass range to determine the corrections to be applied to the MC. 
The CMLLPs considered here have a ${\rm d} E/{\rm d} x$ approximately proportional to $1/{\beta}^2$ which is much higher than the ${\rm d} E/{\rm d} x$ of an identified muon that is essentially a minimum ionizing particle~\cite{bethe}. 
Thus, TOF, which is used to calculate $\beta$, and ${\rm d} E/{\rm d} x$ of a particle can discriminate between CMLLP candidates and muons with ${\beta}\simeq 1$.

The Run IIa analysis, which searched for a pair of CMLLPs, used only the 
TOF of CMLLP candidates to distinguish them from prompt muons. The 
introduction of ${\rm d} E/{\rm d} x$ measurement into the 
Run IIb analyses allowed us to extend the search to events where only one 
CMLLP candidate could be detected. The Run IIa dataset was used to search 
for stau leptons, gaugino-like charginos, and higgsino-like charginos. The Run IIb 
analyses also searched for top squarks. Additional criteria needed for the selection of top squark candidates are described in detail in Sec.~V C. 
 
 Results are presented in this article for stau leptons, gaugino-like charginos, and higgsino-like 
charginos from individual analyses as well as from a combination of Run 
IIa and Run IIb data with a total of 6.3 ${\mathrm{fb}}^{-1}$ integrated luminosity. To avoid double counting of events, the samples used for the two Run IIb 
analyses are constructed to be statistically independent. All events that pass the selection requirements used 
for the search for a pair of CMLLPs are removed from the data used for the single CMLLP search, resulting in an approximately 40\% loss of signal acceptance for the single CMLLP search. The data and the background sample that are used in the search for
 single CMLLPs contain muons that originate mostly from the decays of $W$ 
bosons. The number of such events changes by only about 2\% due to this veto.
Furthermore, we show results for searches for single top squarks and top squark pair production using 5.2 ${\mathrm{fb}}^{-1}$ integrated luminosity. A combination of the two analyses does not improve the result as explained in Sec.~V C, and is not performed here.
\subsection{ Time-of-Flight Measurement \label{tof_section}}
\begin{figure*}
\begin{center}
\scalebox{0.4}{\includegraphics{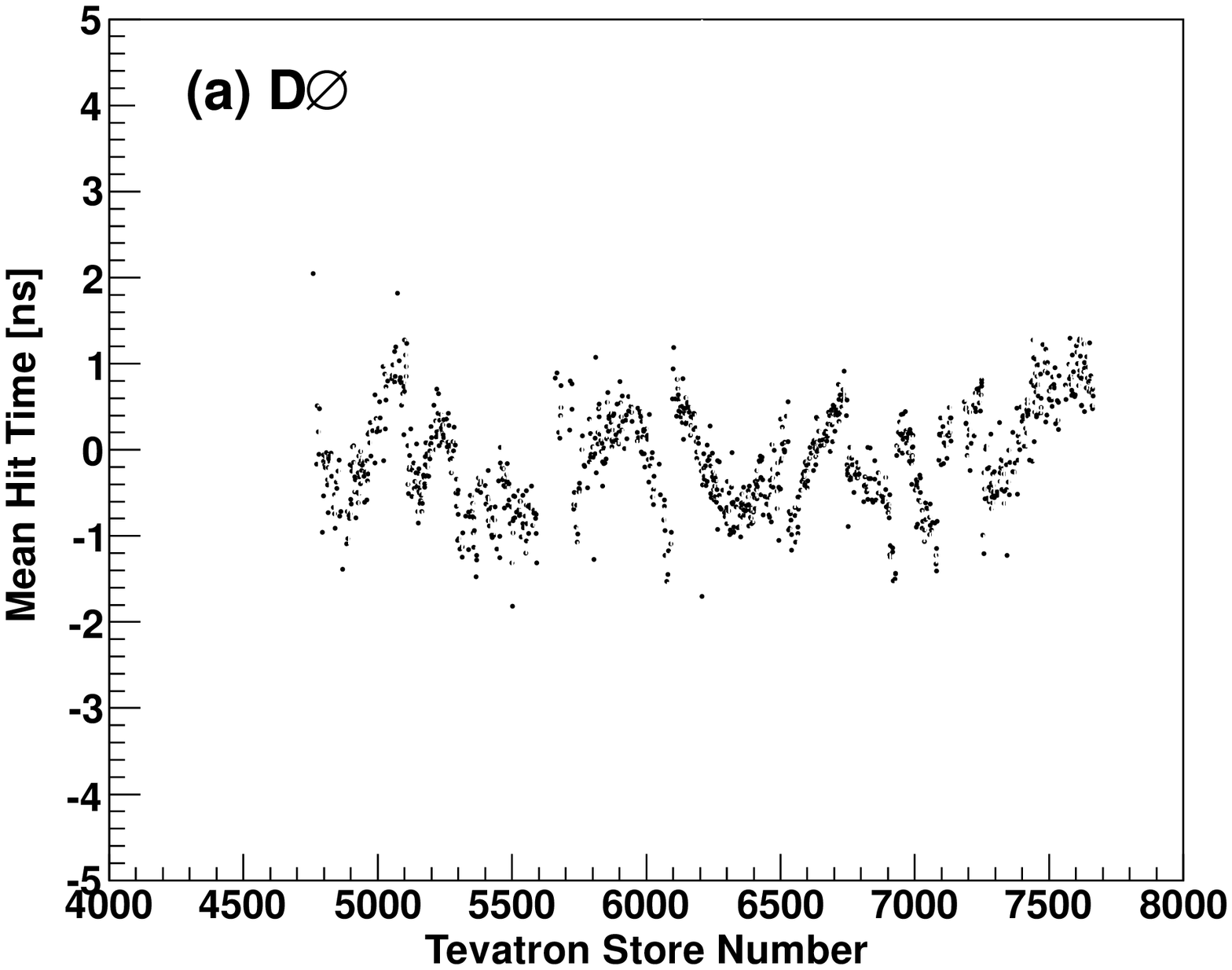}}
\scalebox{0.4}{\includegraphics{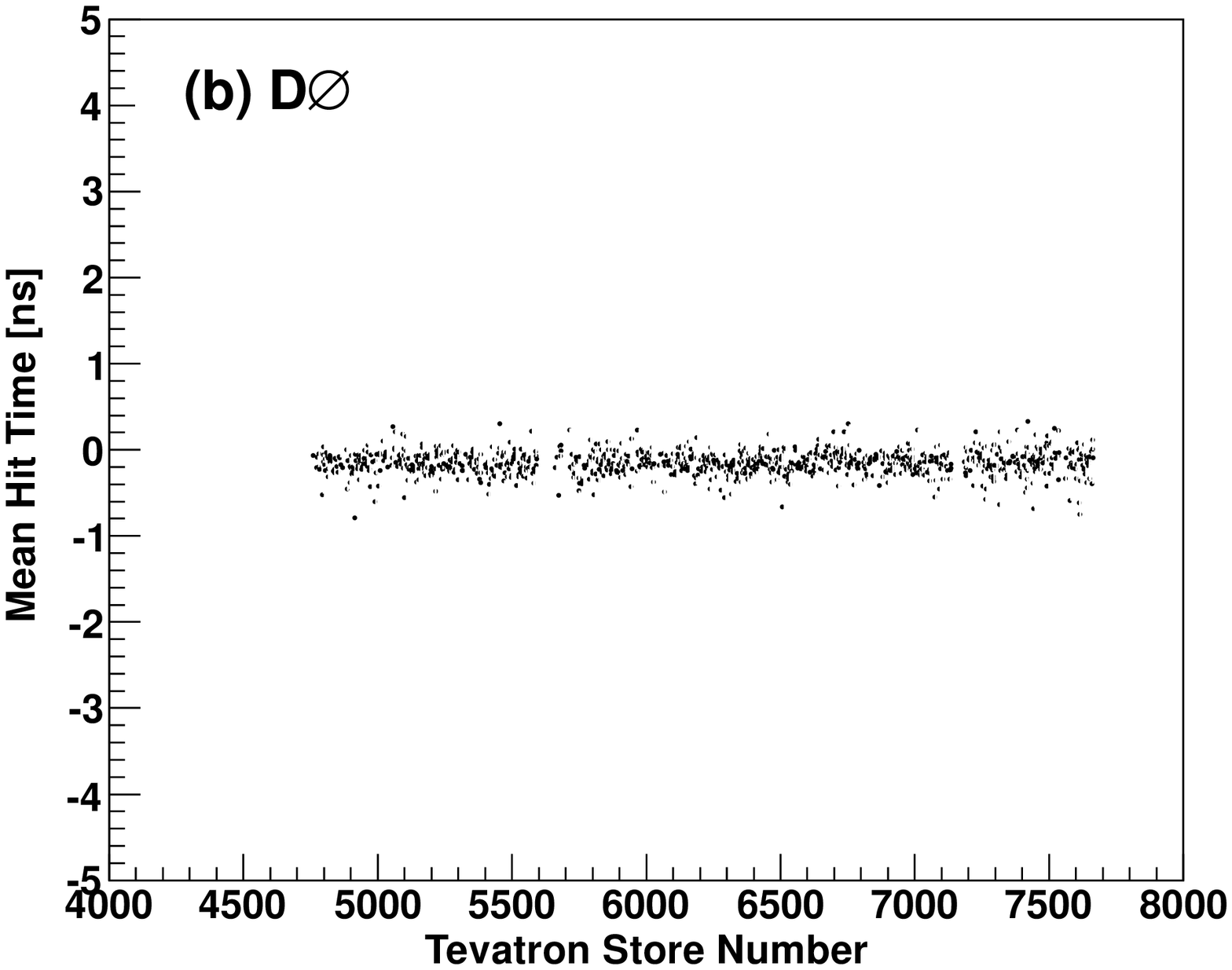}}
\caption{\label{fig:time_store}(color online) Mean hit time as a function of the Tevatron store number in the forward 
region A-layer for muons from the decay of $Z$ bosons in data (a) before correction and 
(b) after correction. A store is the time period for which $p$ and $\bar{p}$ beams 
are circulated in the accelerator after injection. This period is typically 12--24 hours 
for the Tevatron.  When the number of recorded times from $Z\rightarrow \mu\mu$ decays for a given store is large, a Gaussian 
function is fit to those times and the Gaussian mean is used. If the number of recorded 
times for a given store is small, the median of these times is used. These 
distributions cover the data taking period between June 2006 and March 2010.}
\end{center}
\end{figure*}

The TOF of a charged particle reaching the muon system can be calculated from the position of the scintillation counter that is hit and the corresponding time as has been described in Sec.~II. A time offset is determined for each scintillation counter along with its associated cables and time digitizing 
electronics using a sample of muons from experimental data so that a time reading of zero is obtained for particles 
that originate at the center of the detector at the time of a beam-beam interaction and travel 
at the speed of light to the specific scintillation counter. These time offsets are imperfect and have fluctuated with, for example, the 
seasonal variation in the synchronization with the Tevatron accelerator clock ($\approx \pm$1 ns) as shown in Fig.~\ref{fig:time_store}(a). The offsets are corrected by subtracting the relevant amount of deviation from each hit time in each scintillator.
Figure~\ref{fig:time_store}(b) shows the mean of the time distribution of hits 
versus time after the offset correction. The time period for averaging is the duration for which $p$ and $\bar{p}$ beams are circulated in the Tevatron accelerator after injection. This period is typically 12--24 hours. 

\begin{figure*}
\begin{center}
\scalebox{0.35}{\includegraphics{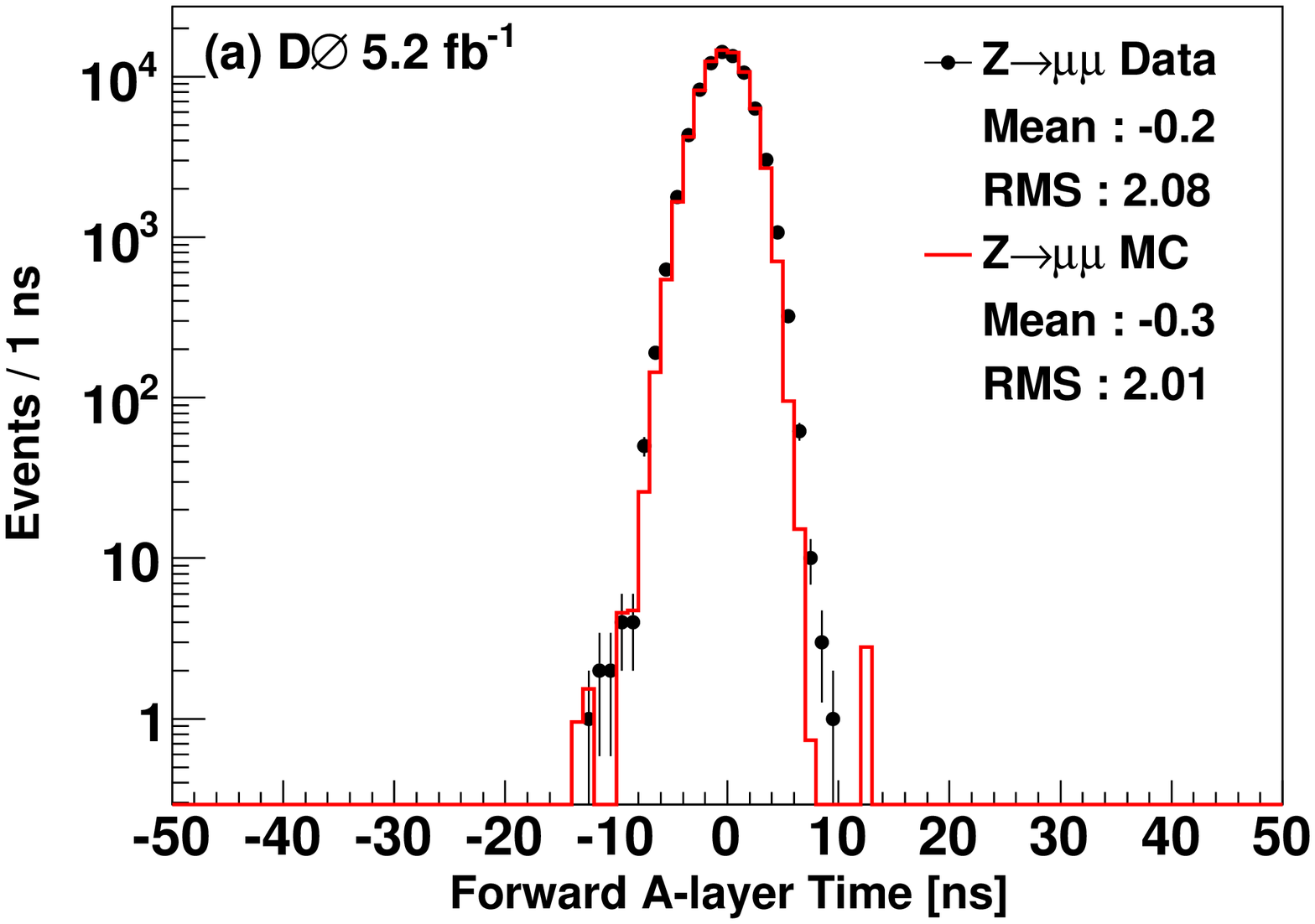}}
\scalebox{0.35}{\includegraphics{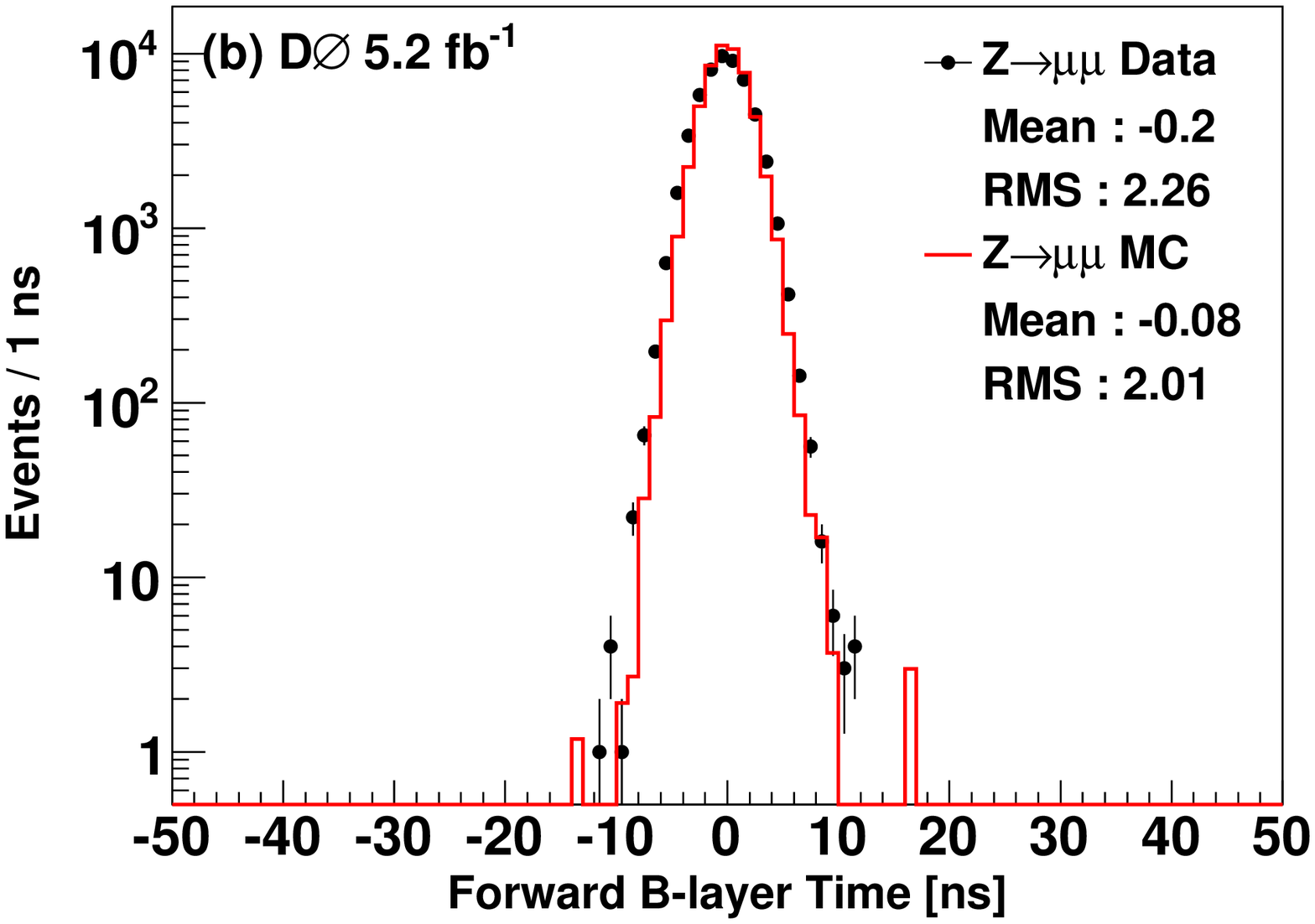}}
\scalebox{0.35}{\includegraphics{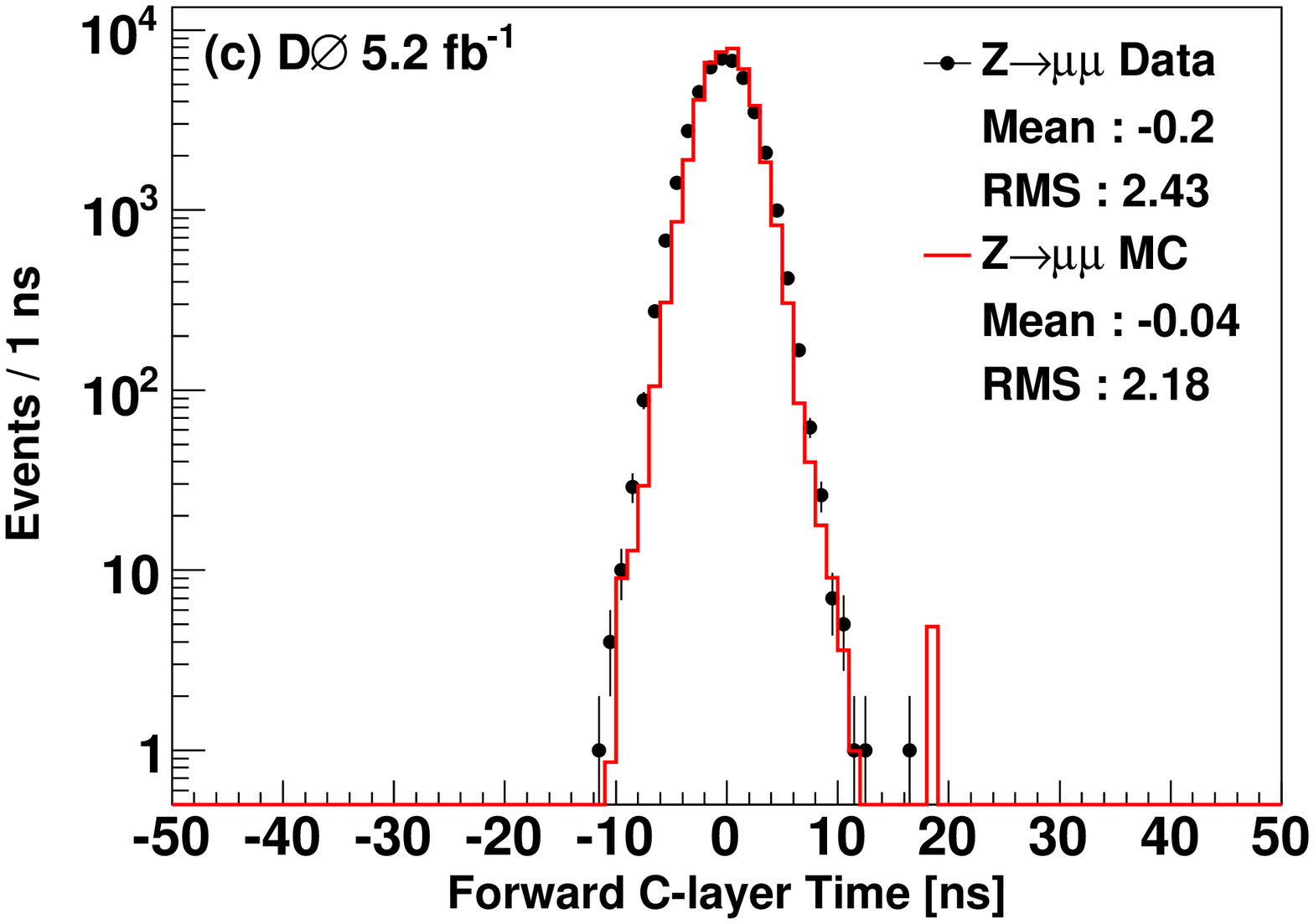}}
\caption{\label{fig:time_forwardA}(color online) Time distribution for scintillation counter layers A, B, and C in the 
forward muon system for times from $Z\rightarrow \mu\mu$ decays for data and MC events.}
\end{center}
\end{figure*}

\begin{figure*}
\begin{center}
\scalebox{0.35}{\includegraphics{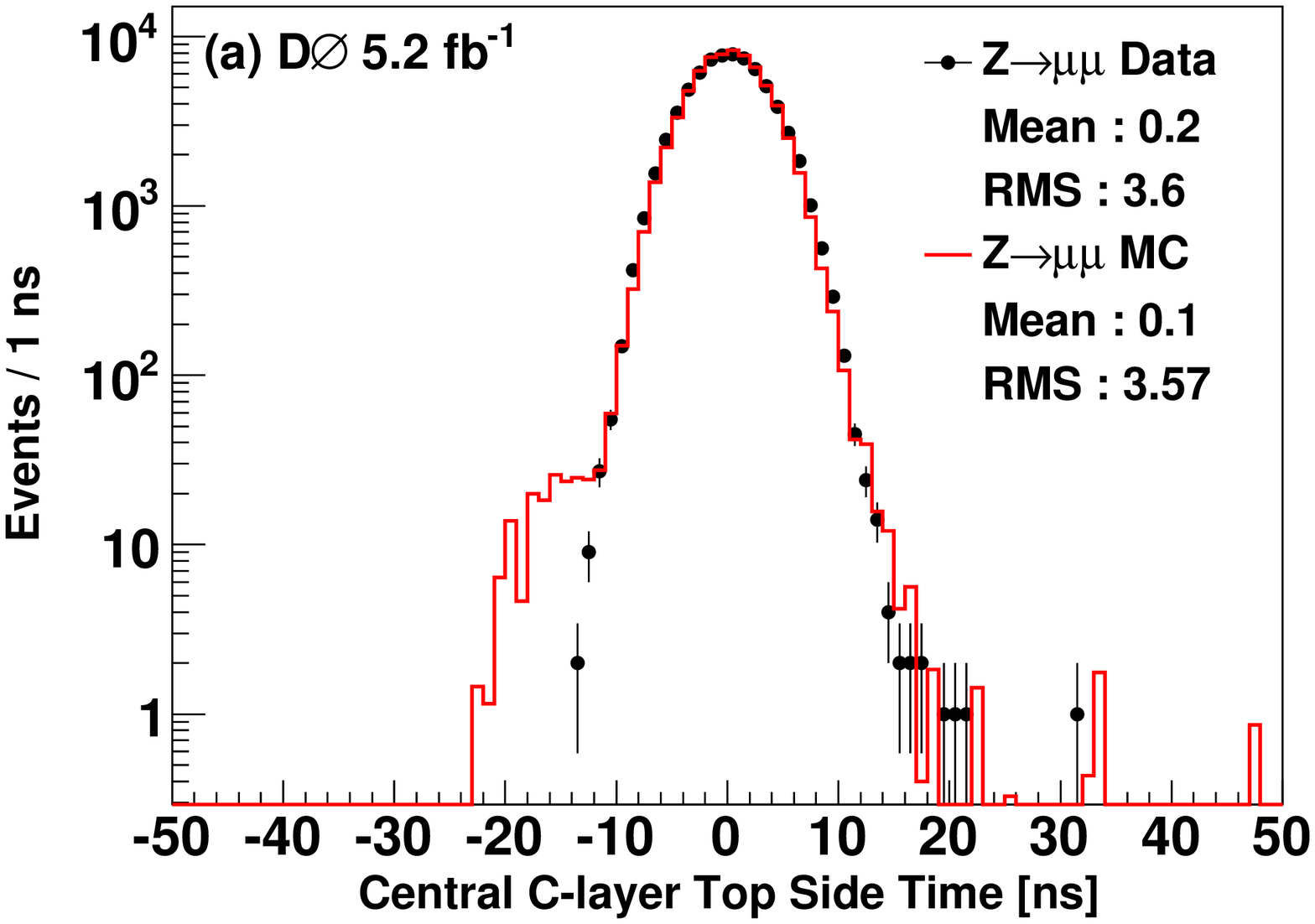}}
\scalebox{0.35}{\includegraphics{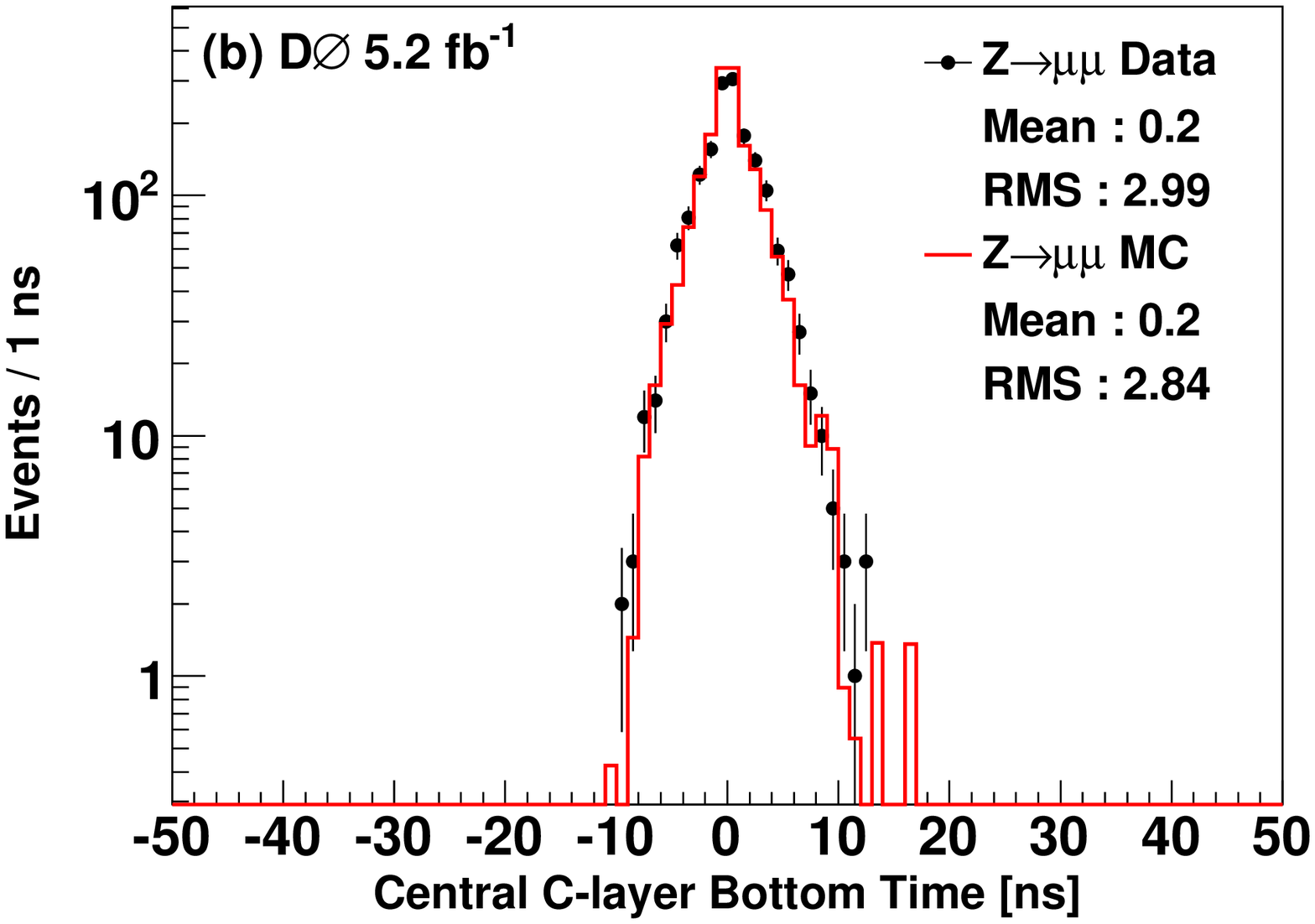}}
\caption{\label{fig:time_centralC}(color online) Time distributions for scintillation counters in layer C, (a) top and sides, (b) bottom, in the central 
muon system for times from $Z\rightarrow \mu\mu$ decays for data and MC events. }
\end{center}
\end{figure*}

\begin{figure*}
\begin{center}
\scalebox{0.35}{\includegraphics{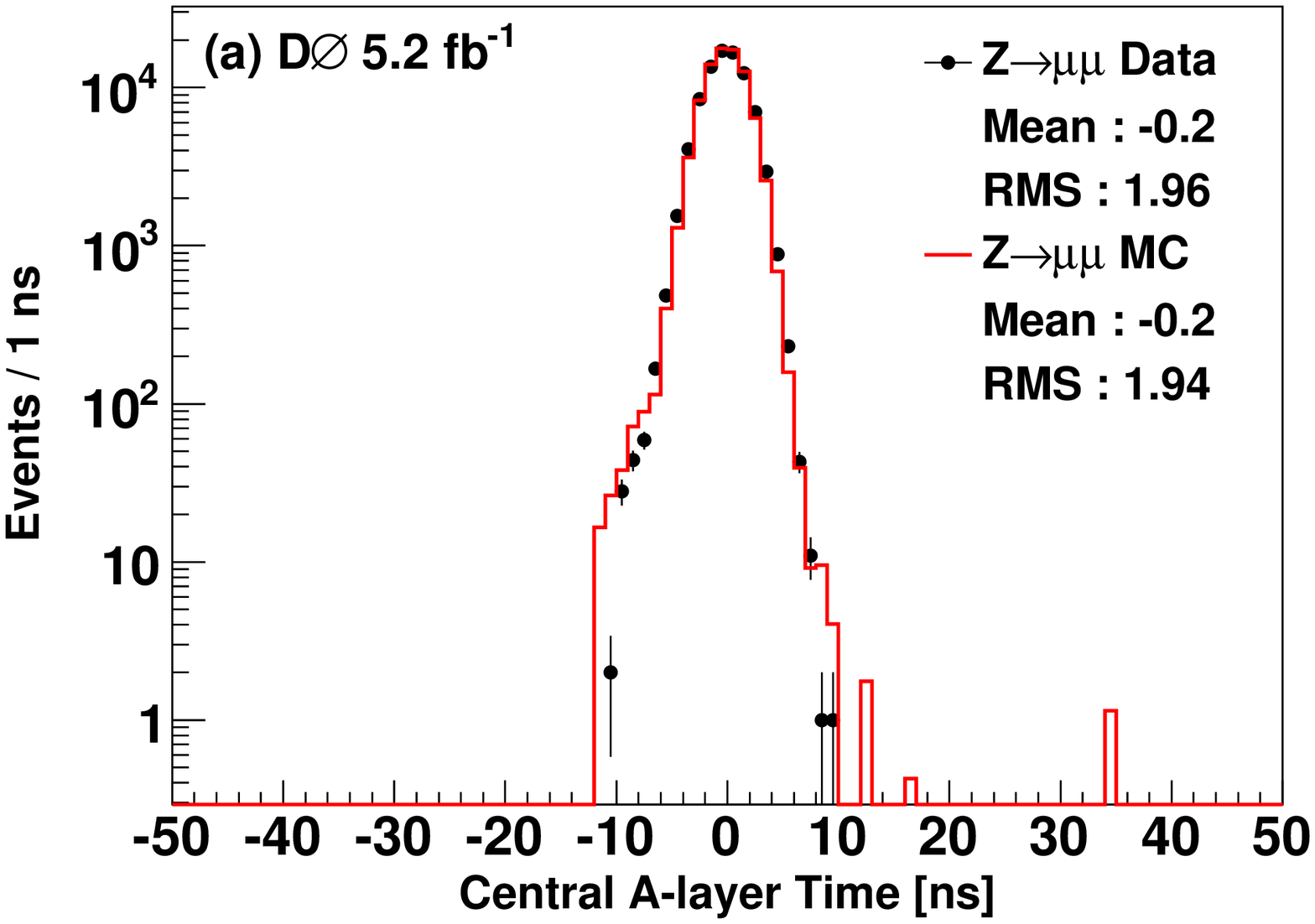}}
\scalebox{0.35}{\includegraphics{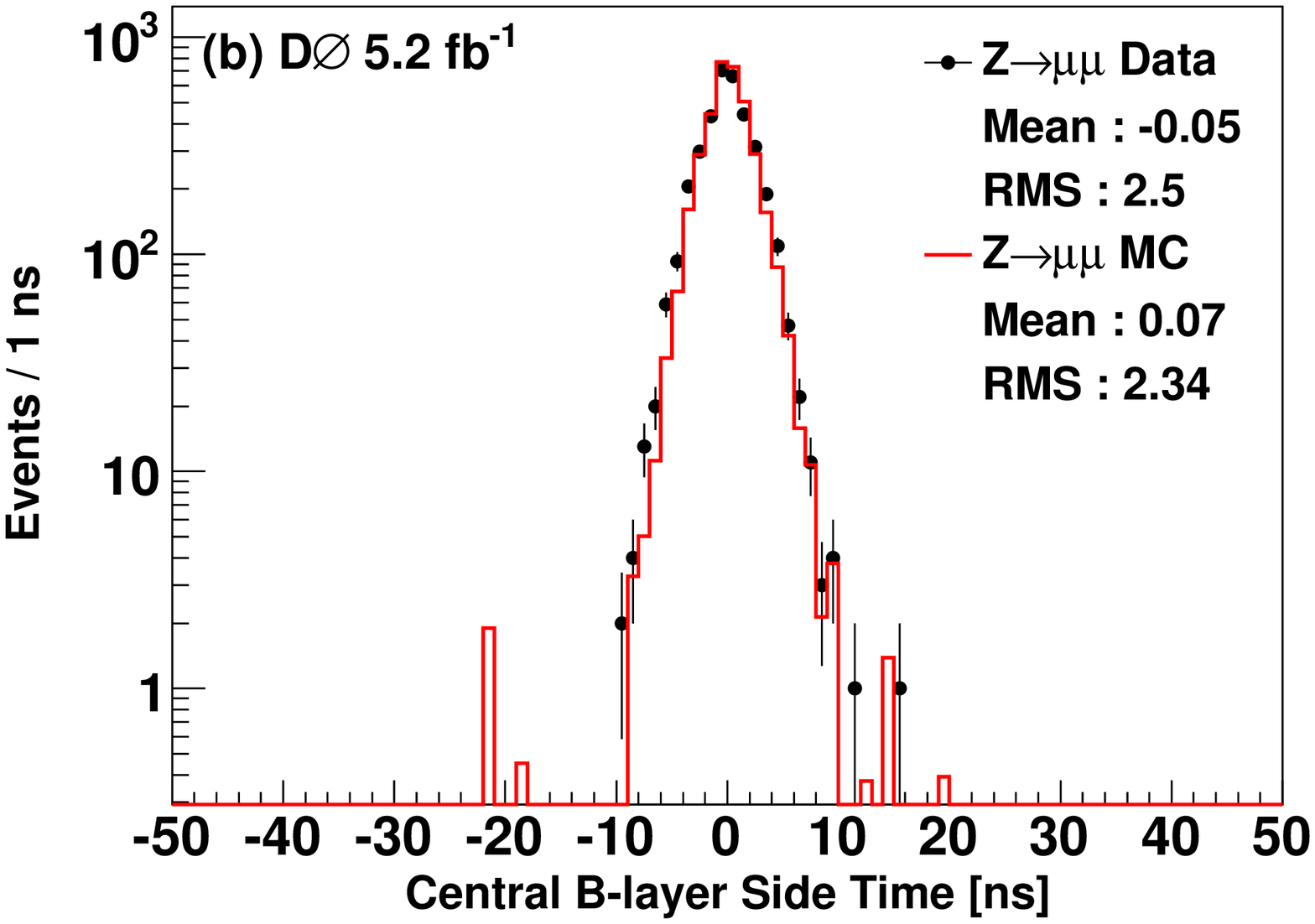}}
\scalebox{0.35}{\includegraphics{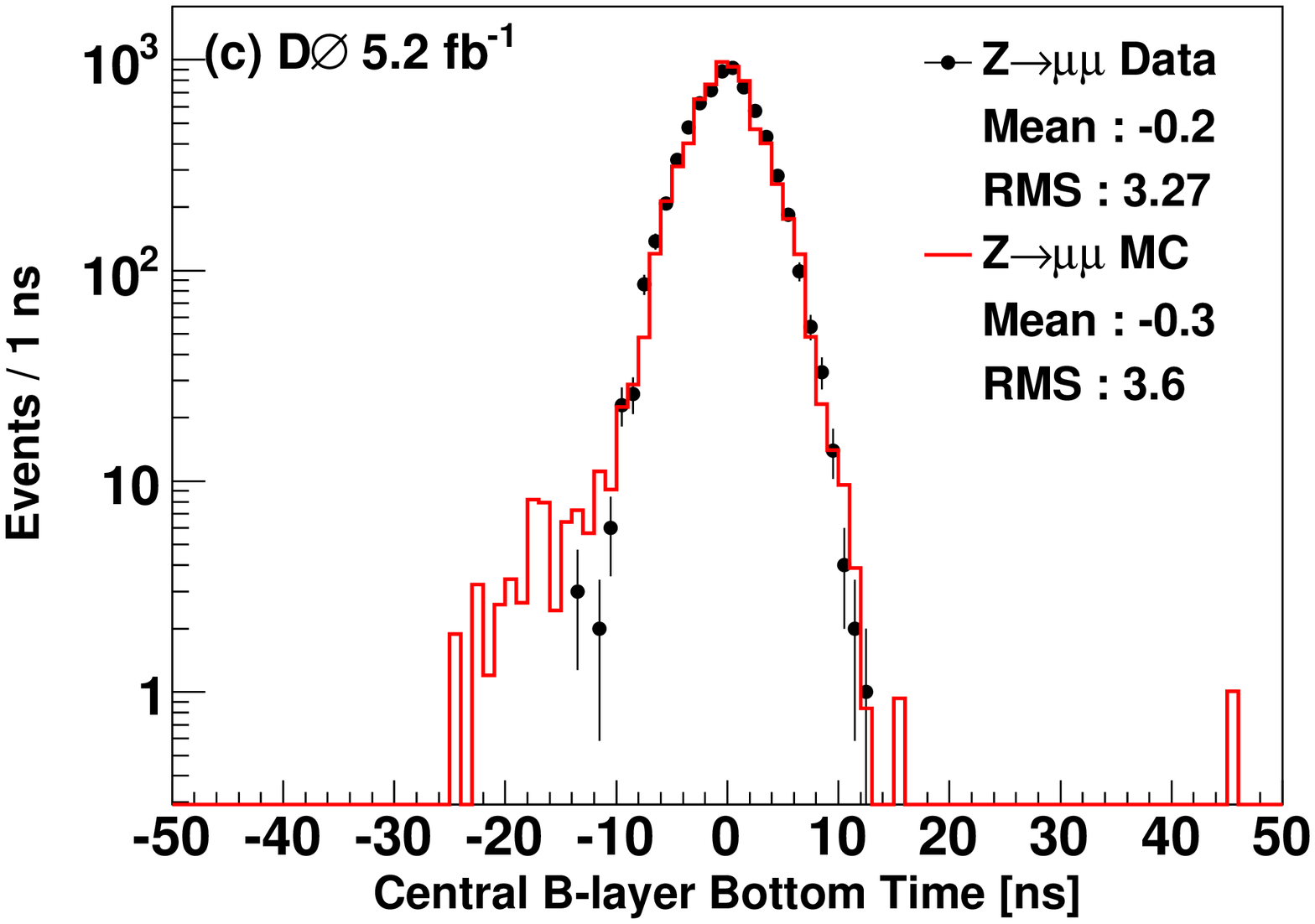}}
\caption{\label{fig:time_centralAB}(color online) Time distributions for scintillation counters in (a) layer A (top, bottom and sides), (b) layer B sides, and (c) layer B bottom, in the central muon system for times from $Z\rightarrow \mu\mu$ decays for data and MC events.}
\end{center}
\end{figure*}

It is observed that the MC simulation gives narrower time distributions than what is observed in data. Therefore, the TOF associated with a muon hit in MC is smeared to reflect the resolution of 
the time distribution of muons from $Z$ boson decays in data. The amount of 
smearing depends on the location of the muon detector because of the 
differing sizes of the scintillation counters. 
Figures~\ref{fig:time_forwardA}-\ref{fig:time_centralAB} 
show the time distributions in selected regions for layers A, B, and C 
in data and MC. There is some mismodeling at early times in the 
central C layer (Fig.~\ref{fig:time_centralC}(a)) and central bottom B layer
(Fig.~\ref{fig:time_centralAB}(c)) arising from the data-driven smearing that has been applied to the hit times of the muons in MC. This mismodeling is not in
our signal region which is at large times and has a negligible effect on the results.


  The value of $\beta$ for a muon-like track is determined from a weighted average of the 
speeds $\beta_i$ determined using times corresponding to individual scintillation counter hits on the track, the weights
being the inverse of the squares of experimentally determined uncertainties.
\begin{figure*}
\begin{center}
\scalebox{0.4}{\includegraphics{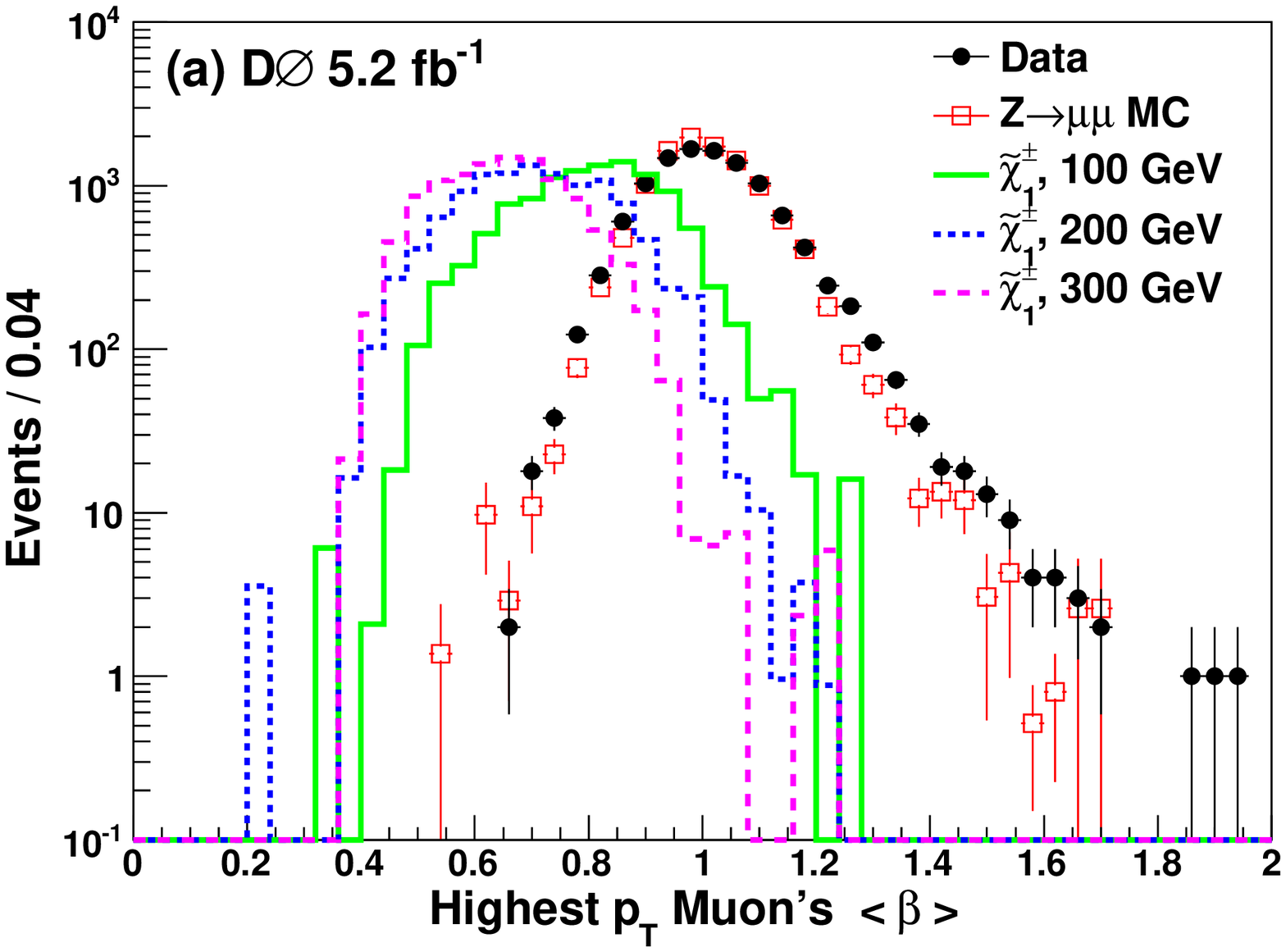}}
\scalebox{0.4}{\includegraphics{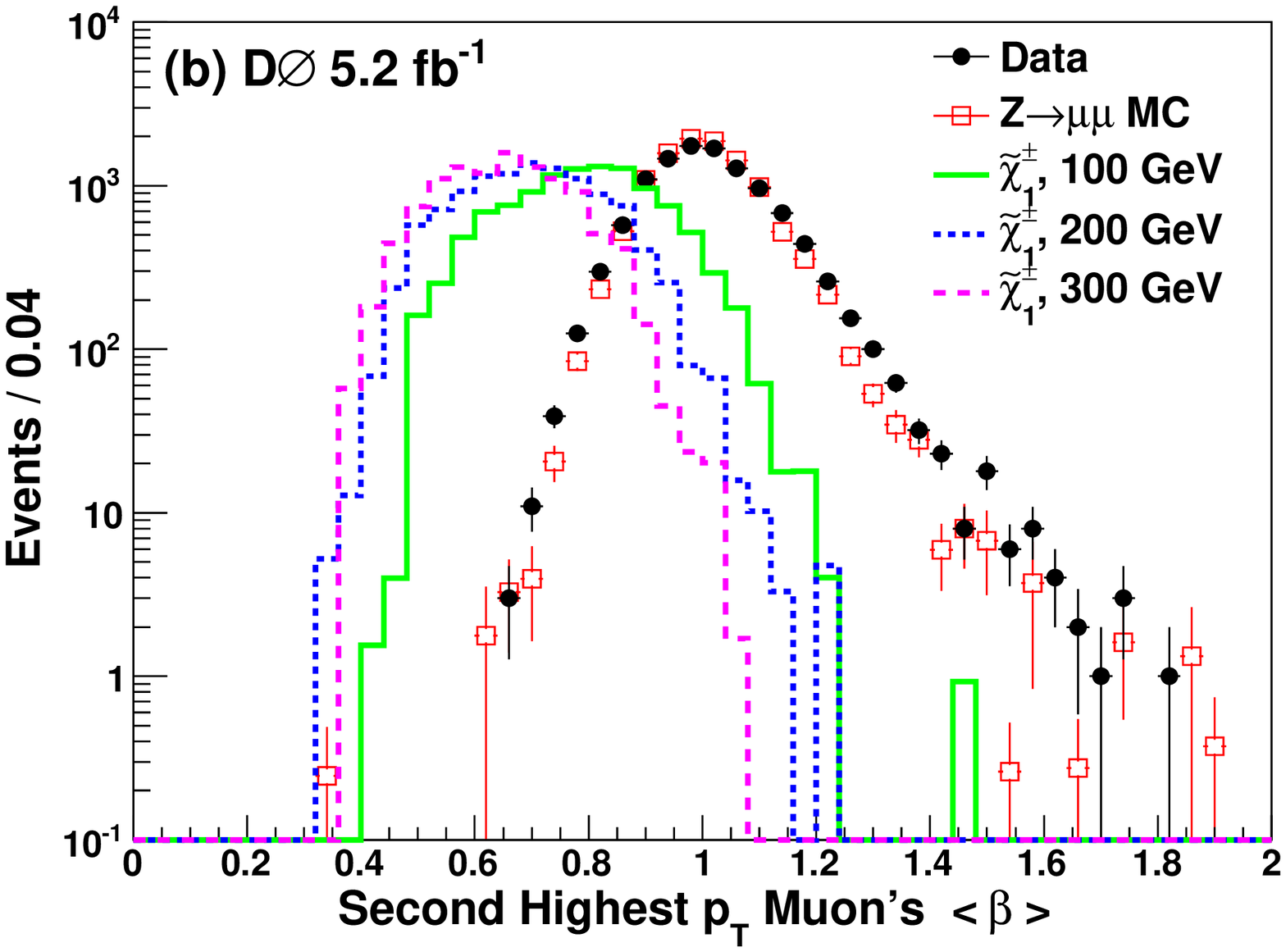}}
\caption{\label{fig:speed_muons}(color online) Distributions of $\langle \beta \rangle$  of the (a) highest $p_T$ and the (b) second-highest 
$p_T$ muons. The distributions are normalized to the same number of events. 
Background is taken from $Z \rightarrow \mu\mu$ MC decays. The selection requirements are identical 
to those used in the pair CMLLP analysis, as described in Sec~\ref{sec:pair}, except that the requirement $\langle \beta \rangle < 1$ is not imposed.}
\end{center}
\end{figure*}
Figure~\ref{fig:speed_muons} shows the $\langle \beta \rangle$ distribution of the highest and
 the second-highest $p_T$ muons in data and in simulated gaugino-like chargino events. 

\begin{figure*}
\begin{center}
\scalebox{0.4}{\includegraphics{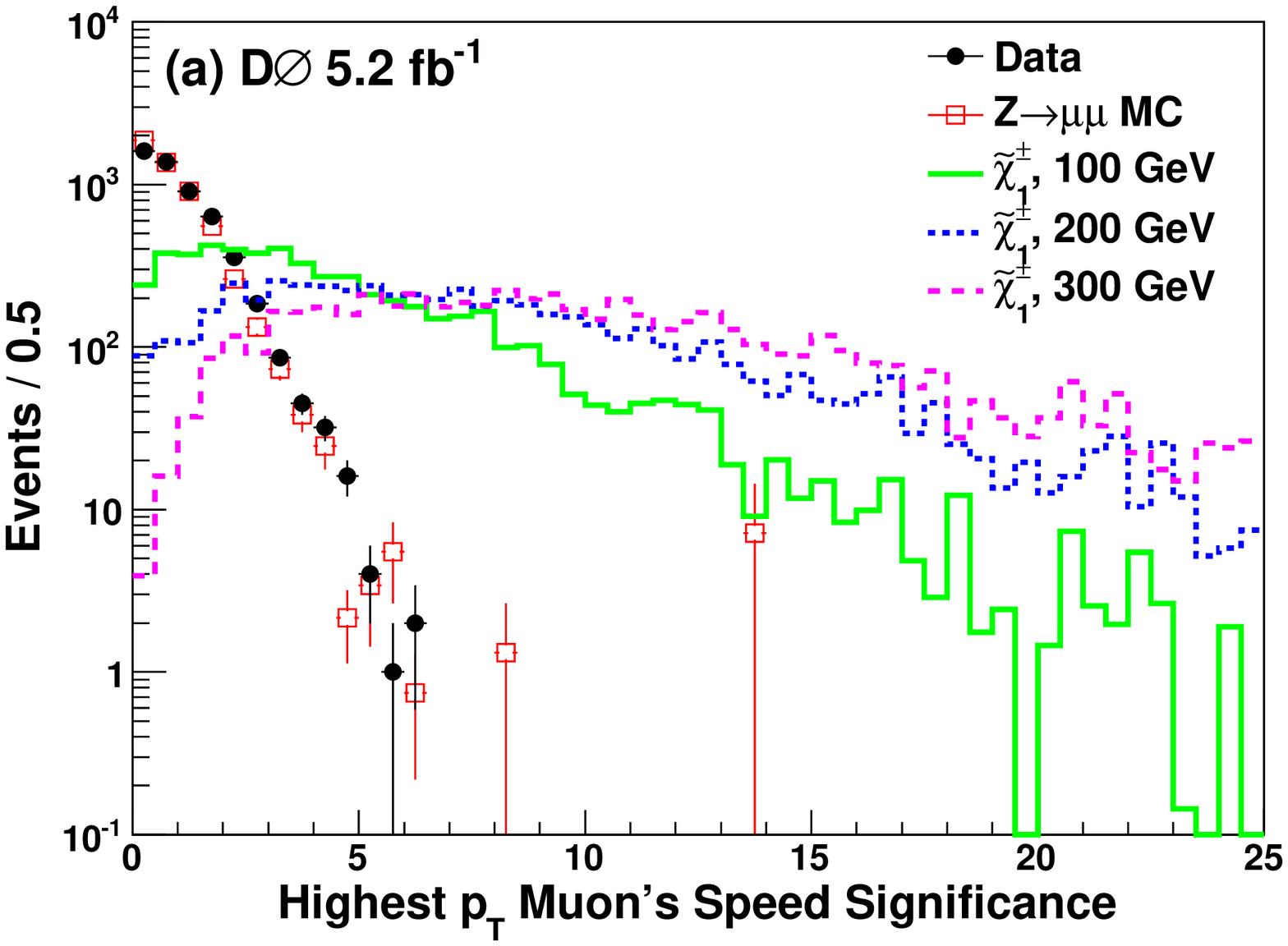}}
\scalebox{0.4}{\includegraphics{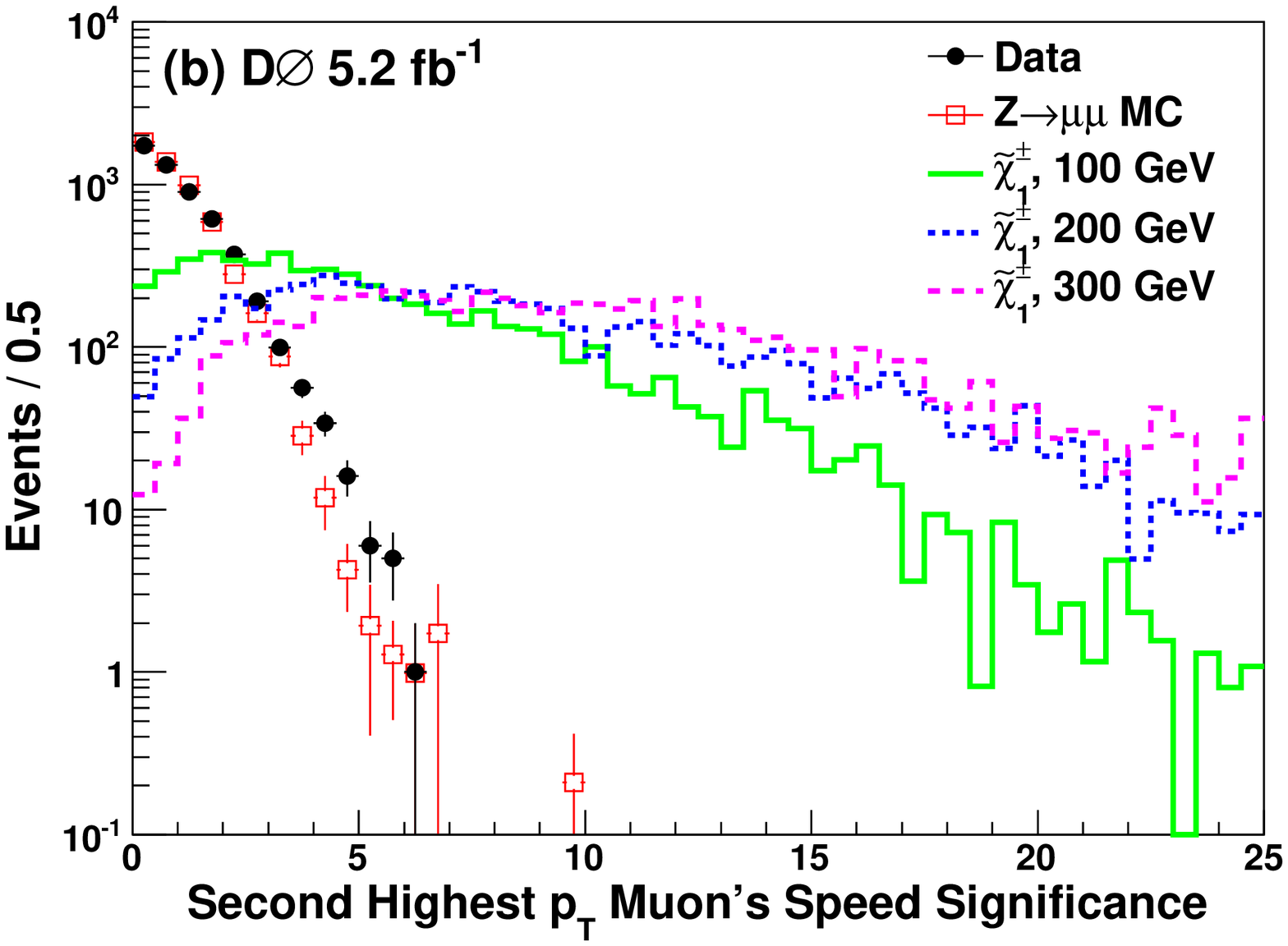}}
\caption{\label{fig:speed_sig}(color online) Speed significance of the (a) highest $p_T$  and the (b) second-highest $p_T$ muons. The distributions are normalized to the same number of events. 
Background is taken from $Z \rightarrow \mu\mu$ MC. The selection requirements are identical to 
those used in the pair CMLLP analysis, as described in Sec~\ref{sec:pair}, except that the 
requirement $\langle \beta \rangle < 1$ is not imposed.}
\end{center}
\end{figure*}

\begin{figure*}
\begin{center}
\scalebox{0.4}{\includegraphics{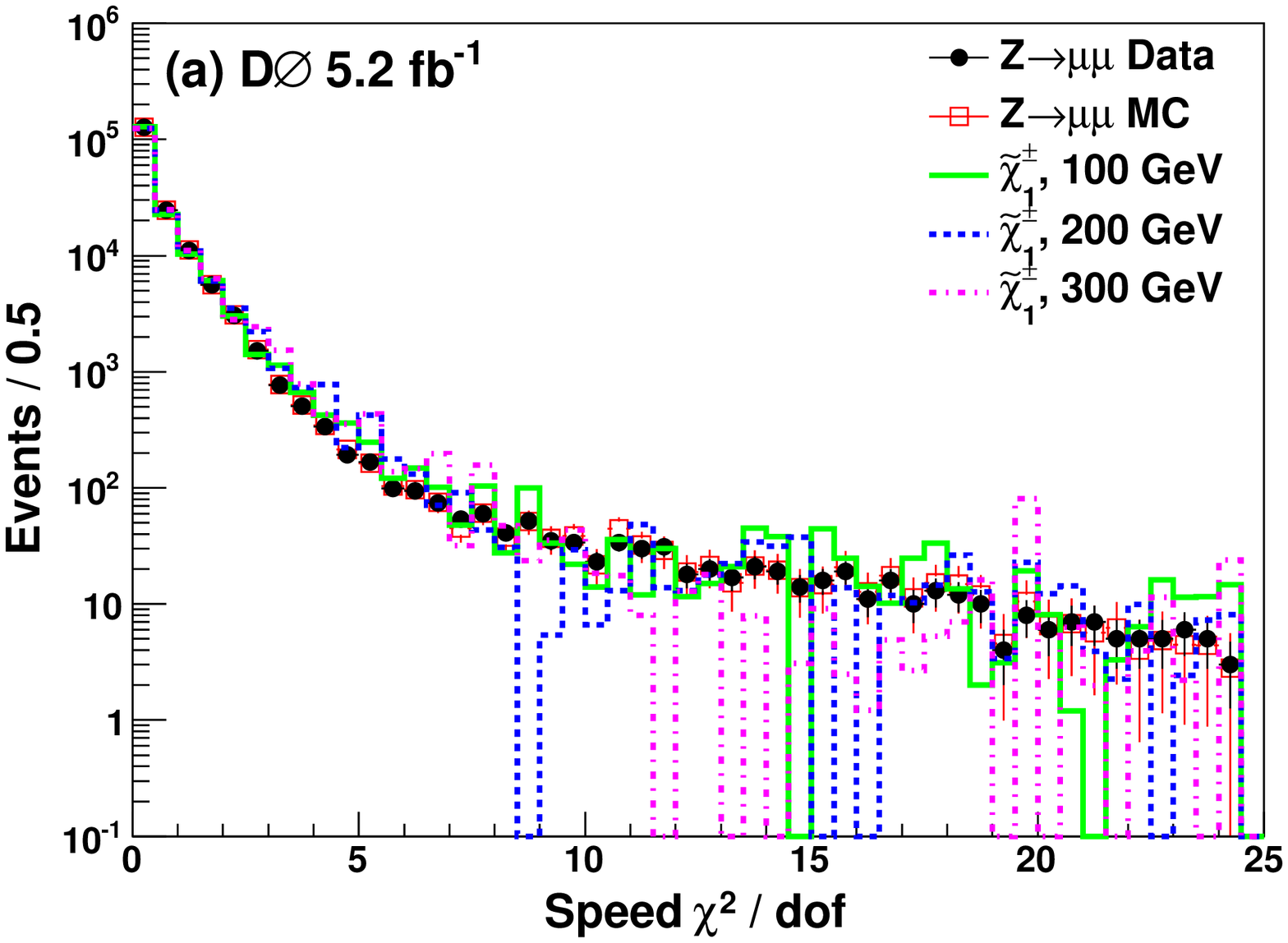}}
\scalebox{0.4}{\includegraphics{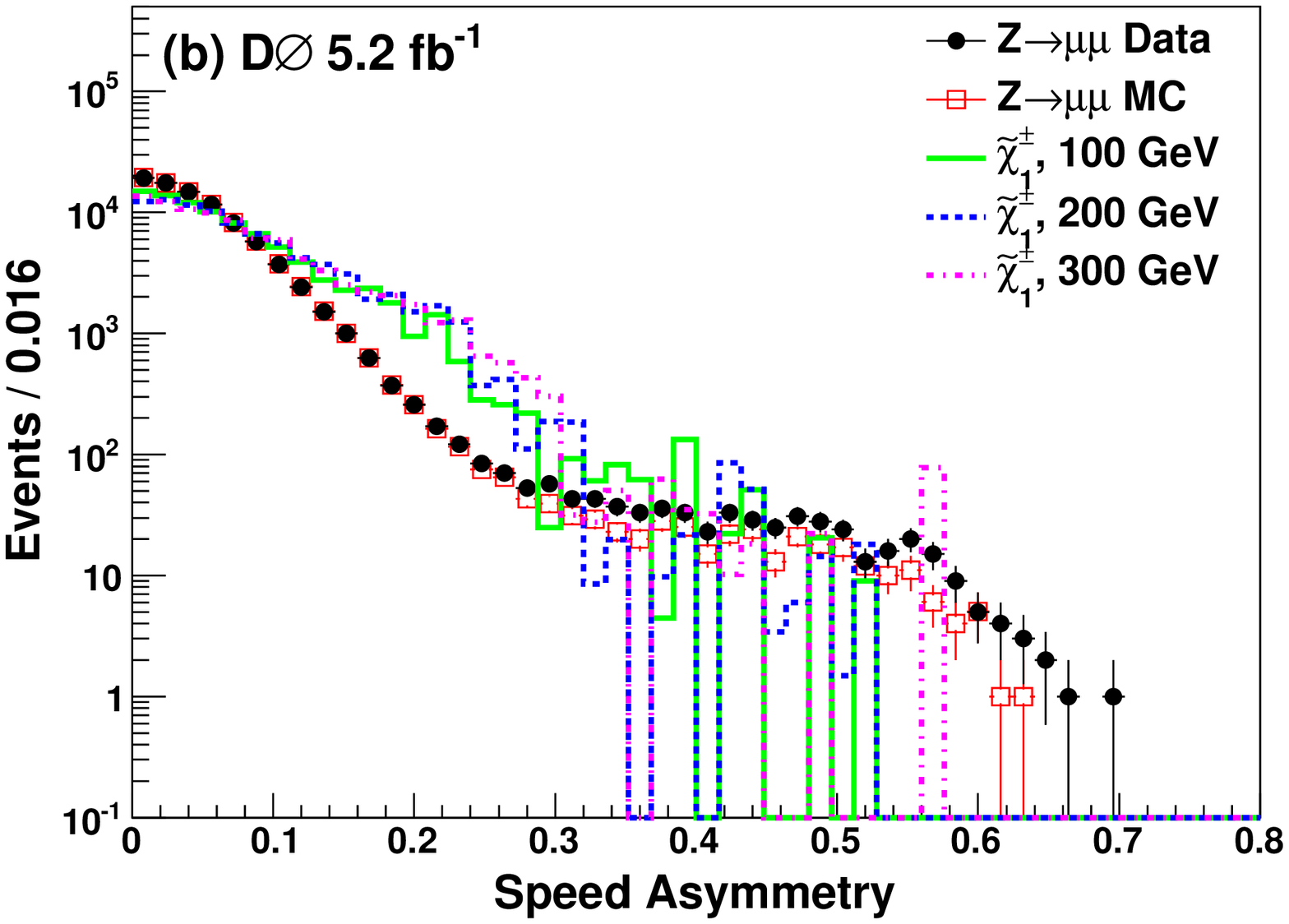}}
\caption{\label{fig:speed_chi_asym}(color online) (a) Distribution of speed ${\chi^2}/{\mathrm{dof}}$ of the highest $p_T$ 
muon in an event, and (b) speed asymmetry distribution of the highest and the second-highest 
$p_T$ muons for $Z\rightarrow\mu\mu$ data, MC and signal (gaugino-like 
chargino with masses of 100 GeV, 200 GeV, and 300 GeV). The distributions are normalized to 
the same number of events. All events are required to have two isolated muons with $p_T>$20 GeV. For $Z\rightarrow\mu\mu$ events, we require 
70 $<M_{\mu\mu}<$ 110 GeV. The muons are required to not be consistent with cosmic ray muons.} 
\end{center}
\end{figure*}

We define the speed significance, which incorporates $\beta$ with its uncertainty ($\sigma_{\mathrm{\langle \beta \rangle}}$) as,
 \begin{equation}
\rm{speed\ significance} = \frac{1-\langle \beta \rangle}{\sigma_{\mathrm{\langle \beta \rangle}}}.
\end{equation}
Figure~\ref{fig:speed_sig} shows speed significance distributions of the highest and the second-highest $p_T$ muons in the event. 
 Prompt muons will have 
speed significance near zero whereas CMLLPs will have positive speed significance. In the Run IIa analysis we use the product of the speed significances of the two muons, which will be positive for a pair of CMLLPs, as an additional criteria to separate signal events from background events. We also calculate the variable, 
\begin{equation}
 \mathrm{speed}~ \chi^2/{\mathrm{dof}} = \frac{1}{N-1}{{\sum}_i}\frac{(\langle \beta \rangle - {\beta}_i)^2}{{\sigma}_i^2} , 
\end{equation}                                                    
where $N$ is the number of scintillation counter hits associated with a muon track, ${\beta}_i$ is the speed of the track corresponding to the hit in the $i^{th}$ scintillation counter and $\langle \beta \rangle$ is the weighted average of the speeds calculated for all scintillation counter hits on the track. For some tracks the measurement ${\beta}_i$ from the hit on a particular layer is inconsistent 
with the ${\beta}_j$ measurements from other layers causing a large contribution to the value of the speed  ${\chi}^2/{\mathrm{dof}}$. To identify and remove this hit from the speed  ${\chi}^2/{\mathrm{dof}}$ calculation, hits on a muon track are removed one at a time and the 
speed ${\chi}^2/{\mathrm{dof}}$ is recalculated with the rest of the hits. The set of hits with the lowest value of speed ${\chi}^2/{\mathrm{dof}}$ is used to recalculate the speed provided it satisfies the qualities of a good muon candidate (described in Sec.~II). Distributions of
speed ${\chi}^2/{\mathrm{dof}}$ for the highest $p_T$ muon in data and
 in simulated gaugino-like chargino events are shown in Fig.~\ref{fig:speed_chi_asym}(a). 

 The speed asymmetry, which is useful in the search for a CMLLP pair, can be defined for the two highest $p_T$ muons in the event as:
\begin{equation}
\rm{speed\ asymmetry} = \frac{|{\beta}_1-{\beta}_2|}{{\beta}_1+{\beta}_2}.
\end{equation}
 The speed asymmetry is near zero for both signal events and events containing two well measured muons, but will be large for events where one of the particles has a speed
that has been mismeasured. Speed asymmetry distributions for data and for simulated gaugino-like chargino events are shown in Fig.~\ref{fig:speed_chi_asym} (b).

We have observed that the disagreements in $\langle \beta \rangle$ distributions from data and MC at small values of $\langle \beta \rangle$,
visible in Fig.~\ref{fig:speed_muons}, can be removed
by applying stringent requirements on either the speed $\chi^2/{\mathrm{dof}}$ or the speed asymmetry distribution. The $\langle \beta \rangle$ distribution extends beyond 1 due to uncertainty in the measurement of velocity. 
There is some mismodeling for $\langle \beta \rangle > 1$, but this is not in our signal region. This disagreement is due to the background coming from mismeasured muons,  
which is characterized by large values of speed $\chi^2/{\mathrm{dof}}$ or speed asymmetry, and not due to signal-like
events, which are characterized by small values of speed $\chi^2/{\mathrm{dof}}$ or speed asymmetry. We apply a requirement on the speed asymmetry of the two candidate CMLLPs in the search for a 
pair of CMLLPs as described in Sec.~VI B. In the search for single CMLLPs we apply a requirement on the speed ${\chi^2}/{\mathrm{dof}}$ of the candidate CMLLP as described in Sec~VII B. 

We correct the mismodeling in the speed $\chi^2/{\mathrm{dof}}$ and the speed asymmetry distributions using a signal-free region (as described in Sec.~VII B) in data. 
We compute the ratio of the speed ${\chi^2}/{\mathrm{dof}}$ (or the speed asymmetry) distributions in data 
and $Z \rightarrow \mu\mu$ MC for events with 70 $<M_{\mu\mu}< $110 GeV, where the potential signal contribution is negligible. The value of this ratio is applied as a weight to all simulated events.

\subsection{The {\boldmath ${\rm d} E/{\rm d} x$} Measurement}
\begin{figure*}
\begin{center}
\scalebox{0.4}{\includegraphics{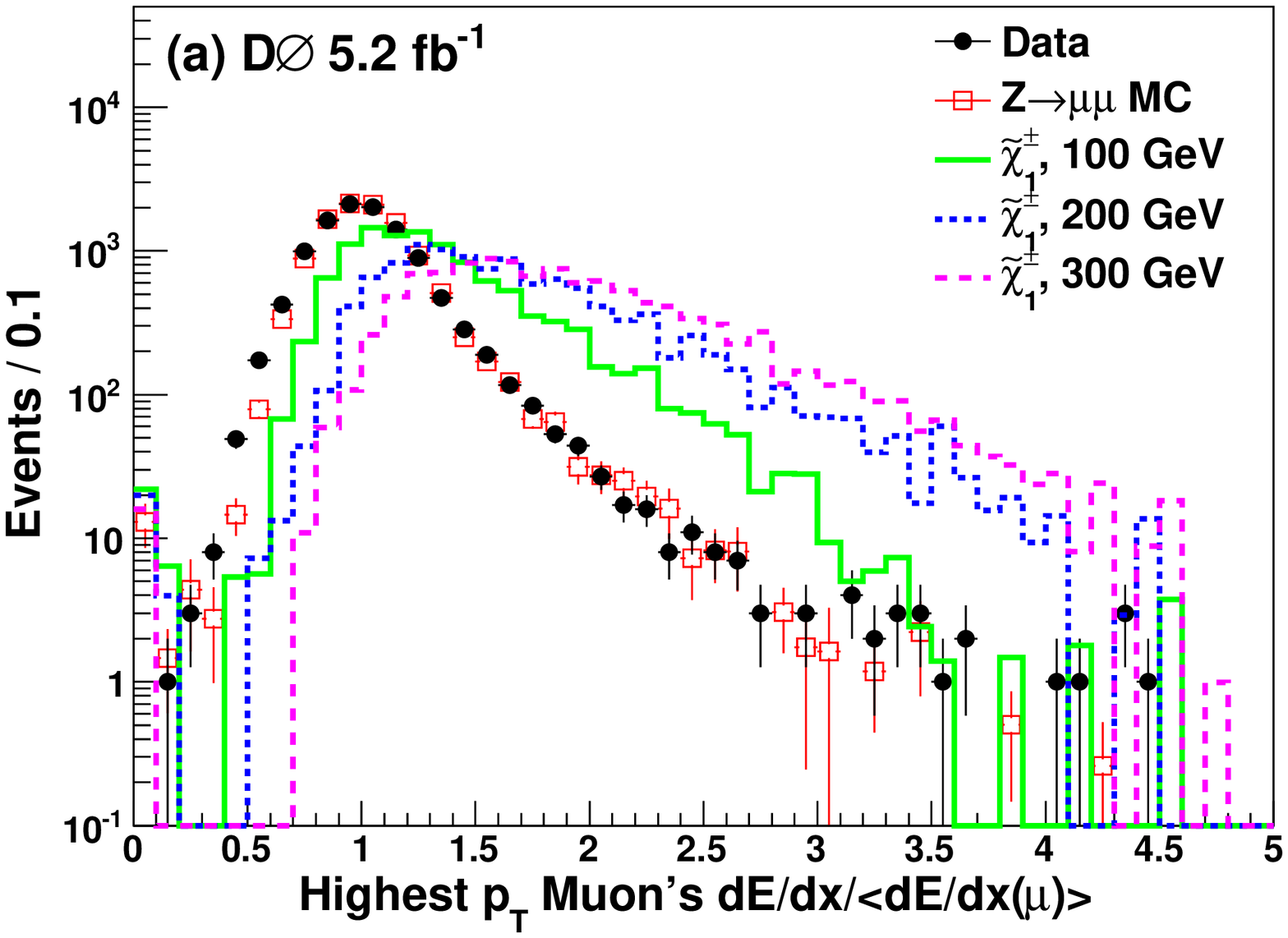}}
\scalebox{0.4}{\includegraphics{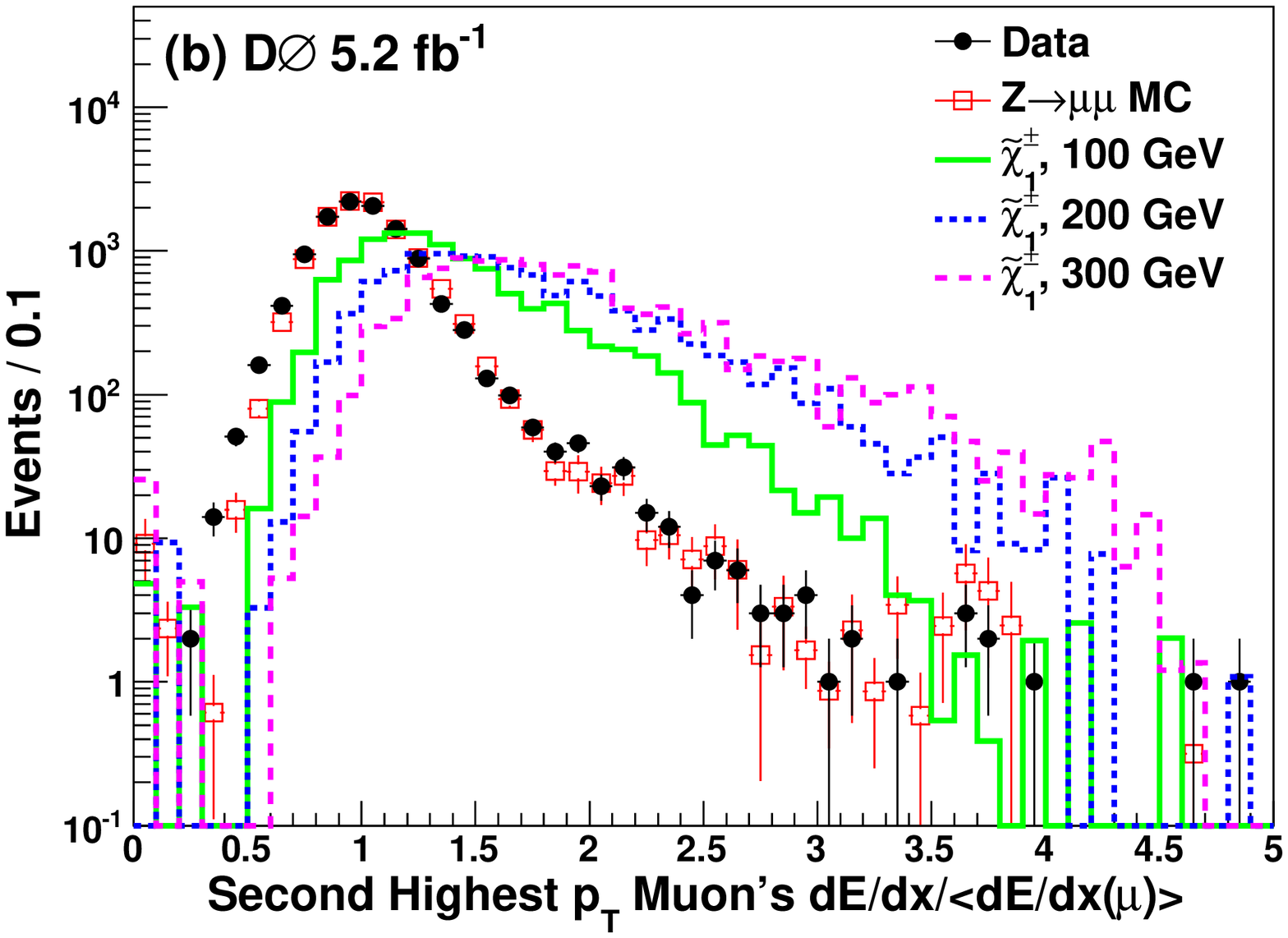}}
\caption{\label{fig:dedx_muons}(color online) Distributions of the adjusted ${\rm d} E/{\rm d} x$ (described in Sec.~V B) of the (a) highest $p_T$ and the (b) second- highest $p_T$ muons in data and the simulated chargino events. The distributions are normalized 
to the same number of events. The Background is taken from $Z \rightarrow \mu \mu$  MC. The selection requirements are identical to those used in the pair CMLLP analysis, as described in Sec~\ref{sec:pair}.}
\end{center}
\end{figure*}
The ${\rm d} E/{\rm d} x$ of a particle is measured in the SMT. 
It is a calibrated average over SMT clusters and is corrected 
for the path length of the particle in the barrel or the disk sensor to reduce the  
dependence on the incident angle of the particle. The calculation of ${\rm d} E/{\rm d} x$ excludes SMT clusters 
with the highest 20\% of ${\rm d} E/{\rm d} x$ values in order to minimize the contribution from Landau tails. 

\begin{figure*}
\begin{center}
\scalebox{0.4}{\includegraphics{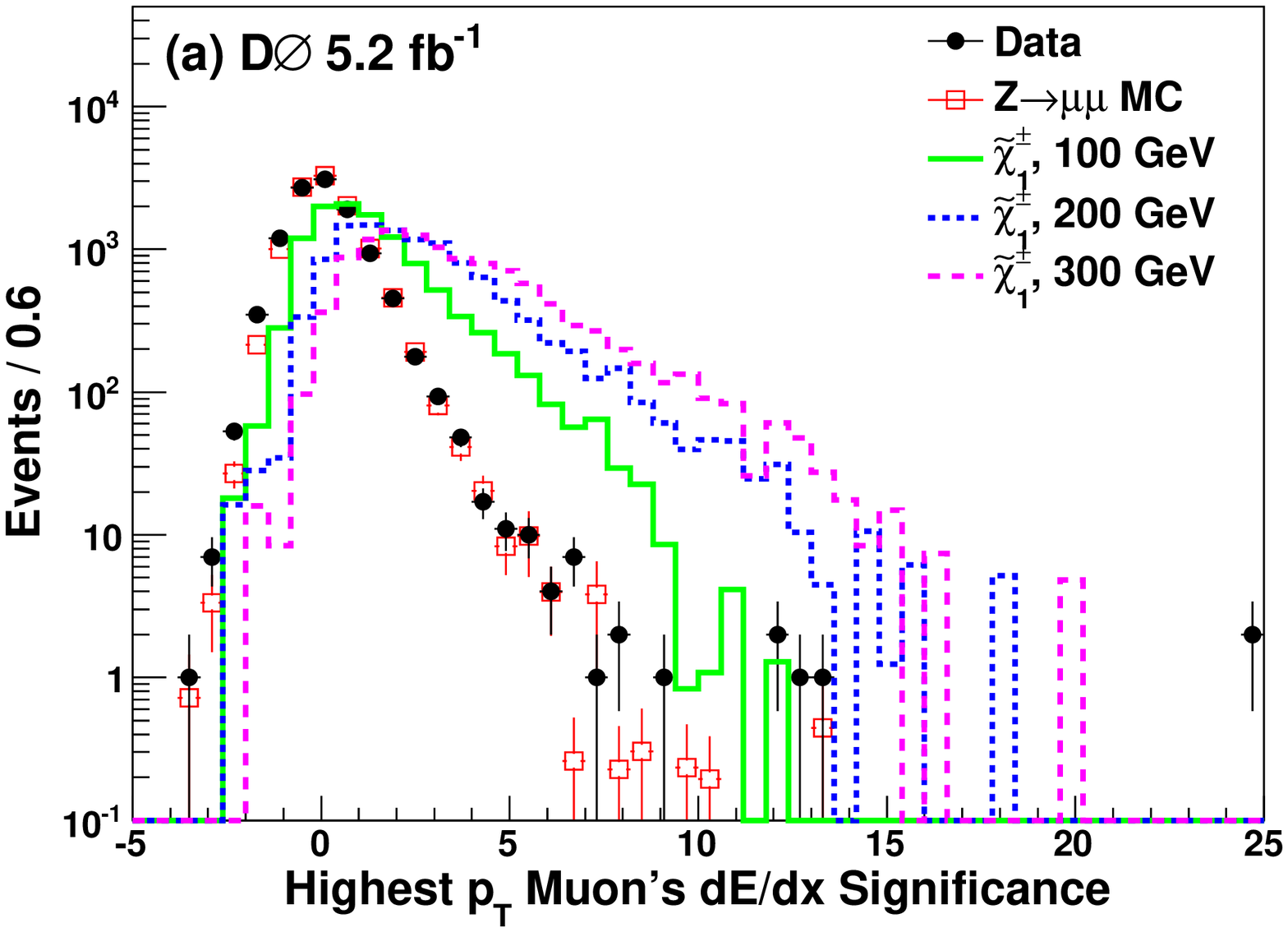}}
\scalebox{0.4}{\includegraphics{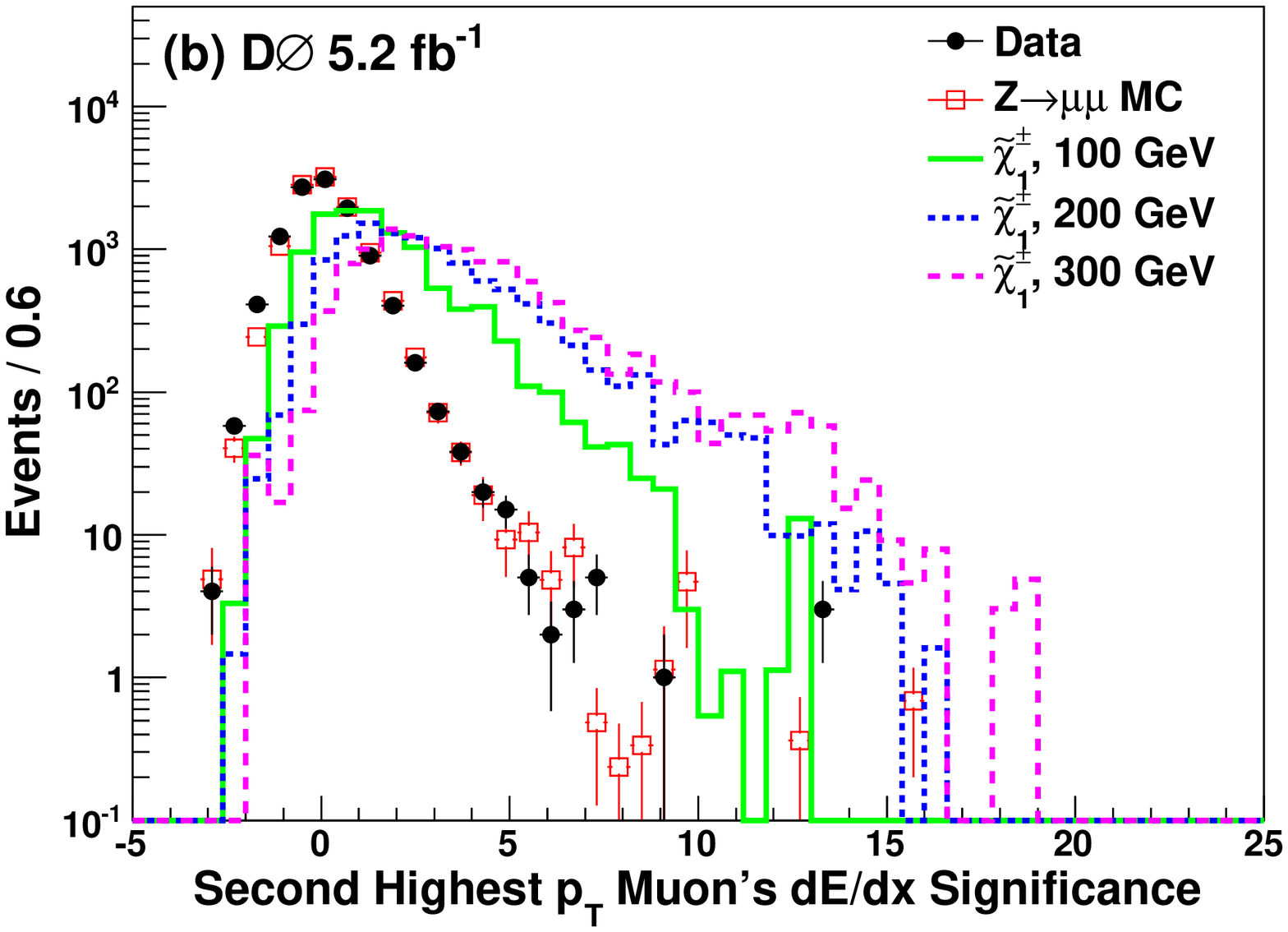}}
\caption{\label{fig:dedx_significance}(color online) Distributions of ${\rm d} E/{\rm d} x$ significance for the (a) highest $p_T$ and the (b) second-highest $p_T$ muons. The distributions are normalized to the same number of events. Background is taken from $Z \rightarrow \mu\mu$ MC. The selection requirements are identical to those used in the pair CMLLP analysis, as described in Sec~\ref{sec:pair}.}
\end{center}
\end{figure*}

   The average value of ${\rm d} E/{\rm d} x$ is observed to decrease with increasing integrated luminosity~\cite{Lum uncert} 
due to radiation damage to the silicon sensors. To correct for this effect, each ${\rm d} E/{\rm d} x$ measurement is divided by the mean ${\rm d} E/{\rm d} x$.
at ${\rm d} E/{\rm d} x =$ 1, to facilitate the comparison between data and MC. The recalibrated ${\rm d} E/{\rm d} x$ distribution is referred to as the ``adjusted ${\rm d} E/{\rm d} x$''.
Figure~\ref{fig:dedx_muons} shows that the adjusted ${\rm d} E/{\rm d} x$ distribution of muons from $Z$
decays is well separated from that of candidate CMLLPs. 
Since the adjusted ${\rm d} E/{\rm d} x$ distribution of muons from $Z$ decays in data does not quite match the adjusted ${\rm d} E/{\rm d} x$ of $Z\rightarrow \mu\mu$ MC events, a Gaussian smearing is applied to the adjusted ${\rm d} E/{\rm d} x$ in MC to better describe the data. 
   The precision of a particle's adjusted ${\rm d} E/{\rm d} x$ depends on the number of SMT clusters used in its calculation. A new variable, ${\rm d} E/{\rm d} x$ significance, is defined to estimate 
this dependence. If $N_c$ is the number of SMT clusters on a track, and  if the spread in the adjusted ${\rm d} E/{\rm d} x$ 
distribution (which is a function of $N_c$) is ${\sigma}(dE/dx)_{N_c}$, then the ${\rm d} E/{\rm d} x$ significance is:
\begin{equation}
{\rm d} E/{\rm d} x\ \mathrm{significance} = \frac{{\rm d} E/{\rm d} x - 1}{{\sigma}({\rm d} E/{\rm d} x)_{N_c}}.
\end{equation}
Figure~\ref{fig:dedx_significance} shows a distribution of the ${\rm d} E/{\rm d} x$ significance for the 
highest $p_T$ and the second-highest $p_T$ muons in data and background, and for simulated gaugino-like charginos of 100--300 GeV masses. The two data points at high values of  ${\rm d} E/{\rm d} x$ significance have $\langle \beta \rangle$ close to 0.98 and are removed by the selection criteria described in Sec.~VI B.
\subsection{Detection of Top Squarks}
About 60\% of top or anti-top squark hadrons will be charged after initial 
hadronization~\cite{stop-fairbairn}. 
 A top squark hadron passing through matter can exchange light quarks through nuclear interactions, changing the charge of 
the hadron. Because the detector has more quarks than antiquarks, after many nuclear interactions 
most of the top squark mesons become baryons, but anti-top squark anti-baryons become anti-mesons. The charge of a top squark baryon can be 0, $+1$ or $+2$; the fractions are model dependent. We have assumed $2/3$ of the stop baryons to be charged after undergoing many interactions in the detector material. Similarly, the charge of an anti-top squark anti-mesons can be 0 or $-1$. We assume anti-top squark anti-mesons to have a probability of $1/2$ of being charged after passing through the detector material
~\cite{stop-mackeprang-thesis, stop-mackeprang-rizzi, stop-mackeprang-milstead}. We also provide results assuming the top and anti-top squark hadrons do not flip charge at all for reference. In this case we only include a factor of 60\% for the initial hadronization.

   As can be seen in Fig.~\ref{fig:D0_detector}, the top squark or the anti-top  squark 
hadron must be measured as charged at three locations while passing through the D0 detector to be reconstructed: after hadronization, after the
calorimeter, and after the muon toroid. Both the calorimeter and the muon toroid 
contain enough material (10 and 6 interaction lengths respectively) to ensure that a top squark or anti-top squark hadron will undergo a 
large number of interactions when passing through them, randomizing its charge. 
Therefore, the probability of a top squark hadron to be charged at all
three locations is 0.6 (at production) $\times$ 0.67 (at the end of
the calorimeter) $\times$ 0.67 (at end of the muon toroid) $=$ 0.27. Likewise, the probability of an anti-top squark hadron to be charged
 at all three locations is 0.6 $\times$ 0.5 $\times$ 0.5 $=$ 0.15.
The probability for a pair of top squark and anti-top squark hadrons both to be charged in all three locations is 0.27 $\times$ 0.15 $=$ 0.04. 
The probability of at least one of the two being charged in all three locations 
is 0.27 $\times$ (1 $-$ 0.15) $+$ 0.15 $\times$ (1 $-$ 0.27) $+$ 0.27 $\times$ 0.15 $=$ 0.38. 
We apply high enough $p_T$ cuts on the reconstructed tracks so that the selected tracks have small curvatures and even the tracks with $+2$ charge are reconstructed and matched. For both the pair and the single CMLLP searches, these estimates of charge survival probabilities are applied as additional factors when the top squark MC is normalized to the expected number of events.
\section{Search for events with a pair of CMLLPs\label{sec:pair}}
With the selection variables described in
Sec.~V to provide discrimination of CMLLP signal over
background, we describe below the selection criteria for the search for a pair of CMLLPs with 5.2 ${\mathrm{fb}}^{-1}$ of Run IIb integrated luminosity.

\subsection{Background Sample}
To model the background, one million events containing muons from decays 
of $Z$ bosons were simulated using {\sc pythia}.
Although a $Z\rightarrow \mu\mu$ MC sample is used to model the background, we do not 
assume that the background is only from $Z\rightarrow \mu\mu$ decays. Any source of 
muons that are not measured correctly contributes to the background. Since the $p_T$ of the background muons have a wide range, we 
minimize the dependence of the analysis on the $p_T$ of the selected  tracks. Other than the initial requirement on the $p_T$ of the tracks, the selection criteria do not depend any further on the $p_T$ of the candidate CMLLPs. The important variables 
in this analysis, $\beta$ and normalized ${\rm d} E/{\rm d} x$, are largely 
independent of the $p_T$ of the muon. The simulation has 
been tuned so that the $\beta$ and the ${\rm d} E/{\rm d} x$ are well modeled as described in Sec.~V A and B. 
\subsection{Event Selection}
\begin{figure*}
\begin{center}
\scalebox{0.4}{\includegraphics{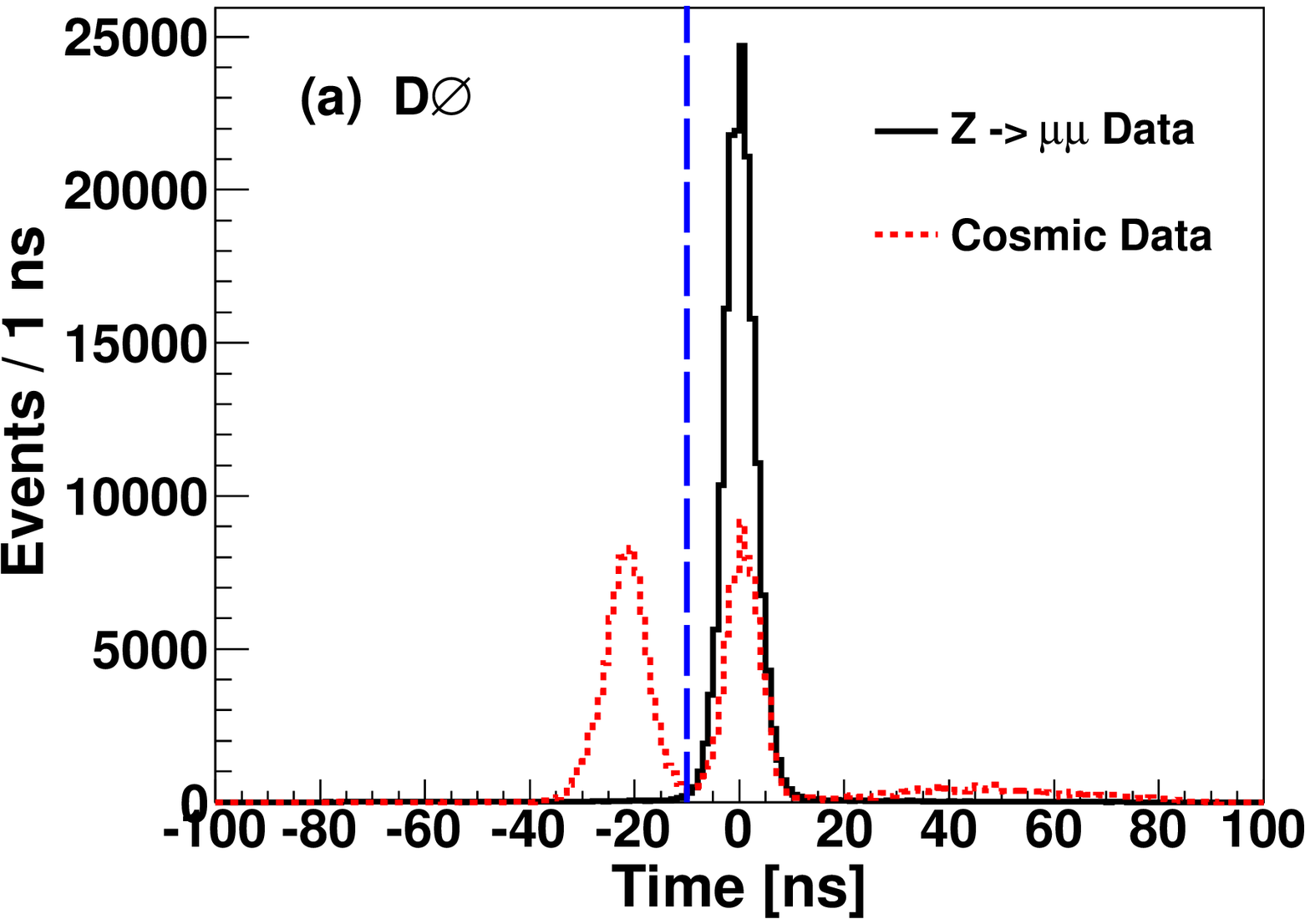}}
\scalebox{0.4}{\includegraphics{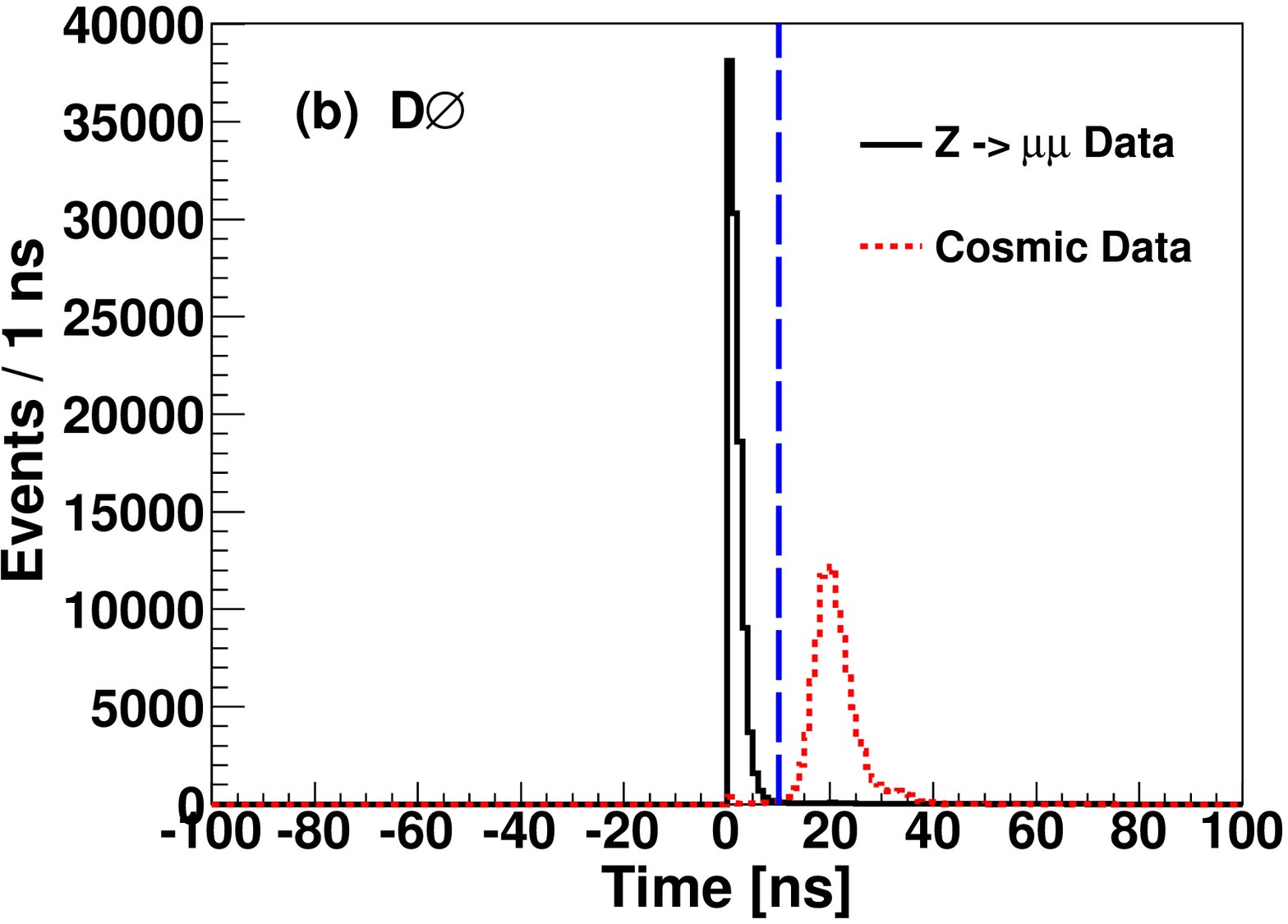}}
\caption{\label{fig:cosmic}(color online) (a) Difference between the A-layer and C-layer times 
for a single muon. There are two cosmic ray peaks for the two possible directions, 
away from or towards the $p\bar{p}$ collision vertex. (b) Absolute value of the difference 
between the A-layer times of the two muons in the event. The times shown in these plots 
are centered at zero for $\langle \beta \rangle = 1$ particles. This cosmic ray data was collected 
when there was no $p$ or $\bar{p}$ beam in the Tevatron collider. Selection requirement on the time difference is shown with a blue vertical line.}
\end{center}
\end{figure*}

Events containing a pair of CMLLP candidates are selected 
by requiring that they contain at least one muon with $p_T >$ 20 GeV and exactly two reconstructed, isolated muons
 of good quality with $|\eta| <$ 2.  The other selection criteria are:
\begin{itemize}
\item Number of SMT clusters, ($N_{SMT})\geq$ 3.
\item $\langle \beta \rangle <$ 1 for the two highest $p_T$ muons.
\item $\langle \beta \rangle$ asymmetry $<$ 0.35.
\item The highest $p_T$ muon has $p_T >$ 55 GeV and the second-highest $p_T$ muon has $p_T >$ 50 GeV.
\item Matching ${\chi}^2$ (between the track in the muon system and the central tracker) $<$ 100.
\end{itemize}
Muons originating from cosmic rays are vetoed by checking the difference in their arrival times at the scintillation counter layers A and C shown in Fig.~\ref{fig:cosmic}(a). The difference in the TOFs from the center of the detector to layer-A of two 
muons (Fig.~\ref{fig:cosmic}(b)) also provides discrimination of  
 muons produced in $p\bar{p}$ collisions from cosmic ray muons. 
The distance of closest approach (DCA) in the $r$-$\phi$ plane of each reconstructed muon track to the beam line, and the pseudo-acolinearity, $\Delta\alpha = |\Delta\phi + \Delta\theta- 2\pi|$, between the two muon tracks are also used to reject cosmic ray muons. These criteria are as follows:
\begin{itemize}
\item DCA in the $r$-$\phi$ plane of each muon $<$ 0.2 cm.
\item  Difference between A-layer and C-layer times of a muon $\ge$ $-10$ ns.
\item  Absolute value of the difference in A-layer times between the two muons $\le$ 10 ns.
\item  $\Delta\alpha \geq$ 0.05.
\end{itemize}
These selection criteria are optimized to produce the best expected limits on the masses of CMLLP candidates.
  
\begin{table*}
\begin{center}
\begin{ruledtabular}
\begin{tabular}{c c c}
Selection Criteria & Data events & Background Acceptance\\
    &      & $Z\rightarrow \mu\mu$ MC (\%) \\
\hline
Initial muon selection & 231487 & 49.6 \\
$N_{\mu}=$ 2 & 178204 & 15.8 \\
Trigger matching & 125662 & \\
Trigger probability & & 8.85\\
Cosmic veto & 106941 & 7.57 \\
L1 trigger gate (MC only), $N_{SMT}$ $\geq 3$ & 98195 & 6.94\\
$\langle \beta \rangle <1 $, $\langle \beta \rangle$ asymmetry $<0.35 $ & 34376 & 2.17\\
$p_T^{\mu1}>$55, $p_T^{\mu2}>$50 GeV & 709 & 0.03\\
Matching ${\chi}^2<$100 & 702& 0.03 \\
\end{tabular}
\end{ruledtabular}
\end{center}
\caption{\label{tab:pair_select}Selection efficiencies for data and background 
events before the application of BDT requirements for the search for a pair of CMLLPs
 with Run IIb data. Initially, events with at least one isolated muon of good
 quality and $p_T > 20$ GeV are required and a match between a muon and a central
 track is required based on ${\chi}^2$. }
\end{table*}
\begin{table*}
\begin{center}
\begin{ruledtabular}
\begin{tabular}{c c c c c c}
 M(stau lepton) in GeV \ &\ 100\ &\ 150\ &\ 200\ &\ 250\ &\ 300\ \\
Selection criteria & & & & &\\
\hline
Initial muon selection& 62.0 & 61.4 & 62.2 & 61.4 & 62.0 \\
$N_{\mu}=$ 2 & 25.0 & 25.2 & 25.6 & 25.6 & 25.5 \\
Trigger probability & 14.4 & 14.6 & 14.6 & 14.5 & 14.4 \\
Cosmic veto & 12.2 & 11.9 & 11.7 & 11.6 & 11.5 \\
L1 trigger gate, $N_{SMT}\geq 3$ & 10.5 & 10.2 & 9.93 & 9.50 & 8.84 \\
$\langle \beta \rangle <$1, $\langle \beta \rangle$ asymmetry$<$0.35 & 9.08 & 9.44 & 9.45 & 9.17 & 8.62 \\
$p_T^{\mu1}>$55, $p_T^{\mu2}>$ 50 GeV & 8.35 & 9.30 & 9.42 & 9.15 & 8.60 \\
Matching ${\chi}^2 <$ 100 & 8.24 & 9.15 & 9.29 & 9.00 & 8.45 \\
\end{tabular}
\end{ruledtabular}
\end{center}
\caption{\label{tab:eff_stau}Selection efficiencies (in \%) before the application of BDT requirements for a pair of stau leptons in simulated events. 
The initial muon selection requires at least one isolated muon of good quality, matched to a central track with $p_T >$ 20 GeV.}
\end{table*}
\begin{table*}
\begin{center}
\begin{ruledtabular}
\begin{tabular}{c c c c c c c c}
M(top squark) in GeV&\ 100\ &\ 150\ &\ 200\ &\ 250\ &\ 300\ &\ 350\ &\ 400\ \\
 Selection criteria & & & & & & &  \\
\hline
Initial muon selection & 50.6 & 54.9 & 57.9 & 59.6 & 58.6 & 58.3 & 57.7\\
$N_{\mu}=$ 2 & 18.1 & 20.7 & 22.3 & 22.9 & 21.9 & 21.9 & 21.2 \\
Trigger probability & 10.4 & 12.0 & 12.9 & 13.2 & 12.5 & 12.5 & 12.0 \\
Cosmic veto & 8.67 & 9.81 & 10.5 & 10.5 & 9.72 & 9.71 & 9.31 \\
L1 trigger gate, $N_{SMT} \geq$ 3 & 7.79 & 8.79 & 9.26 & 8.83 & 7.63 & 6.76 & 5.34 \\
$\langle \beta \rangle <$ 1, $\langle \beta \rangle $\ asymmetry $<0.35$ & 7.16 & 8.28 & 8.93 & 8.63 & 7.45 & 6.65 & 5.24 \\
$p_T^{\mu1}>$ 55, $p_T^{\mu2}>$ 50 GeV & 5.97 & 8.06 & 8.85 & 8.61 & 7.44 & 6.64 & 5.24 \\
Matching ${\chi}^2 <$100 & 5.90 & 7.96 & 8.73 & 8.49 & 7.34 & 6.53 & 5.18 \\
Charge survival efficiency (4\%) & 0.24 & 0.32 & 0.35 & 0.34 & 0.29 & 0.26 & 0.21 \\
\end{tabular}
\end{ruledtabular}
\end{center}
\caption{\label{tab:eff_stop} Selection efficiencies (in \%) before the application of BDT requirements for a pair of top squarks in simulated events. 
The initial muon selection requires at least one isolated muon of good quality, matched to a 
central track with $p_T > 20$ GeV. The top squark charge survival efficiency is 4\%.}
\end{table*}
%
%
\begin{table*}
\begin{center}
\begin{ruledtabular}
\begin{tabular}{c c c c c c}
M(gaugino-like chargino) in GeV  &\ 100\  &\ 150\ &\ 200\  &\ 250\  &\ 300\ \\
Selection criteria & & & & &\\
\hline 
 {Initial muon selection}  &  {46.8}  &  {49.8}  &  {50.3}  &  {49.1}  &  {49.1}\\
 {$N_{\mu}=$ 2}  &  {16.2}  &  {17.9}  &  {18.3}  &  {17.6}  &  {17.7}\\
Trigger probability & 9.32 & 10.4 & 10.6 & 10.3 & 10.3 \\
 {Cosmic veto}  &  {7.73}  &  {8.46}  &  {8.42}  &  {7.94}  &  {7.77}\\
L1 trigger gate, $N_{SMT}\geq 3$ & 6.96 & 7.39 & 7.11 & 6.09 & 5.29  \\
 {$\langle \beta \rangle <$1}, {$\langle \beta \rangle $ asymmetry$<$0.35}  &  {6.36}  &  {6.98}  &  {6.88}  &  {5.95}  &  {5.16}\\
 {$p_{T}^{\mu1}>$55, $p_{T}^{\mu2}>$50 GeV}  &  {5.23}  &  {6.67}  &  {6.77}  &  {5.92}  &  {5.14}\\
 {Matching $\chi^{2} <$100}  &  {5.16}  &  {6.59}  &  {6.67}  &  {5.85}  &  {5.05}\\
\end{tabular}
\end{ruledtabular}
\end{center}
\caption{\label{tab:eff_charginoG} Selection efficiencies (in \%) before the application of BDT requirements for a pair of gaugino-like charginos in simulated  events. The initial muon selection requires at 
least one isolated muon of good quality, matched to a central track with $p_T > 20$ GeV.}
\end{table*}
\begin{table*}
\begin{center}
\begin{ruledtabular}
\begin{tabular}{c c c c c c}
M(Higgsino-like chargino) in GeV &\ 100\  &\ 150\ &\ 200\ &\ 250\  &\ 300 \  \\
Selection criteria & & & & &\\
\hline
Initial muon selection& 47.5 & 50.5 & 51.4 & 51.1 & 49.0 \\
$N_{\mu}=$ 2 &17.1 &18.6 &19.3 &19.0 &18.0\\
Trigger probability & 9.78 & 10.8 & 11.2 & 11.1 & 11.4 \\
Cosmic veto& 8.15 &8.87 &8.88 &8.65 &8.04\\
L1 trigger gate, $N_{SMT}\geq 3$  & 7.29 & 8.13 & 7.56 & 6.83 & 5.69  \\
$\langle \beta \rangle < 1$, $\langle \beta \rangle $\ asymmetry$<$0.35 & 6.57 &7.70 &7.31 &6.68 &5.59\\
$p_T^{\mu1}>$ 55, $p_T^{\mu2}>$ 50 GeV & 5.40 &7.39 &7.23 &6.65 &5.56\\
Matching ${\chi}^2 <100$ & 5.31 & 7.31 & 7.14 & 6.56 & 5.49\\
\end{tabular}
\end{ruledtabular}
\end{center}
\caption{\label{tab:eff_charginoH} Selection efficiencies (in \%) before the application of BDT requirements for a pair of higgsino-like 
charginos in simulated events. The initial muon selection requires at least one isolated 
muon of good quality, matched to a central track with $p_T > 20$ GeV.}
\end{table*}
   The background sample is normalized to data events that pass the conditions described above and have invariant mass of the two 
highest $p_T$ muons in each event within 70 $<M_{\mu\mu}<$ 110 GeV. 
The contribution of potential signal in this region is negligible. The number of data and background events selected using the above 
conditions are listed in Table~\ref{tab:pair_select}, and the efficiencies 
for signal events for various CMLLP candidates are given in Tables~\ref{tab:eff_stau}--~\ref{tab:eff_charginoH}. 

\begin{figure*}
\begin{center}
\scalebox{0.4}{\includegraphics{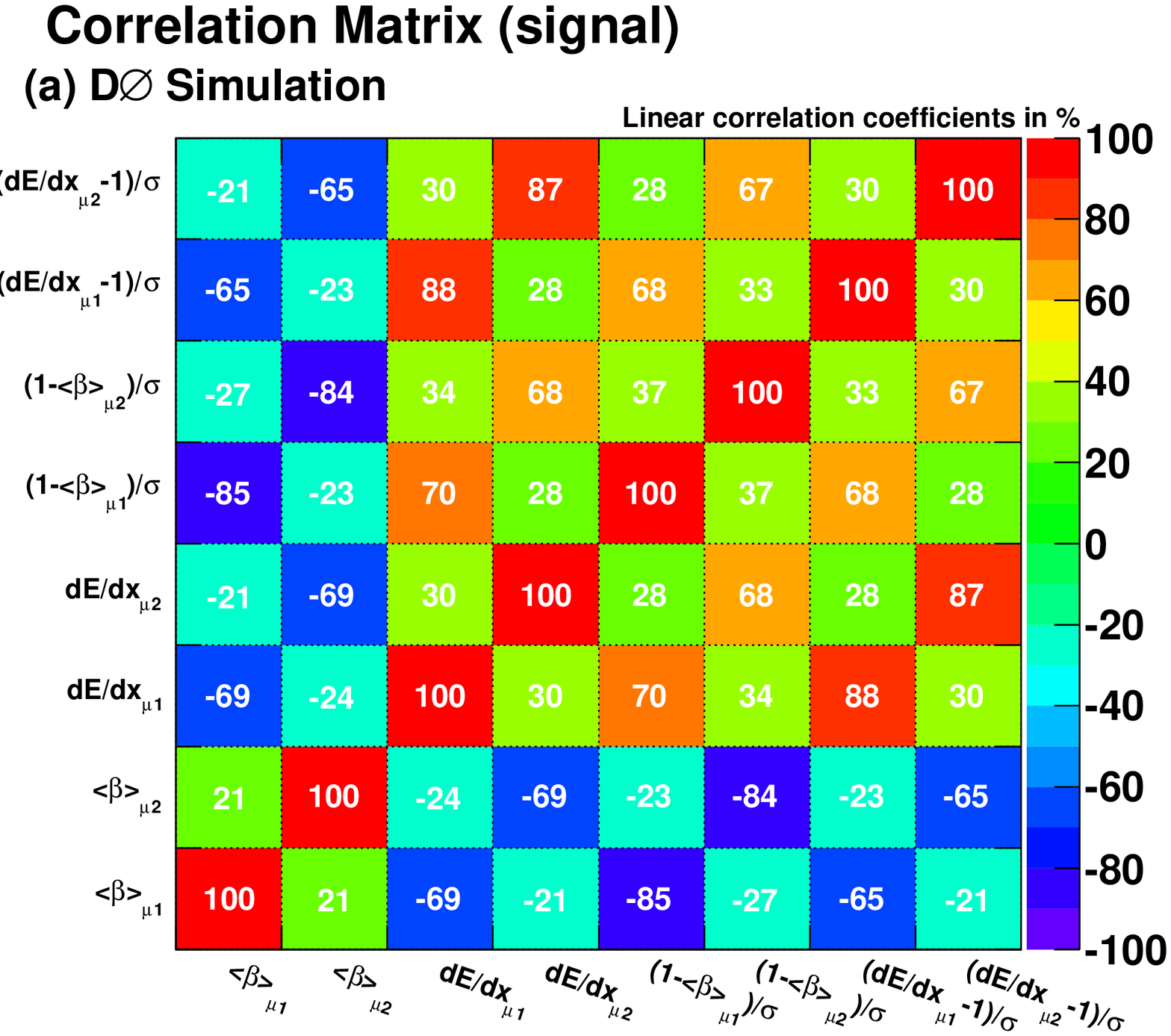}}
\scalebox{0.4}{\includegraphics{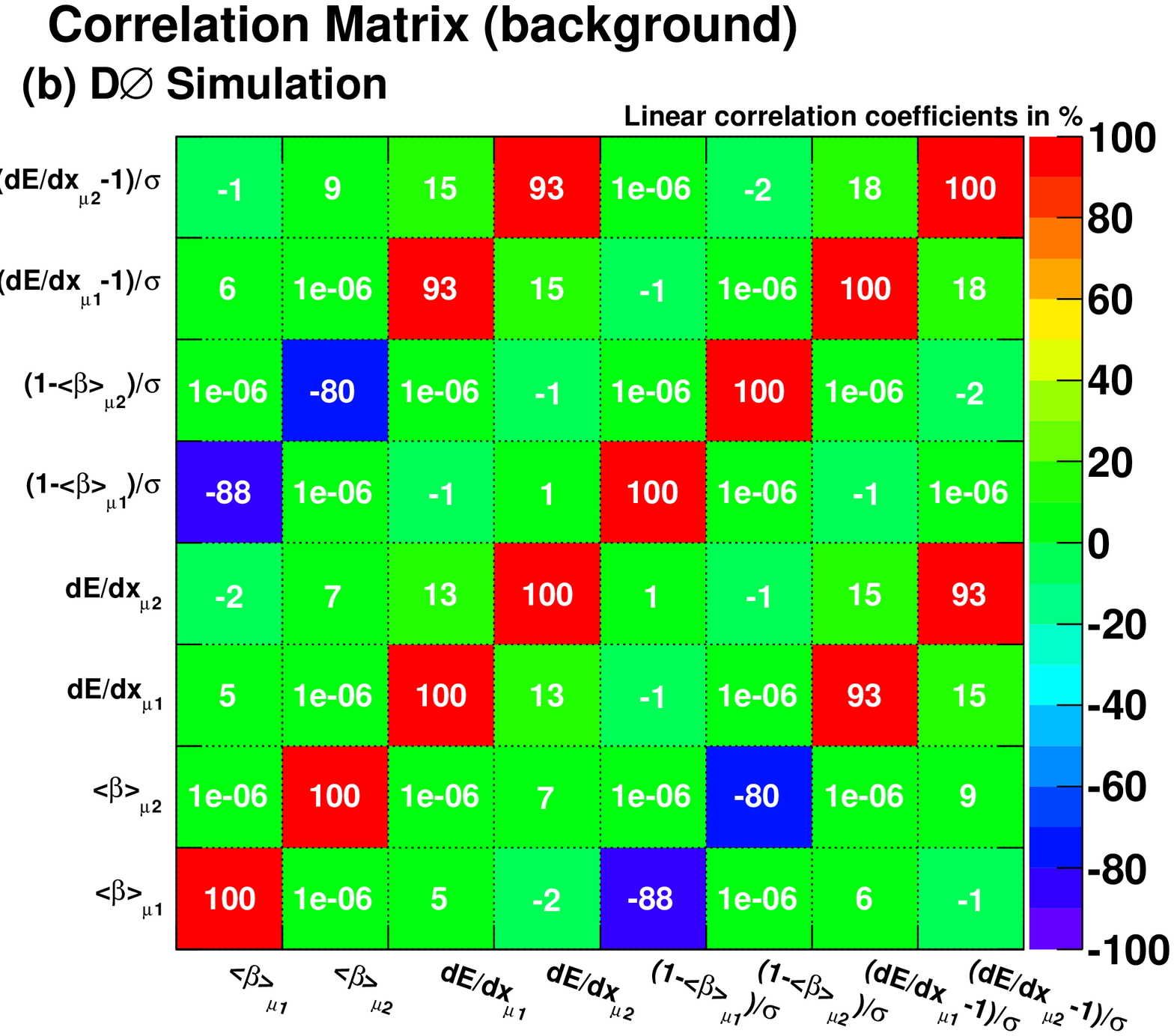}}
\caption{\label{fig:corr_matrix}(color online) Correlation matrix for different kinematic 
variables for (a) stau leptons of 300 GeV mass, and (b) background for the search of a pair of CMLLPs in the Run IIb data.}
\end{center}
\end{figure*}

\begin{figure*}
\begin{center}
\scalebox{0.32}{\includegraphics{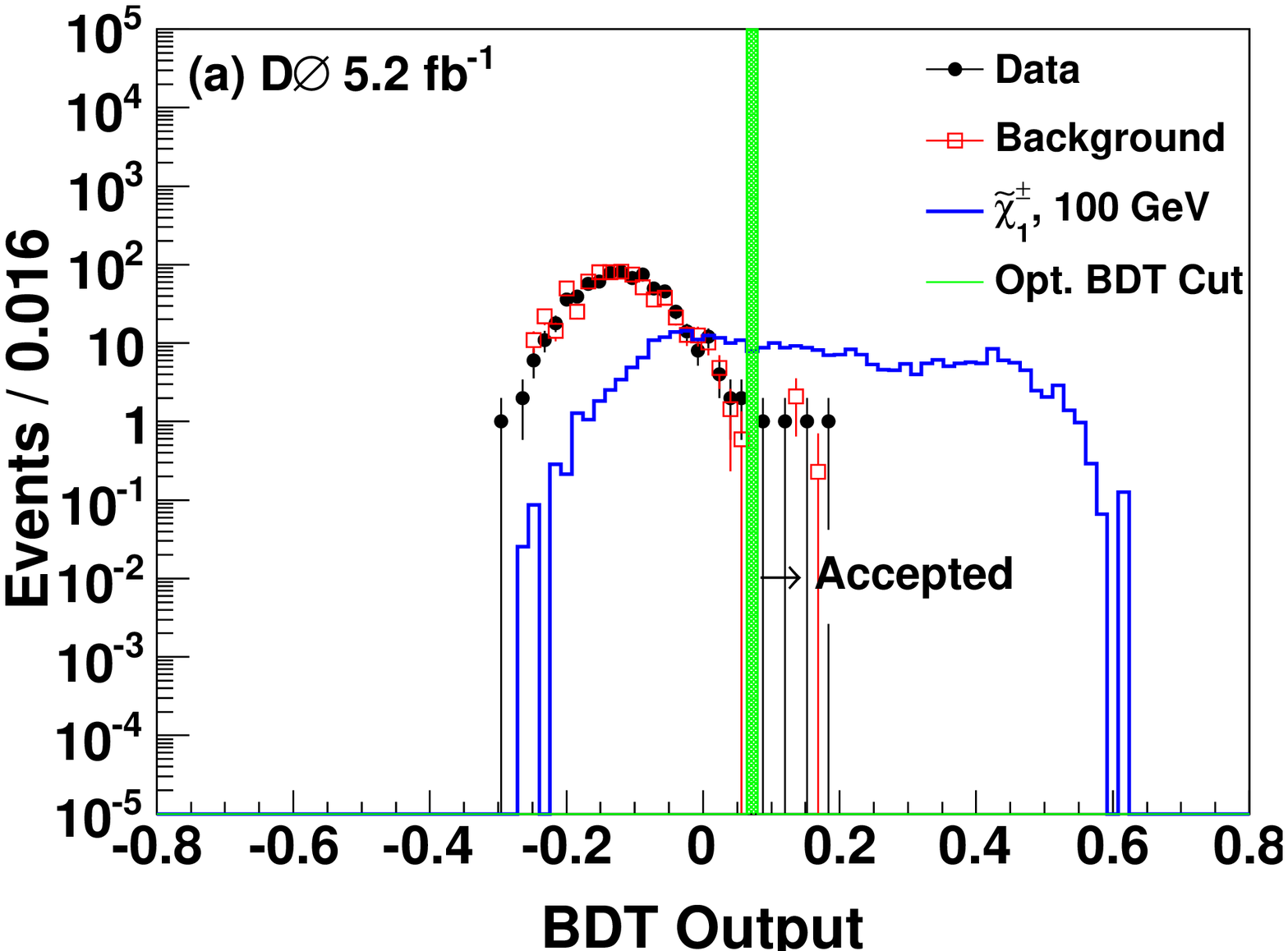}}
\scalebox{0.32}{\includegraphics{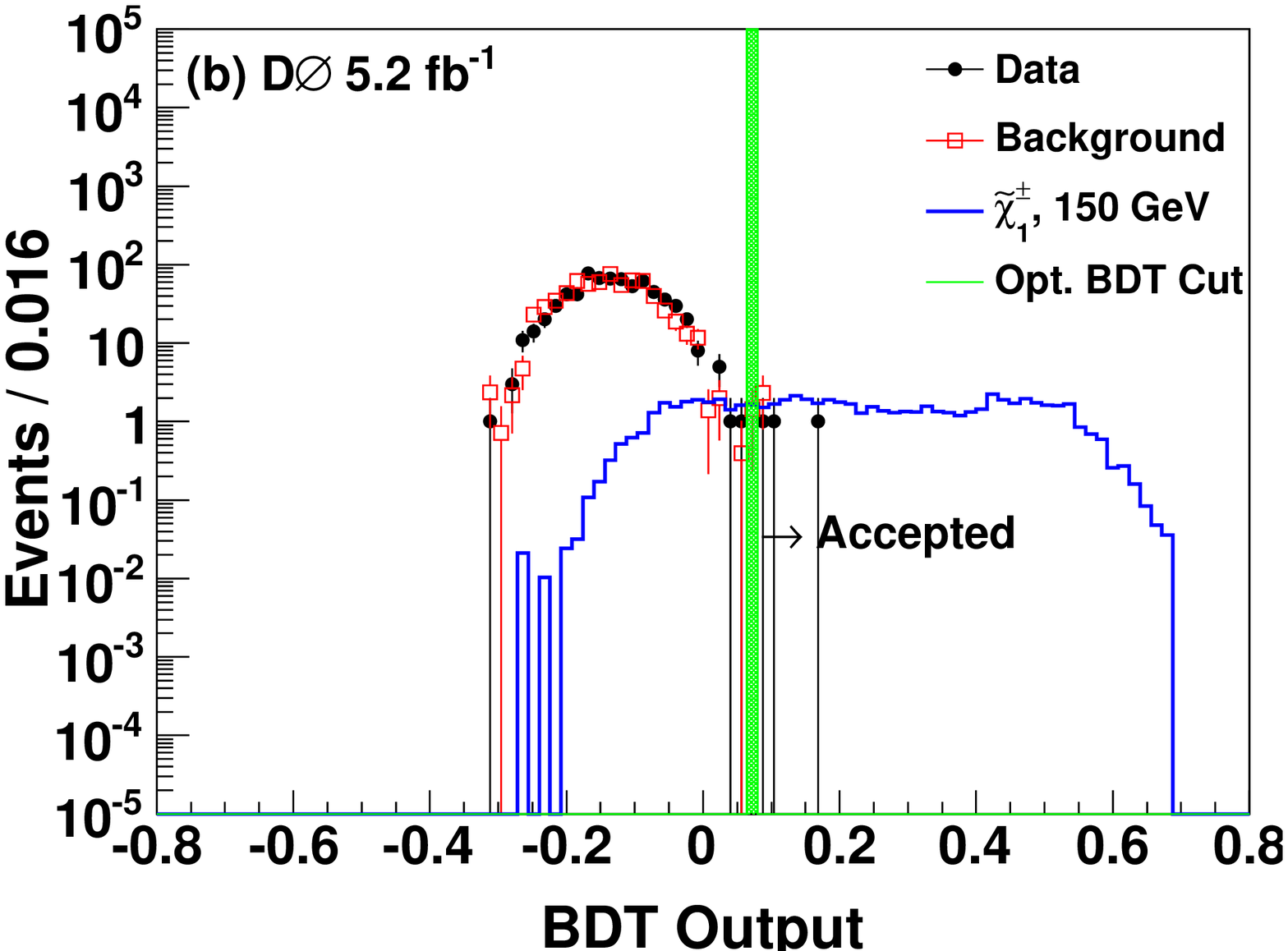}}
\scalebox{0.32}{\includegraphics{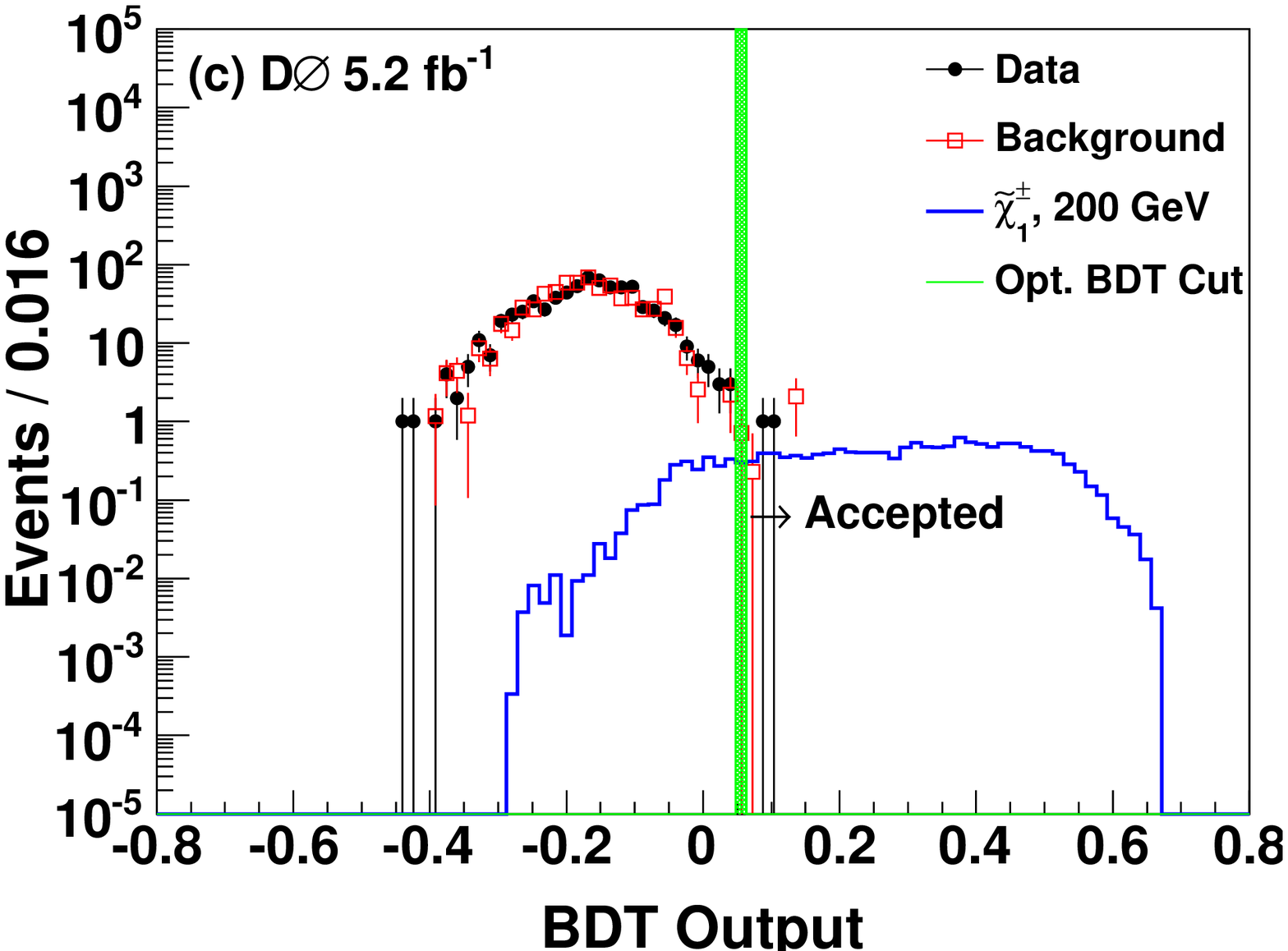}}
\scalebox{0.32}{\includegraphics{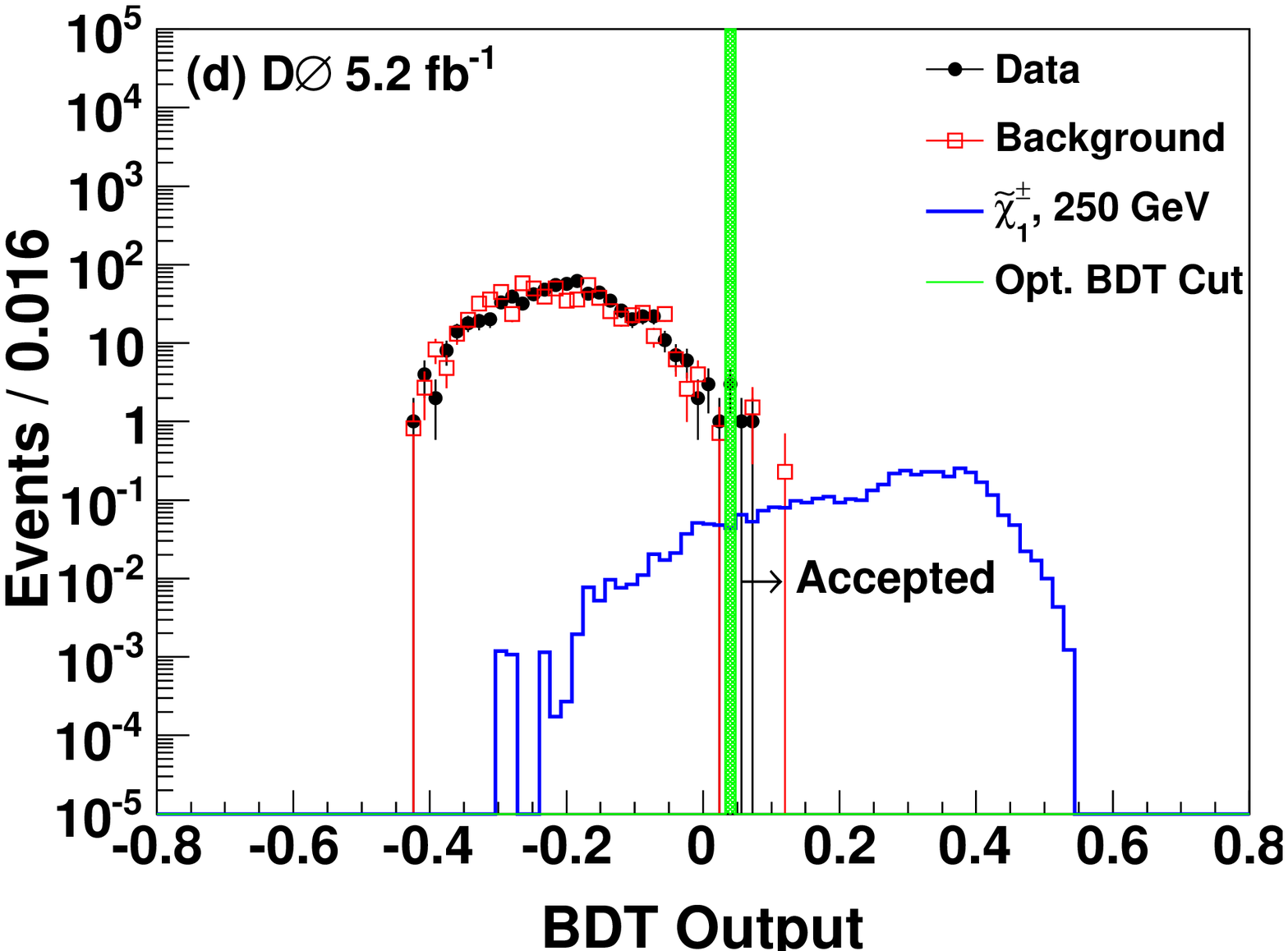}}
\scalebox{0.32}{\includegraphics{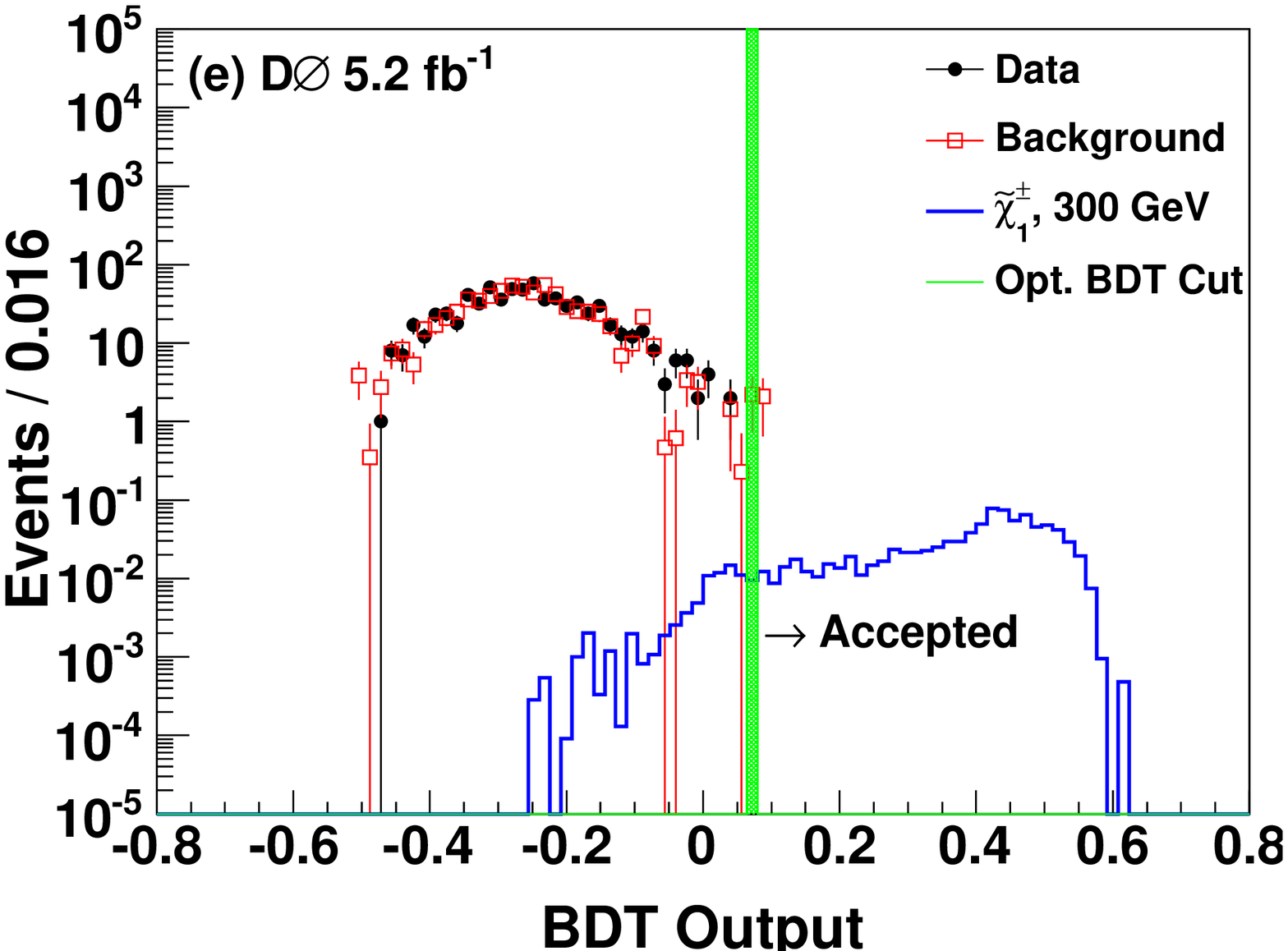}}
\caption{\label{fig:p20pair_BDT_charginoG}(color online) BDT-output distributions for simulated gaugino-like chargino masses of 
100--300 GeV in 50 GeV steps for the search for a CMLLP pair with the 
Run IIb data. Distributions are normalized to the expected number of events. Selection requirement on the BDT value is shown with a green vertical line. Note that a different BDT was constructed for each mass and the BDT
selection requirement optimized separately for each mass.}
\end{center}
\end{figure*}

    A Boosted Decision Tree (BDT), as implemented in Ref.~\cite{BDT}, is used to 
further discriminate signal events from background events. The BDT is trained 
using the expected signal and background distributions from MC, modified as described above. Half of the events are used to train the BDT while the remaining half are used to test the background model and signal response of the BDT.
The variables used as inputs to the BDT are $\beta$, speed significance, normalized ${\rm d} E/{\rm d} x$, 
and ${\rm d} E/{\rm d} x$ significance for the highest $p_T$ and the second-highest $p_T$ muons. 
The distributions for these variables are shown in Figs.~\ref{fig:speed_muons}, 
\ref{fig:speed_sig}, \ref{fig:dedx_muons}, and~\ref{fig:dedx_significance}, where Figs.~\ref{fig:speed_muons} and~\ref{fig:speed_sig} show the 
distributions without the $\langle \beta \rangle < 1$ requirement. The correlation matrices for 
these variables for a stau signal with a mass of 300 GeV and muons from $Z$ boson decays in 
MC are shown in Fig.~\ref{fig:corr_matrix}(a) and (b), respectively. These figures show that there are non-trivial correlations between the variables
and therefore MVA methods such as BDT, which will be able to take the correlations into account, are appropriate for this analysis. An example 
of the BDT output distribution is shown in Fig.~\ref{fig:p20pair_BDT_charginoG} for simulated gaugino-like charginos of different masses. The 
BDT outputs for simulated stau leptons, top squarks, and higgsino-like chargino signals are shown 
in Figs.~\ref{fig:p20pair_BDT_stau}-\ref{fig:p20pair_BDT_charginoH} in Appendix A. The final selection criteria on the BDT output are optimized to yield the best expected cross section limit in the no-signal hypothesis for each mass point, for each signal type.

%
%
\subsection{Systematic Uncertainties\label{sec:pair_syst}}

Systematic uncertainties are included in the estimation of cross section limits as follows. Each input parameter to the BDT distribution used to distinguish between signal and background is varied
 within its one standard deviation uncertainty and a new BDT distribution is produced for each variation for both signal and background models. The new BDT distributions are compared to the nominal ones and the average of the change  
in the occupancy of the BDT bins satisfying the selection requirement is taken as the systematic uncertainty 
due to that parameter. This procedure is applied to the sources of 
systematic uncertainties summarized below. These uncertainties are used in the limit calculation to model the effects of systematics in determining the limits.
 
\begin{itemize}

\item Uncertainty on the muon momentum scale in simulated samples is calculated 
by varying the $p_T$ of the muon track by $\pm$1 standard deviation as measured in data.

\item Time distributions for background and signal are modeled using $Z\rightarrow\mu\mu$ data events. We repeat the time smearing with time distributions of  
muons coming from $W\rightarrow\mu\nu$ decays in data and take the difference as the uncertainty.

\item To account for effects of the calibration of individual scintillation 
counters on the L1 trigger gate, we shift the trigger gate by $\pm$1 ns and calculate the resulting change in signal efficiency.

\item  Uncertainty due to the correction to the ${\rm d} E/{\rm d} x$ measurement of selected tracks to equalize the degrading response due to radiation damage in silicon is evaluated by varying the correction factor by its $\pm$1 standard deviation uncertainty. 

\item ${\rm d} E/{\rm d} x$ modeling uncertainty: the Gaussian smearing function applied to the ${\rm d} E/{\rm d} x$ distribution of muons in MC is derived
  separately using $Z\rightarrow\mu\mu$ and $W\rightarrow\mu\nu$ data events. The difference between the
  BDT distributions obtained with the two methods is taken to be the systematic uncertainty due to ${\rm d} E/{\rm d} x$ modeling.


\item Theoretical values of production cross sections (described in Sec.~IV) depend on the choice of parton distribution functions (PDF). Their effect is estimated by using the 40 CTEQ6.1M error PDFs~\cite{pdf1} for signal and background. 
The variations from each of the error PDF sets and from the renormalization and factorization scale uncertainties are added 
in quadrature.

\end{itemize}

The remaining systematic uncertainties, given below, are added in quadrature to  the uncertainties described above to obtain the total systematic uncertainties on signal acceptance and background prediction.
\begin{itemize}
\item The uncertainty in muon identification is a combination of the  uncertainties due to selection of a muon (1.2\%), central track 
reconstruction (1.4\%), and isolation of the muon (0.9\%). 

\item To determine the uncertainty in background normalization, the mass window for 
the control region is changed from 70 $<M_{\mu\mu}<$ 110 GeV to 60 $<M_{\mu\mu}<$ 120 GeV and 80 $<M_{\mu\mu}<$ 100 GeV. 

\item The MDTs have an asymmetric timing gate with a total 
length of 94 ns, which is not modeled in the MC. 
The signal from the earliest muon arrives at the 
MDTs within 74 ns of the beam crossing from $Z\rightarrow\mu\mu$ data. 
At most 1.2\% of the CMLLPs that we consider will be in the forward muon system 
and will arrive at the MDTs more than 20 ns after a prompt muon. 
We have therefore introduced an additional 1.2\% 
uncertainty on the signal acceptance.

\item Uncertainty due to the speed asymmetry correction (described in Sec.~\ref{tof_section}) is estimated by the change in signal acceptance with and without the correction. This uncertainty is found to vary between 1\% and 10\% depending on the masses and types of CMLLPs.  The value of this uncertainty is 3.6\% for the background sample. 
\end{itemize}
An uncertainty of 6.1\%~\cite{Lum uncert} on the integrated luminosity 
 is applied to the signal efficiency. Systematic uncertainties for signal and background samples are listed in the second columns of Tables \ref{tab:systematics_combination_sig} and \ref{tab:systematics_combination_bkgd}. The
 total systematic uncertainty for the background estimate is 18.2\% and that for the 
signal acceptance is 11.2--15.2\%, depending on the signal model and mass of the CMLLP. 
\subsection{Results}
\begin{figure*}
\begin{center}
\scalebox{0.4}{\includegraphics{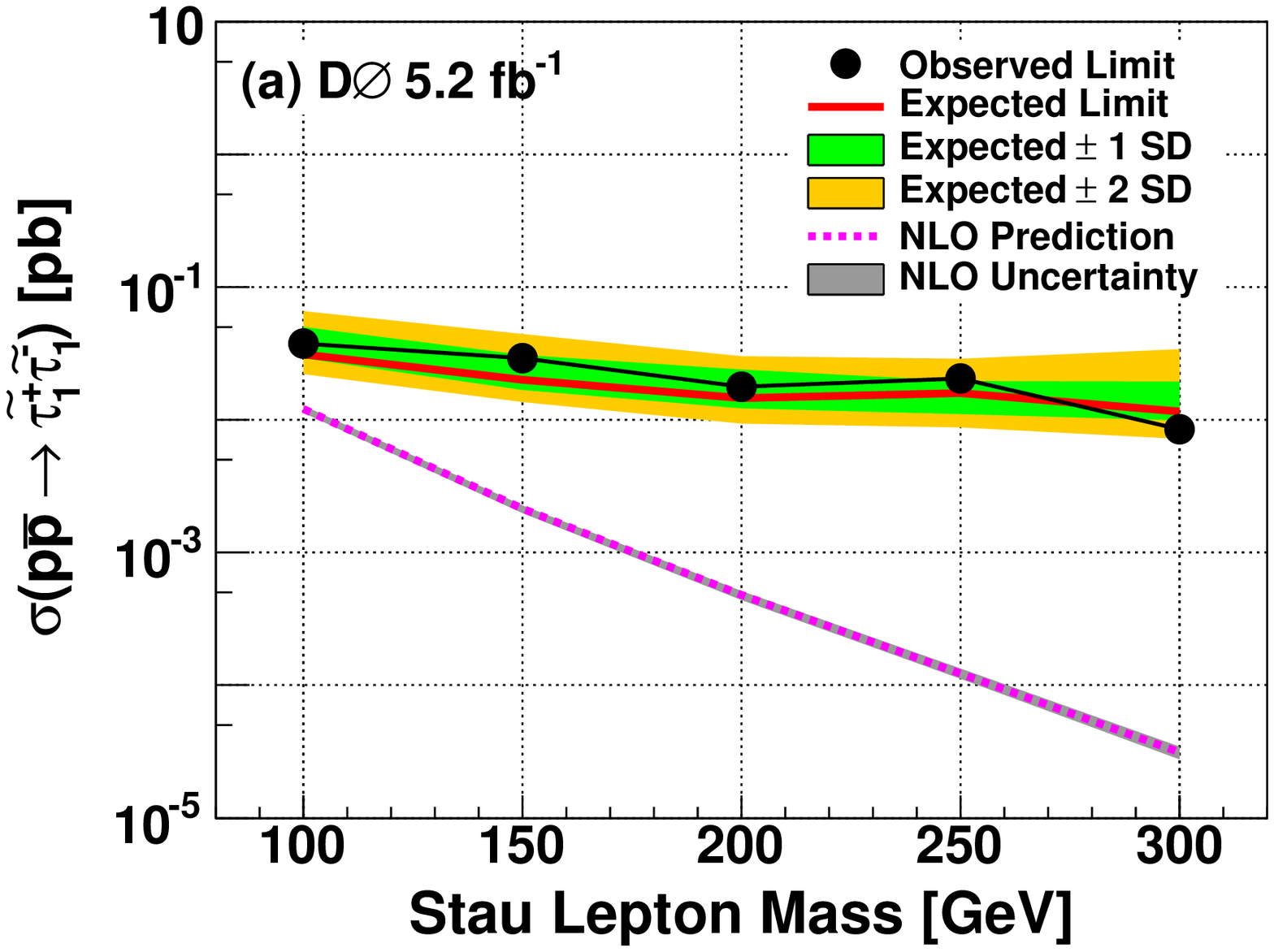}}
\scalebox{0.4}{\includegraphics{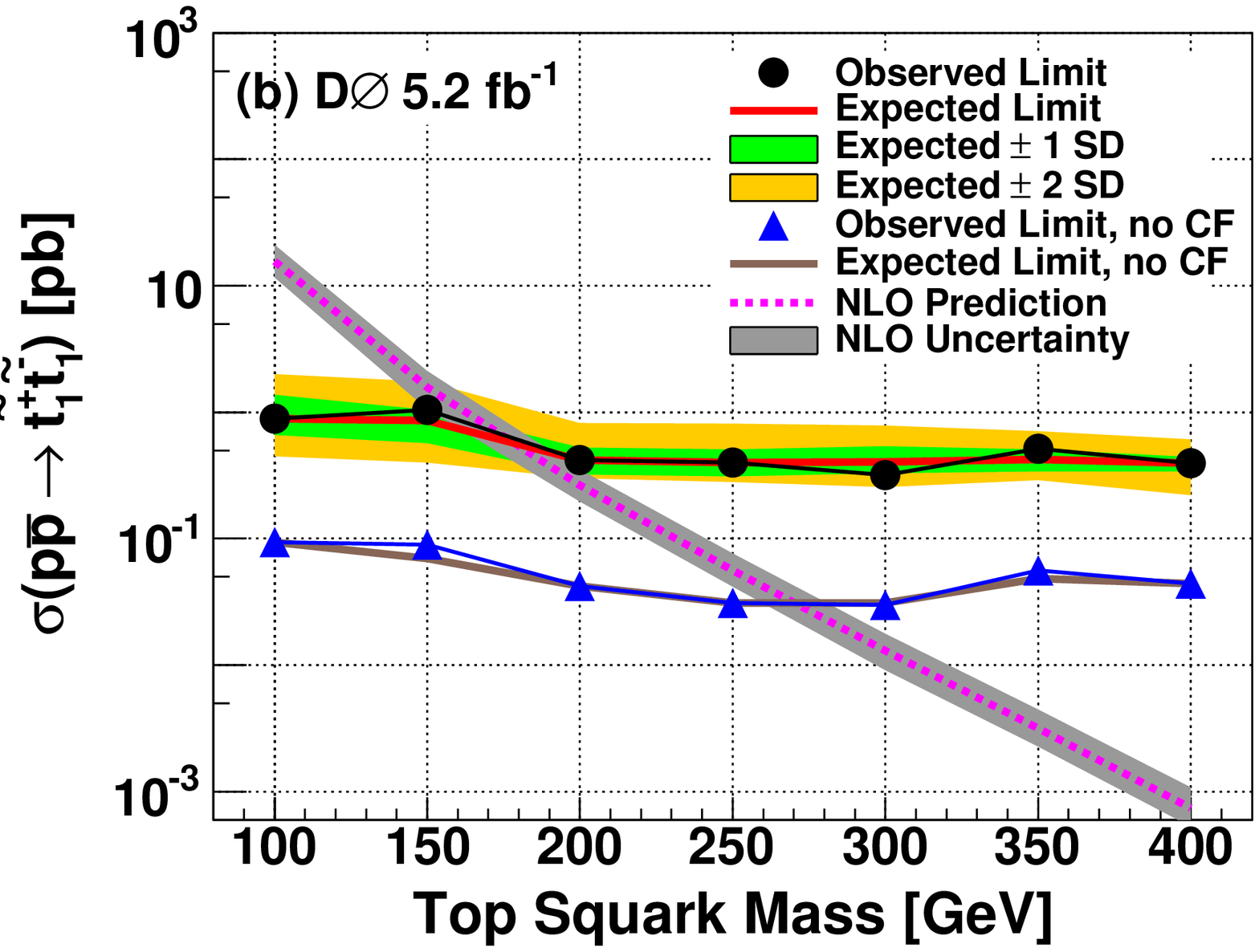}}
\scalebox{0.4}{\includegraphics{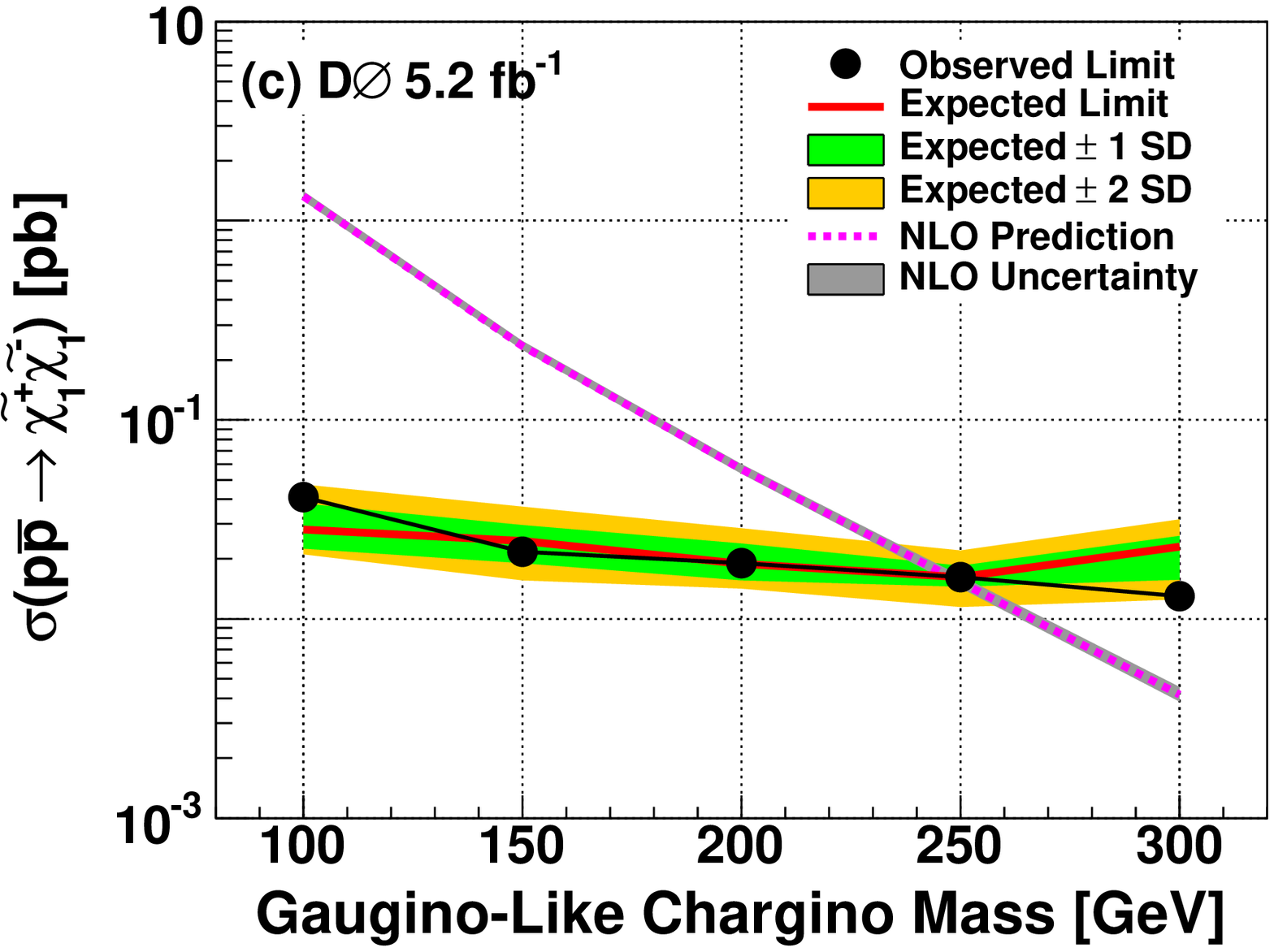}}
\scalebox{0.4}{\includegraphics{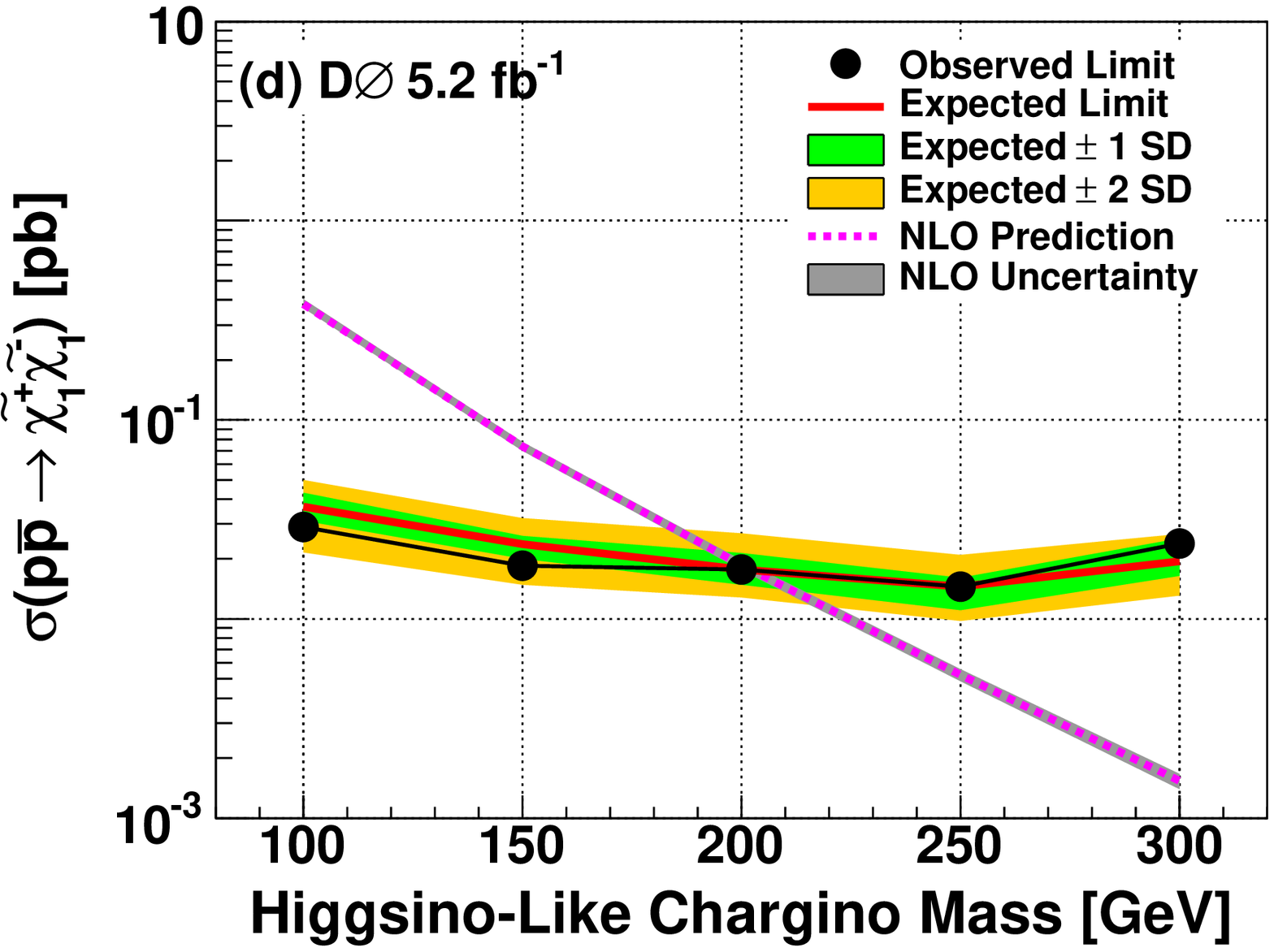}}
\caption{\label{fig:p20pair_limits}(color online) 95\% C.L. limits on production cross sections of a pair of stau leptons, top squarks, gaugino-like charginos, and higgsino-like charginos as a function of their masses from the search for a pair of CMLLPs with Run IIb data.  ``CF'' is the scenario without charge flipping. $\pm 1$ SD and $\pm 2$ SD are the 1 and 2 standard deviation bands respectively around the expected limit curves. }
\end{center}
\end{figure*}

The signal acceptance, the background prediction, and the observed number of events after the BDT requirement are shown in Table \ref{tab:p20pair_sig_accep_bck_pred}. These numbers are used as 
inputs to a modified Bayesian method~\cite{limit_calculator} for calculating the 
limits on production cross sections at 95\% C.L. Theoretical values of the production 
cross sections (described in Sec. IV) and observed and expected values of limits on the production cross 
sections of various CMLLPs are shown in Table \ref{tab:p20pair_limits} and Fig.~\ref{fig:p20pair_limits}. The lower mass limits that are obtained from the cross section limits are 189 GeV for top squarks, 250 GeV for gaugino-like charginos, and 204 GeV for higgsino-like charginos.
The limit on the mass of top squarks would increase to 280 GeV if we would include only the effects of their initial hadronization. If the intersection point of the $-1$ ($+1$) standard deviation band with the NLO cross section  is used, then the mass limits shift down (up) by $\sim$1 GeV for charginos and by $\sim$20 GeV for top squarks.

\begin{table*}
\begin{center} 
\begin{ruledtabular}
\begin{tabular}{c c c c}
{Mass (GeV)}  & {Signal Acceptance (\%)}  & {Predicted Background}  & {No. Observed Data events}\\
\hline
Stau lepton & & & \\
 { 100}  &  {$3.27\pm0.43$}  &  {$2.90\pm1.77$}  &  {$3$}\\
 { 150}  &  {$5.24\pm0.73$}  &  {$2.41\pm1.58$}  &  {$4$}\\
 { 200}  &  {$7.24\pm1.15$}  &  {$2.56\pm1.63$}  &  {$3$}\\
 { 250}  &  {$6.90\pm1.08$}  &  {$2.90\pm1.77$}  &  {$4$}\\
 { 300}  &  {$7.25\pm1.16$}  &  {$1.72\pm1.25$}  &  {$0$}\\
\hline
Top squark & & &  \\
 { 100}  &  {$0.12\pm0.01$}  &  {$2.41\pm1.58$}  &  {$2$}\\
 { 150}  &  {$0.12\pm0.01$}  &  {$2.41\pm1.58$}  &  {$3$}\\
 { 200}  &  {$0.24\pm0.04$}  &  {$2.71\pm1.63$}  &  {$3$}\\
 { 250}  &  {$0.26\pm0.04$}  &  {$2.41\pm1.58$}  &  {$2$}\\
 { 300}  &  {$0.25\pm0.04$}  &  {$2.41\pm1.58$}  &  {$1$}\\
 { 350}  &  {$0.25\pm0.04$}  &  {$2.41\pm1.58$}  &  {$3$}\\
 { 400}  &  {$0.20\pm0.04$}  &  {$1.72\pm1.25$}  &  {$1$}\\
\hline
Gaugino-like chargino & & &  \\
 { 100}  &  {$3.67\pm0.51$}  &  {$2.41\pm1.58$}  &  {$4$}\\
 { 150}  &  {$4.76\pm0.59$}  &  {$2.41\pm1.58$}  &  {$3$}\\
 { 200}  &  {$5.57\pm0.91$}  &  {$2.41\pm1.58$}  &  {$2$}\\
 { 250}  &  {$5.20\pm0.82$}  &  {$1.72\pm1.25$}  &  {$1$}\\
 { 300}  &  {$4.63\pm0.72$}  &  {$2.17\pm1.37$}  &  {$0$}\\
\hline
Higgsino-like chargino & & &  \tabularnewline
 { 100}  &  {$2.79\pm0.31$}  &  {$2.41\pm1.58$}  &  {$1$}\\
 { 150}  &  {$4.36\pm0.45$}  &  {$2.41\pm1.58$}  &  {$0$}\\
 { 200}  &  {$5.74\pm0.66$}  &  {$1.72\pm1.25$}  &  {$2$}\\
 { 250}  &  {$5.62\pm0.71$}  &  {$1.72\pm1.25$}  &  {$1$}\\
 { 300}  &  {$5.29\pm0.64$}  &  {$2.17\pm1.37$}  &  {$3$}\\
\end{tabular}
\end{ruledtabular}
\end{center}
\caption{\label{tab:p20pair_sig_accep_bck_pred} Signal acceptance, 
number of predicted background events, and number of observed events in the search for a pair of CMLLPs with Run IIb data after each BDT selection. The error is a sum in quadrature of statistical and systematic errors.}
\end{table*}
\begin{table*}
\begin{center}
\begin{ruledtabular}
\begin{tabular}{c c c c}
{Mass (GeV)}  & {NLO cross section (pb)}  & ${\sigma}_{95}^{\mathrm{obs}}$ (pb)  & ${\sigma}_{95}^{\mathrm{exp}}$ (pb) \\
\hline

Stau lepton & & &  \\

 { 100}  &  {$0.0120^{+0.0006}_{-0.0008}$}  &  {$0.038$}  
&  {$0.031^{+0.020}_{-0.002}$}\\
 { 150}  &  {$0.0021^{+0.0001}_{-0.0002}$}  &  {$0.029$}  
&  {$0.020^{+0.011}_{-0.010}$}\\
 { 200}  &  {$0.00050^{+0.00003}_{-0.00002}$}  &  {$0.018$}  
&  {$0.015^{+0.009}_{-0.003}$}\\
 { 250}  &  {$0.00010^{+0.00001}_{-0.00001}$}  &  {$0.020$}  
&  {$0.016^{+0.004}_{-0.004}$}\\
 { 300}  &  {$0.000030^{+0.000003}_{-0.000004}$}  &  {$0.009$}  
&  {$0.012^{+0.008}_{-0.002}$}\\
\hline
Top squark & & &   \tabularnewline
  { 100}  &  {$15.60^{+5.40}_{-4.00}$}  &  {$0.89$}  
&  {$0.89^{+0.45}_{-0.23}$}\\
 { 150}  &  {$1.58^{+0.53}_{-0.42}$}  &  {$1.05$}  
&  {$0.86^{+0.24}_{-0.29}$}\\
 { 200}  &  {$0.270^{+0.088}_{-0.068}$}  &  {$0.42$}  
&  {$0.42^{+0.11}_{-0.09}$}\\
 { 250}  &  {$0.056^{+0.020}_{-0.014}$}  &  {$0.40$}  
&  {$0.40^{+0.11}_{-0.09}$}\\
 { 300}  &  {$0.0130^{+0.0048}_{-0.0039}$}  &  {$0.32$}  
&  {$0.40^{+0.14}_{-0.09}$}\\
 { 350}  &  {$0.0032^{+0.0012}_{-0.0009}$}  &  {$0.52$}  
&  {$0.42^{+0.09}_{-0.08}$}\\
 { 400}  &  {$0.00075^{+0.0003}_{-0.0002}$}  &  {$0.40$}  
&  {$0.40^{+0.05}_{-0.06}$}\\
\hline
Gaugino-like chargino  &  &  &  \\
 { 100}  &  {$1.33^{+0.08}_{-0.07}$}  &  {$0.041$}  
&  {$0.028^{+0.010}_{-0.002}$}\\
 { 150}  &  {$0.240^{+0.014}_{-0.010}$}  &  {$0.022$}  
&  {$0.025^{+0.005}_{-0.005}$}\\
 { 200}  &  {$0.0570^{+0.0034}_{-0.0030}$}  &  {$0.019$}  
&  {$0.019^{+0.005}_{-0.003}$}\\
 { 250}  &  {$0.0150^{+0.0011}_{-0.0010}$}  &  {$0.162$}  
&  {$0.016^{+0.002}_{-0.002}$}\\
 { 300}  &  {$0.0042^{+0.0004}_{-0.0003}$}  &  {$0.013$}  
&  {$0.023^{+0.003}_{-0.007}$}\\
\hline
Higgsino-like chargino  & & & \\

  { 100}  &  {$0.380^{+0.023}_{-0.017}$}  &  {$0.029$}  
&  {$0.037^{+0.007}_{-0.005}$}\\
 { 150}  &  {$0.0740^{+0.0040}_{-0.0038}$}  &  {$0.019$}  
&  {$0.024^{+0.003}_{-0.003}$}\\
 { 200}  &  {$0.0190^{+0.0012}_{-0.0010}$}  &  {$0.018$}  
&  {$0.018^{+0.004}_{-0.003}$}\\
 { 250}  &  {$0.0053^{+0.0004}_{-0.0004}$}  &  {$0.015$}  
&  {$0.015^{+0.002}_{-0.005}$}\\
 { 300}  &  {$0.0015^{+0.0001}_{-0.0001}$}  &  {$0.024$}  
&  {$0.020^{+0.006}_{-0.003}$}\\

\end{tabular}
\end{ruledtabular}
\end{center}
\caption{\label{tab:p20pair_limits} NLO cross section and 95\% C.L. limits (${\sigma}_{95}$) for the search for a pair of CMLLPs in the Run IIb data. The top squark charge survival efficiency is 4\% (see text).}
\end{table*}
%
\section{search for events with a single CMLLP}
The following section describes the search for a single CMLLP 
in 5.2 ${\mathrm{fb}}^{-1}$ of integrated luminosity. More details can be found in Ref.~\cite{D0 CMSP PRL Run IIb}.
\subsection{Background Sample}
\begin{figure}
\begin{center}
\scalebox{0.45}{\includegraphics{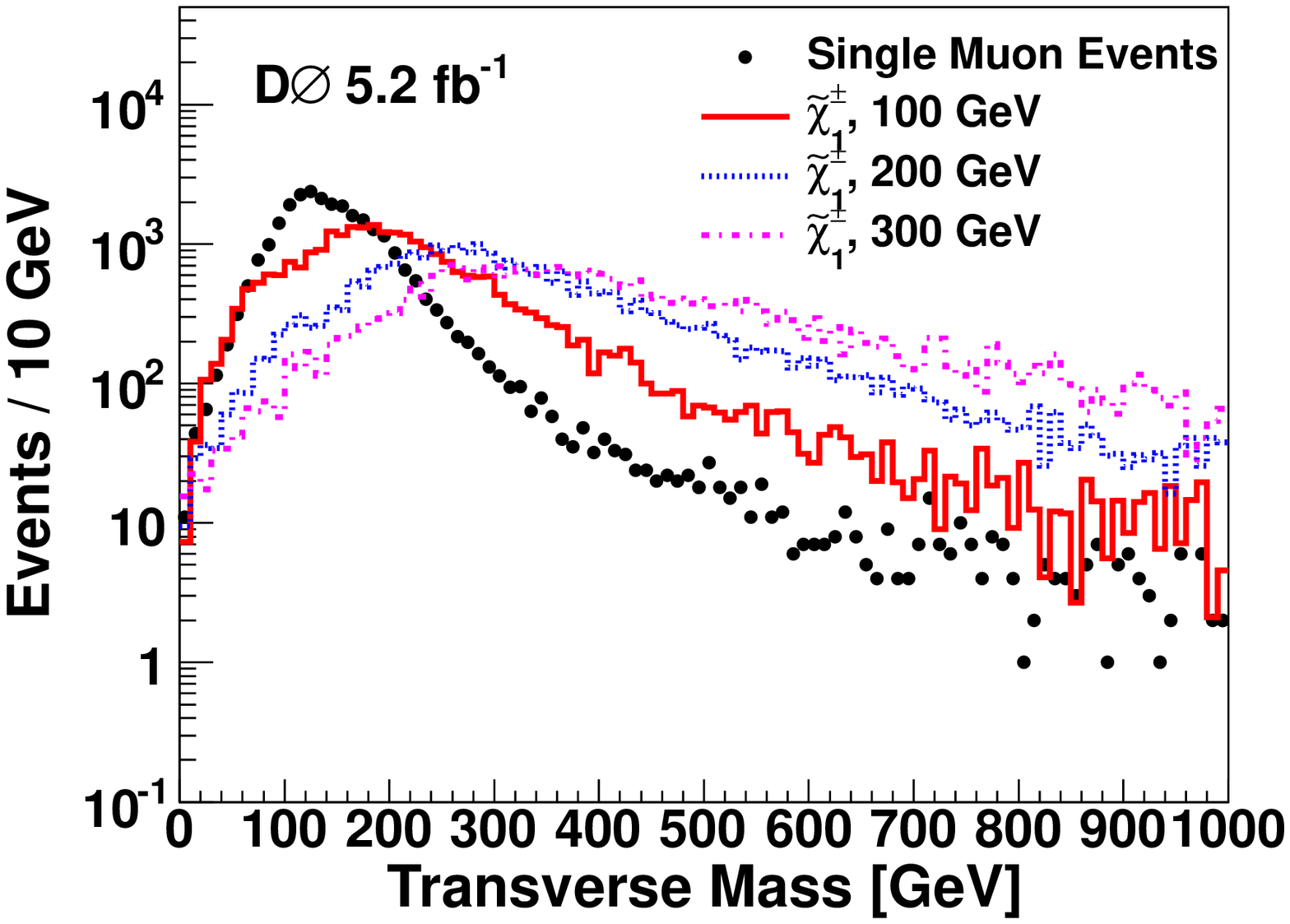}}
\caption{\label{fig:transverse_mass}(color online) The transverse mass $M_T$ for single 
muon events and  higgsino-like chargino events with chargino masses 
100 GeV, 200 GeV, and 300 GeV. The single muon event sample satisfies all 
of the selection criteria described in Table \ref{tab:p20single_cutflow_data}
except the $M_T$ cut, which separates the data and background samples.
All the selection criteria described in Table \ref{tab:p20single_cutflow_charginoH} 
have been applied to the charginos except the $M_T >$ 200 GeV cut. 
The distributions have been normalized to the same number of events.}
\end{center}
\end{figure}

The dominant background in the single CMLLP search is muons from the decays 
of $W$ bosons, which is modeled with data. To define independent data 
and background samples, we select events using the transverse mass, $M_T$, of the $W$ boson given by
\begin{equation}
M_{T}=\sqrt{({p_{T}}^{\mu}+{\not}E_{T})^{2}-(p_{x}+{\not}E_{x})^{2}-(p_{y}+{\not}E_{y})^{2}}.
\end{equation}
Here ${p_T}^{\mu}$ is the transverse momentum of the muon and ${\not}E_T$ is the total unbalanced momentum transverse to the beamline as measured in the calorimeter and corrected for the muons. Events with $M_T \leq$ 200 GeV and $\langle \beta \rangle <1$ are selected for the background sample and events with $M_T >$ 200 GeV and $\langle \beta \rangle <1$ constitute the search sample. Figure~\ref{fig:transverse_mass} shows a distribution of $M_T$ for single muon events from data and for higgsino-like chargino MC events. 
\subsection{Event Selection}
The criteria to select events with one or more CMLLPs are 
similar to those used in the search for a pair of CMLLPs (Sec.~VI B). 
Events satisfying a suite of single muon triggers are required to 
contain an isolated muon of good quality (described in Sec.~VI B) 
within $|\eta| <$ 1.6. The muon must originate at the $p\bar{p}$ interaction vertex and must 
satisfy the following criteria:   
\begin{itemize}
\item $p_T >$ 60 GeV.
\item speed ${\chi}^2/{\mathrm{dof}} <$ 2.
\end{itemize}
If there is more than one such muon in the event, 
only the highest $p_T$ muon is considered as the CMLLP candidate. 
To ensure that selected muons do not originate from cosmic rays, we require the DCA in the $r$-$\phi$ plane of the selected muon track to the beam line to be less than 0.2 cm and the difference between A-layer and C-layer times to be $\ge-$10 ns.

 If there is a second muon passing all the selection criteria, conditions to remove the cosmic ray events are the same as in the search for CMLLP pairs (described in Sec.~VI B).
The selection criteria and the corresponding efficiencies for number of events in data 
and the CMLLP signals are given in Tables~\ref{tab:p20single_cutflow_data} --~\ref{tab:p20single_cutflow_charginoH}. 

%
\begin{table*}
\begin{center}
\begin{ruledtabular}
\begin{tabular}{c c c}
Selection Criteria & \multicolumn{2}{c}{Number of Events}\\
\hline
Initial Muon Selection & \multicolumn{2}{c}{345173}\\

Isolated muon & \multicolumn{2}{c}{141008}\\

Trigger matching  & \multicolumn{2}{c}{79983} \\

Cosmic veto & \multicolumn{2}{c}{79096}\\

$N_{SMT}\geq$ 3  & \multicolumn{2}{c}{57532}\\

$p_{T} >$ 60 GeV & \multicolumn{2}{c}{56466}\\


$\langle \beta \rangle $ $<$ 1, Speed $\chi^{2}/{dof} <$ 2 & \multicolumn{2}{c}{27876}\\

Matching $\chi^{2}\leq$ 100 & \multicolumn{2}{c}{27742}\\

\cline{2-3}

& Background ($M_{T} \leq 200$ GeV) & Data ($M_{T}>200$ GeV)\\

\cline{2-3}

 & 22368 & 5374\\
\end{tabular}
\end{ruledtabular}
\end{center}
\caption{\label{tab:p20single_cutflow_data}Selection efficiencies for 
data in the search for single CMLLPs before applying the BDT selection
requirements. Initially, events with at 
least one isolated muon of good quality and $p_T > 20$ GeV are required and a match
 between a muon and a central track is required based on ${\chi}^2$.}
\end{table*}
\begin{table*}
\begin{center}
\begin{ruledtabular}
\begin{tabular}{c c c c c c}
M(stau lepton) in GeV
&\ 100\ &\ 150\ &\ 200\ &\ 250\ &\ 300\ \\
Selection criteria&&&&&\\
\hline
Initial muon selection& 68.9 & 67.9 & 68.9 & 68.1 & 68.5\\

Isolated muon & 59.1 & 59.5 & 60.6 & 60.2 & 60.8\\

Trigger probability & 31.5 & 31.2 & 31.4 & 31.0 & 31.0 \\

Cosmic veto & 27.9 & 26.2 & 25.3 & 27.3 & 26.1\\

L1 trigger gate, $N_{SMT}\geq$ 3& 22.2 & 19.6 & 18.0 & 18.0 & 15.6\\


$p_{T} >$ 60 GeV & 20.3 & 19.4 & 18.0 & 18.0 & 15.6\\


$\langle \beta \rangle$ $<$ 1, Speed $\chi^{2}/{\mathrm{dof}} <$ 2 & 16.9 & 16.5 & 15.6 & 15.7 & 13.7\\

Matching $\chi^{2} \leq$ 100 & 16.8 & 16.4 & 15.5 & 15.6 & 13.6\\

$M_{T} >$ 200 GeV & 14.4 & 15.6 & 15.2 & 15.4 & 13.6\\
\end{tabular}
\end{ruledtabular}
\end{center}
\caption{\label{tab:p20single_cutflow_stau}Selection efficiencies (in \%) for a 
single stau lepton in MC events. Initially, events with at least one isolated 
muon of good quality and $p_T > 20$ GeV are required and a match between a muon
 and a central track is required based on ${\chi}^2$.}
\end{table*}
\begin{table*}
\begin{center}
\begin{ruledtabular}
\begin{tabular}{c c c c c c c c}
M(top squark) in GeV
&\ 100\ &\ 150\ &\ 200\ &\ 250\ &\ 300\ &\ 350\ &\ 400\ \\
Selection criteria &&&&&&&\\
\hline
Initial muon selection& 56.9 & 62.3 & 65.6 & 56.6 & 67.2 & 66.9 & 66.5\\
Isolated muon & 46.0 & 52.0 & 55.8 & 47.0 & 57.3 & 57.3 & 56.6\\

Trigger probability & 24.9 & 27.6 & 29.3 & 25.3 & 29.4 & 29.3 & 28.8 \\

Cosmic veto & 21.1 & 23.3 & 26.2 & 24.3 & 24.6 & 23.3 & 24.6\\

L1 trigger gate, $N_{SMT}\geq$ 3 & 15.7 & 16.8 & 18.0 & 15.3 & 14.3 & 11.8 & 10.6\\


$p_{T} >$ 60 GeV & 13.0 & 16.3 & 18.0 & 15.3 & 14.2 & 11.8 & 10.6\\


$\langle \beta \rangle$ $<$ 1, Speed $\chi^{2}/{dof} <$ 2 & 11.1 & 14.3 & 15.7 & 13.5 & 12.5 & 10.4 & 9.3\\

Matching $\chi^{2} \leq$ 100 & 11.0 & 14.1 & 15.6 & 13.4 & 12.4 & 10.4 & 9.3\\

$M_{T} >$ 200 GeV& 8.6 & 13.2 & 15.2 & 13.3 & 12.4 & 10.3 & 9.3\\


Charge survival efficiency (38\%) & 3.3 & 5.0 & 5.8 & 5.0 & 4.7 & 3.9 & 3.5 \\
\end{tabular}
\end{ruledtabular}
\end{center}
\caption{\label{tab:p20single_cutflow_stop}Selection efficiencies (in \%) for a single top squark in MC events. 
Initially, events with at least one isolated muon of
good quality and $p_T > 20$ GeV are required and a match between a muon and a
 central track is required based on ${\chi}^2$.}
\end{table*}
\begin{table*}
\begin{center}
\begin{ruledtabular}
\begin{tabular}{c c c c c c}
M(gaugino-like chargino) in GeV
&\ 100\  &\ 150\ &\ 200\ &\ 250\ &\ 300\ \\
Selection criteria & & & & & \\
\hline

Initial muon selection& 51.7 & 54.7 & 55.4 & 54.3 & 53.8\\

Isolated muon & 41.3 & 44.6 & 45.7 & 44.7 & 44.9\\

Trigger probability & 22.7 & 24.3 & 24.7 & 24.2 & 24.1 \\

Cosmic veto & 21.0 & 20.8 & 21.0 & 19.9 & 19.7\\

L1 trigger gate, $N_{SMT}\geq$ 3 & 15.5 & 13.8 & 12.4 & 10.0 & 8.4 \\


$p_{T} >$ 60 GeV & 12.2 & 13.1 & 12.3 & 10.0 & 8.4\\


$\langle \beta \rangle$ $<$ 1, Speed $\chi^{2}/{dof} <$ 2 & 10.4 & 11.7 & 11.0 & 8.9 & 7.4\\

Matching $\chi^{2}\leq$ 100 & 10.3 & 11.6 & 10.9 & 8.9 & 7.4\\

$M_{T} >$ 200 GeV & 7.8 & 10.6 & 10.6 & 8.8 & 7.4\\

\end{tabular}
\end{ruledtabular}
\end{center}
\caption{\label{tab:p20single_cutflow_charginoG}Selection efficiencies (in \%) for a single gaugino-like chargino in simulated events. Initially, events with at least one
isolated muon of good quality and $p_T > 20$ GeV are required and a match between
 a muon and a 
central track is required based on ${\chi}^2$.}
\end{table*}
\begin{table*}
\begin{center}
\begin{ruledtabular}
\begin{tabular}{c c c c c c}
M(higgsino-like chargino) in GeV
&\ 100\ &\ 150\ &\ 200\ &\ 250\  &\ 300\ \\
Selection criteria &&&&&\\
\hline
Initial muon selection & 52.4 & 55.6 & 56.6 & 56.1 & 53.9\\

Isolated muon & 42.2 & 45.8 & 47.0 & 47.0 & 45.3\\

Trigger probability & 23.1 & 25.0 & 25.3 & 25.3 & 24.2\\ 

Cosmic veto & 21.1 & 23.4 & 23.7 & 21.3 & 20.1\\

L1 trigger gate, $N_{SMT}\geq$ 3  & 15.9 & 15.9 & 14.5 & 11.4 & 9.1\\


$p_{T} >$ 60 GeV & 12.7 & 15.2 & 14.4 & 11.3 & 9.1\\


$\langle \beta \rangle$ $<$ 1, Speed $\chi^{2}/{dof} <$ 2 & 10.8 & 13.4 & 12.7 & 10.1 & 8.2\\

Matching $\chi^{2}\leq$100 & 10.8 & 13.4 & 12.7 & 10.1 & 8.1\\

$M_{T} >$ 200 GeV & 8.3 & 12.2 & 12.4 & 10.0 & 8.1\\

\end{tabular}
\end{ruledtabular}
\end{center}
\caption{\label{tab:p20single_cutflow_charginoH}Selection efficiencies (in \%) for a 
single higgsino-like chargino in MC events. Initially, events with at least 
one isolated muon of good quality and $p_T > 20$ GeV are required and a match 
between a muon and 
a central track is required based on ${\chi}^2$.}
\end{table*}
%

Since the background is modeled using data, it 
is necessary to normalize it to data in a signal-free region. 
Events containing muons with measured $\langle \beta \rangle \geq$ 1 are used to define 
a signal-free region. Signal-free control events contain muons with $\langle \beta \rangle \geq$ 1 and $M_T \leq$ 200 GeV, and 
signal-free data contain muons with $\langle \beta \rangle >$ 1 and 
$M_T >$ 200 GeV. If the number of background events is $N_B$, the number of 
signal-free control events is $N_{SFC}$, and the number of signal-free data 
events is $N_{SFD}$, then the number of normalized background events, $N_{NB}$, 
can be expressed as
\begin{equation} 
N_{NB} = N_B \frac{N_{SFD}}{N_{SFC}}.
\end{equation}

\begin{figure*}
\begin{center}
\scalebox{0.35}{\includegraphics{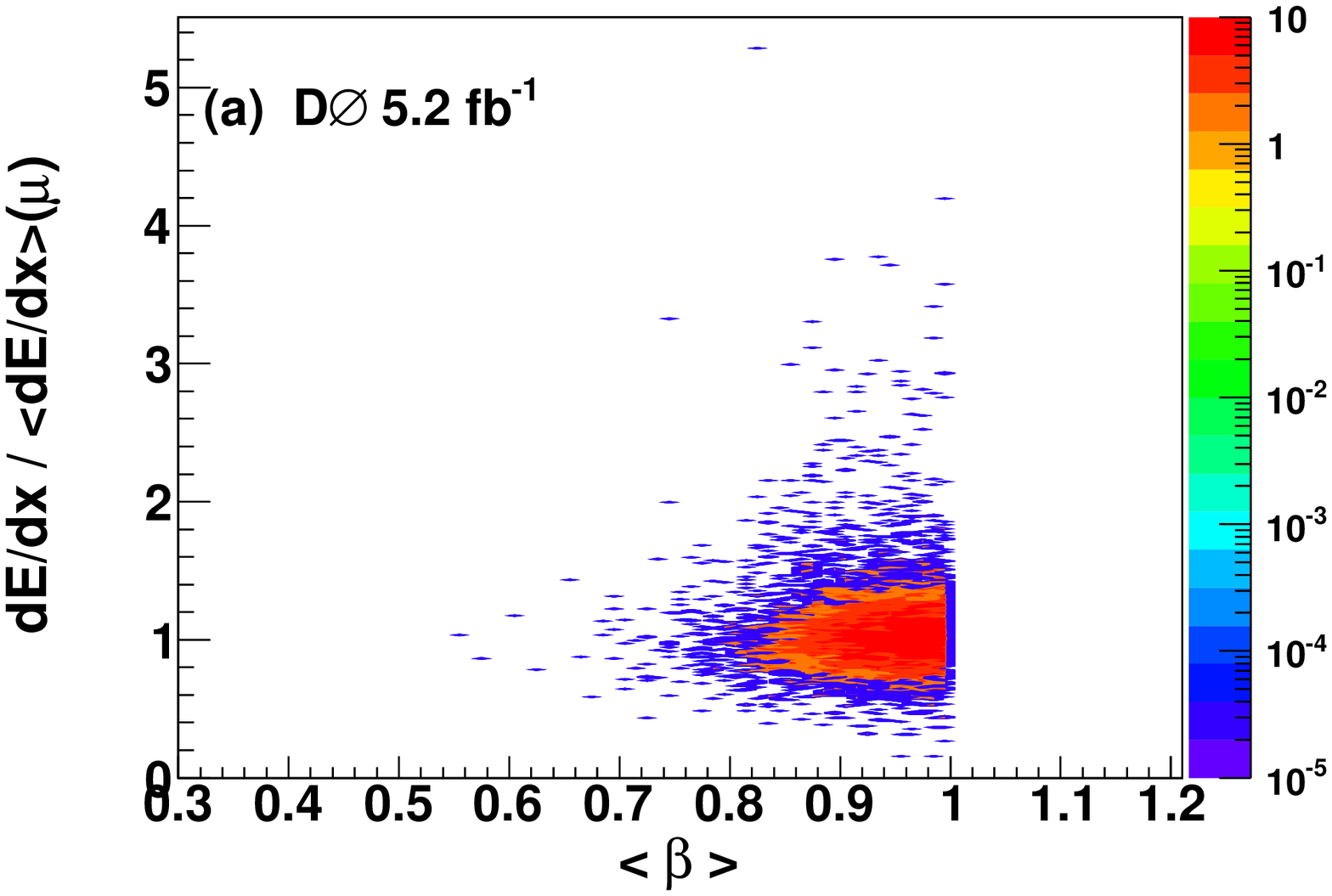}}
\scalebox{0.35}{\includegraphics{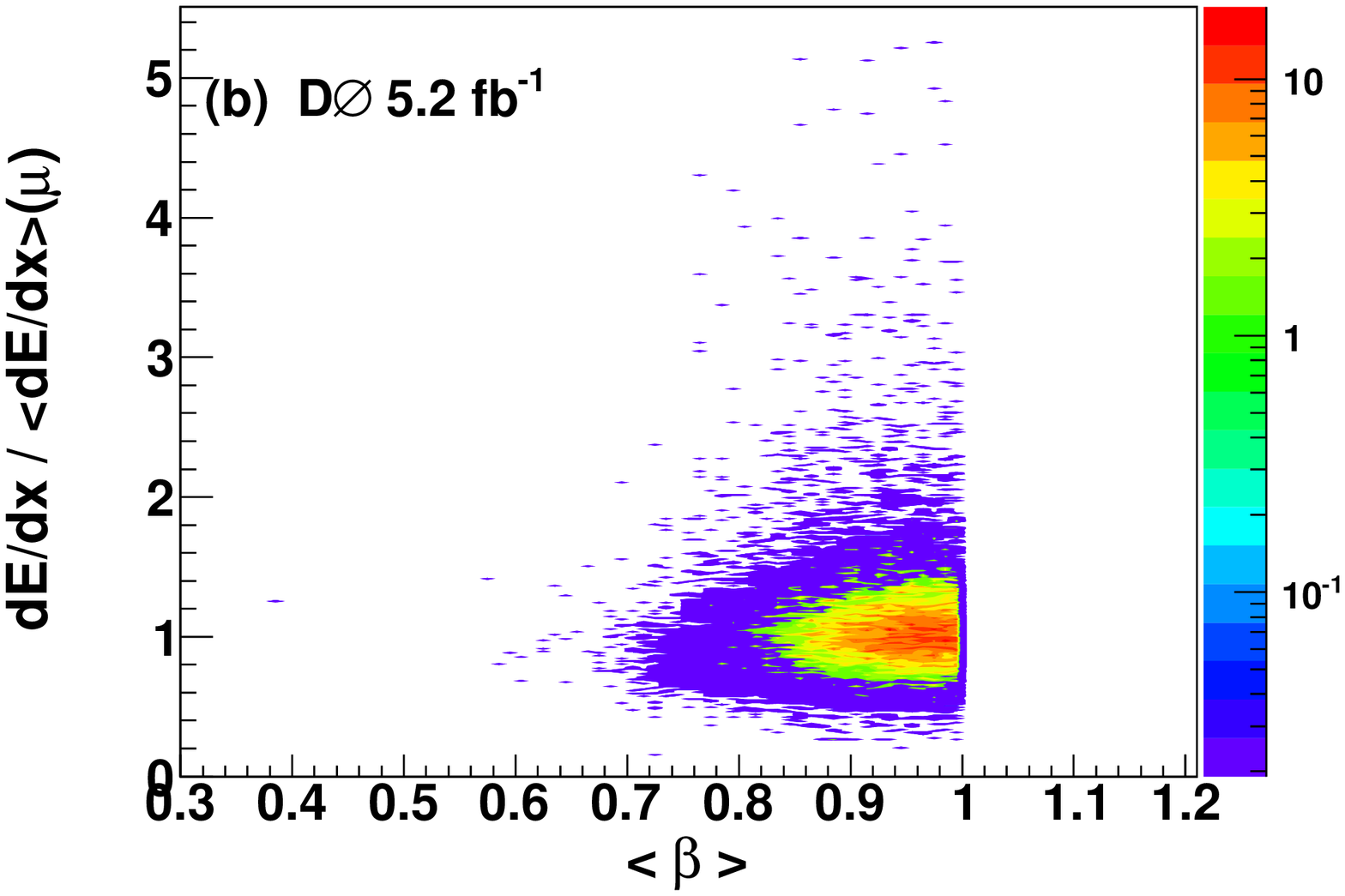}}
\scalebox{0.35}{\includegraphics{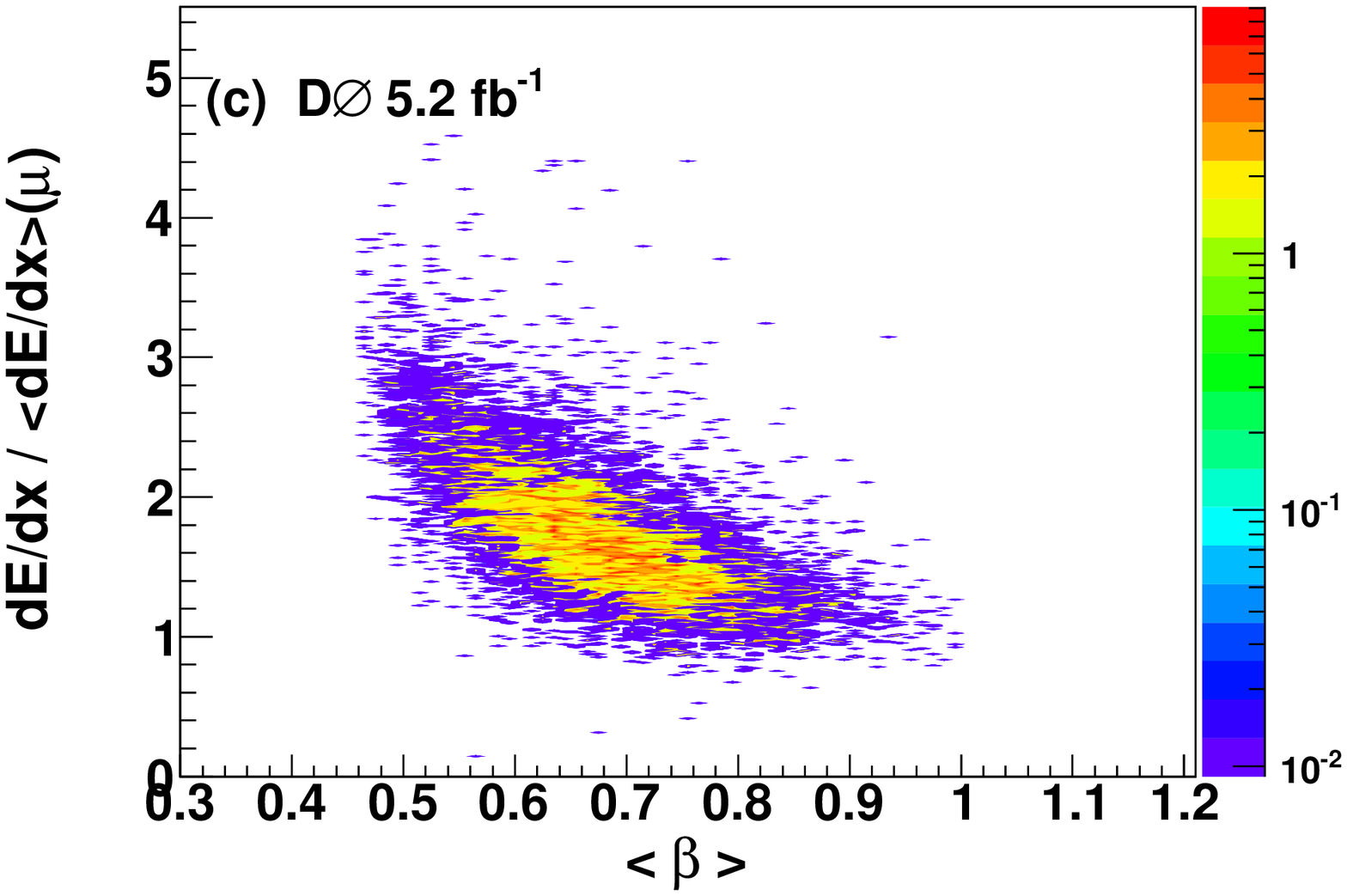}}
\caption{\label{fig:dedx_beta}(color online) Adjusted ${\rm d} E/{\rm d} x$ versus $\langle \beta \rangle$ for (a) data events, (b) background (data events with $M_T \leq$ 
200 GeV), (c) MC sample for gaugino-like charginos with mass of 300 GeV.}
\end{center}
\end{figure*}

\begin{figure*}
\begin{center}
\scalebox{0.35}{\includegraphics{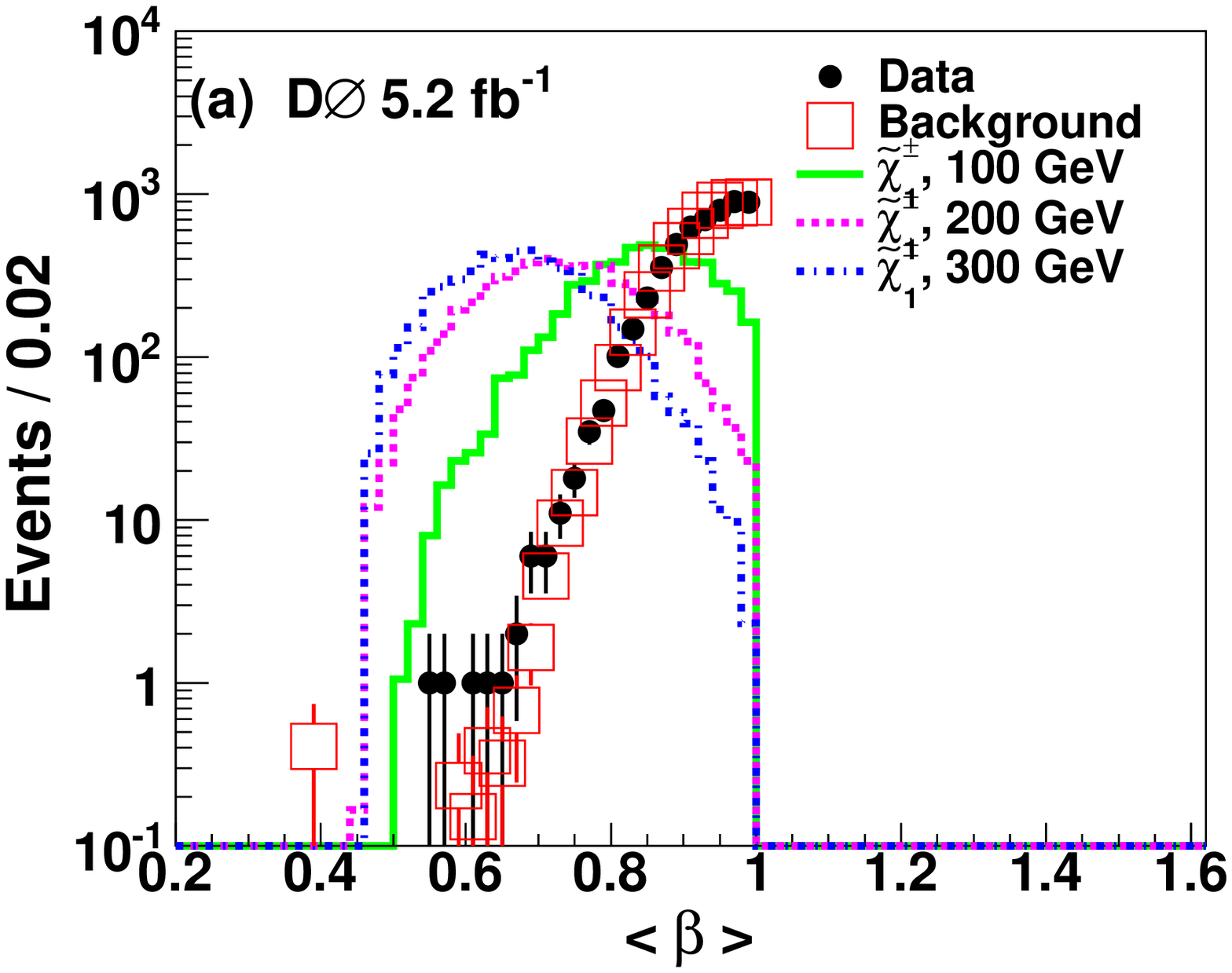}}
\scalebox{0.35}{\includegraphics{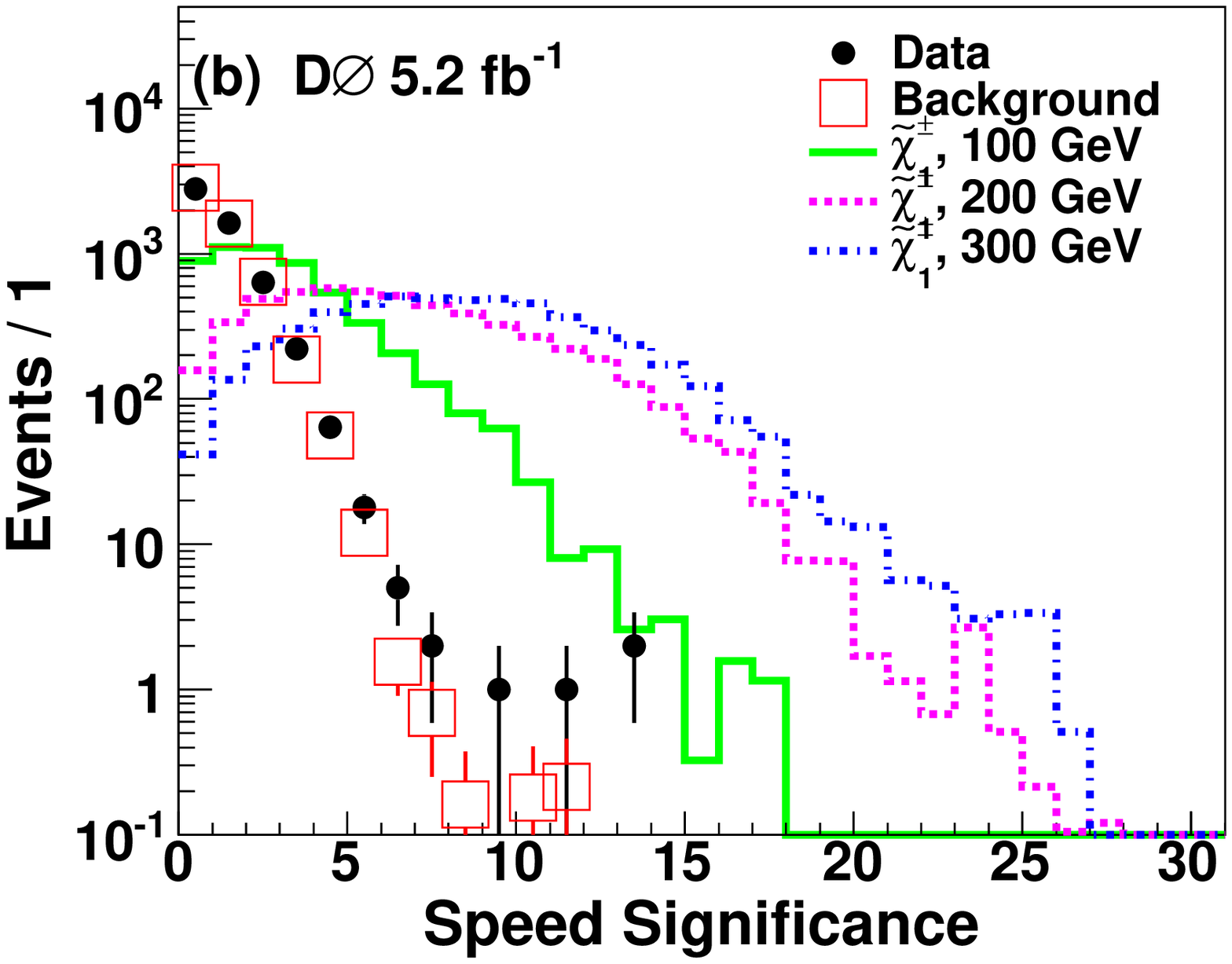}}
\scalebox{0.35}{\includegraphics{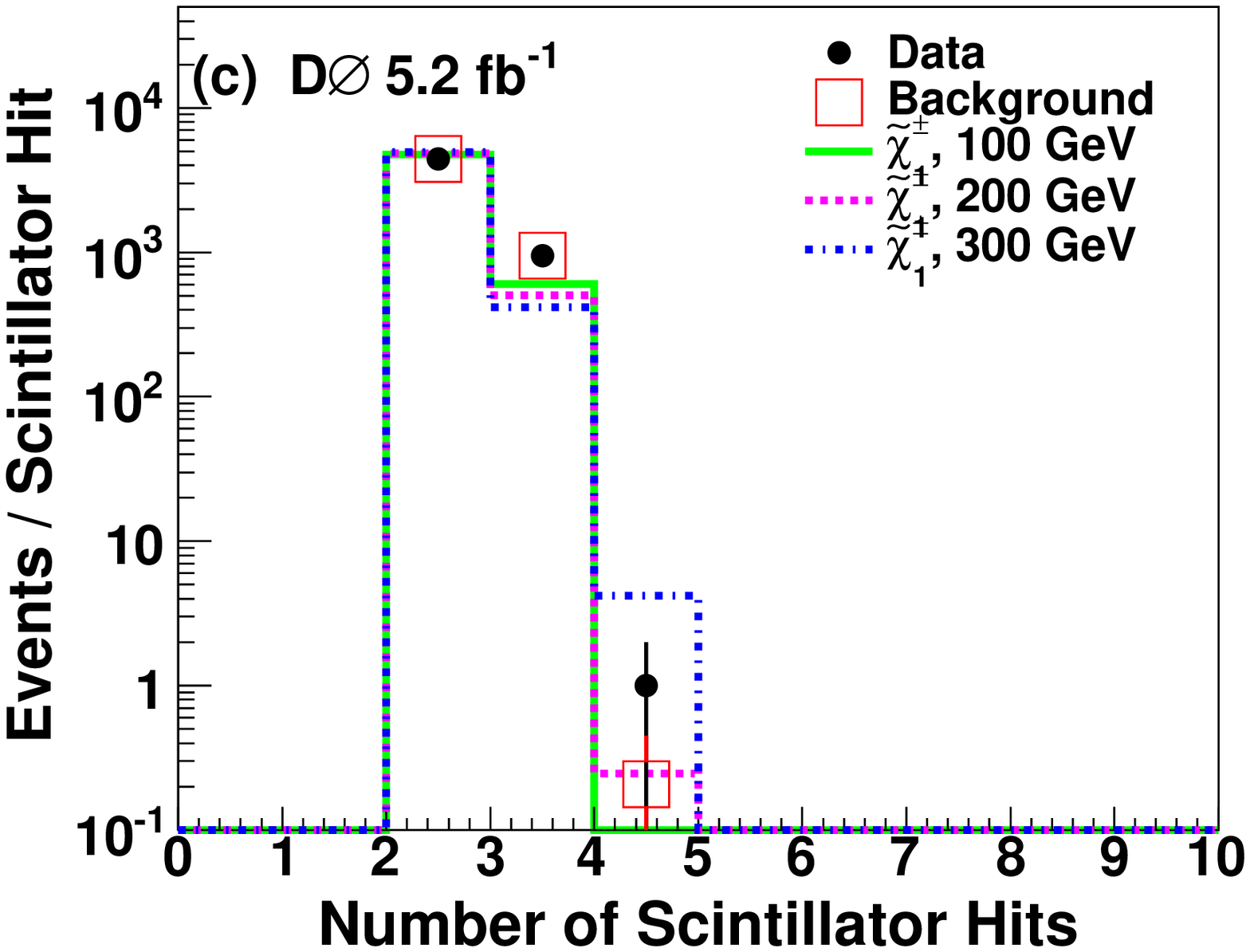}}
\caption{\label{fig:p20single_speed_BDT_input}(color online) Speed related input distributions 
to BDTs for the search for single CMLLPs. Background comes from events containing 
a muon with $\langle \beta \rangle <1$ and $M_T \leq 200$ GeV. Signal is gaugino-like charginos of 
masses 100 GeV, 200 GeV, and 300 GeV. The distributions are normalized to the same 
number of events.}
\end{center}
\end{figure*}

\begin{figure*}
\begin{center}
\scalebox{0.35}{\includegraphics{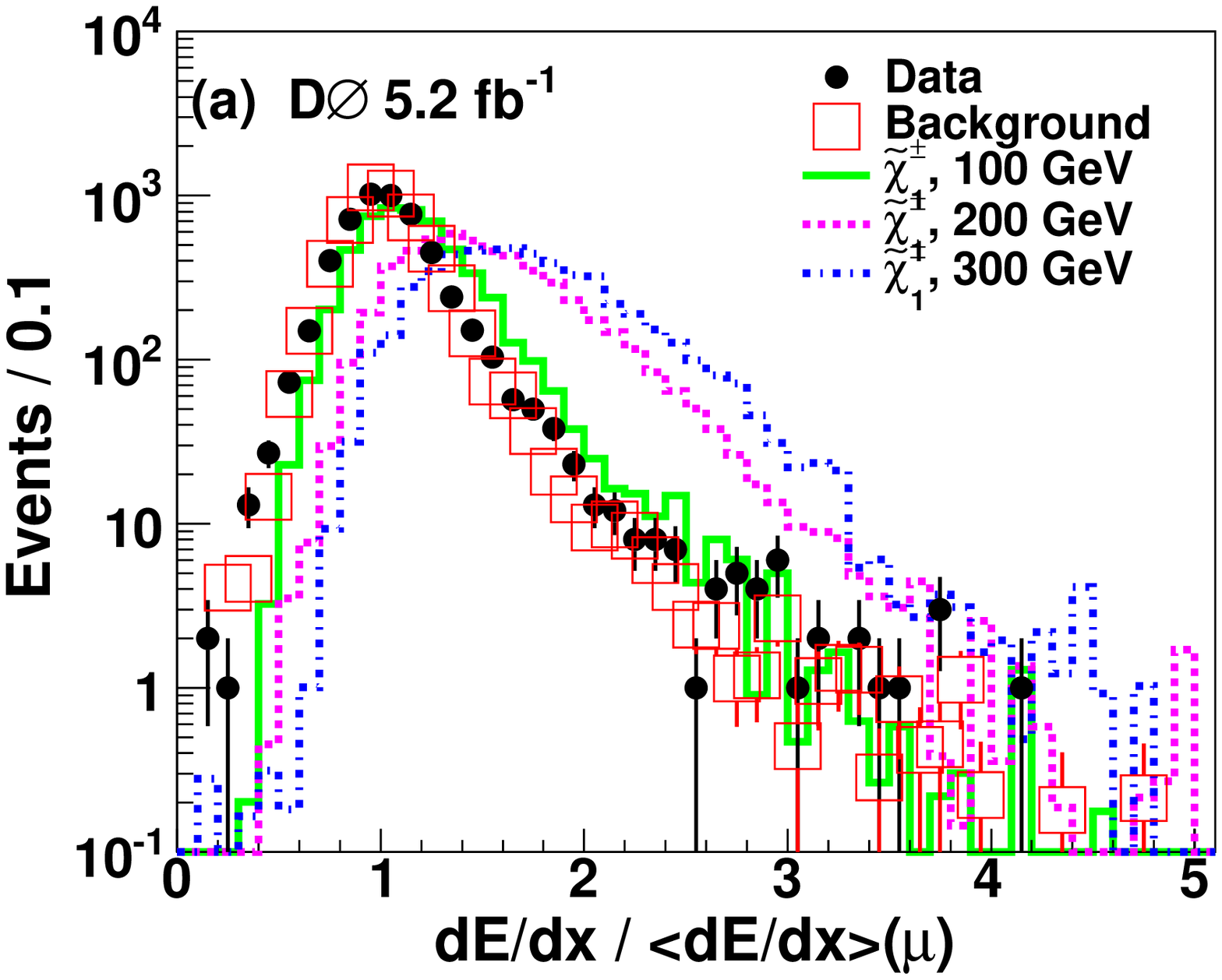}}
\scalebox{0.35}{\includegraphics{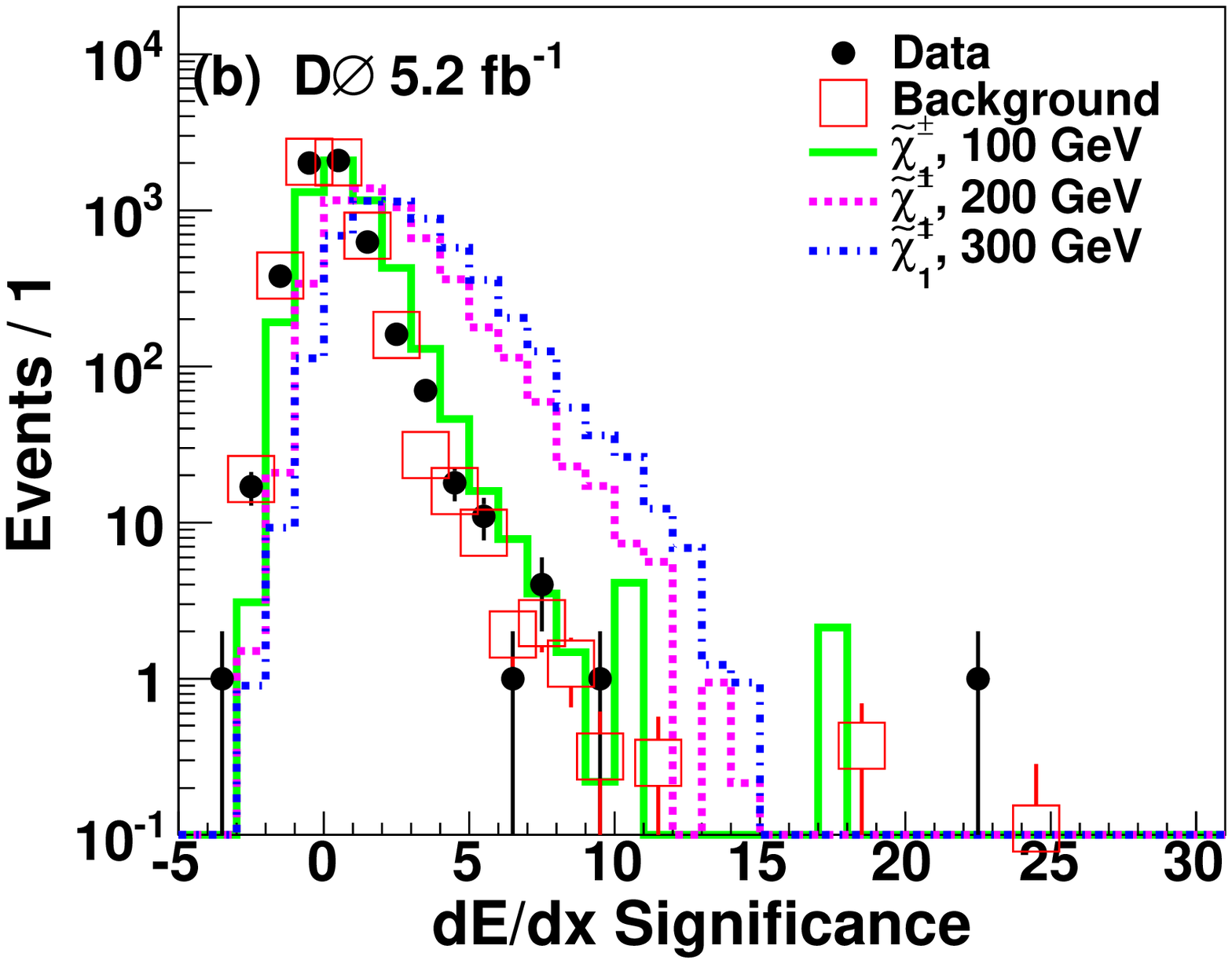}}
\scalebox{0.35}{\includegraphics{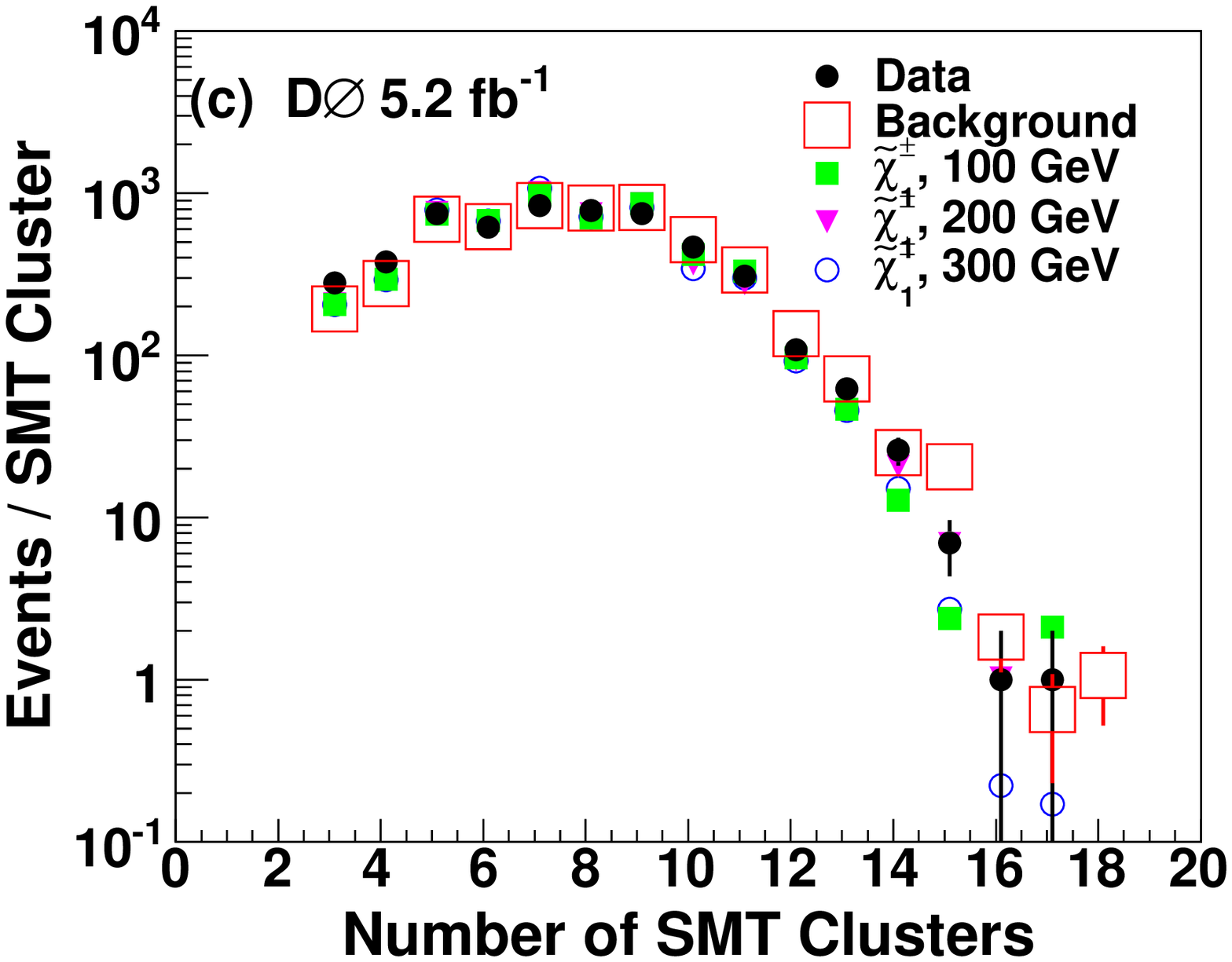}}
\caption{\label{fig:p20single_dedx_BDT_input}(color online) Distributions related to ${\rm d} E/{\rm d} x$ used in the BDT for the search for single CMLLPs. Background comes from events containing 
a muon with $\langle \beta \rangle <1$ and $M_T \leq 200$ GeV. Signal is gaugino-like charginos of masses 
100 GeV, 200 GeV, and 300 GeV. The distributions are normalized to the same number 
of events. }
\end{center}
\end{figure*}

  The key variables used for discrimination between signal and background are $\langle \beta \rangle$ and ${\rm d} E/{\rm d} x$. These variables are anti-correlated for candidate tracks originating from signal, but not for those originating from background. Figure~\ref{fig:dedx_beta} shows the adjusted 
${\rm d} E/{\rm d} x$ as a function of $\langle \beta \rangle$ for simulated gaugino-like charginos, 
background, and data. The variables $\langle \beta \rangle$, speed significance, number of scintillation counter hits, 
${\rm d} E/{\rm d} x$, ${\rm d} E/{\rm d} x$ significance, and $N_{SMT}$ are used as inputs to a BDT. The BDT is trained to distinguish between signal and background events using signal events from MC and background events from
data. Half the input events are used for training, while the
other half are used as a test sample to model the background and signal response of the BDT. The distributions of the BDT input variables are 
shown in Figs.~\ref{fig:p20single_speed_BDT_input} and \ref{fig:p20single_dedx_BDT_input}.

\begin{figure*}
\begin{center}
\scalebox{0.35}{\includegraphics{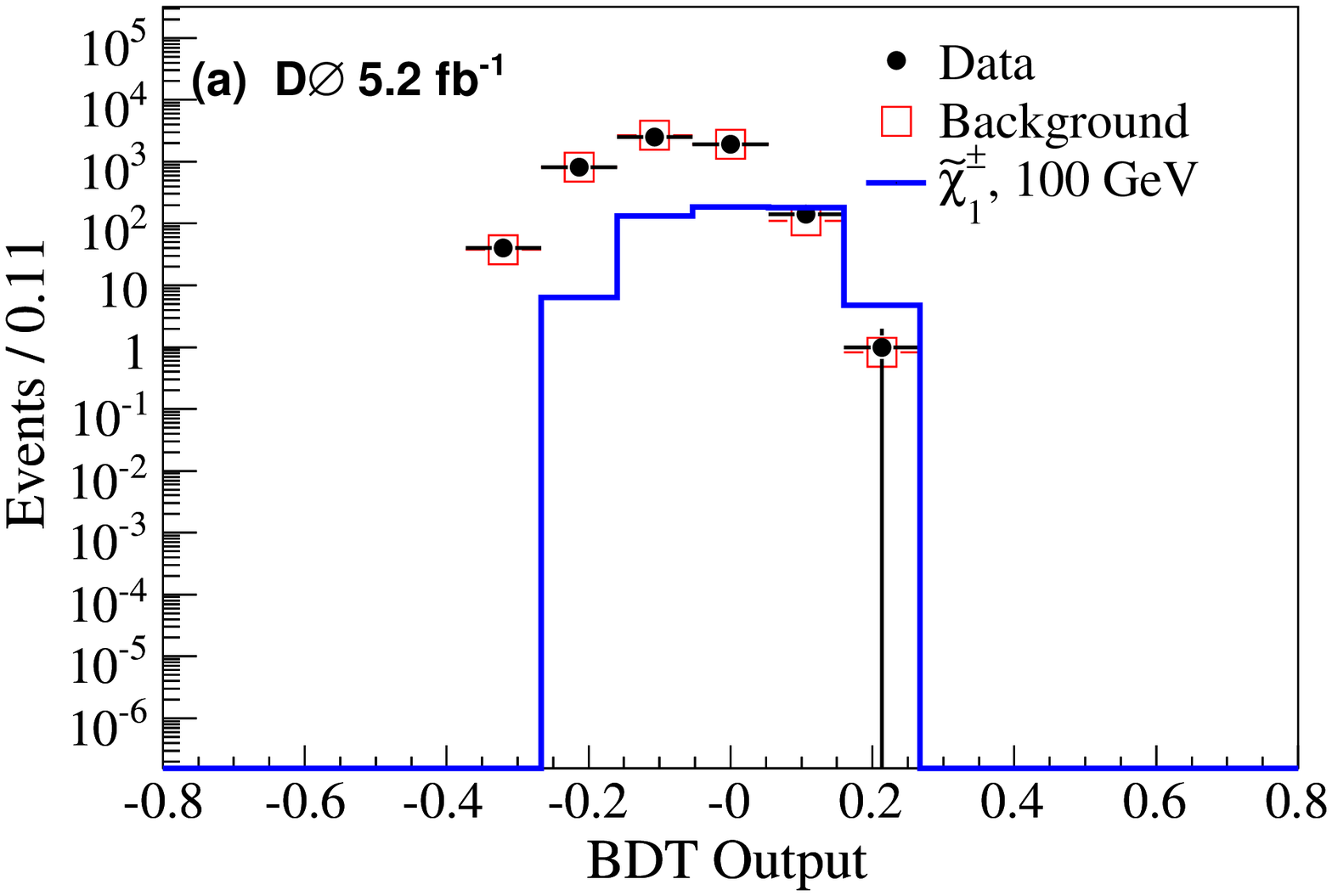}}
\scalebox{0.35}{\includegraphics{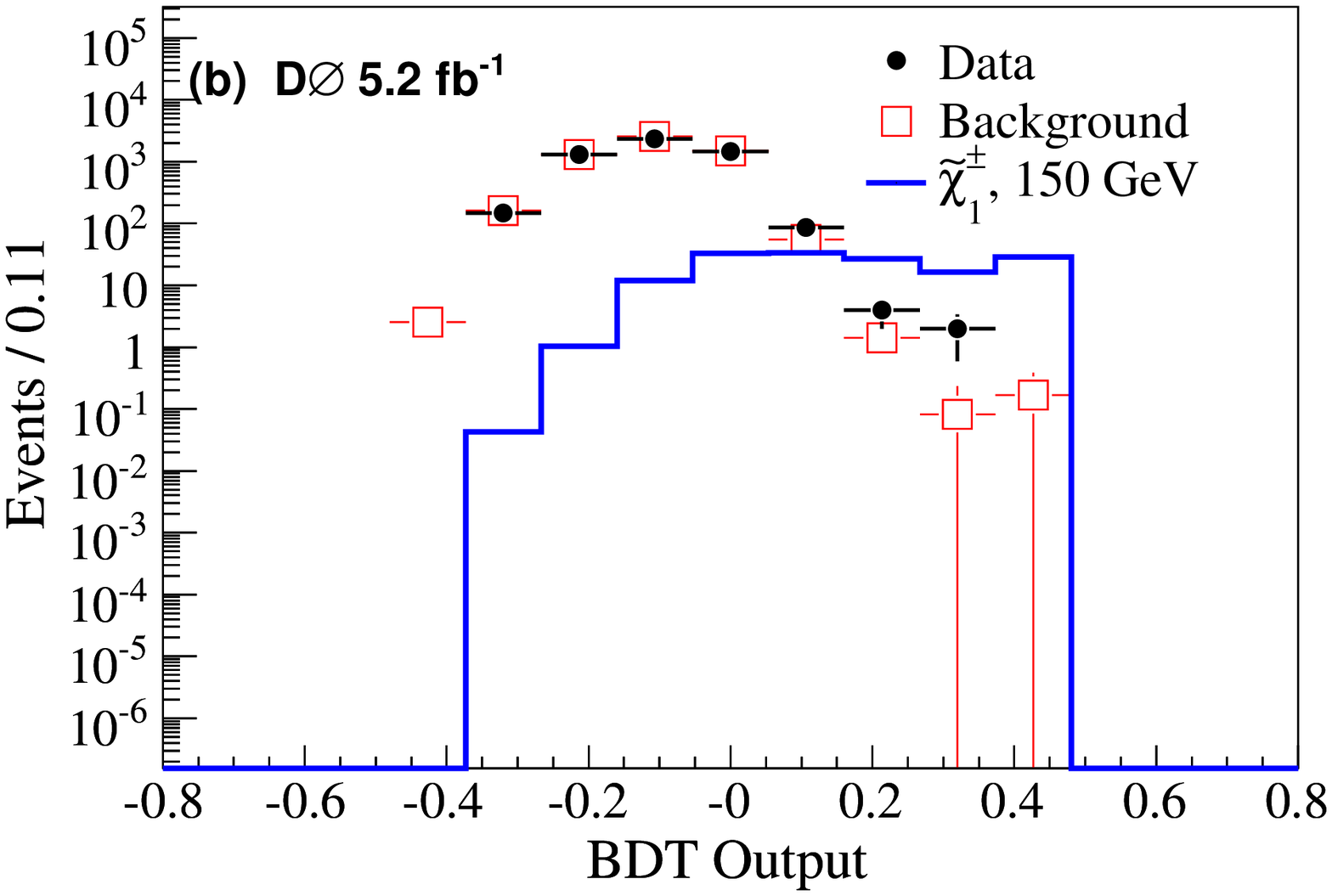}}
\scalebox{0.35}{\includegraphics{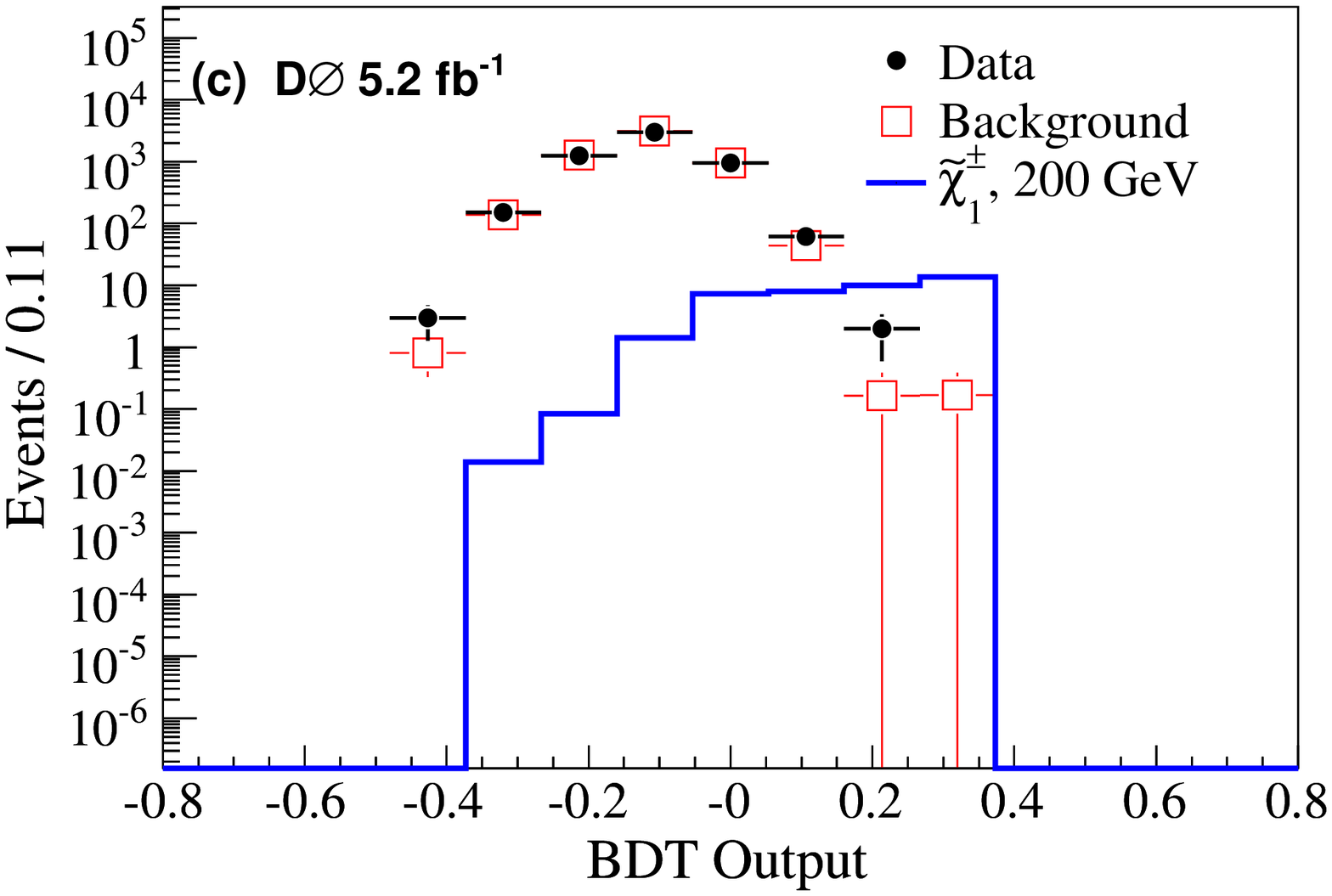}}
\scalebox{0.35}{\includegraphics{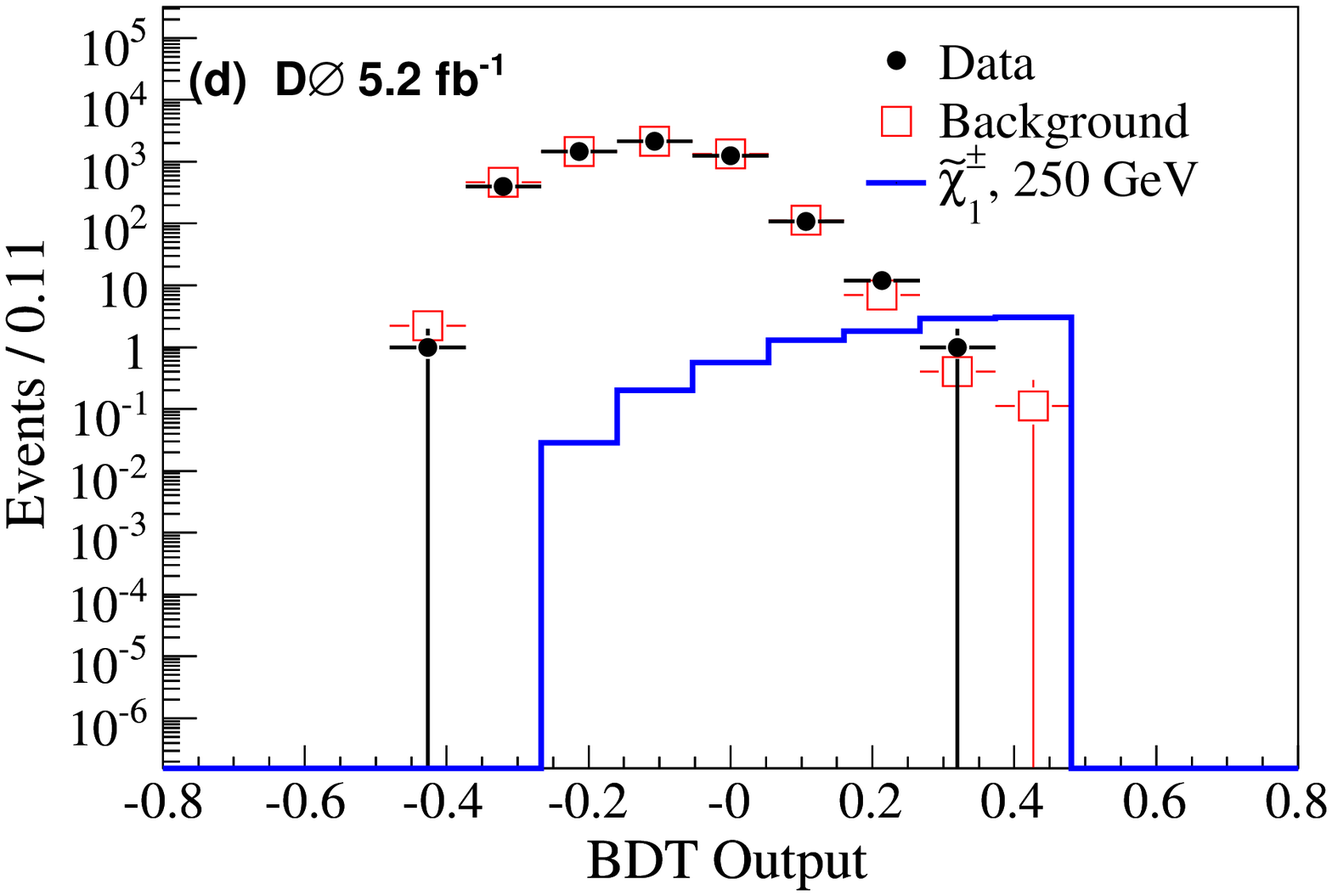}}
\scalebox{0.35}{\includegraphics{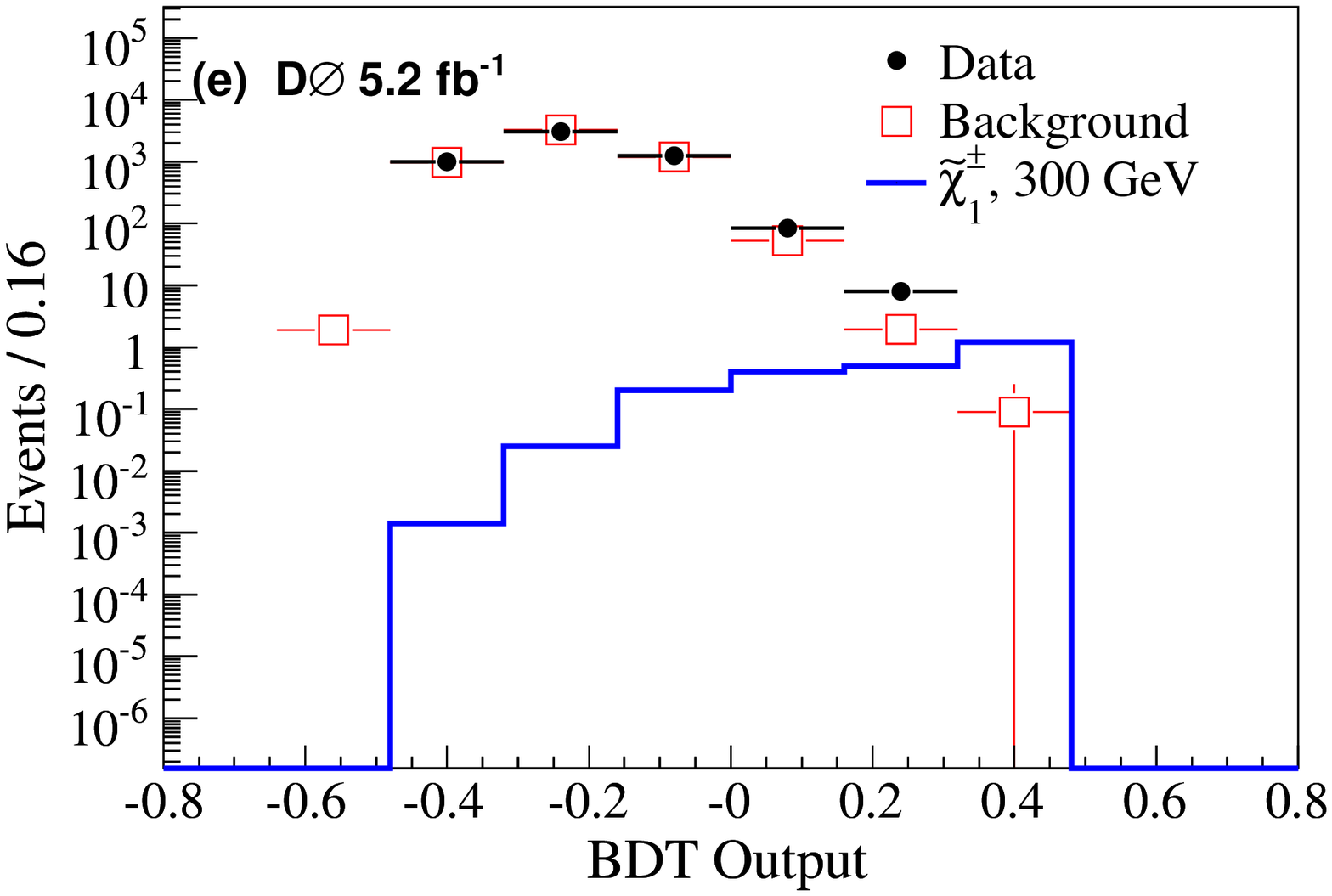}}
\caption{\label{fig:p20single_BDTout_charginoG}(color online) BDT-output distributions for simulated gaugino-like chargino masses 100--300 GeV in 50 GeV steps in 
the search for single CMLLPs. The distributions are normalized to the expected number of events.}
\end{center}
\end{figure*}
An example of the BDT-output distribution is shown in Fig.~\ref{fig:p20single_BDTout_charginoG} for simulated gaugino-like charginos for masses 100--300 GeV.  
The BDT-output distributions for simulated stau lepton, top squark, and higgsino-like 
charginos, after being normalized to the expected number of events, are shown in Figs.~\ref{fig:p20single_BDTout_stau}-\ref{fig:p20single_BDTout_charginoH} in Appendix A.
These BDT distributions are used as input to a modified frequentist limit calculation  
employing a log-likelihood ratio (LLR) test statistic~\cite{COLLIE, CLS-Wade} 
to obtain 95\% C.L. cross section limits. To minimize the degrading effects of 
systematic uncertainties on the search sensitivity, the signal and background 
are fitted to the observed data by maximizing a likelihood function over the 
systematic uncertainties for both the background-only 
and the signal-plus-background hypotheses. 

\subsection{Systematic Uncertainties\label{sec:single_syst}}

The sources of systematic uncertainties and their estimation are similar to the search for CMLLP pairs, described in Sec. VI C. A parameter is varied within its uncertainty and the change is propagated through the entire analysis to produce a BDT distribution. 
The systematic uncertainties can be divided into two categories, ``normalization'' and ``shape'' uncertainties. Normalization uncertainties affect only the overall event rate whereas the shape uncertainties can also change the differential distribution.

The systematic uncertainty due to uncertainty in luminosity (6.1\%) and in muon 
reconstruction efficiency (2.1\%) are normalization uncertainties. The other sources of normalization uncertainties are: the background normalization 
uncertainty due to the choice of the cuts on $\langle \beta \rangle$ (7.2\%) and $M_T$ (2.2\%), 
the muon $p_T$ resolution uncertainty (0.2\%), the ${\rm d} E/{\rm d} x$ correction uncertainty 
($<$ 0.1\%), the ${\rm d} E/{\rm d} x$ smearing uncertainty (0.2\%), and the speed ${\chi}^2/{\mathrm{dof}}$ 
correction uncertainty (0.4\%). The systematic uncertainty for the speed $\chi^2/{\mathrm{dof}}$ correction is determined by
 repeating the data to MC correction (described in Sec.~\ref{tof_section}) with a MC sample
 of $W\rightarrow \mu\nu$ decays and then taking the difference between the two corrections as the
 uncertainty. The uncertainties due to the choice of PDF, and the 
${\rm d} E/{\rm d} x$ correction are the same as in Sec.~VI C. The systematic uncertainties due 
to the width of the L1 timing gate, and the timing simulation change with 
the output value of the BDT and therefore are shape systematics. 
The average uncertainty for the L1 timing gate width is 4\% and that for the timing simulation is 7\%. These uncertainties are summarized in the 
third columns of Tables 
\ref{tab:systematics_combination_sig}, and \ref{tab:systematics_combination_bkgd}.
\subsection{Results}
\begin{figure*}
\begin{center}
\scalebox{0.4}{\includegraphics{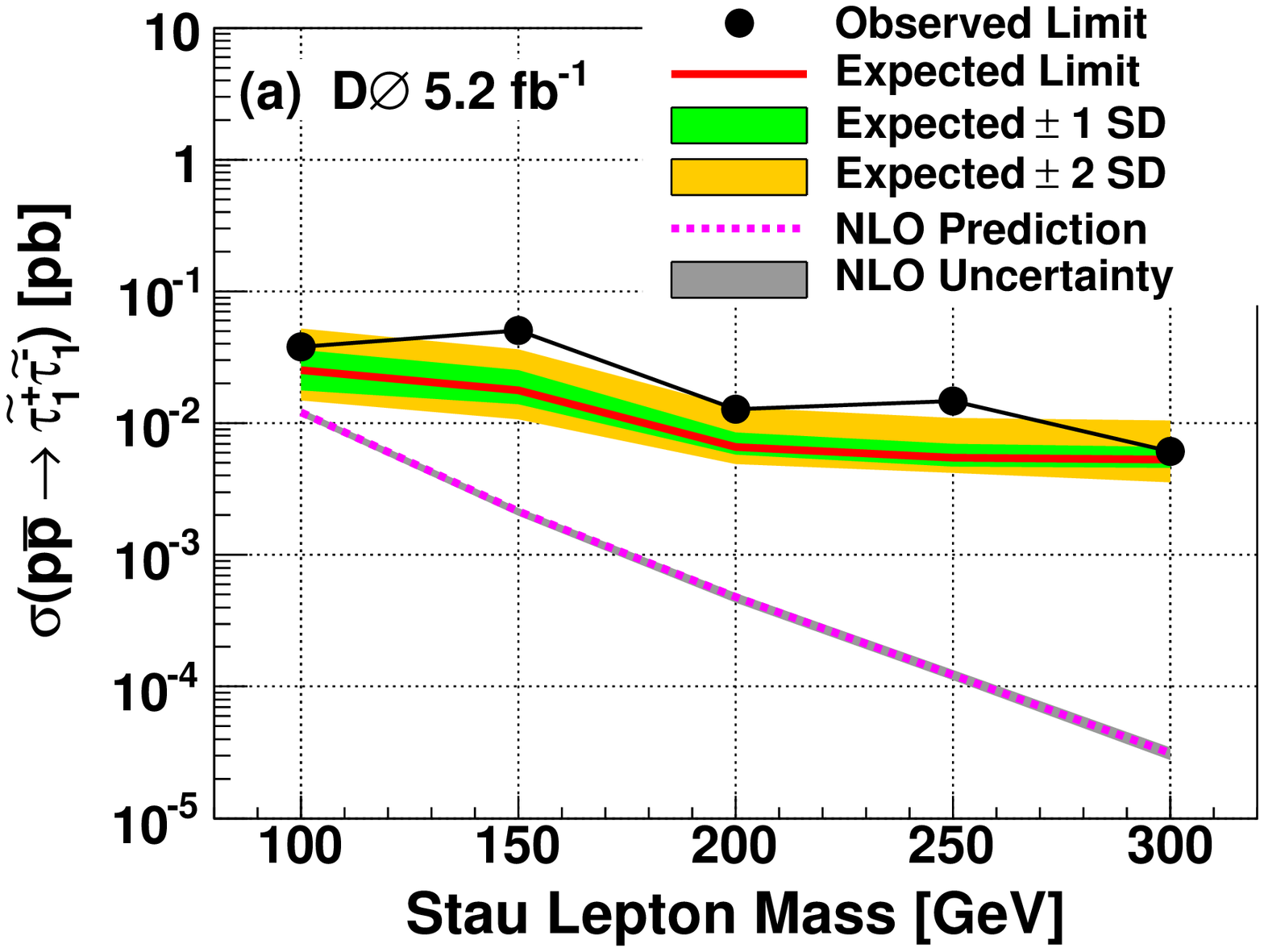}}
\scalebox{0.4}{\includegraphics{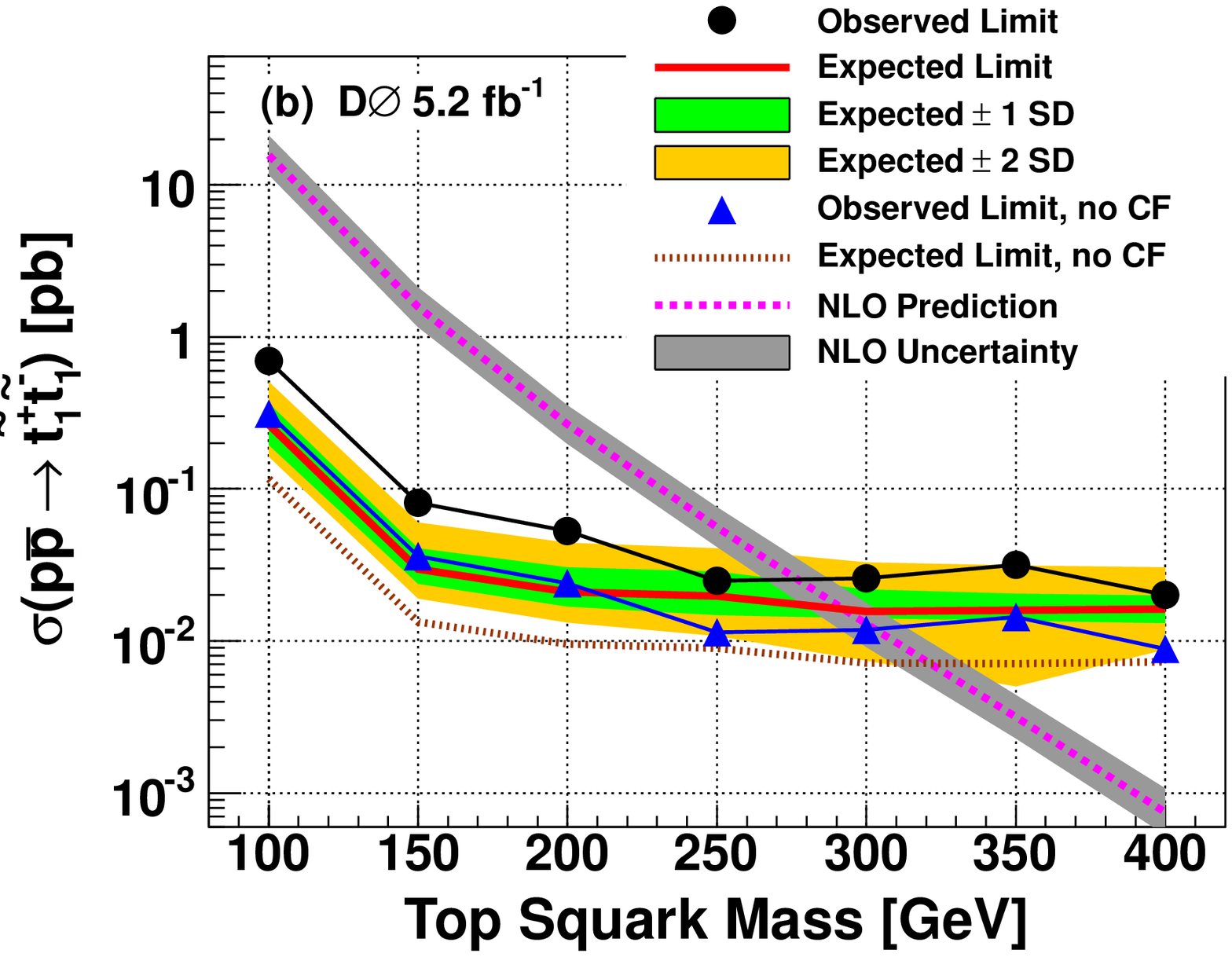}}
\scalebox{0.4}{\includegraphics{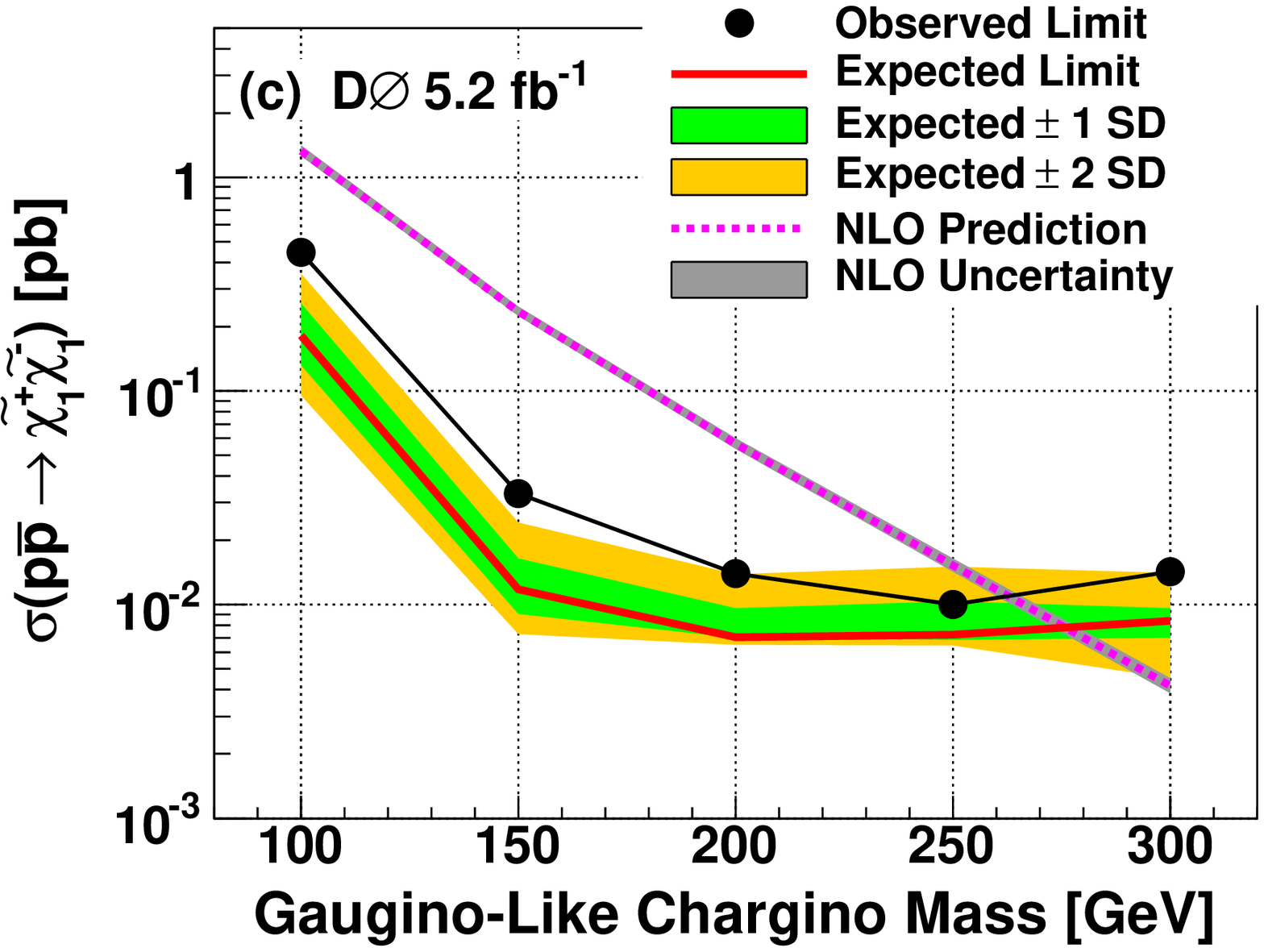}}
\scalebox{0.4}{\includegraphics{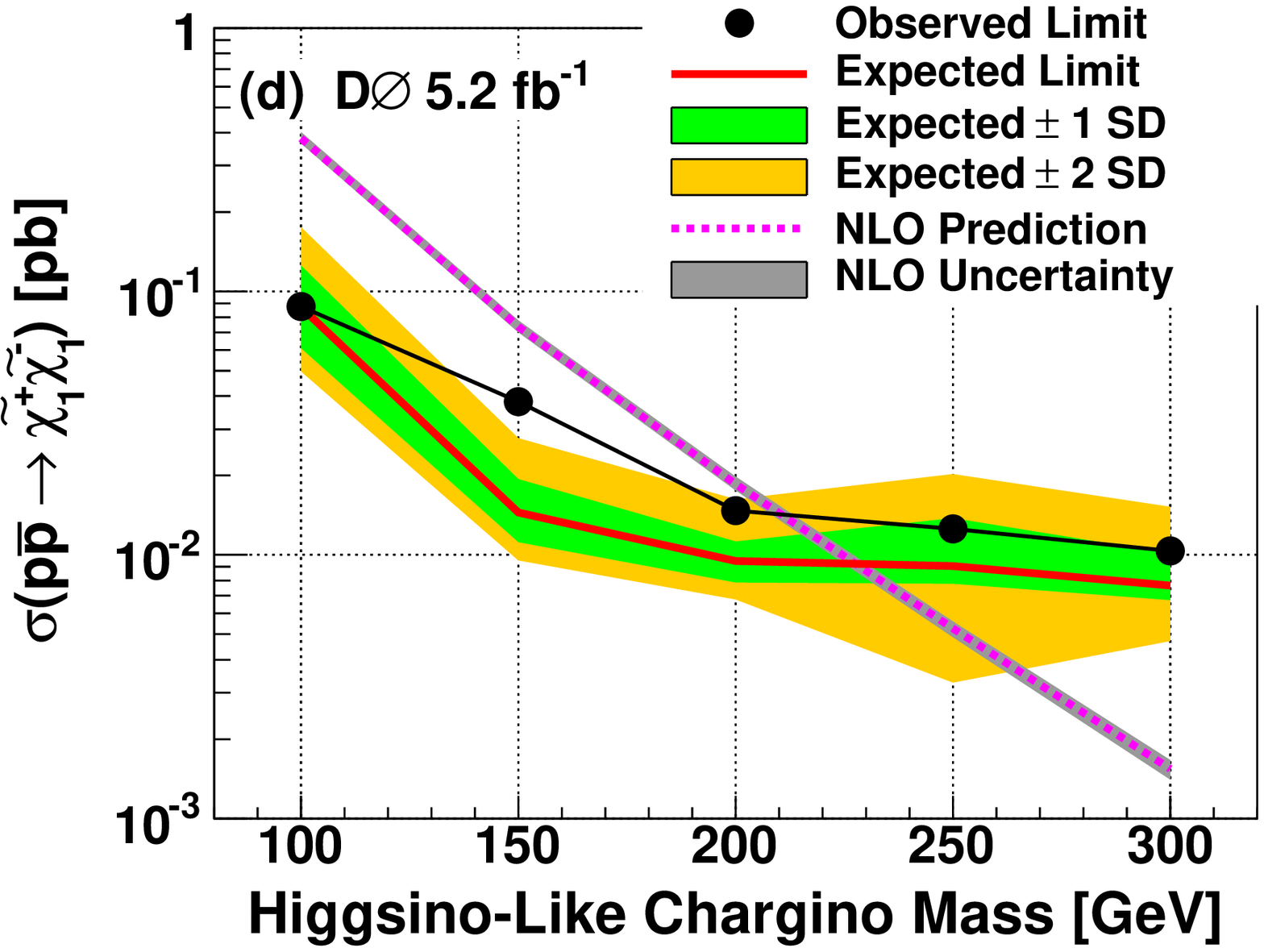}}
\caption{\label{fig:p20single_limits}(color online) 95\% C.L. limits on production cross sections of a pair of stau leptons, top squarks, gaugino-like charginos, and higgsino-like charginos, as a function 
of their masses from the search for one or more CMLLPs with Run IIb data. $\pm 1$ SD and $\pm 2$ SD are the 1 and 2 standard deviation bands respectively around the expected limit curves.}
\end{center}
\end{figure*}

For the single CMLLP search with 5.2 ${\mathrm{fb}}^{-1}$ integrated luminosity, 95\% C.L. upper limits on production cross sections for stau leptons, top squarks, 
gaugino-like, and higgsino-like charginos are shown in Table \ref{tab:p20single_limits} and in Fig.~\ref{fig:p20single_limits}. Limits are set on 
production cross section of stau leptons from 0.05 pb to 0.006 pb, for stau lepton masses 
in the range between 100 GeV and 300 GeV. Pair-produced long-lived top squarks are 
excluded below masses of 285 GeV. If we only include the effects of initial 
hadronization of top squarks and do not include the effects of charge flipping 
during their passage through the detector, the lower limit on long-lived 
top squark mass is found to be 305 GeV. Pair-produced long-lived gaugino-like charginos 
are excluded below masses of 267 GeV, and higgsino-like charginos below masses 
of 217 GeV.  They are identical to those presented 
in~\cite{D0 CMSP PRL Run IIb}. 
Using the intersection of the $-1$ ($+1$) standard deviation band with the NLO 
cross section, the mass limits shift down (up) by 1 GeV for charginos and 
by 10 GeV for top squarks.
\begin{table*}
\begin{center} 
\begin{ruledtabular}
\begin{tabular}{c c c c}
Mass (GeV) & NLO cross section [pb]  & ${\sigma}_{95}^{obs}$ [pb]  & ${\sigma}_{95}^{exp} $  [pb]\\ 
\hline
Stau lepton& & & \\

 { 100}  &  {$0.0120^{+0.0006}_{-0.0008}$}  &  {$0.038$}  &  {$0.025^{+0.011}_{-0.0075}$}\\
 { 150}  &  {$0.0021^{+0.0001}_{-0.0002}$}  &  {$0.050$}  &  {$0.018^{+0.0076}_{-0.0038}$}\\
 { 200}  &  {$0.00050^{+0.00003}_{-0.00002}$}  &  {$0.013$}  &  {$0.0066^{+0.0020}_{-0.0008}$}\\
 { 250}  &  {$0.00010^{+0.00001}_{-0.00001}$}  &  {$0.015$}  &  {$0.0055^{+0.0015}_{-0.0008}$}\\
 { 300}  &  {$0.000030^{+0.000003}_{-0.000004}$}  &  {$0.006$}  &  {$0.0053^{+0.0013}_{-0.0007}$}\\
\hline
Top squark  & & & \\

 { 100}  &  {$15.6^{+5.4}_{-4.0}$}  &  {$0.70$}  &  {$0.26^{+0.094}_{-0.078}$}\\
 { 150}  &  {$1.58^{+0.53}_{-0.42}$}  &  {$0.081$}  &  {$0.030^{+0.011}_{-0.0063}$}\\
 { 200}  &  {$0.27^{+0.088}_{-0.068}$}  &  {$0.053$}  &  {$0.021^{+0.0096}_{-0.0043}$}\\
 { 250}  &  {$0.056^{+0.020}_{-0.014}$}  &  {$0.025$}  &  {$0.020^{+0.0088}_{-0.0049}$}\\
 { 300}  &  {$0.013^{+0.0048}_{-0.0039}$}  &  {$0.026$}  &  {$0.016^{+0.0061}_{-0.0016}$}\\
 { 350}  &  {$0.0032^{+0.0012}_{-0.0009}$}  &  {$0.032$}  &  {$0.016^{+0.0046}_{-0.0024}$}\\
 { 400}  &  {$0.0008^{+0.0003}_{-0.0002}$}  &  {$0.020$}  &  {$0.016^{+0.0036}_{-0.0031}$}\\
\hline
Gaugino-like chargino  & & & \\

 { 100}  &  {$1.33^{+0.08}_{-0.07}$}  &  {$0.44$}  &  {$0.180^{+0.076}_{-0.051}$}\\
 { 150}  &  {$0.240^{+0.014}_{-0.010}$}  &  {$0.033$}  &  {$0.0120^{+0.0047}_{-0.0028}$}\\
 { 200}  &  {$0.0570^{+0.0034}_{-0.0030}$}  &  {$0.014$}  &  {$0.0070^{+0.0026}_{-0.00006}$}\\
 { 250}  &  {$0.0150^{+0.0011}_{-0.0010}$}  &  {$0.010$}  &  {$0.0072^{+0.0031}_{-0.0004}$}\\
 { 300}  &  {$0.0042^{+0.0004}_{-0.0003}$}  &  {$0.014$}  &  {$0.0084^{+0.0012}_{-0.0014}$}\\
\hline
Higgsino-like chargino  & & & \\

 { 100}  &  {$0.380^{+0.023}_{-0.017}$}  &  {$0.088$}  &  {$0.087^{+0.038}_{-0.026}$}\\
 { 150}  &  {$0.0740^{+0.0040}_{-0.0038}$}  &  {$0.038$}  &  {$0.015^{+0.0049}_{-0.0033}$}\\
 { 200}  &  {$0.0190^{+0.0012}_{-0.0010}$}  &  {$0.015$}  &  {$0.0095^{+0.0018}_{-0.0017}$}\\
 { 250}  &  {$0.0053^{+0.0004}_{-0.0004}$}  &  {$0.013$}  &  {$0.0091^{+0.0047}_{-0.0013}$}\\
 { 300}  &  {$0.0015^{+0.0001}_{-0.0001}$}  &  {$0.010$}  &  {$0.0077^{+0.0025}_{-0.0009}$}\\
\end{tabular}
\end{ruledtabular}
\end{center}
\caption{\label{tab:p20single_limits} 95\% C.L. cross section 
limits for stau lepton, top squark, gaugino-like, and higgsino-like charginos in the search for single CMLLPs in Run IIb data.}
\end{table*}

\section{Run II\lowercase{a} Result} 
The search for a pair of CMLLPs in 1.1 ${\mathrm{fb}}^{-1}$ of Run IIa integrated luminosity utilizes 
the TOF measurement in addition to other kinematic variables to select events with 
candidate CMLLPs~\cite{D0 CMSP PRL Run IIa}. The CMLLP candidates in this analysis are staus, gaugino-like, and higgsino-like charginos.   
The principal background comes from mismeasured muons from the 
decays of the $Z$ boson. Table \ref{tab:p17pair_sample} presents the signal 
acceptance, the number of predicted background events, and the number of observed events in this analysis.
\begin{table*}
\begin{center} 
\begin{ruledtabular}
\begin{tabular}{c c c c}
Mass (GeV)  & Signal Acceptance (\%)  & Predicted Background  & Observed Data\tabularnewline
\hline
Stau lepton & & & \\

 { 100}  &  {$5.56\pm0.11\pm0.41$}  &  {$1.55\pm0.49\pm0.30$}  &  {$1$}\\
 { 150}  &  {$12.3\pm0.16\pm1.27$}  &  {$1.70\pm0.51\pm0.15$}  &  {$1$}\\
 { 200}  &  {$13.9\pm0.17\pm1.11$}  &  {$1.70\pm0.51\pm0.51$}  &  {$1$}\\
 { 250}  &  {$13.3\pm0.16\pm1.25$}  &  {$1.70\pm0.51\pm0.31$}  &  {$1$}\\
 { 300}  &  {$11.7\pm0.15\pm1.34$}  &  {$1.86\pm0.54\pm0.15$}  &  {$2$}\\
\hline
Gaugino-like chargino  & & & \\

 { 100}  &  {$4.63\pm0.10\pm0.35$}  &  {$1.55\pm0.49\pm0.31$}  &  {$1$}\\
 { 150}  &  {$8.51\pm0.13\pm0.88$}  &  {$1.24\pm0.44\pm0.11$}  &  {$1$}\\
 { 200}  &  {$8.89\pm0.13\pm0.71$}  &  {$1.86\pm0.54\pm( < 0.01)$}  &  {$1$}\\
 { 250}  &  {$7.40\pm0.12\pm0.70$}  &  {$1.70\pm0.51\pm0.31$}  &  {$1$}\\
 { 300}  &  {$5.88\pm0.11\pm0.68$}  &  {$1.70\pm0.51\pm0.14$}  &  {$2$}\\
\hline
Higgsino-like chargino  & & & \\

 { 100}  &  {$4.94\pm0.10\pm0.37$}  &  {$1.55\pm0.49\pm0.31$}  &  {$1$}\\
 { 150}  &  {$8.91\pm0.13\pm0.92$}  &  {$1.39\pm0.46\pm0.13$}  &  {$1$}\\
 { 200}  &  {$9.56\pm0.14\pm0.76$}  &  {$1.86\pm0.54\pm( < 0.01)$}  &  {$1$}\\
 { 250}  &  {$8.13\pm0.13\pm0.76$}  &  {$1.70\pm0.51\pm0.31$}  &  {$1$}\\
 { 300}  &  {$6.36\pm0.11\pm0.73$}  &  {$1.70\pm0.51\pm0.14$}  &  {$1$}\\
\hline
\end{tabular}
\end{ruledtabular}
\end{center}
\caption{\label{tab:p17pair_sample} The signal acceptance, number 
of predicted background events, and the number of observed events from the search 
for a pair of CMLLPs with 1.1 ${\mathrm{fb}}^{-1}$ of Run IIa integrated luminosity. The first error is statistical and the second one is systematic.}
\end{table*}
In the absence of any signal, limits of 206 GeV and 171 GeV are set 
on masses of gaugino-like charginos and higgsino-like charginos,  respectively. Limits 
on the production cross section of stau leptons are set from 0.31 pb to 0.04 pb for the stau lepton mass range of 60 to 300 GeV.

\section{Combination of Results}

In the absence of observed signal in all three analyses, the searches 
for a pair of CMLLPs in 1.1 ${\mathrm{fb}}^{-1}$ of Run IIa, in 5.2 ${\mathrm{fb}}^{-1}$ of Run IIb, and the search for a single CMLLP 
in 5.2 ${\mathrm{fb}}^{-1}$ of Run IIb integrated luminosity, we combine the results to find limits on the production cross sections of stau leptons, gaugino-like 
charginos, and higgsino-like charginos. Due to the effect of hadronization and 
charge-flipping, the sensitivity of the search for single CMLLPs for top squarks is 
much better than the sensitivity of the search for a pair of CMLLPs. As a result, the top squark mass limits for the combination of the single CMLLP (with a veto for the events common with the pair search) and the pair CMLLP analyses do 
not show a significant improvement over the results from the single CMLLP analysis on its own and hence is not performed.
\subsection{Method of Combination}
In the Run IIb search for a pair of CMLLPs, the 
value of the requirement on the BDT output is optimized for each signal mass point. 
A similar procedure was used in the search for a pair of CMLLPs with Run IIa data. In the search for single CMLLPs in Run IIb, 
the entire BDT distribution is used as input to a \emph{CL$_s$} limit setting  method. 
To obtain results from the combination of the three analyses, the signal acceptance, 
the number of predicted background events, and the number of observed events 
for each signal mass, for the two pair analyses (Tables \ref{tab:p20pair_sig_accep_bck_pred}, 
\ref{tab:p17pair_sample}) along with the BDT distributions from the search for 
single CMLLPs (Figs.~\ref{fig:p20single_BDTout_charginoG}, and \ref{fig:p20single_BDTout_stau}-\ref{fig:p20single_BDTout_charginoH}) (see Appendix A) are used as inputs to the same \emph{CL$_s$} method. 
To avoid double counting of events, the datasets
used for the two Run IIb analyses are made statistically independent by removing the events that have been selected for the search of CMLLP pairs from the dataset used to search for single CMLLPs. The different analyses are combined by summing the log-likelihood ratios over all the bins and all the analyses. 
\subsection{Systematic Uncertainties}
Systematic uncertainties are treated as Gaussian distributions and are applied to
the expected number of signal and background events. 
Various sources of systematic uncertainties along with their values for 
the four types of CMLLP signals that have been studied are listed in 
Table~\ref{tab:systematics_combination_sig} and the systematic uncertainties 
on the background sample are listed in Table~\ref{tab:systematics_combination_bkgd}.
\begin{table*}
\begin{center} 
\begin{ruledtabular}
\begin{tabular}{c c c}
Pair (1.1 ${\mathrm{fb}}^{-1}$)  & Pair (5.2 ${\mathrm{fb}}^{-1}$)  & Single (5.2 ${\mathrm{fb}}^{-1}$)\tabularnewline
\hline
 {Luminosity ($\pm$6.1\%)}  &  {Luminosity ($\pm$6.1\%)}  &  {Luminosity ($\pm$6.1\%)}\tabularnewline
 {Muon Reco. ($\pm$0.7\%)}  &  {Muon Reco. ($\pm$2.1\%)}  &  {Muon Reco. ($\pm$2.1\%)}\tabularnewline
 {PDF ($\pm 0.1$--2.7\%)}  &  {PDF ($\pm$0.2\%)}  &  {PDF ($\pm$0.2\%)}\tabularnewline
 {Timing gate ($\pm 2.8$--13\%)}  &  {$p_{T}$ resolution ($\pm$2.8\%)}  &  {$p_{T}$ resolution ($\pm$0.2\%)}\tabularnewline
 {Time simulation ($\pm 6$--13\%)}  &  {Timing gate ($\pm$2.4\%)}  &  {Timing gate (shape)}\tabularnewline
 {}  &  {Time simulation ($\pm$2.8\%)}  &  {Time simulation (shape)}\tabularnewline
 {}  &  {${\rm d} E/{\rm d} x$ corr. ($\pm$0.1\%)}  &  {${\rm d} E/{\rm d} x$ corr. ($\pm$0.02\%)}\tabularnewline
 {}  &  {${\rm d} E/{\rm d} x$ smearing ($\pm$0.6\%)}  &  {${\rm d} E/{\rm d} x$ smearing ($\pm$0.2\%)}\tabularnewline
 {}  &  {MDT Timing gate ($\pm$1.2\%)}  &  {Speed $\chi^2/{\mathrm{dof}}$ Corr. ($\pm$0.4\%)}\tabularnewline
 {}  &  {Speed Asym. Corr. ($\pm 1.06$--10.1\%)}  &  {}\tabularnewline
\end{tabular}
\end{ruledtabular}
\end{center}
\caption{\label{tab:systematics_combination_sig} Systematic uncertainties for signals for all three analyses.}
\end{table*}
%
\begin{table*}
\begin{center} 
\begin{ruledtabular}
\begin{tabular}{ c c c}
Pair (1.1 ${\mathrm{fb}}^{-1}$)  & Pair (5.2 ${\mathrm{fb}}^{-1}$)  & Single (5.2 ${\mathrm{fb}}^{-1}$)\tabularnewline
\hline
 {Bkgd. Norm. ($\pm$9-28\%)}  &  {Luminosity ($\pm$6.1\%)}  &  {${\rm d} E/{\rm d}x$ Corr. Uncertainty ($\pm$0.02)}\tabularnewline
 {}  &  {Muon Reco. ($\pm$2.1\%)}  &  {Bkgd. Norm.-$\langle \beta \rangle$ ($\pm$7.2\%)}\tabularnewline
 {}  &  {PDF ($\pm$0.3\%)}  &  {Bkgd. Norm.-$M_{T}$ ($\pm$2.2\%)}\tabularnewline
 {}  &  {$p_{T}$ resolution ($\pm$11.0\%)}  &  {}\tabularnewline
 {}  &  {Timing gate ($\pm$3.8\%)}  &  {}\tabularnewline
 {}  &  {Time simulation ($\pm$9.5\%)}  &  {}\tabularnewline
 {}  &  {${\rm d} E/{\rm d} x$ Corr. ($\pm$1.5\%)}  &  {}\tabularnewline
 {}  &  {${\rm d} E/{\rm d} x$ smearing ($\pm$4.9\%)}  &  {}\tabularnewline
 {}  &  {Bkgd. Norm. ($\pm$2.2\%)}  &  {}\tabularnewline
 {}  &  {Speed Asym. Corr. ($\pm$3.6\%)}  &  {}\tabularnewline
\end{tabular}
\end{ruledtabular}
\end{center}
\caption{\label{tab:systematics_combination_bkgd} Systematic uncertainties 
for background events for all three analyses.}
\end{table*}
All systematic uncertainties except the luminosity uncertainty~\cite{Lum uncert} 
for the two searches for pairs of CMLLPs are treated as uncorrelated. The recent 
search for CMLLP pairs and the search for a single CMLLP are based on the same 
dataset. Therefore, the systematic uncertainties for these two analyses are correlated except for the uncertainties on background normalizations (the background samples are different in the two analyses). The systematic uncertainties for the 
single CMLLP dataset after removal of the events containing CMLLP pairs are the same as those in the search for a single CMLLP analysis described in Sec.~VII D. The shape systematic uncertainties however are updated after the removal of the common events. 

\subsection{Results}

\begin{figure*}
\begin{center}
\scalebox{0.4}{\includegraphics{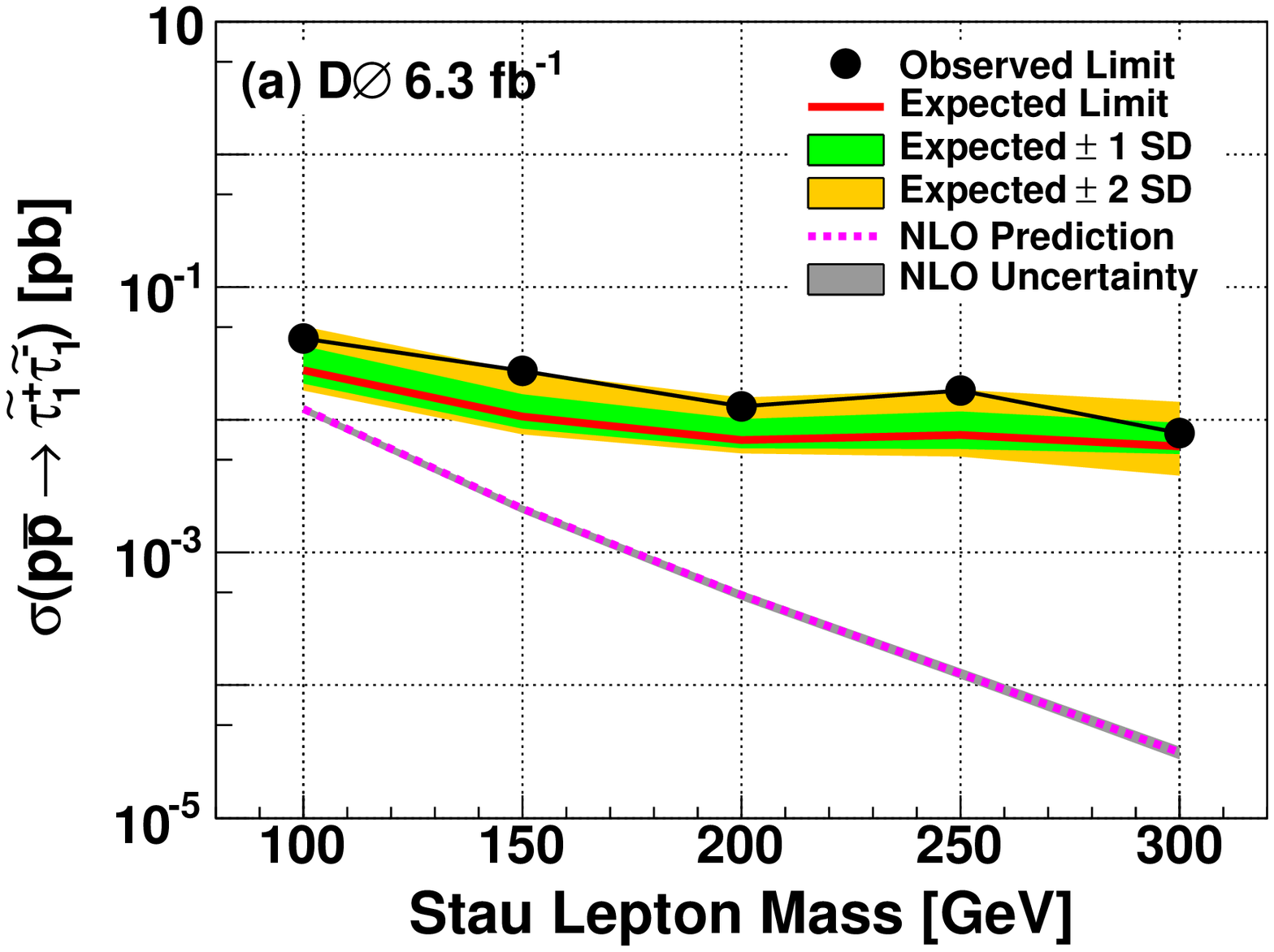}}
\scalebox{0.4}{\includegraphics{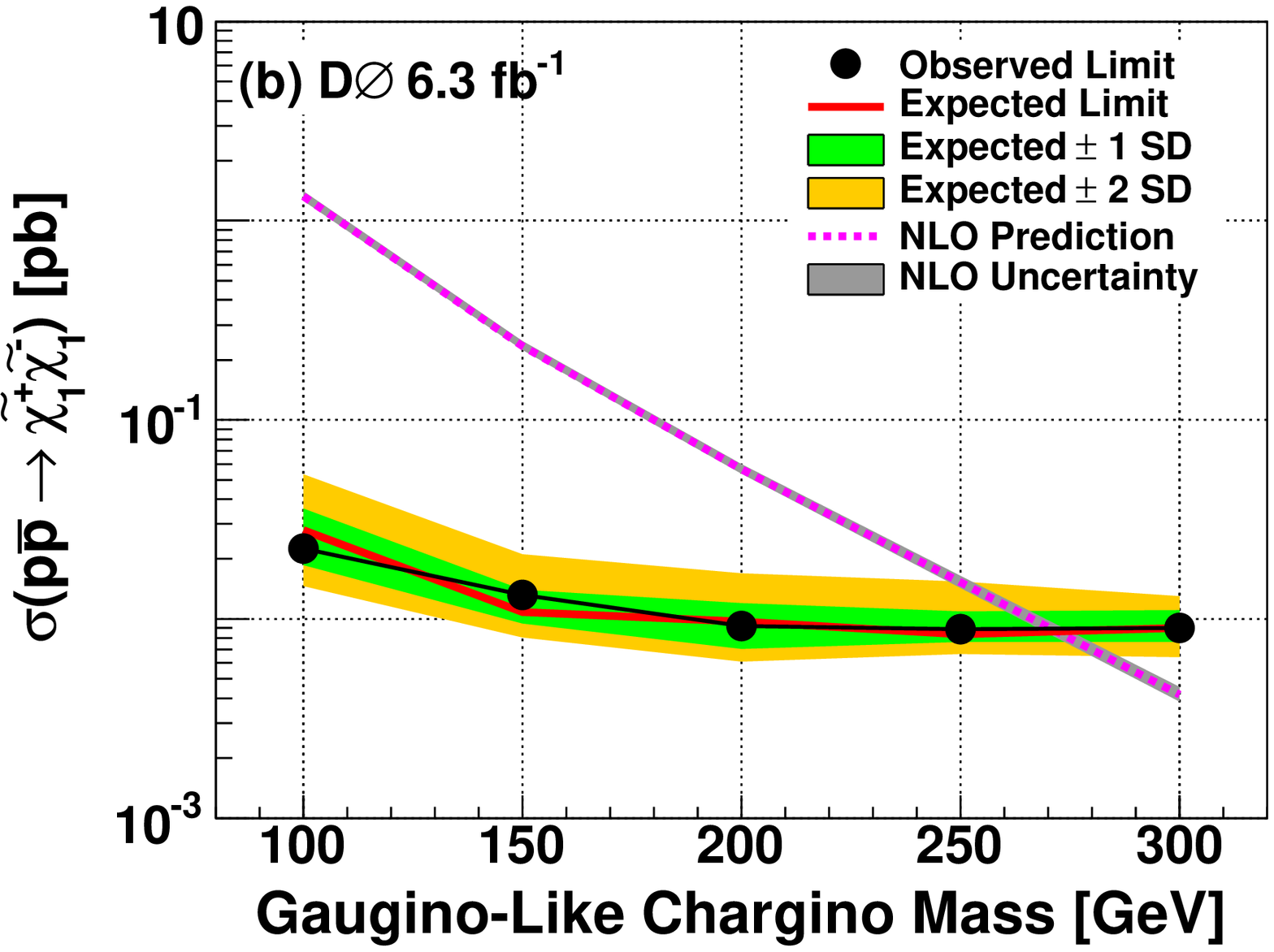}}
\scalebox{0.4}{\includegraphics{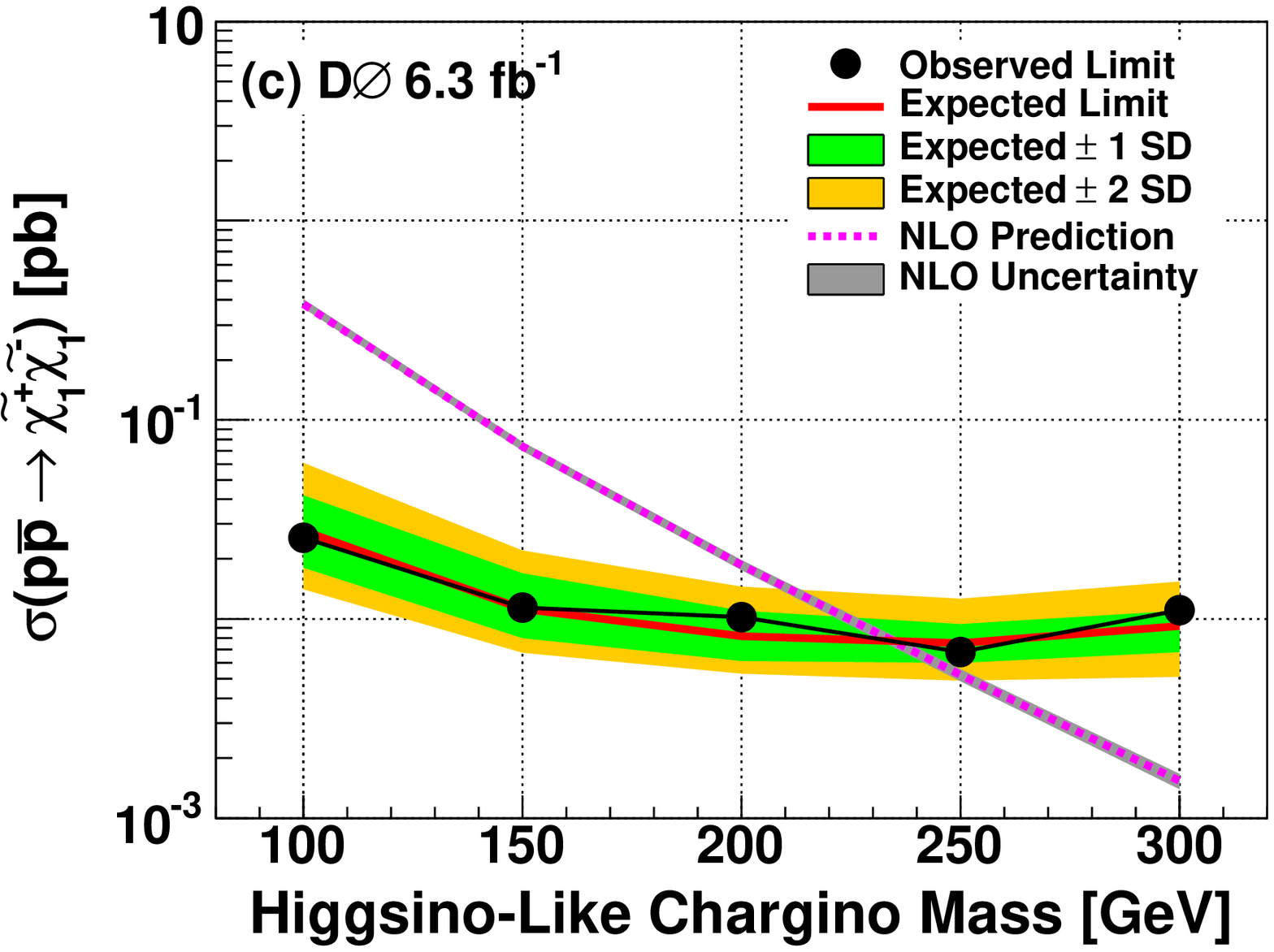}}
\caption{\label{fig:combined_limits}(color online) Combined limits at 95\% C.L. on production cross 
sections of a pair of stau leptons, gaugino-like charginos, and higgsino-like charginos as a function of their masses with Run IIa and Run IIb data. $\pm 1$ SD and $\pm 2$ SD are the 1 and 2 standard deviation bands respectively around the expected limit curves.}
\end{center}
\end{figure*}

Combined 95\% C.L. cross section limits for stau leptons, gaugino-like,  and higgsino-like charginos are shown in 
Table~\ref{tab:Combination_LimitSummary}, and Fig.~\ref{fig:combined_limits}. Using the observed cross section and the
theoretical NLO cross section, we set mass limits of 278 GeV for 
gaugino-like charginos and 244 GeV for higgsino-like charginos. 
Using the intersection of the $-1$ ($+1$) standard deviation ($\sigma$) band on the NLO 
cross section shifts the mass limits down (up) by $\sim$1 GeV for the charginos. 
We do not have enough sensitivity to set a limit on the 
stau lepton mass and therefore we set an upper limit on production cross sections of 
stau leptons to be 0.04 pb to 0.008 pb for the stau lepton mass range of 100 to 300 GeV.
\begin{table*}
\begin{center} 
\begin{ruledtabular}
\begin{tabular}{c c c c}
Mass (GeV)  & NLO cross section (pb)  
& ${\sigma}_{95}^{obs}$ (pb) & ${\sigma}_{95}^{exp}$ (pb)\tabularnewline
\hline
Stau lepton & & & \\

  { 100}  &  {$0.0120^{+0.0006}_{-0.0008}$}  &  {$0.041$}  
&  {$0.024^{+0.013}_{-0.006}$}\tabularnewline
 { 150}  &  {$0.0021^{+0.0001}_{-0.0002}$}  &  {$0.023$}  
&  {$0.011^{+0.004}_{-0.003}$}\tabularnewline
 { 200}  &  {$0.00050^{+0.00003}_{-0.00002}$}  &  {$0.013$}  
&  {$0.008^{+0.003}_{-0.001}$}\tabularnewline
 { 250}  &  {$0.00010^{+0.00001}_{-0.00001}$}  &  {$0.017$}  
&  {$0.008^{+0.004}_{-0.001}$}\tabularnewline
 { 300}  &  {$0.000030^{+0.000003}_{-0.000004}$}  &  {$0.008$}  
&  {$0.006^{+0.003}_{-0.002}$}\tabularnewline

Gaugino-like chargino  & & & \\

  { 100}  &  {$1.33^{+0.08}_{-0.07}$}  &  {$0.023$}  
&  {$0.028^{+0.008}_{-0.010}$}\tabularnewline
 { 150}  &  {$0.240^{+0.014}_{-0.010}$}  &  {$0.013$}  
&  {$0.011^{+0.004}_{-0.001}$}\tabularnewline
 { 200}  &  {$0.0570^{+0.0034}_{-0.0030}$}  &  {$0.009$}  
&  {$0.010^{+0.003}_{-0.002}$}\tabularnewline
 { 250}  &  {$0.0150^{+0.0011}_{-0.0010}$}  &  {$0.009$}  
&  {$0.008^{+0.003}_{-0.001}$}\tabularnewline
 { 300}  &  {$0.0042^{+0.0004}_{-0.0003}$}  &  {$0.009$}  
&  {$0.009^{+0.002}_{-0.001}$}\tabularnewline

Higgsino-like chargino  & & & \\

  { 100}  &  {$0.380^{+0.023}_{-0.017}$}  &  {$0.026$}  
&  {$0.028^{+0.014}_{-0.009}$}\tabularnewline
 { 150}  &  {$0.074^{+0.0040}_{-0.0038}$}  &  {$0.011$}  
&  {$0.011^{+0.005}_{-0.003}$}\tabularnewline
 { 200}  &  {$0.0190^{+0.0012}_{-0.0010}$}  &  {$0.010$}  
&  {$0.008^{+0.003}_{-0.002}$}\tabularnewline
 { 250}  &  {$0.00530^{+0.00035}_{-0.0004}$}  &  {$0.007$}  
&  {$0.008^{+0.002}_{-0.001}$}\tabularnewline
 { 300}  &  {$0.0015^{+0.0001}_{-0.0001}$}  &  {$0.011$}  
&  {$0.009^{+0.001}_{-0.001}$}\tabularnewline
\end{tabular}
\end{ruledtabular}
\end{center}
\caption{\label{tab:Combination_LimitSummary} Combined 95\% C.L. cross 
section limits for stau leptons, gaugino-like, and higgsino-like charginos from the three search strategies.}
\end{table*}

\section{Summary}

A search for CMLLPs has been performed with the D0 detector with 5.2 
${\mathrm{fb}}^{-1}$ integrated luminosity using two different strategies: a search for a pair 
of identified CMLLPs and a search for a single identified CMLLP in events expected to contain a pair of CMLLP's. These two searches 
are combined with the earlier search for CMLLP pairs with 1.1 ${\mathrm{fb}}^{-1}$ 
integrated luminosity. 
 We use the central value of the theoretical cross section predictions to set 
 95\% C.L. lower limits on the masses of top squarks and charginos and on the cross section of stau leptons.

Using the combination of the three searches we set mass limits of 278 GeV 
for gaugino-like charginos and 244 GeV for higgsino-like charginos.
 For stau leptons we set an upper limit of 0.04 pb to 0.008 pb on the production cross section for the mass range of 100 to 300 GeV.

In the search for single CMLLPs we exclude  
top squarks with masses below 285 GeV with a charge flipping probability of 38\%.
A combination of the analyses is not performed for the top squarks since  
improvement in the stop limit by combining the searches 
is negligible. 

 Limits on the chargino cross 
sections are the most restrictive limits to date, with 
about an order of magnitude improvement over the previous D0 result with  1.1 ${\mathrm{fb}}^{-1}$ integrated luminosity~\cite{D0 CMSP PRL Run IIa}. The improvement in both the pair and the single CMLLP 
searches over the previous results is due to the increased luminosity 
as well as the additional use of another key variable, the measured ${\rm d} E/{\rm d} x$ of the tracks.
\begin{acknowledgments}
We thank the staffs at Fermilab and collaborating institutions,
and acknowledge support from the
DOE and NSF (USA);
CEA and CNRS/IN2P3 (France);
MON, Rosatom and RFBR (Russia);
CNPq, FAPERJ, FAPESP and FUNDUNESP (Brazil);
DAE and DST (India);
Colciencias (Colombia);
CONACyT (Mexico);
NRF (Korea);
FOM (The Netherlands);
STFC and the Royal Society (United Kingdom);
MSMT and GACR (Czech Republic);
BMBF and DFG (Germany);
SFI (Ireland);
The Swedish Research Council (Sweden);
and
CAS and CNSF (China).
\end{acknowledgments}

\newpage

\appendix{\section{BDT distributions}}

\begin{figure*}
\begin{center}
\scalebox{0.35}{\includegraphics{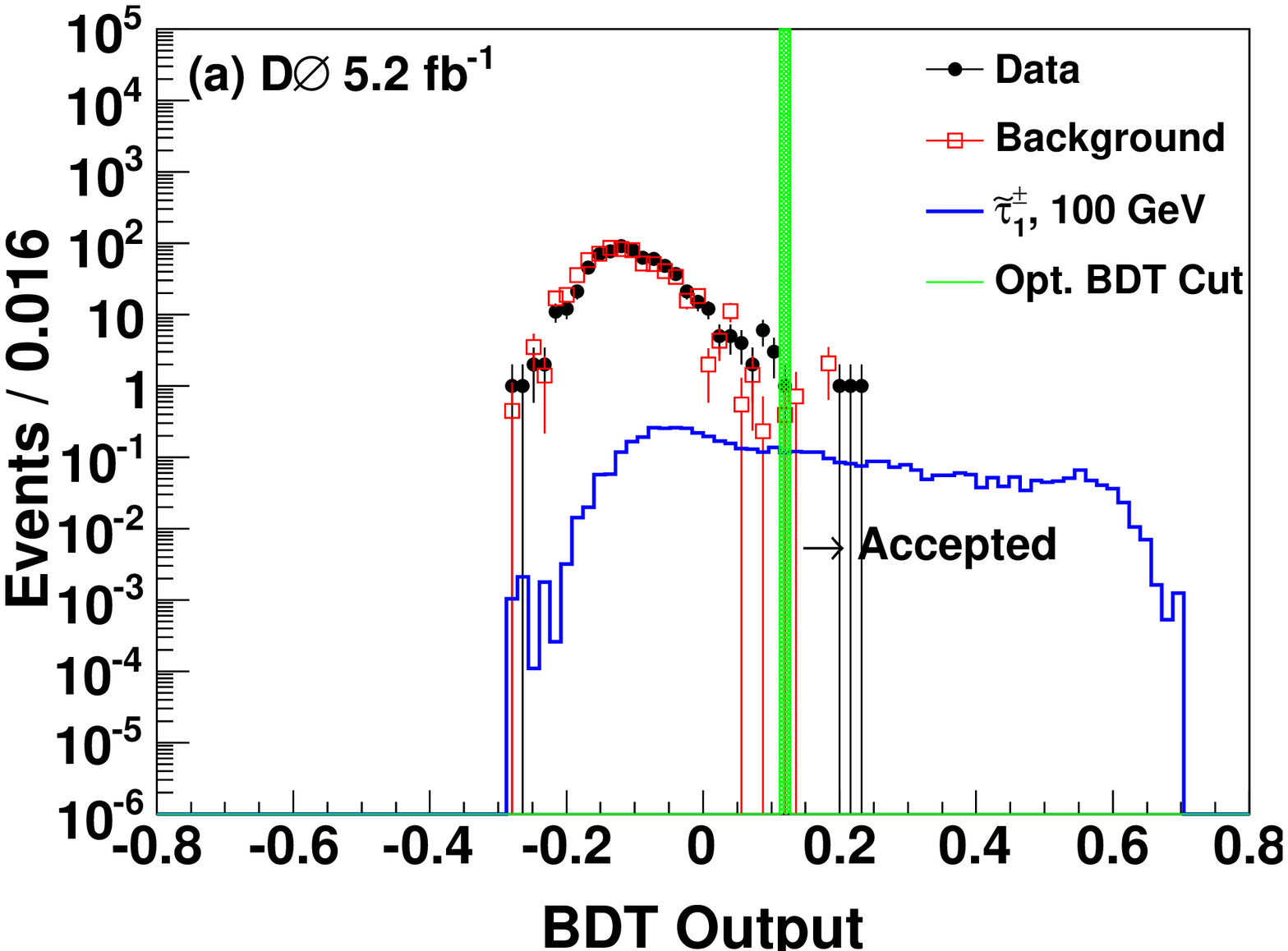}}
\scalebox{0.35}{\includegraphics{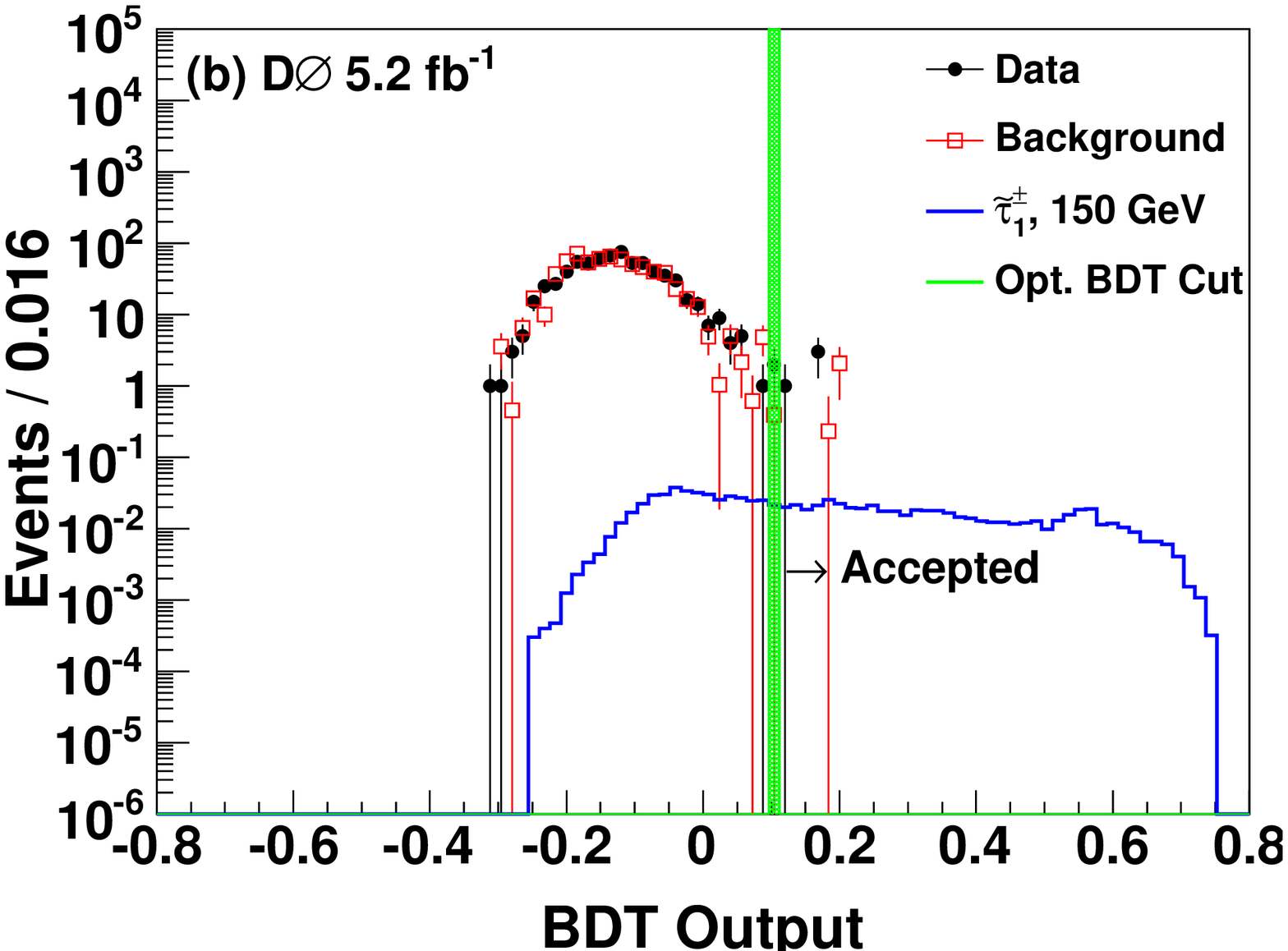}}
\scalebox{0.35}{\includegraphics{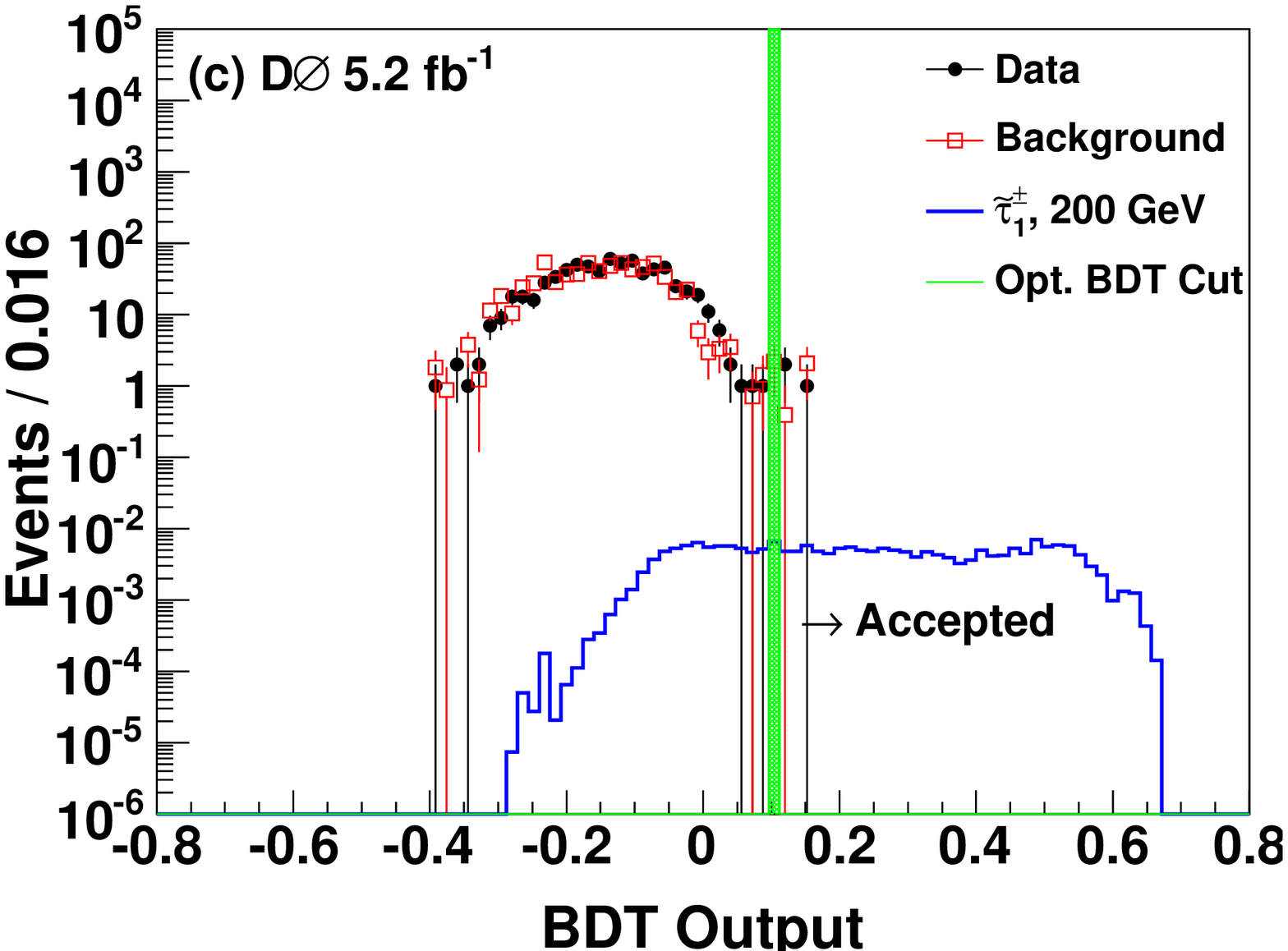}}
\scalebox{0.35}{\includegraphics{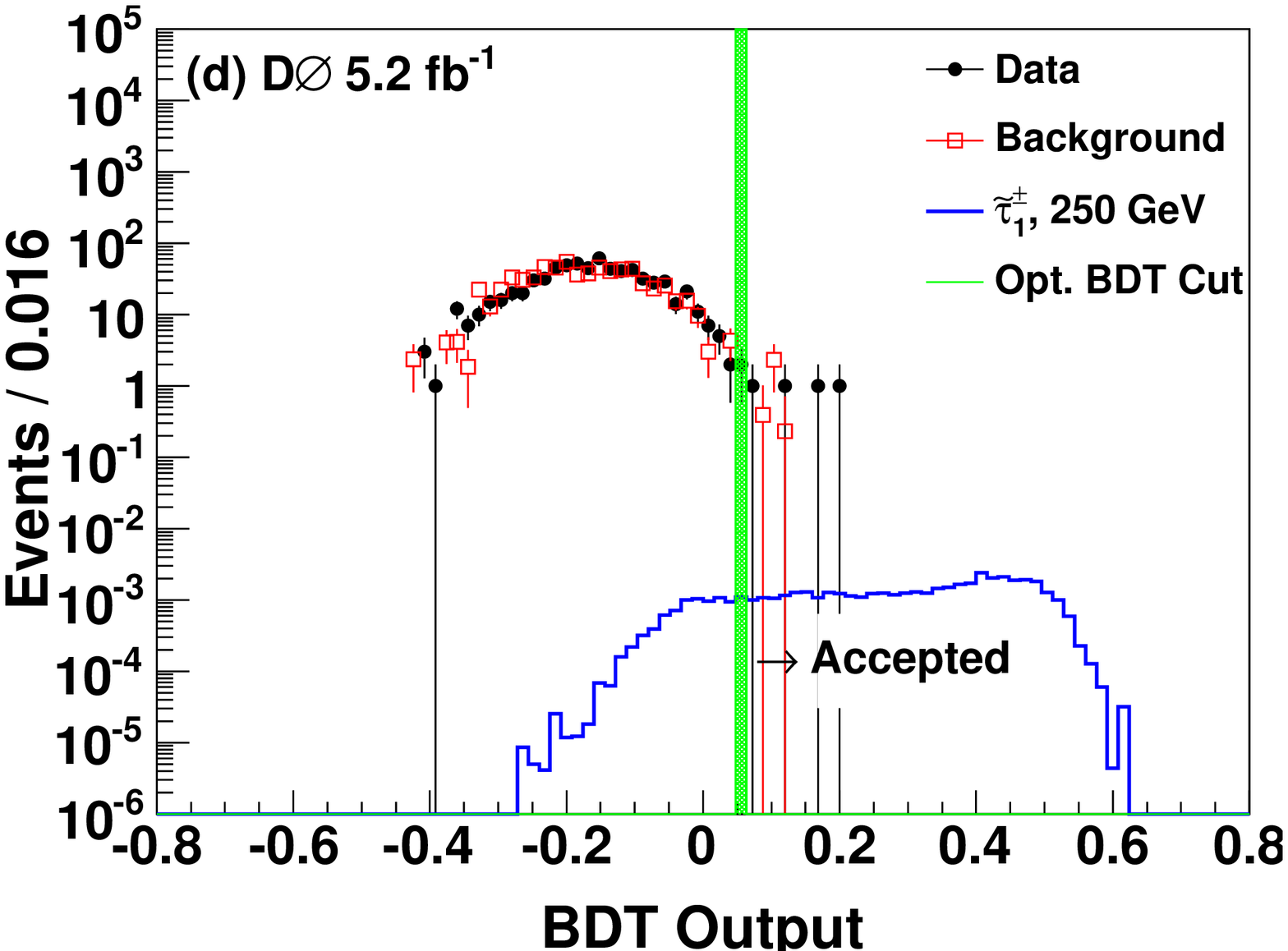}}
\scalebox{0.35}{\includegraphics{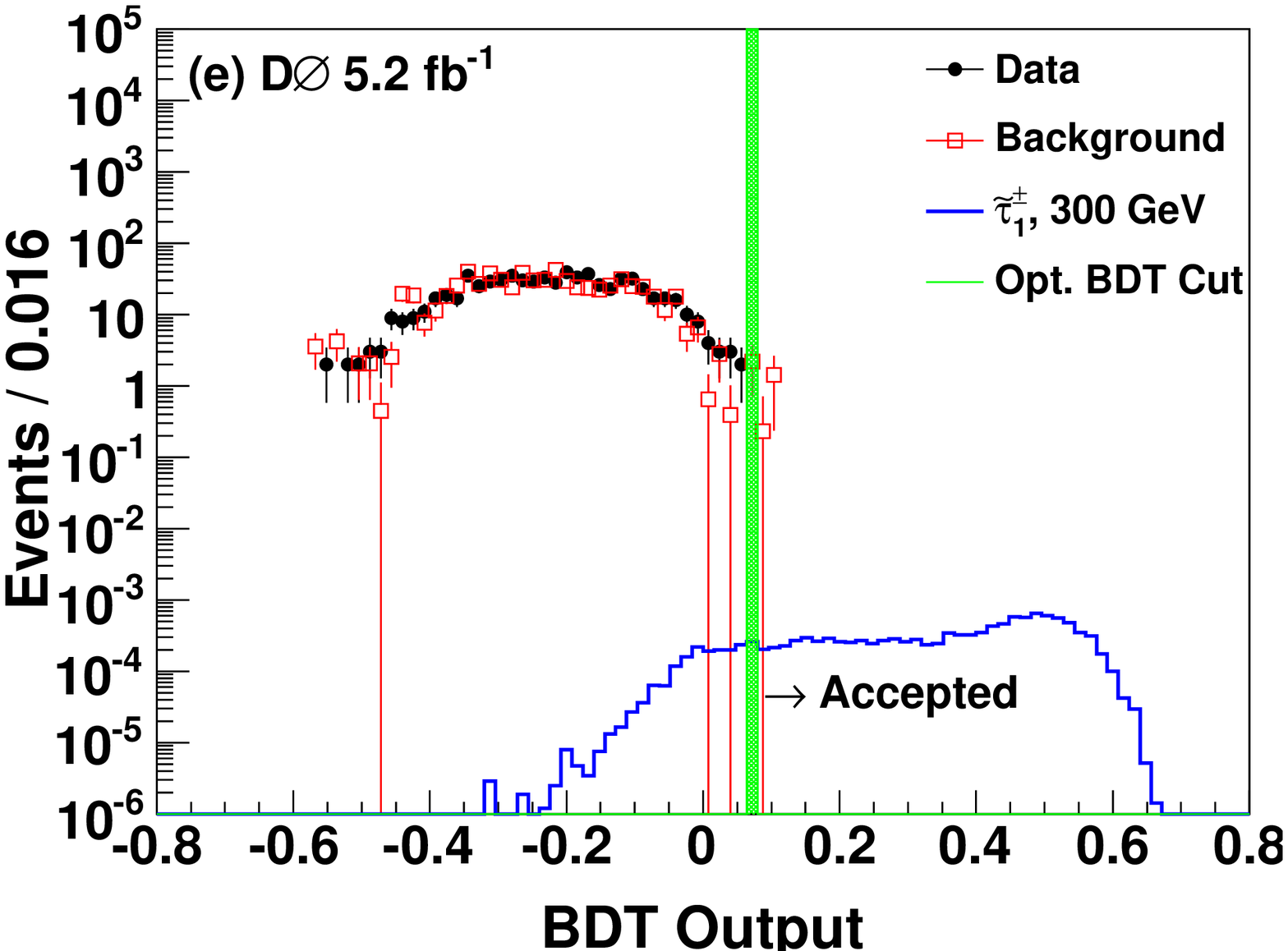}}
\caption{\label{fig:p20pair_BDT_stau}(color online) BDT-output distributions for stau masses 100-300 GeV in 50 GeV steps for the search for a CMLLP pair with Run IIb data. 
The distributions are normalized to the expected number of events. The selection requirement on the BDT value is shown with a green vertical line.}
\end{center}
\end{figure*}

\begin{figure*}
\begin{center}
\scalebox{0.35}{\includegraphics{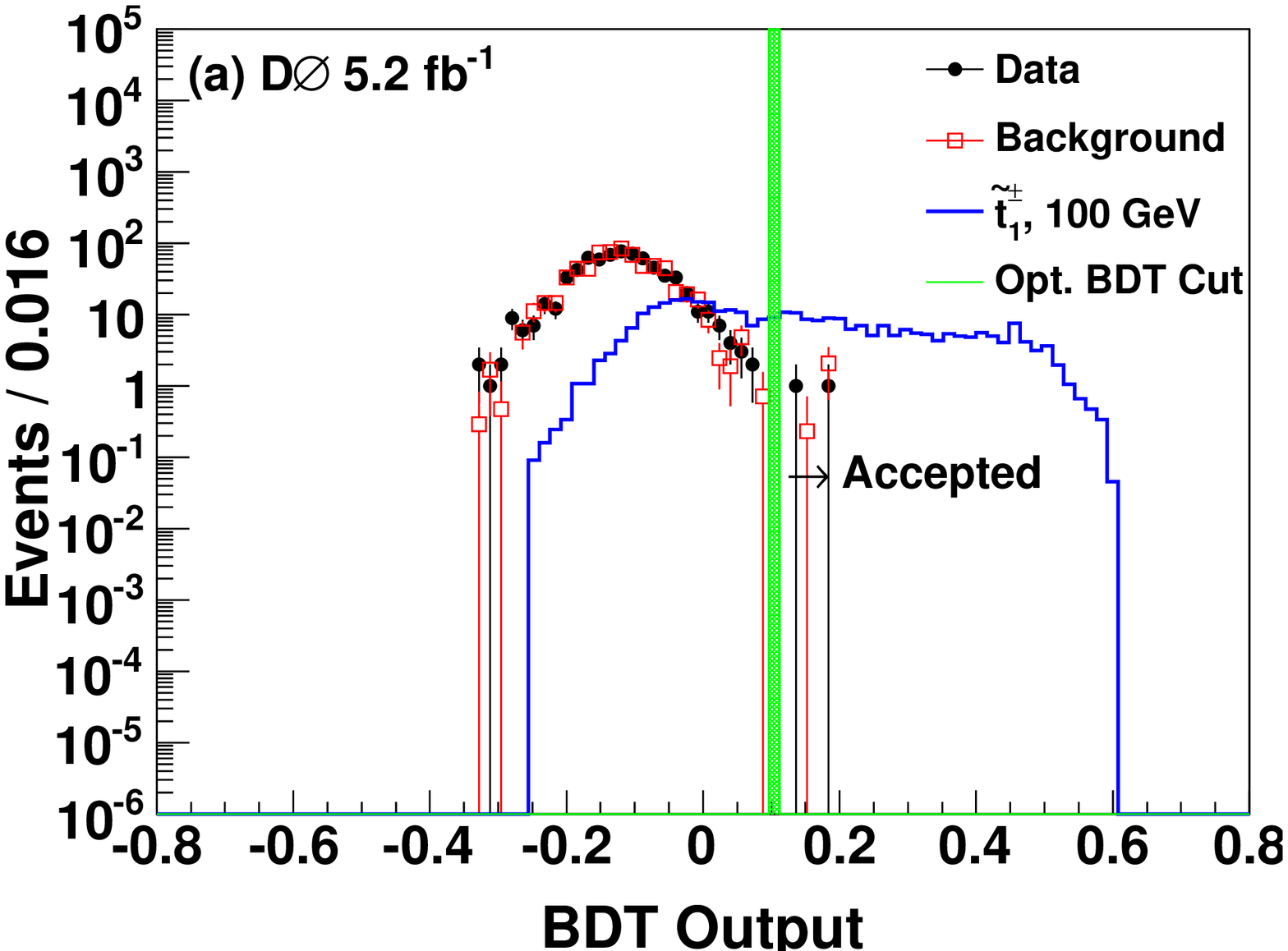}}
\scalebox{0.35}{\includegraphics{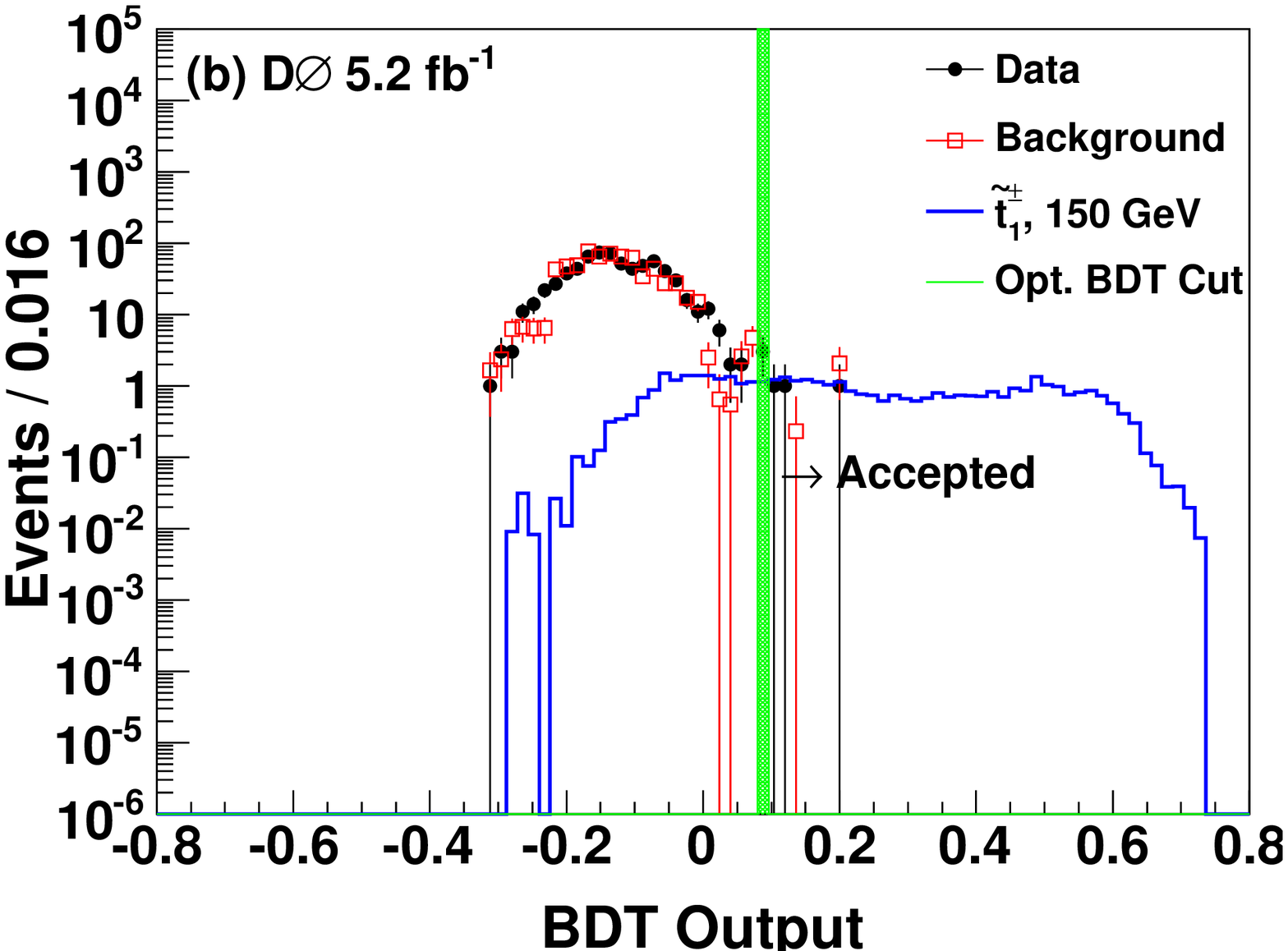}}
\scalebox{0.35}{\includegraphics{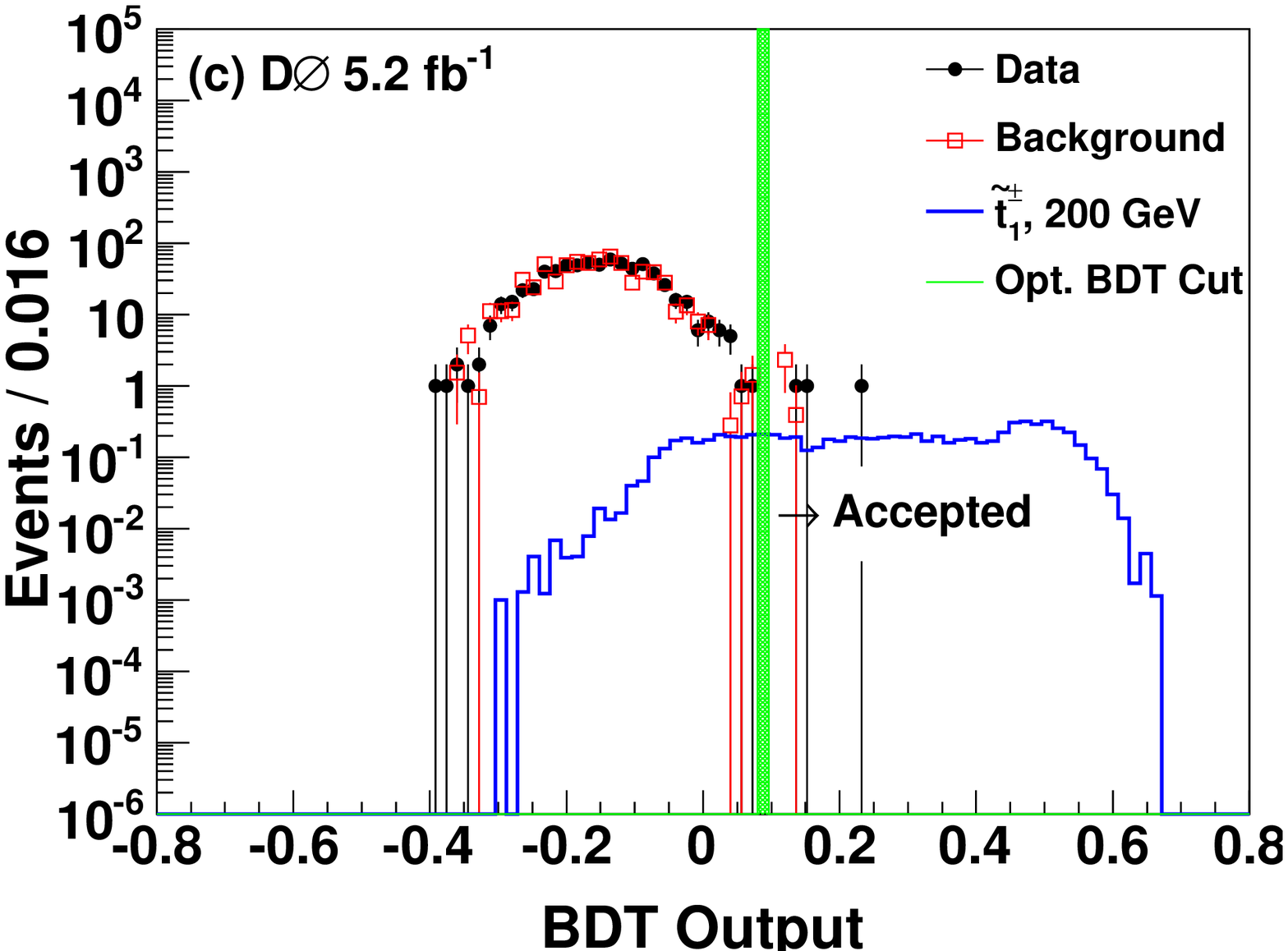}}
\scalebox{0.35}{\includegraphics{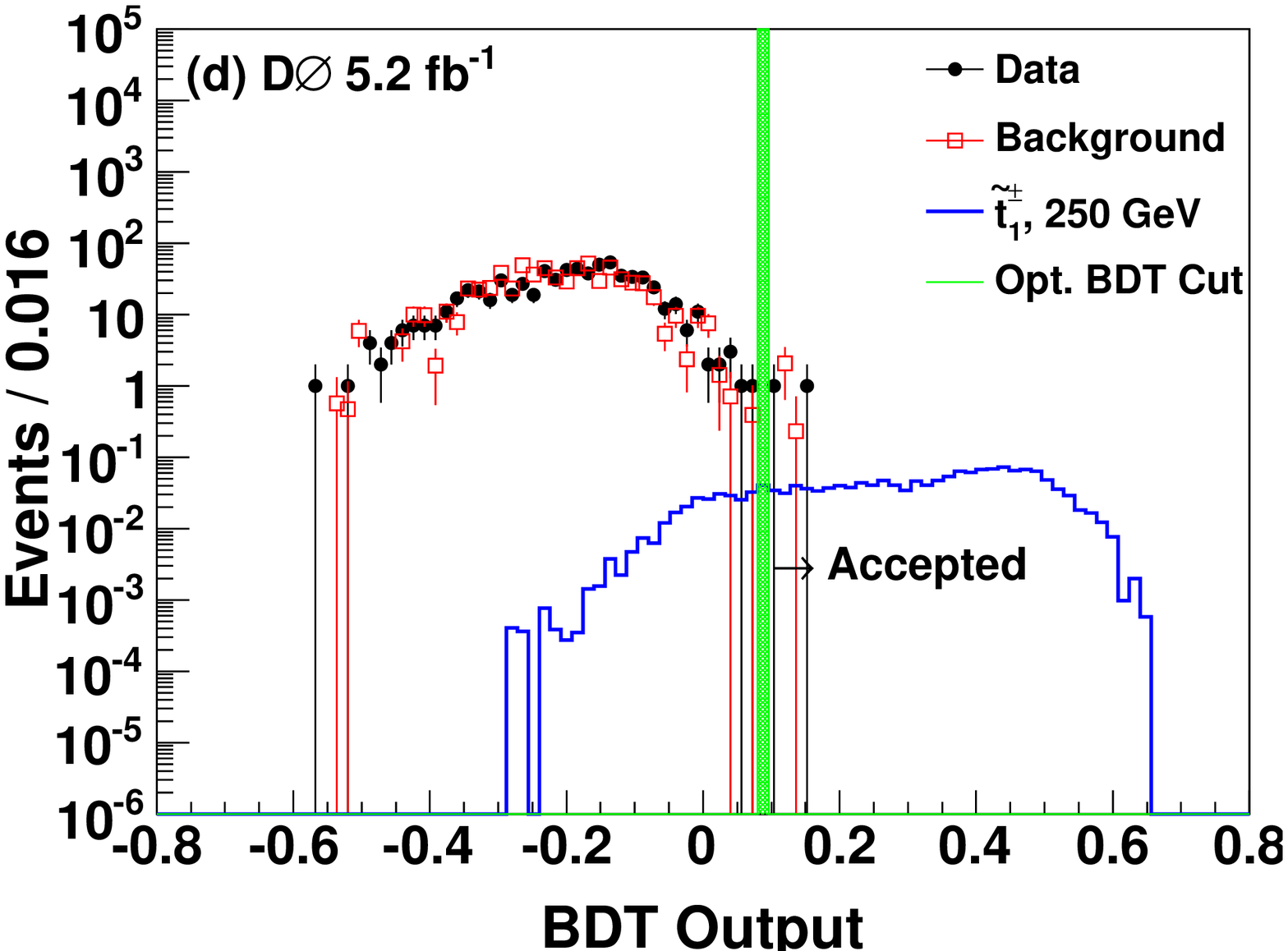}}
\scalebox{0.35}{\includegraphics{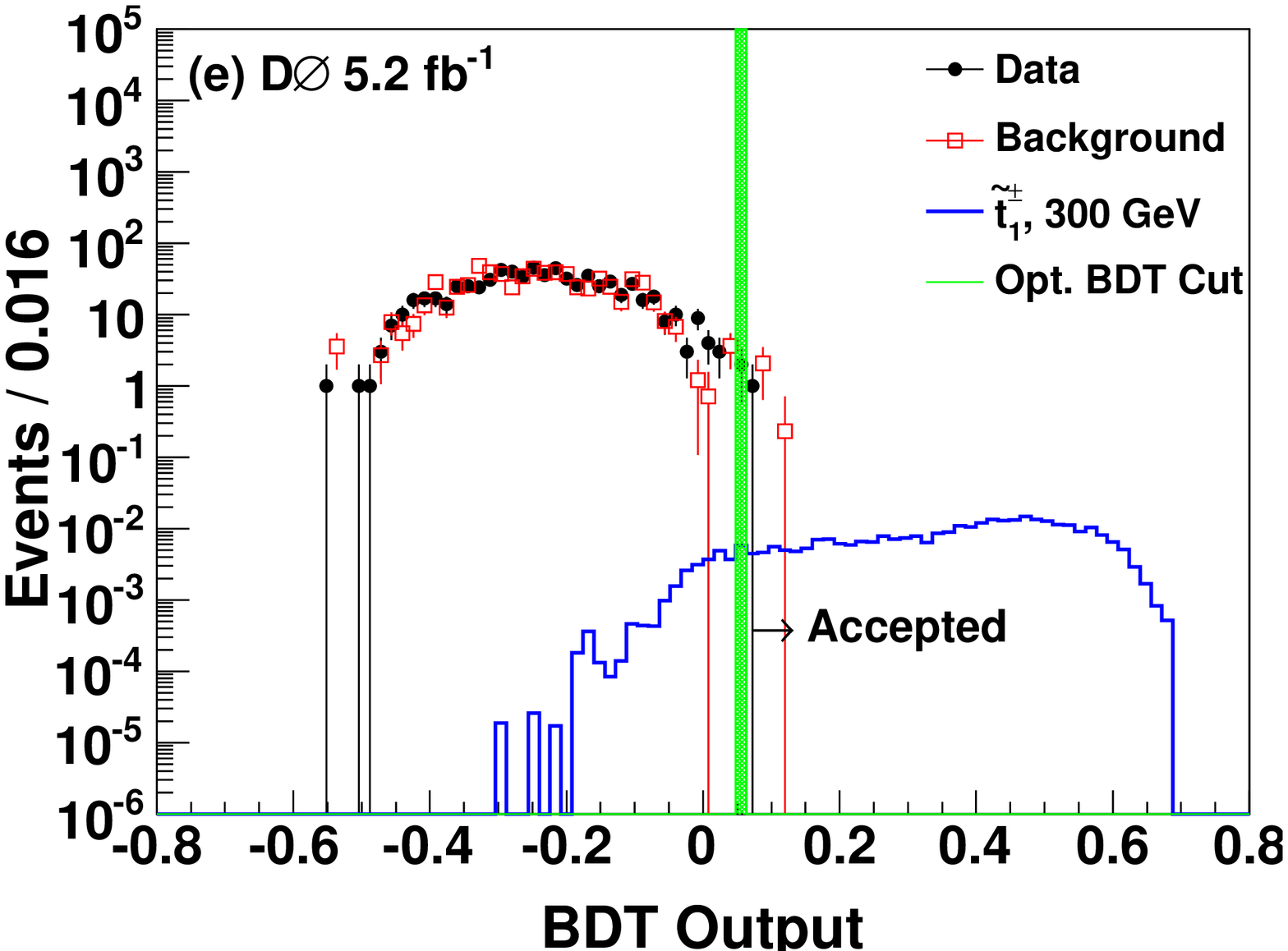}}
\scalebox{0.35}{\includegraphics{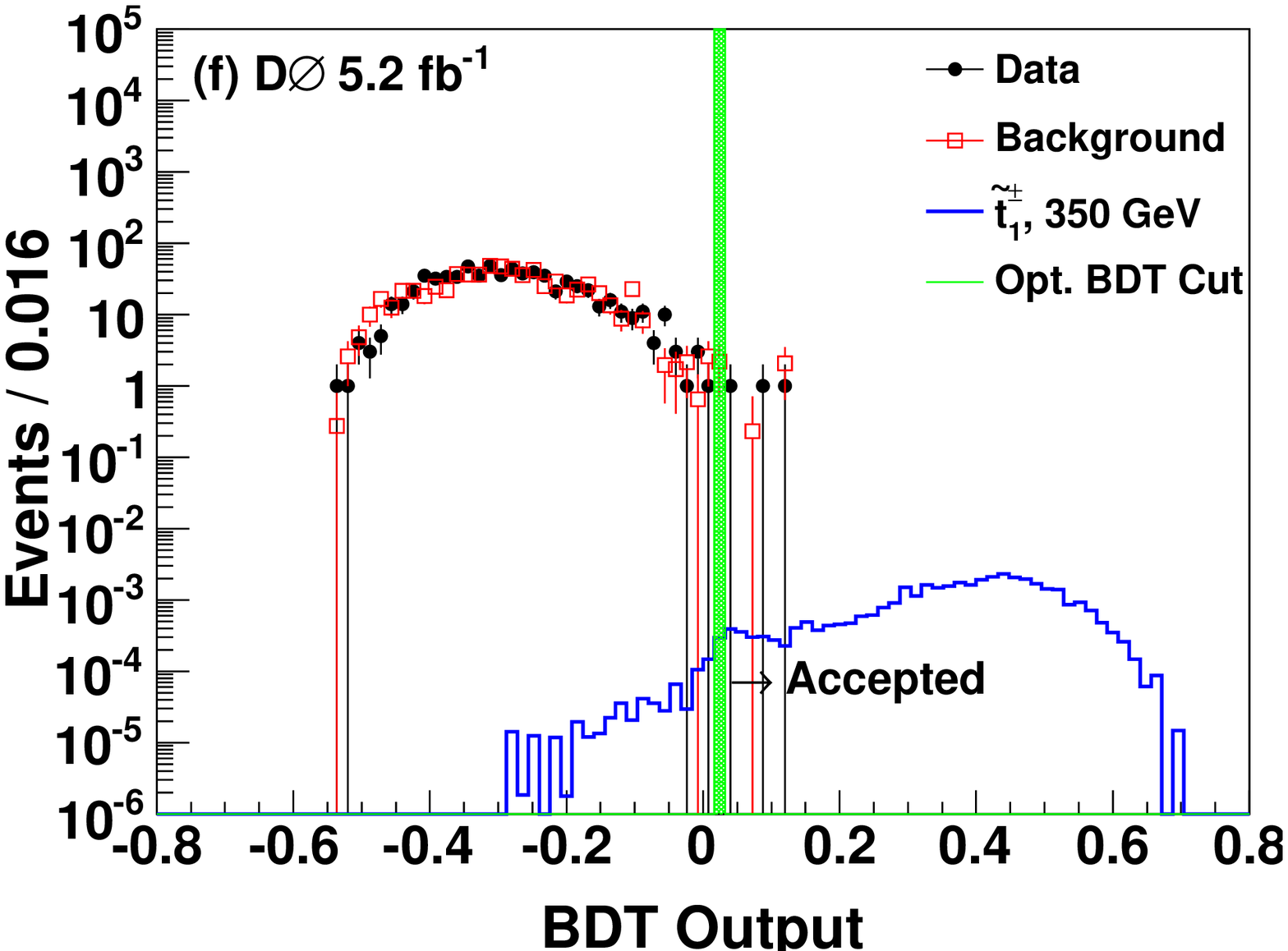}}
\scalebox{0.35}{\includegraphics{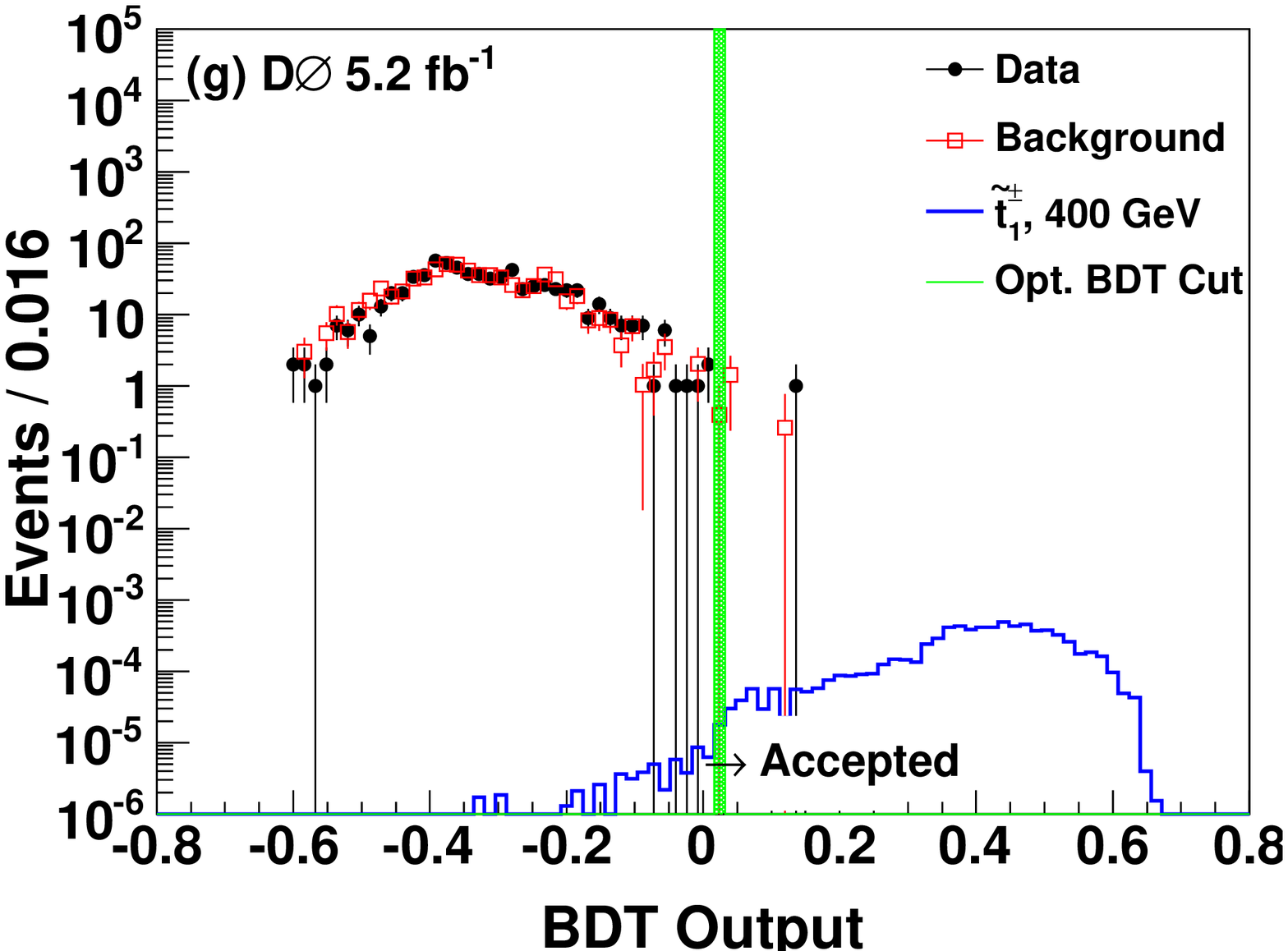}}
\caption{\label{fig:p20pair_BDT_stop}(color online) BDT-output distributions for top squark masses 100-400 GeV in 50 GeV steps for the search for a CMLLP pair with 
the Run IIb data. Distributions are normalized to the expected number of events.
Selection requirement on the BDT value is shown with a green vertical line.}
\end{center}
\end{figure*}

\begin{figure*}
\begin{center}
\scalebox{0.35}{\includegraphics{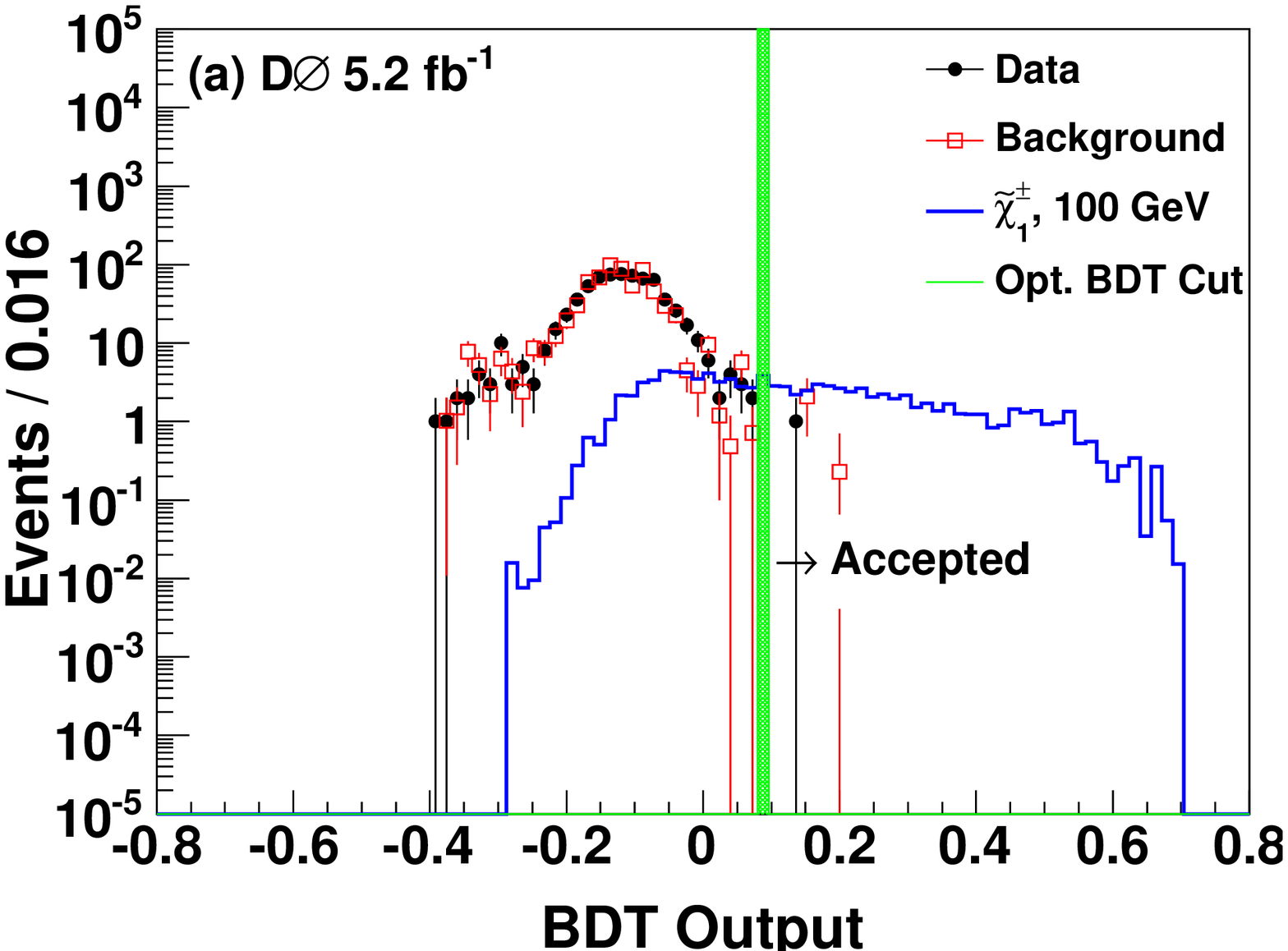}}
\scalebox{0.35}{\includegraphics{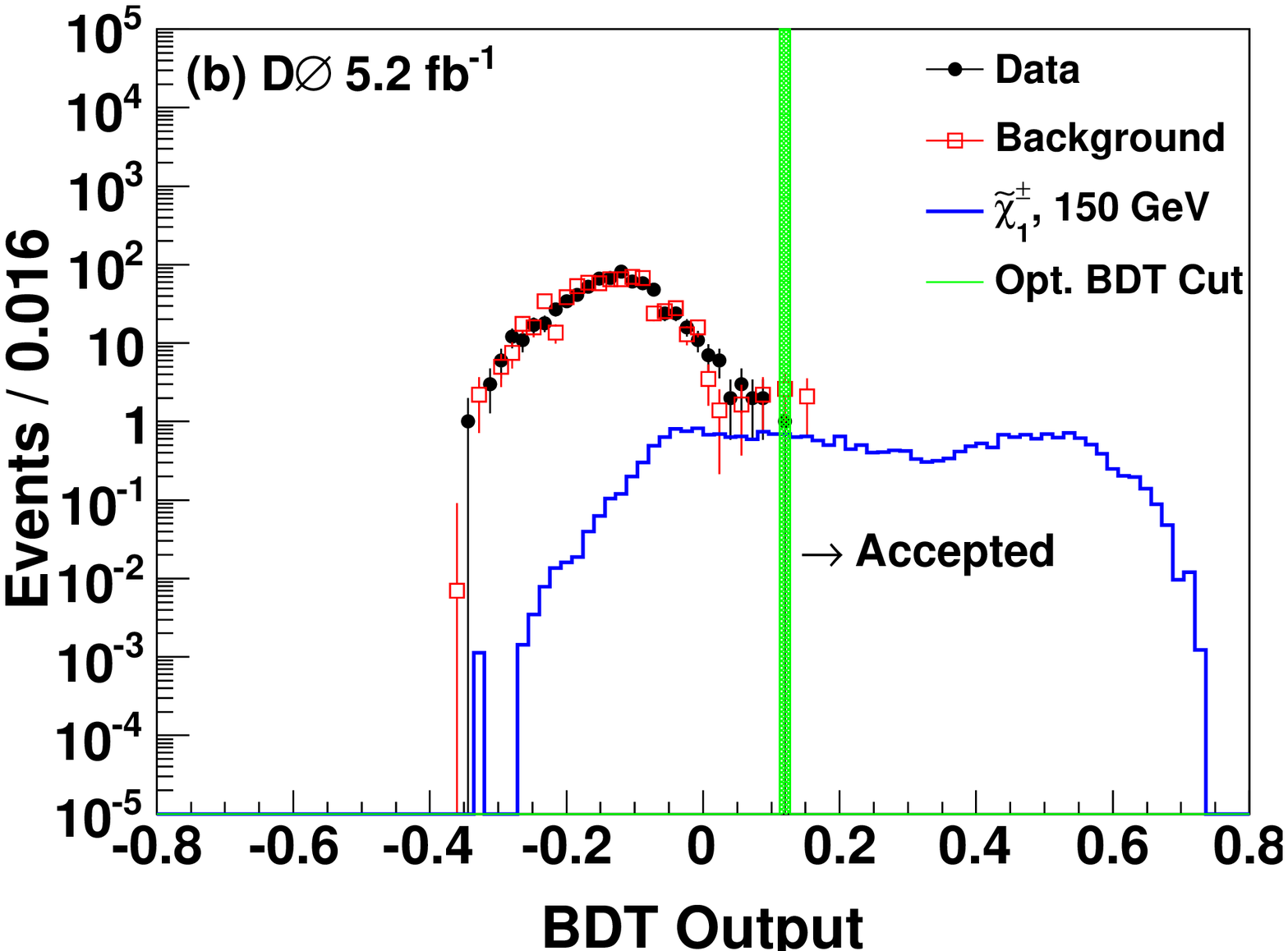}}
\scalebox{0.35}{\includegraphics{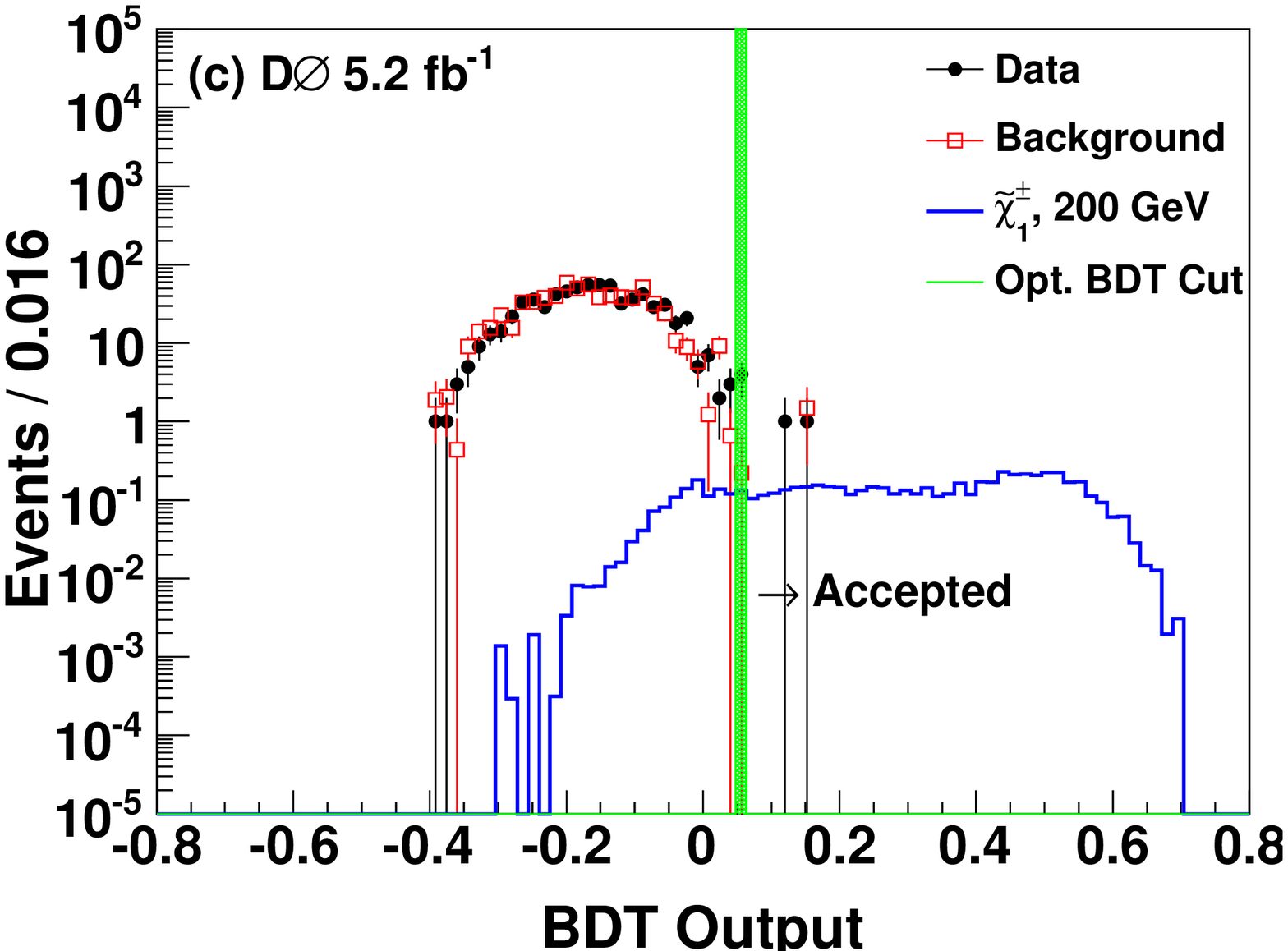}}
\scalebox{0.35}{\includegraphics{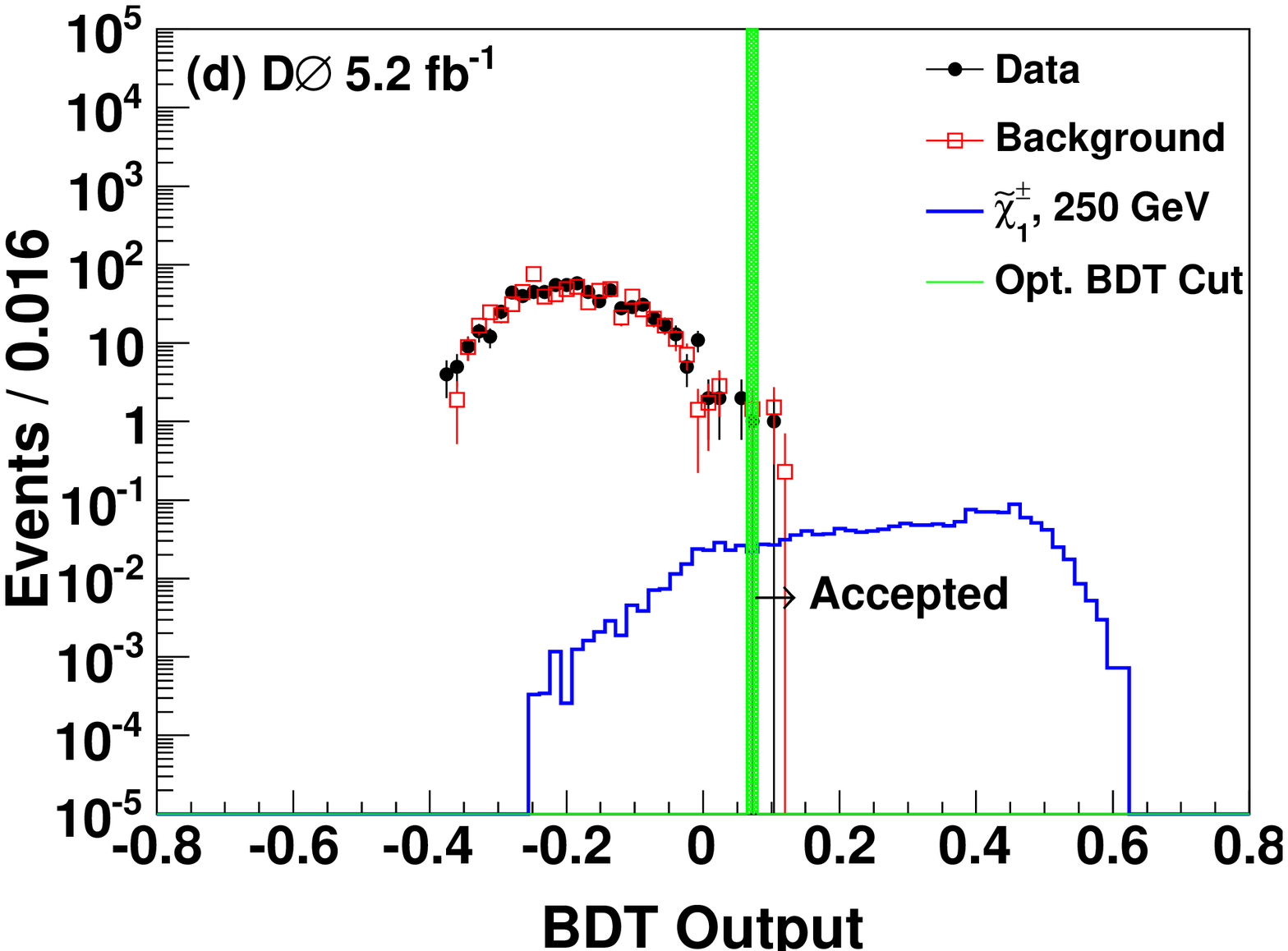}}
\scalebox{0.35}{\includegraphics{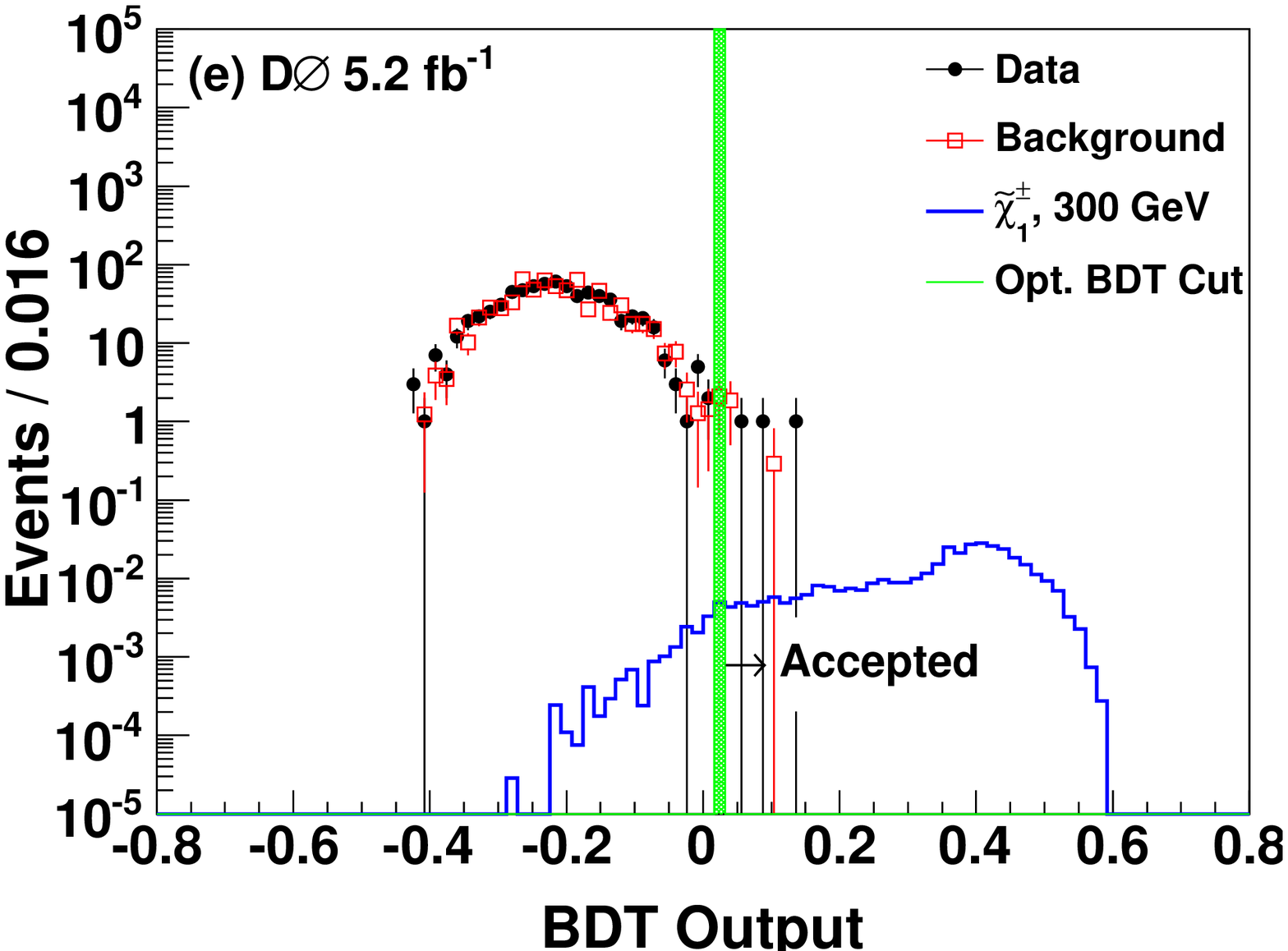}}
\caption{\label{fig:p20pair_BDT_charginoH}(color online) BDT-output distributions for higgsino-like chargino masses 100-300 GeV in 50 GeV steps for the search for a CMLLP pair with the 
Run IIb data. Distributions are normalized to the expected number of events.
Selection requirement on the BDT is shown with a green vertical line.}
\end{center}
\end{figure*}

\begin{figure*}
\begin{center}
\scalebox{0.35}{\includegraphics{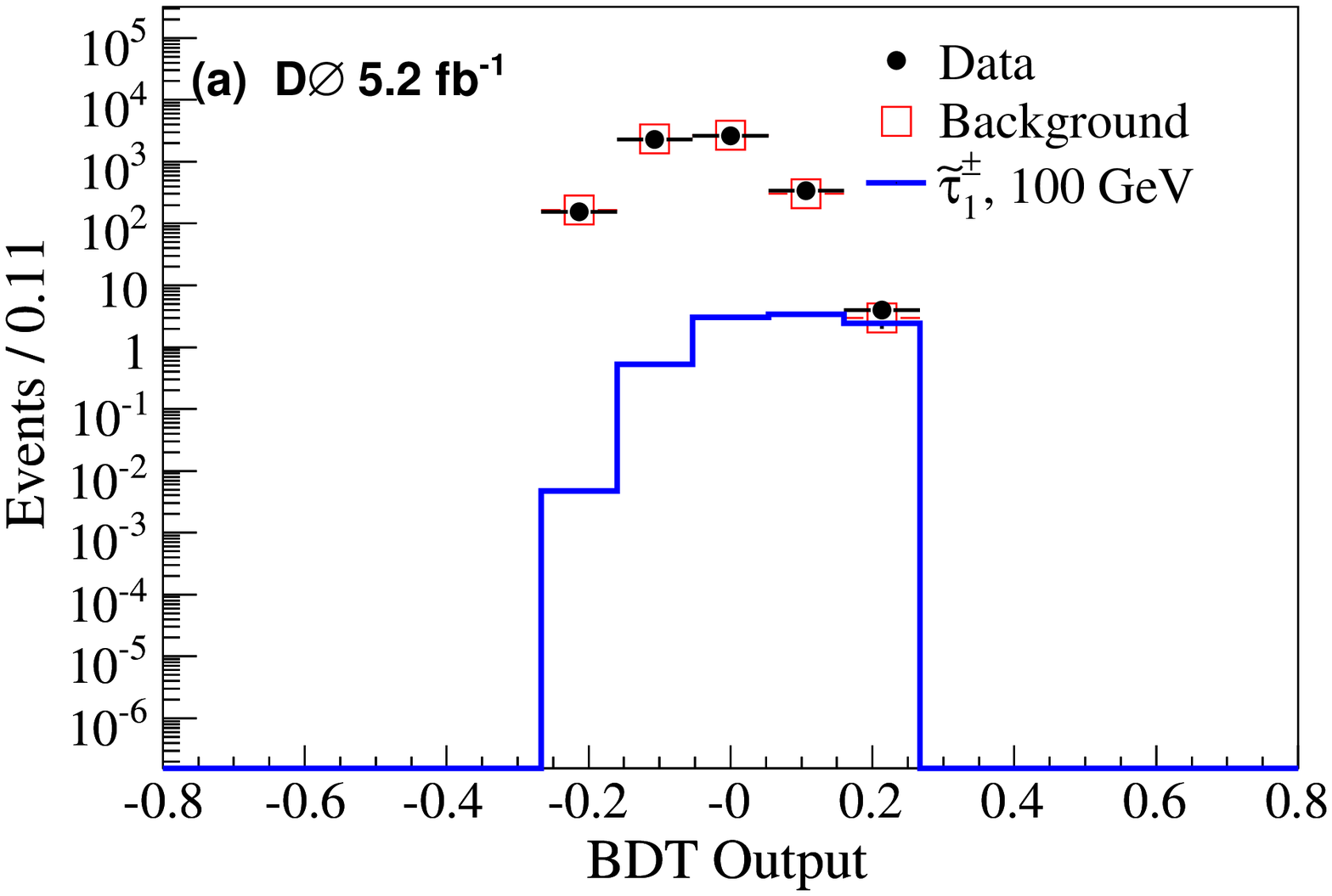}}
\scalebox{0.35}{\includegraphics{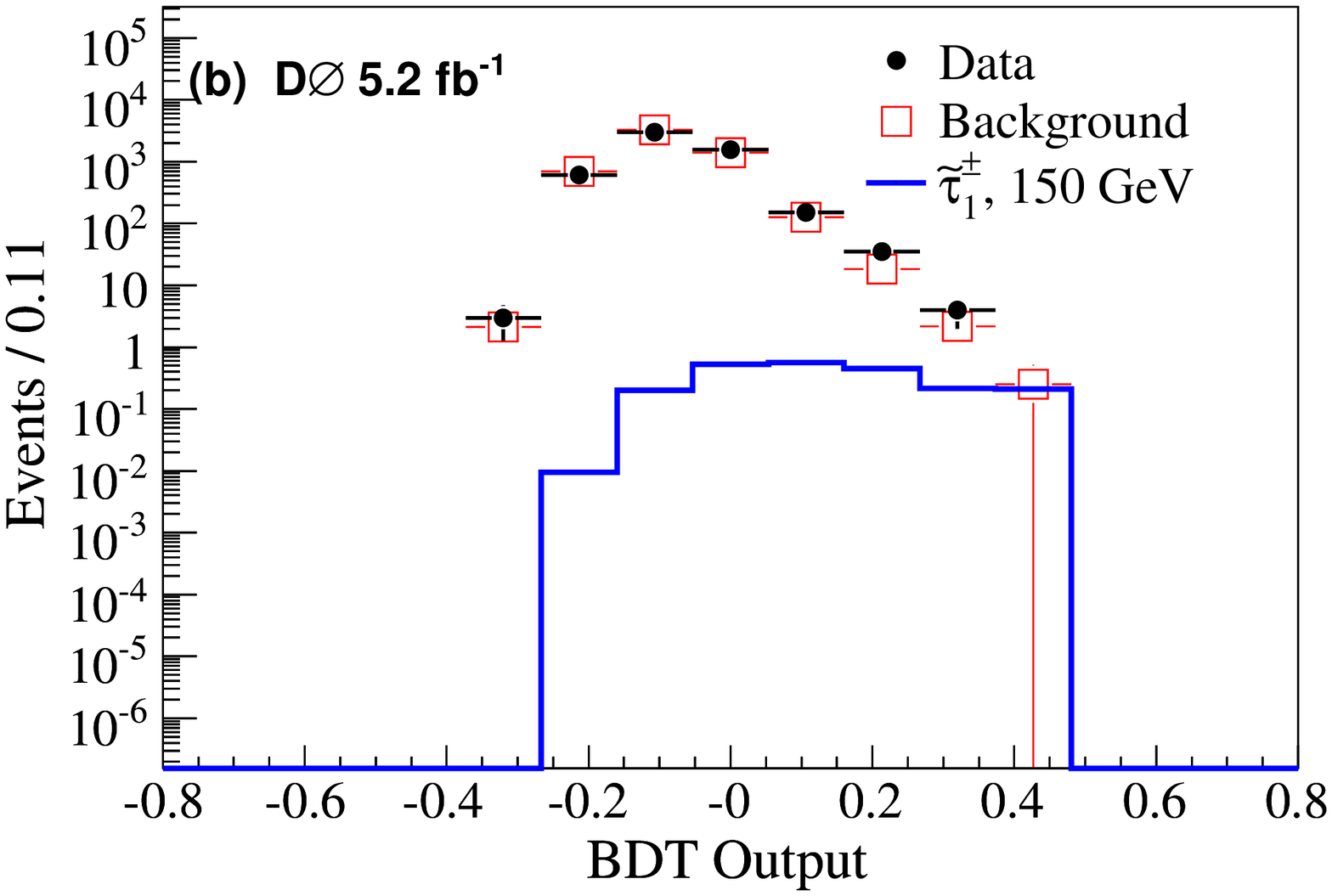}}
\scalebox{0.35}{\includegraphics{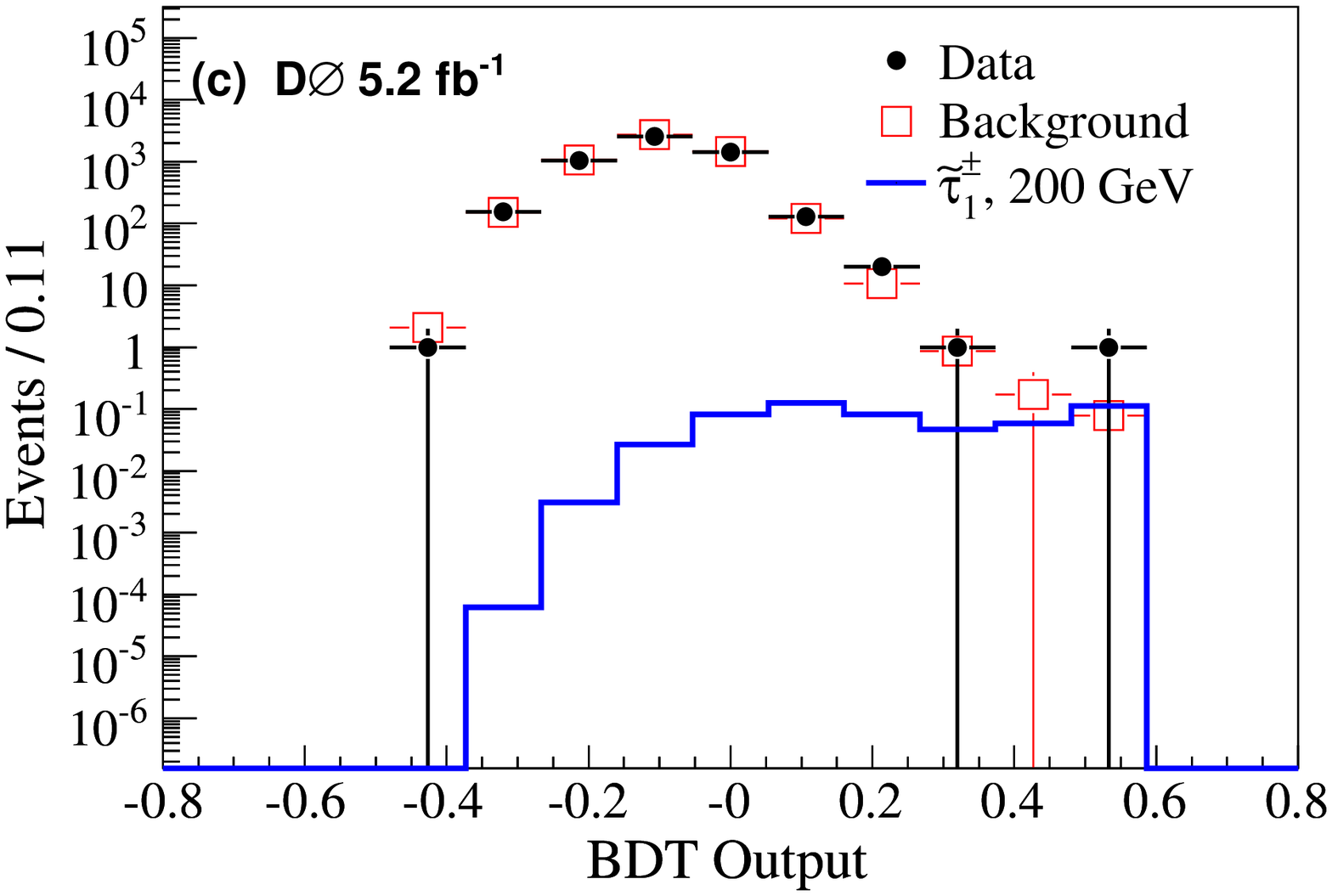}}
\scalebox{0.35}{\includegraphics{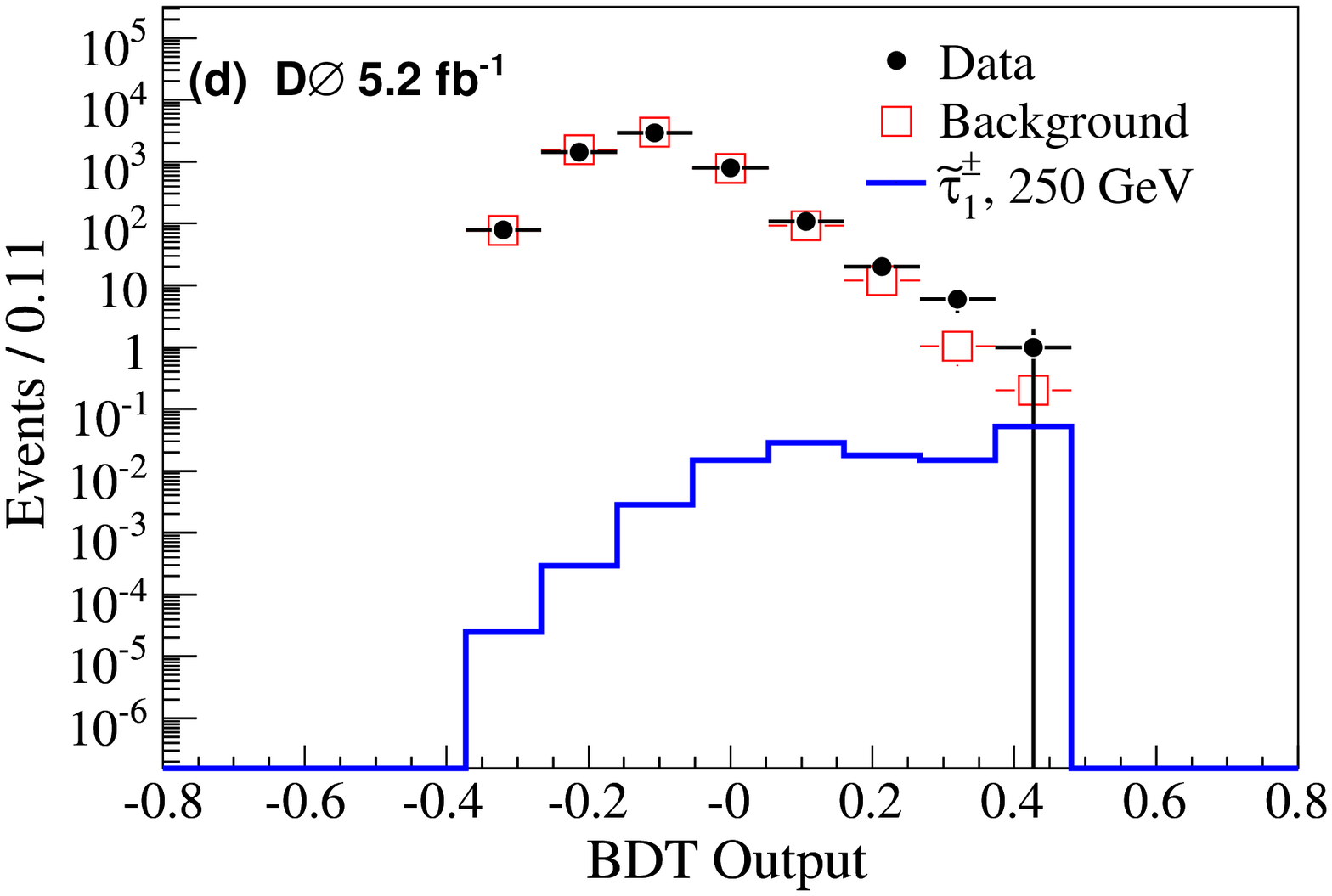}}
\scalebox{0.35}{\includegraphics{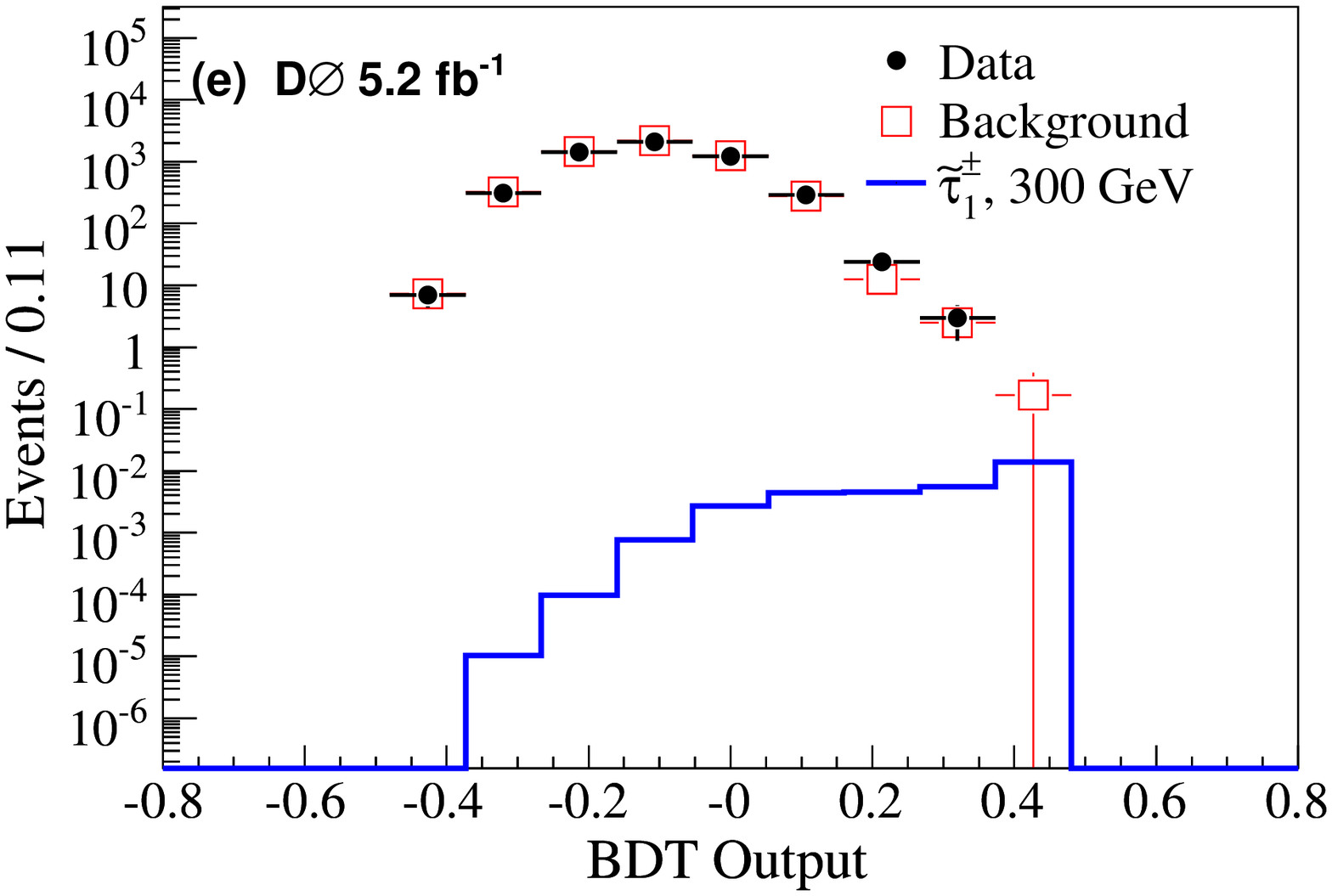}}
\caption{\label{fig:p20single_BDTout_stau}(color online) BDT-output distributions for stau 
masses 100-300 GeV in 50 GeV steps in the search for single 
CMLLPs. The distributions are normalized to the expected number of events.}
\end{center}
\end{figure*}

\begin{figure*}
\begin{center}
\scalebox{0.35}{\includegraphics{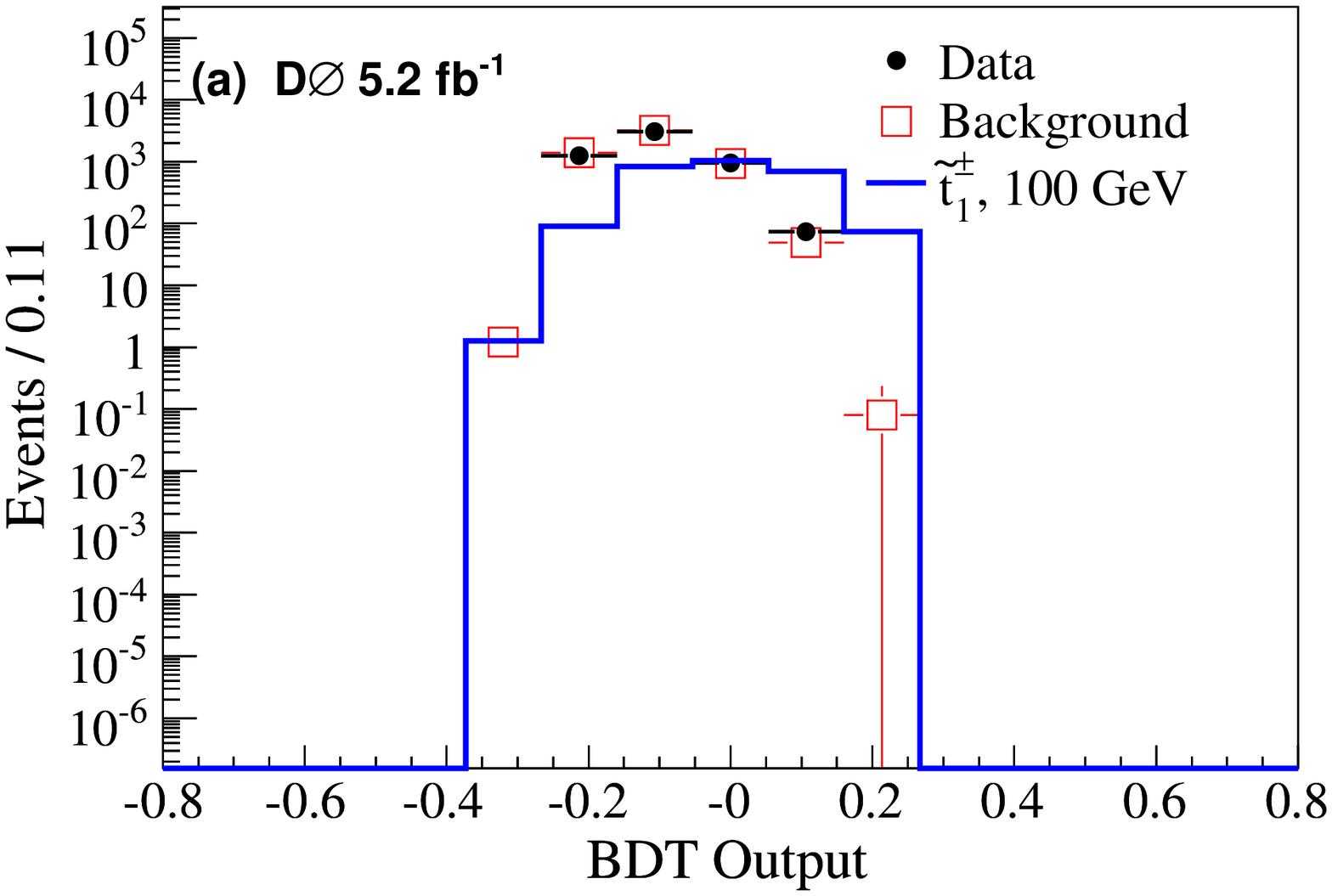}}
\scalebox{0.35}{\includegraphics{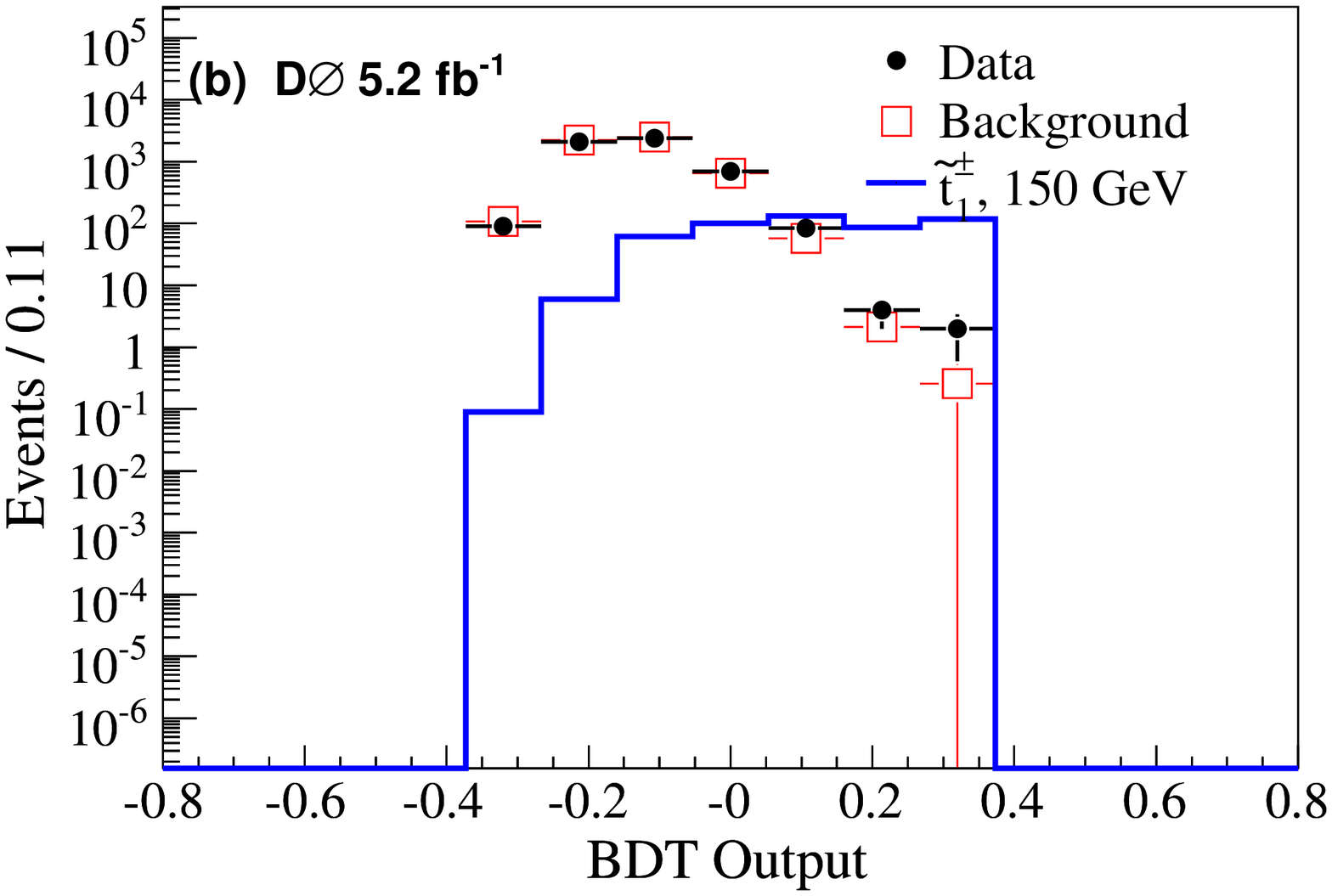}}
\scalebox{0.35}{\includegraphics{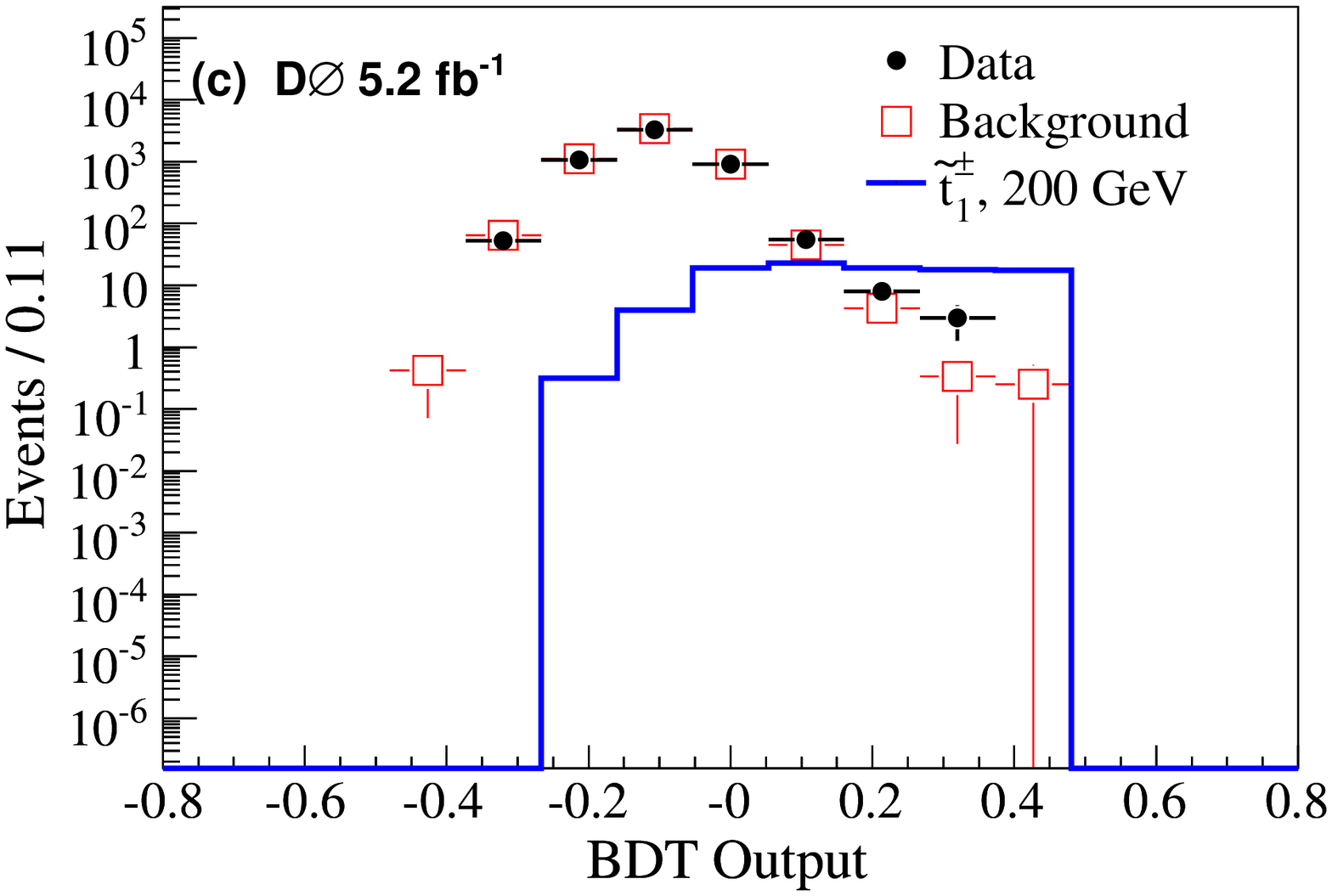}}
\scalebox{0.35}{\includegraphics{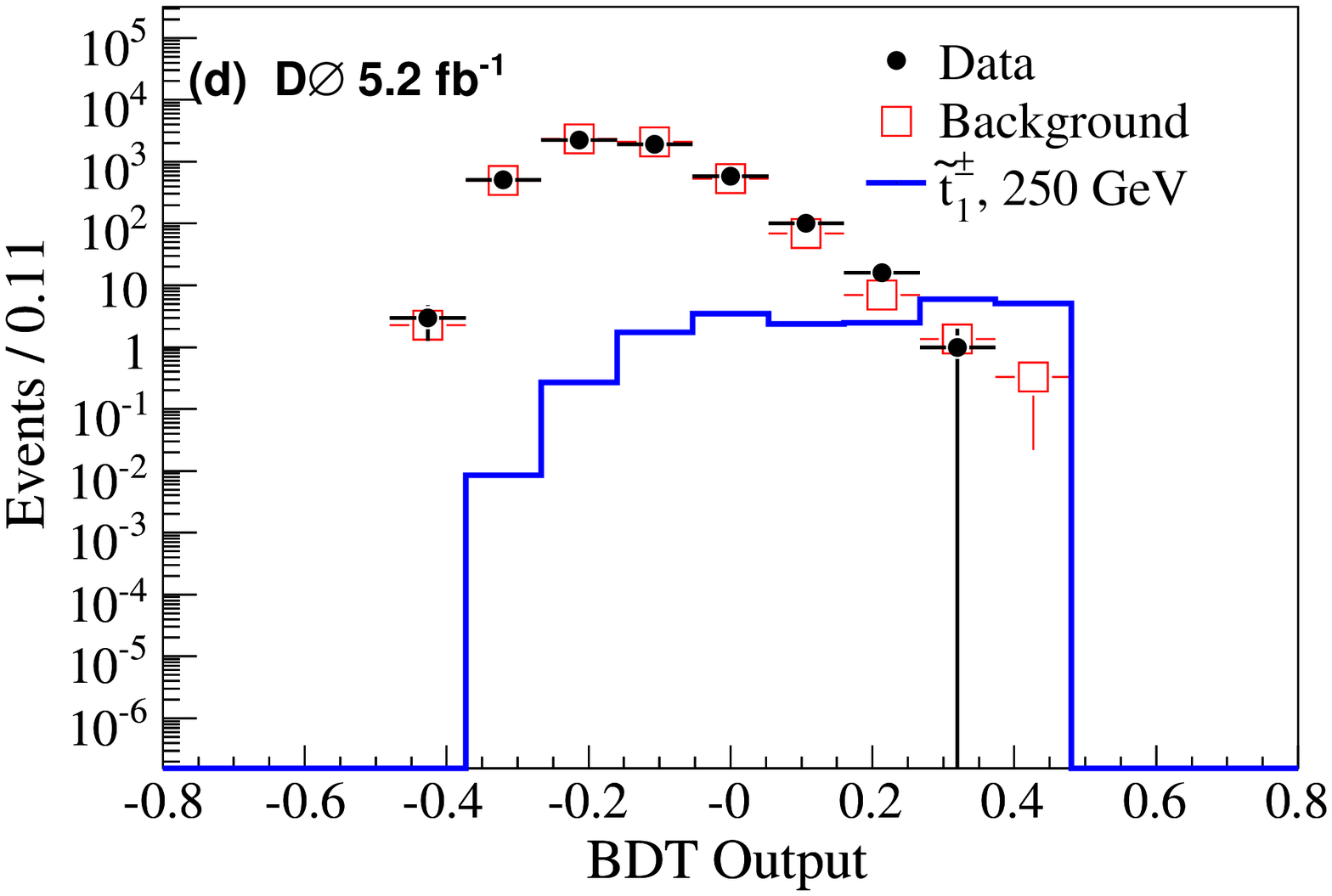}}
\scalebox{0.35}{\includegraphics{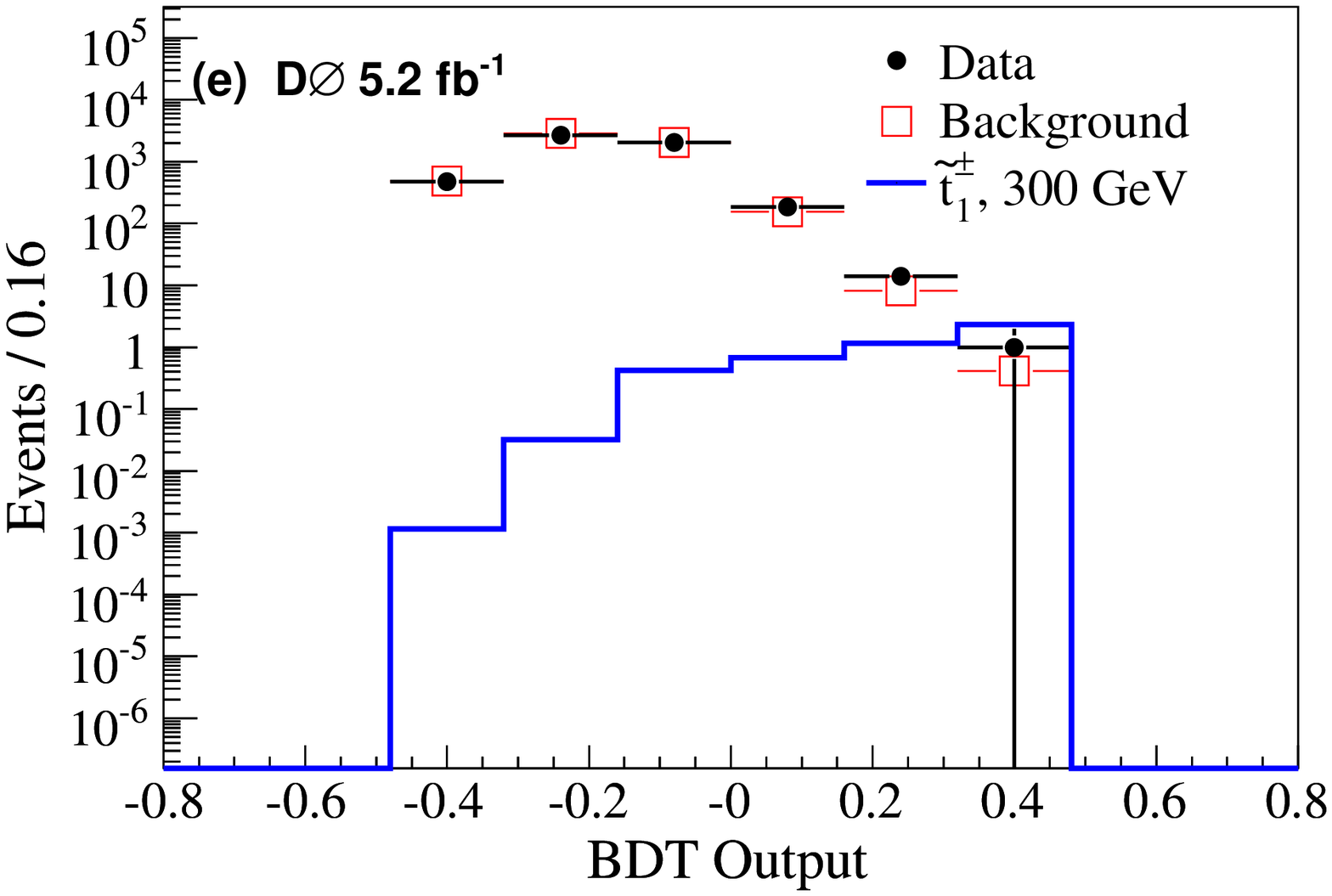}}
\scalebox{0.35}{\includegraphics{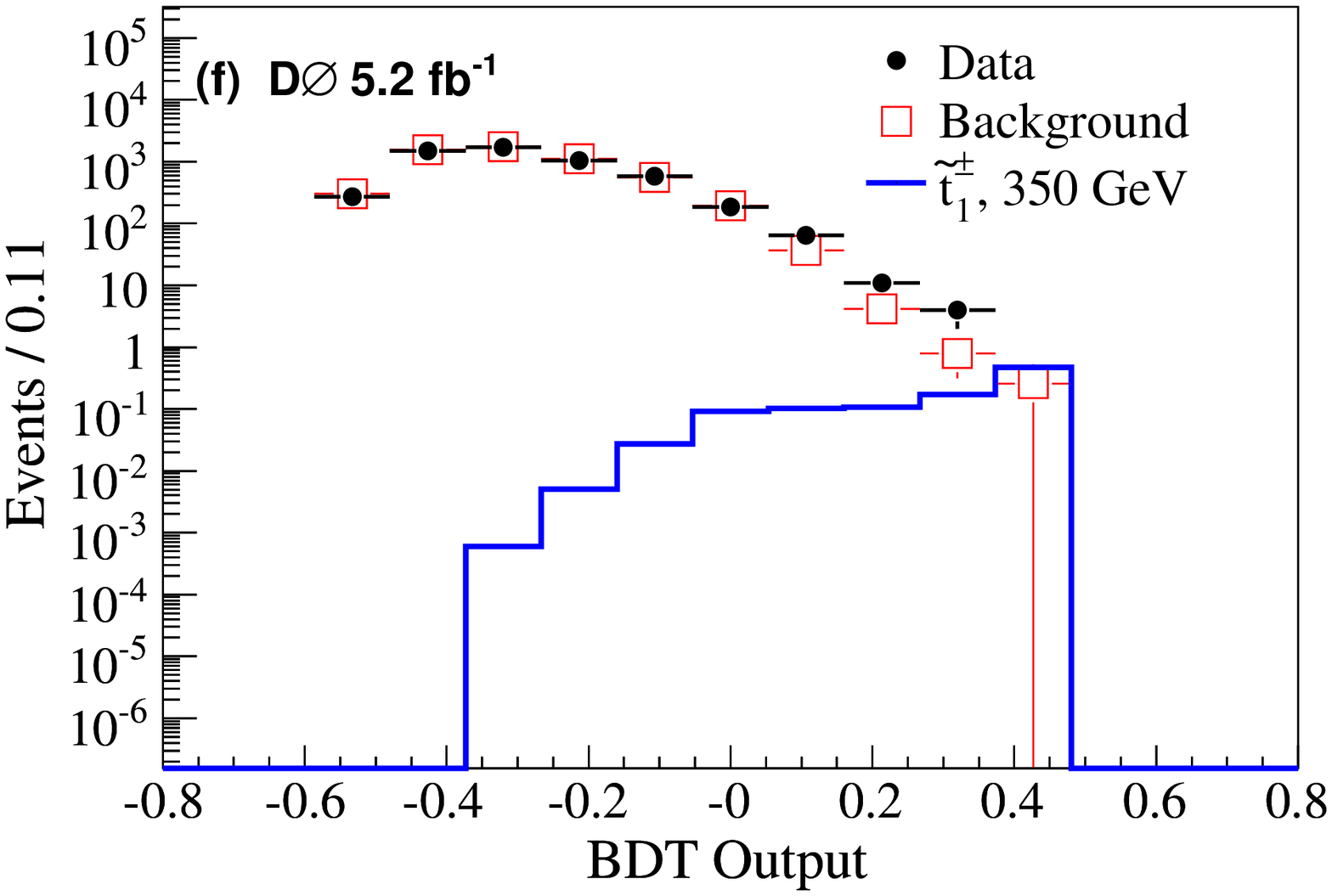}}
\scalebox{0.35}{\includegraphics{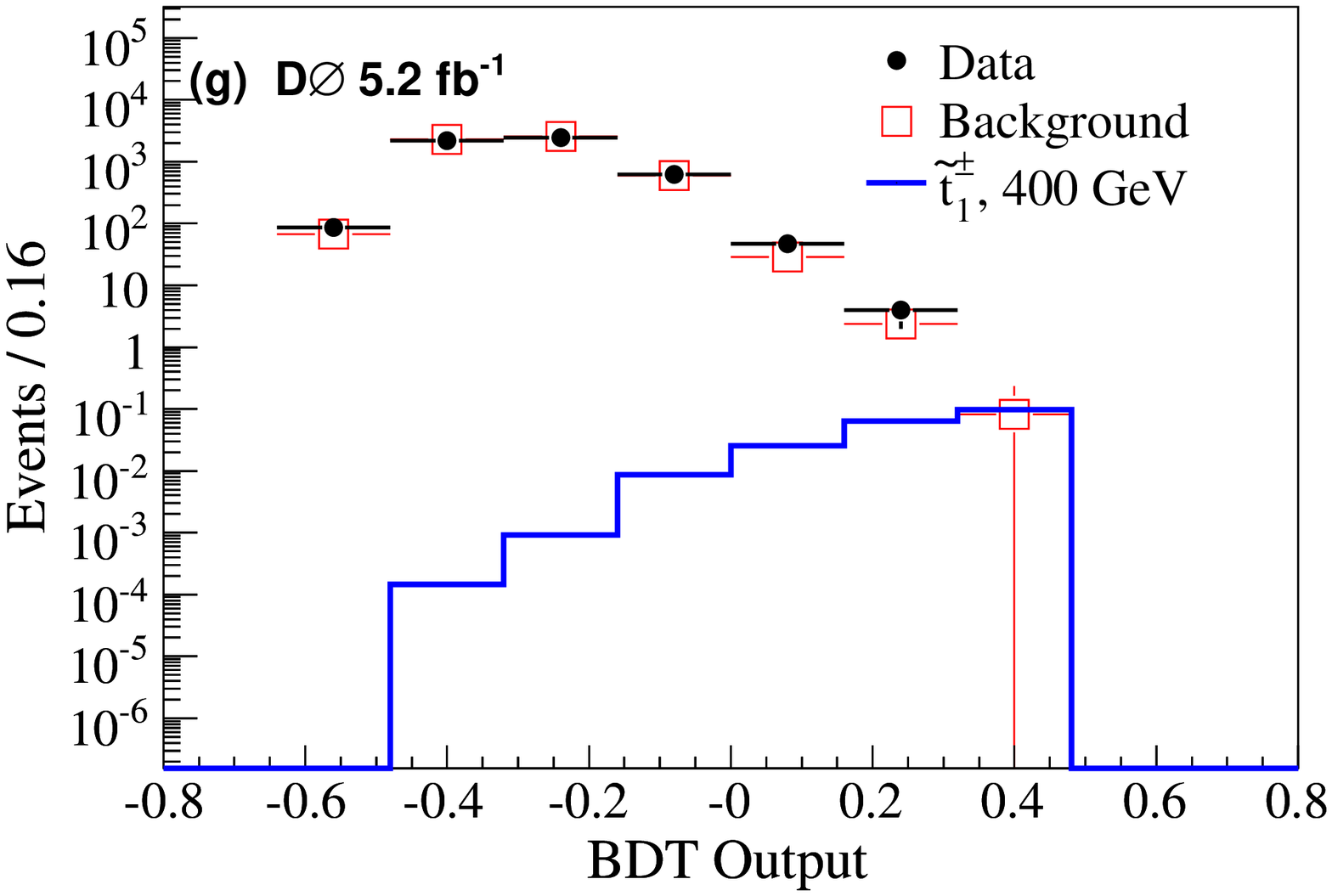}}
\caption{\label{fig:p20single_BDTout_stop}(color online) BDT-output distributions for top squark 
masses 100-400 GeV in 50 GeV steps in the search for single CMLLPs. The distributions are 
normalized to the expected number 
of events.}
\end{center}
\end{figure*}

\begin{figure*}
\begin{center}
\scalebox{0.35}{\includegraphics{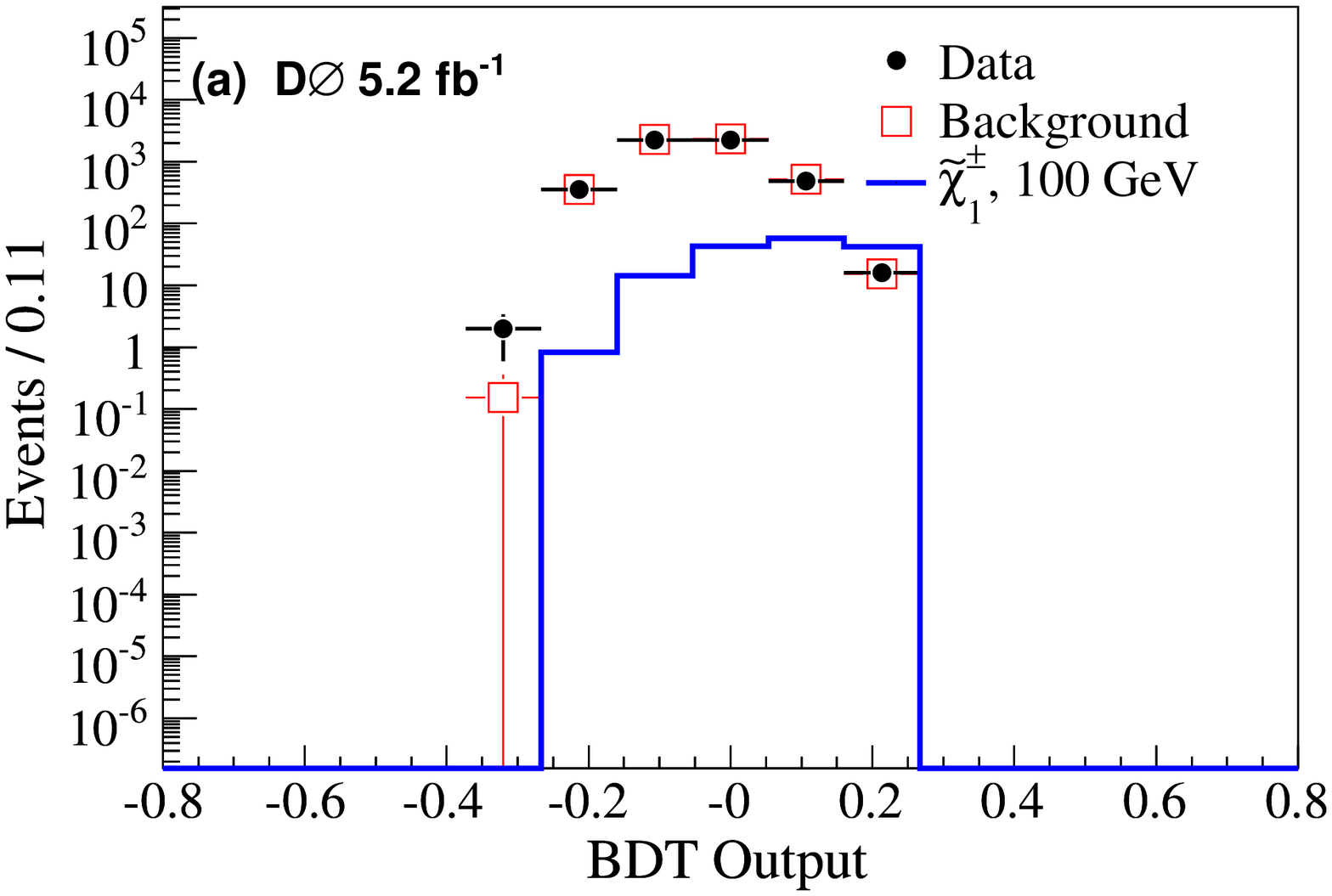}}
\scalebox{0.35}{\includegraphics{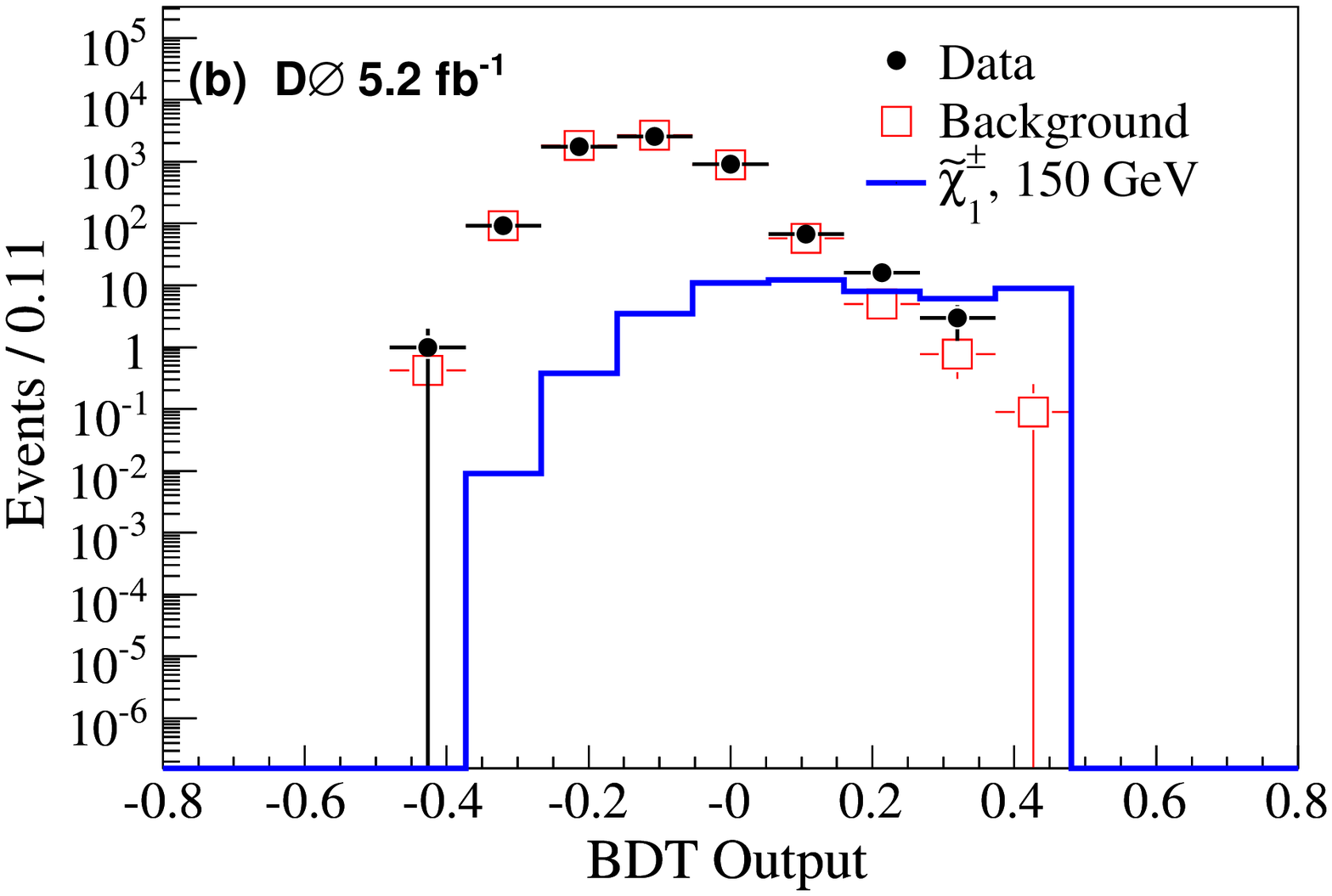}}
\scalebox{0.35}{\includegraphics{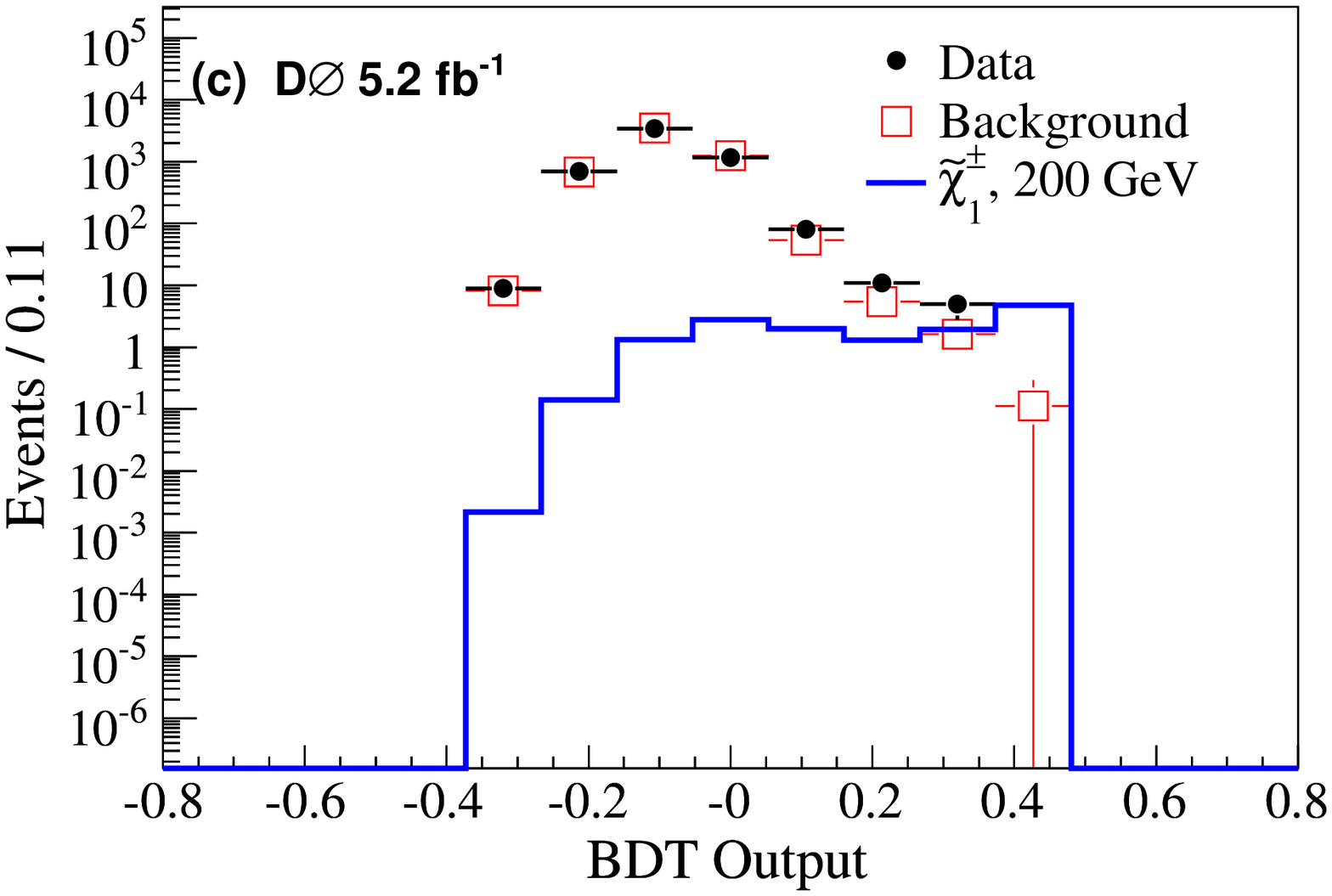}}
\scalebox{0.35}{\includegraphics{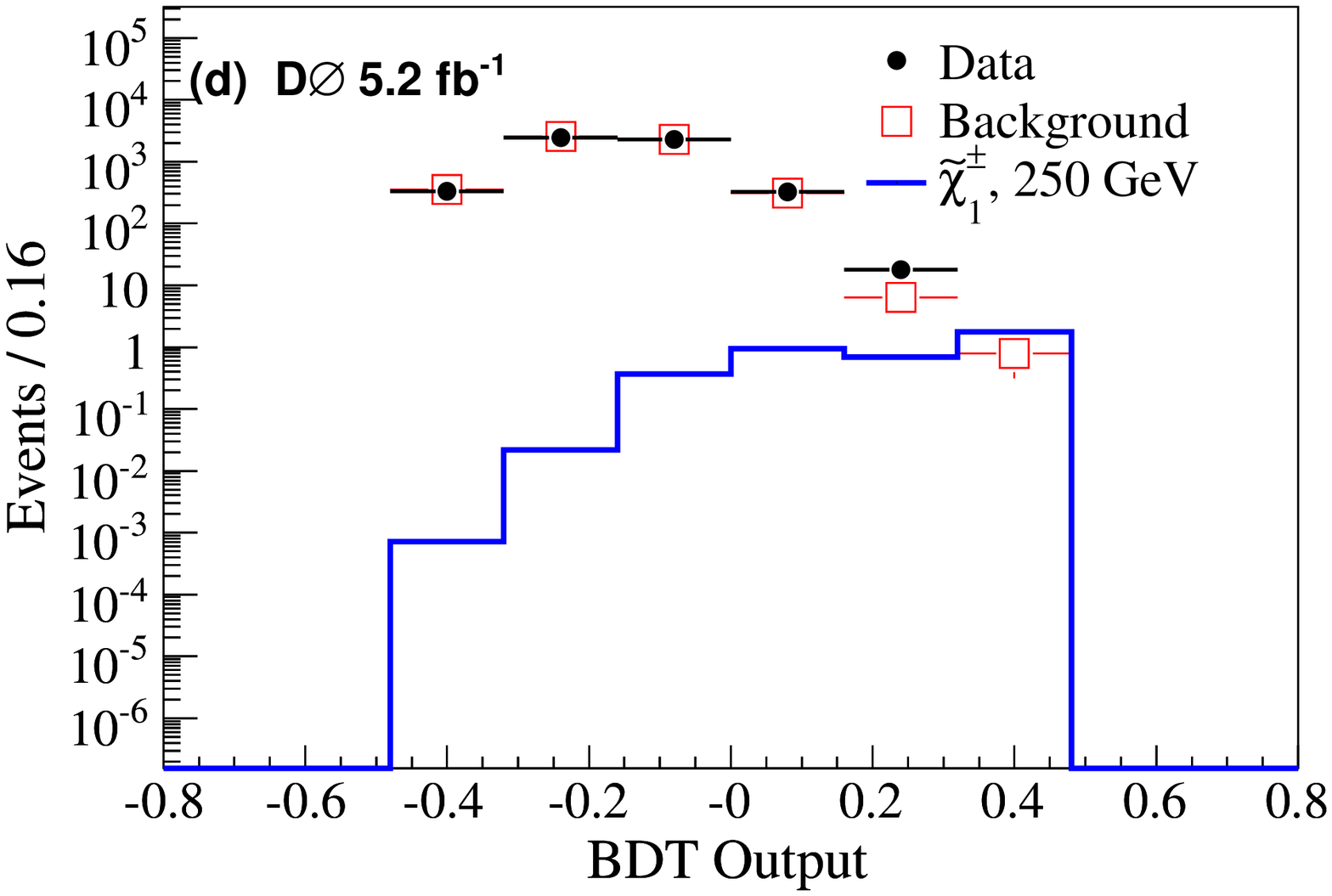}}
\scalebox{0.35}{\includegraphics{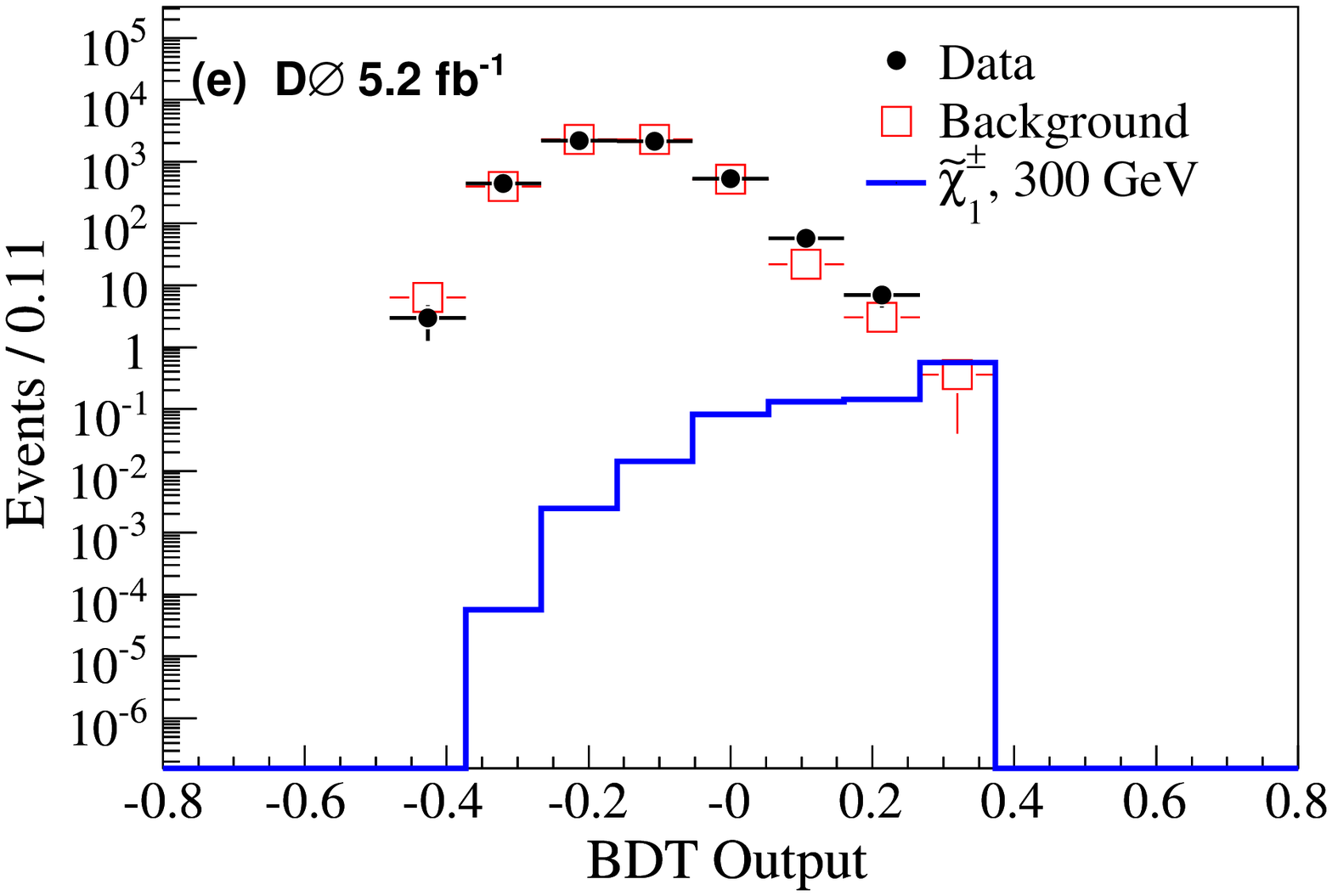}}
\caption{\label{fig:p20single_BDTout_charginoH}(color online) BDT-output distributions for 
higgsino-like chargino masses 100-300 GeV in 50 GeV steps in  
the search for single CMLLPs. The distributions are normalized to the expected number 
of events. }
\end{center}
\end{figure*}

 The BDT-output distributions for stau, top squark, and higgsino-like 
charginos, after being normalized to the expected number of events, for the search of a pair of CMLLPs 
are shown in Figs.~\ref{fig:p20pair_BDT_stau}--\ref{fig:p20pair_BDT_charginoH}. The BDT-output distributions for stau, top squark, and higgsino-like charginos, after being normalized to the expected number of events, for the search of a single CMLLP are shown in Figs.~\ref{fig:p20single_BDTout_stau}--\ref{fig:p20single_BDTout_charginoH}.


\end{document}